\documentclass[preprintnumbers,amsmath,11pt,amssymb,floatfix,superscriptaddress,nofootinbib]{revtex4}
\usepackage{graphicx}
\usepackage{subcaption}
\usepackage{epsfig}
\usepackage{bm}
\usepackage{amsfonts}

\input epsf

\begin{document}

\title{Generalized Holographic and Ricci Dark Energy: Cosmological Diagnostics and Scalar Field Realizations}
\author{Antonio Pasqua}

\email{toto.pasqua@gmail.com}

\date{\today}
\newpage


\begin{abstract}
In this work, we present two generalized formulations of the Holographic and Ricci Dark Energy (DE) models, given by $
\rho_{GH} = 3c^2M^{2}_{pl} \left[ 1-\epsilon\left(1-\frac{R}{H^2}\right) \right]H^2$ and 
$\rho_{GR} = 3c^2M^{2}_{pl}\left[ 1-\eta\left(1-\frac{H^2}{R}\right) \right]R$ where $H$ and $R$ denote the Hubble parameter and the Ricci scalar, while $\epsilon$ and $\eta$ are model parameters related by $\epsilon = 1 - \eta$.
We derived explicit analytical expressions for key cosmological quantities, including the Hubble parameter, the DE density $\rho_D$, the DE pressure $p_D$, the equation of state parameter of DE $\omega_D$ and the deceleration parameter $q$.
The analysis was carried out for four distinct cases: (i) the standard model in its original formulation; (ii) the inclusion of spatial curvature; (iii) the addition of interactions between the dark sectors; and (iv) the presence of both interaction and curvature. Moreover, we also considered the limiting case of a DE Dominated Universe.

To further characterize the dynamical features of the models, we investigated several diagnostic tools, namely the statefinder parameters, the $Om(z)$ diagnostic, the squared speed of the sound $v_s^2$, the cosmographic parameters and the age of the present Universe. 
Moreover, we established a correspondence between the DE models we studied and some scalar field theories, including tachyon, k-essence, dilaton, quintessence, Dirac-Born-Infeld, Yang-Mills and Nonlinear Electrodynamics (NLED) fields.
\end{abstract}

\maketitle
\tableofcontents

\section{Introduction}
Observational evidence from the Wilkinson Microwave Anisotropy Probe (WMAP) \cite{cmb1,cmb2}, the Supernova Cosmology Project \cite{sn2,sn4}, the Sloan Digital Sky Survey (SDSS) \cite{sds1,sds3,sds4}, the Planck satellite mission \cite{planck}, and X-ray measurements \cite{xray} has firmly established that our Universe is undergoing an accelerated expansion. To explain this phenomenon, an exotic component known as dark energy (DE)—characterized by a negative equation of state—has been introduced. While the Cosmological Constant $\Lambda_{CC}$ remains the simplest and most widely supported candidate for DE, it suffers from theoretical challenges, such as the cosmological constant problem and the coincidence problem, motivating the investigation of alternative models \cite{cosm3,cosm4,cosm5}.

The Cosmological Constant can be interpreted as an additional term in Einstein’s field equations. However, predictions from Quantum Field Theory, employing cut-offs at the Planck or electroweak scales, yield a vacuum energy density that exceeds the observed value by factors of $10^{123}$ or $10^{55}$, respectively. The absence of a protective symmetry that could naturally reduce $\Lambda_{CC}$ to a small value underlies the so-called Cosmological Constant problem. Furthermore, the coincidence problem questions why the energy densities of matter and DE are comparable precisely in the current epoch. Comprehensive discussions of these foundational issues can be found in \cite{cosm3,cosm4,cosm5}, while a detailed treatment of the early Universe’s inflationary phase is presented in \cite{rev2}.

Within the framework of General Relativity, DE is believed to constitute roughly two-thirds of the present cosmic energy density $\rho_{\text{tot}}$ \cite{twothirds}, with the remainder mainly composed of Dark Matter (DM) and a smaller fraction of baryonic matter. Despite extensive observational and theoretical studies, the fundamental nature of DE remains unknown.

One alternative to the cosmological constant is a class of dynamical DE models, characterized by a time-dependent equation of state parameter $\omega_D$. Analyses of Type Ia Supernovae data indicate that these models may fit observations better than a constant-$\Lambda$ scenario. Proposed theoretical frameworks include scalar field models such as quintessence \cite{quint1,quint4,quint5}, k-essence \cite{kess3,kess4,kess5}, tachyon fields \cite{tac2,tac3,tac4}, phantom energy \cite{pha2,pha5,pha6}, dilaton fields \cite{dil1,dil2}, and quintom models \cite{qui3,qui8,qui10,qui12}. In addition, interacting DE scenarios have been explored, including models based on the Chaplygin gas \cite{cgas1,cgas2,cgas3}, as well as Agegraphic Dark Energy (ADE) and its extension, New ADE (NADE) \cite{ade1,ade2}. A phase-space analysis of multifluid cosmologies considering complex forms of the EoS parameter can be found in \cite{rev2}.

Beyond scalar fields and Chaplygin gas models, another class of dynamical DE theories arises from the holographic principle  \cite{holo1,holo2,holo5}. Among these, the Holographic Dark Energy (HDE) model, first proposed by Li \cite{li}, has attracted considerable attention and has been extensively studied in several papers \cite{nood1,nood2,nood3,nood4,nood5,nood6,nood7}. According to the holographic principle—relevant in black hole thermodynamics and string theory—the entropy of a system scales with the area $A$ of its boundary, rather than the volume $V$, i.e., $S \sim A \sim L^2$, where $L$ is a characteristic length scale.

In Li's formulation, the HDE density is given by:
\begin{equation}
\rho_\Lambda = 3\alpha M_p^2 L^{-2},
\end{equation}
where $\alpha$ is a dimensionless parameter and $M_p = \left(8\pi G_N\right)^{-1/2}$ is the reduced Planck mass, with $G_N$ being Newton’s gravitational constant. The concept of holographic DE was first introduced by Cohen et al.~\cite{coh1}, who suggested that the DE energy density should be limited by the Universe’s size to avoid black hole formation. Their model, $\rho_\Lambda \propto H^2$, however, failed to produce cosmic acceleration, leading to an effective equation of state $\omega = 0$.

To address this, it was proposed using the future event horizon as the infrared (IR) cutoff, yielding a model consistent with late-time acceleration ~\cite{li} . Subsequent modifications include the Holographic Ricci Dark Energy  model, where the IR cut-off scales as $L \propto R^{-1/2}$ with $R$ representing the Ricci scalar \cite{gaoprimo}, and the Granda–Oliveros generalization, where the DE density depends on both the Hubble parameter $H$ and its first derivative $\dot{H}$ \cite{go1,go4,go5}. These holographic models have been extensively tested against Type Ia supernovae, cosmic microwave background data, and baryon acoustic oscillations \cite{cons1,cons2,cons3,cons4,cons5,cons6,cons7,cons8,cons9}. For broader overviews, see also \cite{hde1,hde2,hde7,hde10,hde12,hde13,hde17,hde18,hde19,
hde22,hde23,hde24,hde26,hde28,hde30,hde32,hde33,hde34,hde35,saridakis11,saridakis22,sc1,sc2,sc3}.

In this work, we focus on a recently proposed model introduced in Xu et al \cite{xu2009}. The main features and implications of this model are discussed in the following sections.

The paper is structured as follows.\\
In Section 2, we establish the cosmological framework that serves as the foundation of this study. 
In Section 3, 4, 5 and 6, we introduce the model we are studying and we derive some important cosmological quantities, like the Hubble parameter $H$, the fractional energy density of DE $\Omega_D$ and its evolutionary form $\Omega'_D$, the dark energy density $\rho_D$, the dark energy pressure $p_D$, as well as explicit expressions for the equation of state (EoS) parameter $\omega_D$ and the deceleration parameter $q$, all expressed as functions of the parameter $x=\ln a$ and the redshift $z$. We also evaluate these quantities in the limiting case of a Dark Energy Dominated Universe, i.e. for $\Omega_m =\Omega_k= 0$ and $\Omega_D = 1$. We perform this analysis for four different cases: the general case, introduced in the original work, then we consider the presence of curvature; in the third case, we consider the presence of interaction between Dark Sectors and in the last case we consider the presence of both interactin and curvature. 
In Section 7, we carry out the Statefinder diagnostic for the models under consideration. 
In Section 8, we apply the Om diagnostic to further investigate their dynamical behavior. 
In Section 9, we evaluate the squared speed of the sound $v_s^2$
to examine the stability of the models. 
In Section 10, we analyze the cosmographic parameters, providing a model-independent characterization of the cosmic expansion. 
In Section 11, we estimate the present age of the Universe within the framework of the models considered. 
In Section 12, we establish a correspondence between the proposed models and various scalar field theories, in particular the Tachyon, the k-essence, the dilaton, the quintessence, the DBI, the Yang-Mills and the Non-Linear Electro-Dynamics scalar fields models. 
Finally, in Section 13, we present the Conclusions of this work. 

\section{Cosmological Framework}
We consider a spatially homogeneous and isotropic universe described by the Friedmann–Lemaitre-Robertson–Walker (FLRW) metric, filled with cold dark matter (CDM) and a dark energy (DE) component. In this work, the dark energy is modeled through two distinct frameworks: Holographic Dark Energy (HDE) and Ricci dark energy (RDE), including their respective generalizations. The line element of the FLRW spacetime is given by
\begin{equation}
ds^2 = -dt^2 + a(t)^2 \left[ \frac{dr^2}{1 - k r^2} + r^2 d\theta^2 + r^2 \sin^2\theta\, d\phi^2 \right],
\end{equation}
where $a(t)$ is the scale factor and $k = 0, \pm1$ denotes the spatial curvature of the universe (corresponding to a flat, closed or open Universe, respectively). Moreover, $r$ is the radial component of the metric while $\theta$ and $\phi$ are the angular components of the metric, with $0 \leq \theta \leq \pi$ and $0 \leq \phi < 2\pi$.

The evolution of the universe is governed by the Friedmann equations:
\begin{eqnarray}
H^2 &=& \frac{8\pi G}{3} \left( \rho_m + \rho_D \right) - \frac{k}{a^2}, \label{genfri}\\
    \dot{H} + 2H^2 &=& \left(\frac{8\pi G}{6}\right)\, p_D -\frac{k}{a^2} , \label{7fri2}
\end{eqnarray}
where $H = \dot{a}/a$ is the Hubble parameter, and $\rho_m$ and $\rho_D$ denote the energy densities of cold dark matter and dark energy, respectively. Moreover, $G$ is the gravitational constant.

 In this study, we focus on the spatially flat case $k = 0$, which is well supported by current observational data.

As shown in~\cite{evadu1}, when the Hubble horizon is adopted as the infrared (IR) cut-off, the holographic dark energy (HDE) density takes the form
\begin{equation}
\rho_D = 3c^2 M_{pl}^2 H^2,
\end{equation}
where $c$ is a dimensionless constant and $M_{pl}$ is the reduced Planck mass. However, it has been shown~\cite{evadu1} that this choice of cut-off does not lead to a late-time accelerated expansion of the Universe, making the model incompatible with current observational data.\\
Gao et al.~\cite{gaoprimo} proposed that the Ricci scalar can be employed as the infrared (IR) cutoff in the holographic framework, leading to a model known as Ricci dark energy (RDE), in which the dark energy density is proportional to the Ricci scalar. The Ricci scalar in a FLRW universe is given by
\begin{equation}
R = -6\left( \dot{H} + 2H^2 + \frac{k}{a^2} \right),
\end{equation}
and the corresponding RDE density takes the form
\begin{equation}
\rho_D = 3c^2 M_{pl}^2 R = 3c^2 M_{pl}^2\left( \dot{H} + 2H^2 + \frac{k}{a^2} \right),
\end{equation}
where $R$ denotes the positive contribution of the Ricci scalar, and the numerical coefficient has been absorbed into the constant $c^2$.

The Ricci DE model has been widely and well studied in many different papers \cite{riccide1,riccide3,riccide4,riccide5,riccide6,riccide7,riccide8,riccide10,riccide12,riccide13,riccide14,riccide15,riccide16,riccide17,riccide18}

The fractional energy densities corresponding to the matter component and the dark energy component are defined as follows:
\begin{eqnarray}
    \Omega_m &=& \frac{\rho_m}{\rho_{\text{cr}}} = \frac{\rho_m}{3M^2_p H^2}, \label{8} \\
    \Omega_D &=& \frac{\rho_D}{\rho_{\text{cr}}} = \frac{\rho_D}{3M^2_p H^2}, \label{10} \\
    \Omega_k &=&\frac{k}{a^2 H^2} \label{fridk}.
\end{eqnarray}
where $\rho_{\text{cr}}$ is the critical energy density required for a spatially flat universe, given by
\begin{equation}
    \rho_{\text{cr}} = 3M^2_p H^2.
\end{equation}

Using Eqs.~(\ref{8}), (\ref{10}) and (\ref{fridk}), the Friedmann equation given in  Eq.~(\ref{genfri}) can be rewritten in a dimensionless form in the following way:
\begin{equation}
    \Omega_m + \Omega_D = 1 - \Omega_k. \label{11}
\end{equation}

To ensure consistency with the Bianchi identity—or, equivalently, the local conservation of the energy-momentum tensor, the total energy density $\rho_{\text{tot}}$ must satisfy the continuity equation:
\begin{equation}
    \dot{\rho}_{\text{tot}} + 3H\left( \rho_{\text{tot}} + p_{\text{tot}} \right) = 0, \label{12old}
\end{equation}
where $\rho_{\text{tot}}$ and $p_{\text{tot}}$ denote the total energy density and total pressure of the cosmic fluid, respectively. These are defined as:
\begin{eqnarray}
    \rho_{\text{tot}} &=& \rho_m + \rho_D+\rho_k, \\
    p_{\text{tot}} &=& p_D,
\end{eqnarray}
since cold dark matter (CDM) is assumed to be pressureless.

The continuity equation in Eq.~(\ref{12old}) can also be expressed in terms of the total equation of state (EoS) parameter, $\omega_{\text{tot}} = p_{\text{tot}} / \rho_{\text{tot}}$, as:
\begin{equation}
    \dot{\rho}_{\text{tot}} + 3H\left(1 + \omega_{\text{tot}} \right) \rho_{\text{tot}} = 0. \label{12}
\end{equation}

Assuming that the energy densities of dark matter and dark energy are conserved separately (i.e., in the non-interacting case), the total conservation equation can be decomposed into two independent continuity equations, one for DE and one for matter:
\begin{eqnarray}
    \dot{\rho}_D + 3H\left( 1 + \omega_D \right)\rho_D &=& 0, \label{12deold} \\
    \dot{\rho}_m + 3H\rho_m &=& 0. \label{12dm}
\end{eqnarray}

Using the general definition of the EoS parameter for dark energy,
\begin{equation}
    \omega_D = \frac{p_D}{\rho_D}. \label{mammud}
\end{equation}
Eq.~(\ref{12deold}) can be equivalently written as:
\begin{equation}
    \dot{\rho}_D + 3H\left( p_D + \rho_D \right) = 0. \label{12de}
\end{equation}

Moreover, Eqs.~(\ref{12deold}), (\ref{12dm}), and (\ref{12de}) can be rewritten as functions of the new variable $x = \ln a$, by using the relation $\frac{d}{dt} = H \frac{d}{dx}$. Denoting derivatives with respect to $x$ by a prime $'$, i.e. $' = \frac{d}{dx}$, the equations become:
\begin{eqnarray}
    \rho'_D + 3\left( 1 + \omega_D \right)\rho_D &=& 0, \label{12deoldprime} \\
    \rho'_D + 3\left( p_D + \rho_D \right) &=& 0, \label{12deprime} \\
    \rho'_m + 3\rho_m &=& 0. \label{12dmprime}
\end{eqnarray}

In this paper, we consider two generalized versions of
the Holographic and Ricci Dark Energy models, which were introduced in Xu et al \cite{xu2009}. They are given in generalized forms:
\begin{eqnarray}
\rho_{GH}&=&3c^2M^{2}_{pl} f\left(\frac{R}{H^2}\right)H^2,\label{eq:GHD}\\
\rho_{GR}&=&3c^2M^{2}_{pl}g\left(\frac{H^2}{R}\right)R,
\label{eq:GRD}
\end{eqnarray}
where $f(x)$ and $g(y)$ are functions of the dimensionless variables
$x=R/H^2$ and $y=H^2/R$ respectively. \\
In addition, we will analyze several different scenarios. First, we consider the non-interacting case without curvature. Next, we study the non-interacting case including curvature. Finally, we examine the interacting case both with and without curvature.\\

We will express all quantities as functions of the redshift $ z $ or equivalently as function of the parameter $x = \ln a$.\\
We must consider some useful relations which will be useful in order to change variables. Using:
\begin{eqnarray}
a = \frac{1}{1+z},
\end{eqnarray}
and
\begin{eqnarray}
 \frac{d}{dt} = -H(1+z) \frac{d}{dz},
\end{eqnarray}
then the time derivative of the Hubble parameter becomes:
\begin{eqnarray}
\dot{H} = -H(1+z) \frac{dH}{dz}. \label{change}
\end{eqnarray}

We now aim to derive an expression for the quantity $R/H^2$ in both the flat and curved cases.\\
We start with the flat case. \\
In this case, we have that
\begin{eqnarray}
    R\propto \dot{H}+2H^2.
\end{eqnarray}
Using the result of Eq. (\ref{change}), we can write
 \begin{eqnarray}
R = -H(z)(1+z) \frac{dH(z)}{dz} + 2H^2(z) .
  \end{eqnarray}
Dividing by $H^2$, we obtain:
 \begin{eqnarray}
\frac{R(z)}{H^2(z)} = - (1+z) \frac{1}{H(z)} \frac{dH(z)}{dz} + 2 .
  \end{eqnarray}
Using the relation:
 \begin{eqnarray}
\frac{d \ln H^2(z)}{dz} = \frac{1}{2H(z)} \frac{dH^2(z)}{dz},
  \end{eqnarray}
we can write:
\begin{eqnarray}
\frac{R(z)}{H^2(z)}&=&2-\frac{(1+z)}{2}\frac{d\ln H^2(z)}{dz}.
\end{eqnarray}

WE now consider the case with spatial curvature. In this case, the Ricci scalar can be written as:
\begin{eqnarray}
    R \propto \dot{H}+2H^2+\frac{k}{a^2}.
\end{eqnarray}
Following the same procedure that we did for the previous case, we obtain:
\begin{eqnarray}
\frac{R(z)}{H^2(z)} =2 - \frac{(1+z)}{2} \frac{d \ln H^2(z)}{dz}  + \frac{k(1+z)^2}{H^2(z)},
\end{eqnarray}
which is equivalent to:
\begin{eqnarray}
\frac{R(z)}{H^2(z)} =2 - \frac{(1+z)}{2} \frac{d \ln H^2(z)}{dz} + H_0^2\Omega_{k0} (1+z)^2,
\end{eqnarray}
where $\Omega_{k0}$ represents the present-day value of $\Omega_k$.\\
Given that DE constitutes approximately two-thirds of the current total energy content of the Universe, while its contribution was negligible during the early stages of cosmic evolution, it is reasonable to consider models in which the dark energy density increases with the expansion of the Universe. Since the Hubble parameter $H$ encodes the expansion rate, it is natural to assume that the energy density of dark energy depends on $H$ and its time derivatives.\\
It is straightforward to see that the Holographic and the Ricci Dark Energy models are recovered when the functions $ f(x) = g(y) \equiv 1 $. Moreover, in the limiting cases of $ f(x) = x $ and $ g(y) = y $, the Holographic and Ricci Dark Energy models effectively exchange their roles. Clearly, the functions can be written as
\begin{eqnarray}
f\left(\frac{R}{H^2}\right)=1-\epsilon\left(1-\frac{R}{H^2}\right),\label{eq:epsilonGHD}\\
g\left(\frac{H^2}{R}\right)=1-\eta\left(1-\frac{H^2}{R}\right),\label{eq:epsilonGRD}
\end{eqnarray}
where $ \epsilon $ and $ \eta $ are the two parameters that characterize the models we consider. In this context, the above description can be interpreted in the following way: when $ \epsilon = 0 $ (i.e., $ \eta = 1 $) or $ \epsilon = 1 $ (i.e., $ \eta = 0 $), the generalized dark energy density reduces to the Holographic (Ricci) and Ricci (holographic) dark energy densities, respectively. 

Furthermore, when the function satisfies the relation $ f(x) = x g(1/x) $, where the variable $ x $ is defined as $ x = R/H^2 $, the holographic and Ricci dark energy models become equivalent to the generalized ones.

In the parameterized forms given by Eqs.~(\ref{eq:epsilonGHD}) and (\ref{eq:epsilonGRD}), the following relation holds:
\begin{eqnarray}
    \epsilon=1-\eta .\label{etaepsi}
\end{eqnarray}
In general, when $\epsilon\neq 0$ (i.e. when $\eta\neq 0$) or, equivalently, when $\epsilon\neq 1$ (i.e. when $\eta\neq 1$), they are hybrid ones.

If $ \epsilon = 0 $ (or equivalently $ \eta = 1 $), the conclusion would favor the holographic dark energy model. Otherwise, the Ricci-like model would be preferred. In fact, in these two parameterizations, one may expect the relation $ \epsilon = 1 - \eta $, since both models are constructed as combinations of $ \dot{H} $ and $ H^2 $. The parameters $ \epsilon $ and $ \eta $ simply regulate the balance between these two terms. By fitting the models to observational data, the preferred orientation — whether holographic-like or Ricci-like — can be identified.

In the following sections, we will consider Eq.~(\ref{eq:epsilonGHD}) and Eq.~(\ref{eq:epsilonGRD}) as simple examples to explore the properties of generalized dark energy models for some different scenarios.In particular, we begin by examining the general case. Next, we incorporate the effects of spatial curvature. Finally, we extend the analysis to include interactions between the Dark Sectors in both of the previously considered scenarios.\\
Moreover, we will consider in the calculations some specific values of the parameters involved.\\
In particular, for $\epsilon$ we will assume the value of $\epsilon = 1.312^{+0.353}_{-0.293}$, as it was derived in the original paper of the model we are studying: using Eq. (\ref{etaepsi}), we derive that $\eta= 1-\epsilon = -0.312$.
Moreover, we consider three  different values for $c^2$:
\begin{itemize}
    \item $c^2 = 0.46$, as it was found in the original paper of Gao
    \item $c=0.579^{+0.030}_{-0.029}$, as found in \cite{xu2009}, which leads to $c^2 = 0.335241$
    \item for a flat (i.e. for $k = 0$) Universe $c = 0.818^{+0.113}_{-0.097}$, which implies $c^2=0.669124$, and in the case of a non-flat (i.e. for $k \neq 0$) Universe we have $c = 0.815^{+0.179}_{-0.139}$, which implies $c^2=0.664225$, as it was found in \cite{lic1,lic2}
\end{itemize}

\section{Generalized Holographic and Ricci Dark Energy Models}
\subsection{Case 1: $f\left(\frac{R}{H^2}\right)=1-\epsilon\left(1-\frac{R}{H^2}\right)$}
We begin our analysis by considering the functional form
\begin{eqnarray}
f\left(\frac{R}{H^2}\right) = 1 - \epsilon \left(1 - \frac{R}{H^2}\right).
\end{eqnarray}
This expression corresponds to a linear modification of the standard Ricci Dark Energy (RDE) model, parametrized by the dimensionless quantity $\epsilon$.  
  
In this framework, the case $\epsilon = 0$ recovers the standard $\Lambda$CDM prescription, while $\epsilon > 0$ ($\epsilon < 0$) enhances (suppresses) the relative contribution of the other term.

Under this assumption, the corresponding dark energy density takes the form:
\begin{eqnarray}
\rho_{GH}&=&3c^2M^{2}_{pl} \left[ 1-\epsilon\left(1-\frac{R}{H^2}\right) \right]H^2.\label{eq:GHD}
\end{eqnarray}
In this case, we obtain that the Friedmann equation given in Eq. (\ref{genfri}) can be rewritten in the following way:
\begin{eqnarray}
H^2&=&\frac{1}{3M^{2}_{pl}}\left(\rho_{m}+\rho_{GH}\right)\nonumber\\
&=&H^2\Omega_{m}+c^2\left[1-\epsilon\left(1-\frac{R}{H^2}\right)\right]H^2\label{eq:FREGH},
\end{eqnarray}
where $\Omega_{m}=\rho_{m}/(3M^{2}_{pl}H^2)$ is the dimensionless
energy density of DM. Instead, the dimensionless
energy density  of the generalized holographic dark energy $\Omega_{D_{GH}}$ is given by:
\begin{eqnarray}
\Omega_{D_{GH}}&=&\frac{\rho_{D_{GH}}}{3M^{2}_{pl}H^2}\nonumber\\
&=&c^2\left[1-\epsilon\left(1-\frac{R}{H^2}\right)\right]\nonumber\\
&=&c^2\left[1+\epsilon-\frac{\epsilon(1+z)}{2}\frac{d \ln
H_{GH}^2}{dz}\right]\nonumber \\
&=& c^2 \left( 1 + \epsilon + \frac{\epsilon}{2h_{GH}^2} \frac{d h_{GH}^2}{dx} \right).\label{eq:OmegaGH1}
\end{eqnarray}
We now want to obtain the expression of the fractional energy density for DM. \\
From the continuity equation for DM given in Eq. (\ref{12dmprime}), we obtain the following expression for $\rho_m$:
\begin{eqnarray}
    \rho_m = \rho_{m0}(1+z)^{3}  = \rho_{m0}e^{-3x}, \label{rhoMM}
\end{eqnarray}
which leads to
\begin{eqnarray}
    \rho'_m =3\rho_{m0}(1+z)^{3}  = -3\rho_{m0}e^{-3x} ,\label{rhoM}
\end{eqnarray}
where $\rho_{m0}$ represents the present-day value of $\rho_m$.\\
We can also write:
\begin{eqnarray}
\rho_m &=&  \Omega_{m0}e^{-3x}, \\
\rho'_m &=& -3\Omega_{m0}e^{-3x},\label{boh1}
\end{eqnarray}
where we used the fact that $\rho_{m0}=\Omega_{m0}$ since we are using normalized quantities.\\
Using the expression of $\Omega_{D_{GH}}$ obtained in Eq. (\ref{eq:OmegaGH1}) along with the expression of $\rho_m$ given in Eq. (\ref{rhoMM}), the Friedmann equation obtained in Eq. (\ref{genfri}) can be rewritten as a differential equation of $H_{GH}^2(z)$ in the following way:
\begin{equation}
H_{GH}^2(z)\left\{1-c^2\left[1+\epsilon-\frac{\epsilon(1+z)}{2}\frac{d \ln
H_{GH}^2(z)}{dz}\right]\right\}=H^2_0\Omega_{m0}(1+z)^3,
\end{equation}
which has the following solution:
\begin{equation}
h^2_{GH}(z)=\frac{2\Omega_{m0}(1+z)^3}{2+c^2(\epsilon-2)}+\left[1-\frac{2\Omega_{m0}}{2+c^2(\epsilon-2)}\right](1+z)^{2(1-\alpha)},
\end{equation}
where we defined the parameter $\alpha$ as follows:
\begin{eqnarray}
\alpha = \frac{1}{\epsilon} \left( \frac{1}{c^2} - 1 \right),
\end{eqnarray}
and we put $h^2_{GH}(z) = \frac{H^2_{GH}(z)}{H_0^2}$.\\
Moreover, we have that $\Omega_{m0} = \frac{\rho_{m0}}{3M_{pl}^2H_0^2}$ is the present day value of $\Omega_m$.\\
We can also write $h_{GH}^2(z)$ as:
\begin{equation} \label{st2}
h_{GH}^2(z) = \Omega_{m0} (1+z)^3 
+ \left[1 - \frac{2 \Omega_{m0}}{2 + c^2(\epsilon - 2)} \right] (1+z)^{2\left(1 - \alpha\right)} 
-  \frac{c^2 (\epsilon - 2) \Omega_{m0}(1+z)^3}{2 + c^2(\epsilon - 2)}.
\end{equation}
The form obtained in Eq. (\ref{st2}) is useful because it allows to have the expression of $h_{GH}^2(z)$ divided in two parts, one produced by DM, i.e. $\Omega_{m0} (1+z)^3$, and one which is therefore produced by DE. Therefore, we derive the following expression for the energy density fo DE $\rho_{D_{GH}}(z)$:
\begin{equation}
\rho_{D_{GH}}(z)=  \left[1 - \frac{2 \Omega_{m0}}{2 + c^2(\epsilon - 2)} \right] (1+z)^{2\left(1 - \alpha\right)} 
-  \frac{c^2 (\epsilon - 2)\Omega_{m0}(1+z)^3}{2 + c^2(\epsilon - 2)} .
\end{equation}
We can now write the expressions of $h_{GH}^2$ and $\rho_{D_{GH}}$ we derived as functions of the parameter $x = \ln a$:
\begin{eqnarray}
h_{GH}^2(x) &=& \Omega_{m0} e^{-3x} 
+ \left[1 - \frac{2 \Omega_{m0}}{2 + c^2(\epsilon - 2)} \right] e^{-2\left(1 - \alpha\right)x} 
-  \frac{c^2 (\epsilon - 2)\Omega_{m0} e^{-3x}}{2 + c^2(\epsilon - 2)} ,\\
\rho_{D_{GH}}(x)&=& \left[1 - \frac{2 \Omega_{m0}}{2 + c^2(\epsilon - 2)} \right] e^{-2\left(1 - \alpha\right)x} 
-  \frac{c^2 (\epsilon - 2) \Omega_{m0} e^{-3x}}{2 + c^2(\epsilon - 2)}.\label{density1}
\end{eqnarray}
We now want to calculate the final expressions of other important cosmological quantities, like the fractional energy density of dark energy $\Omega_{D_{GH}}$, the evolutionary form of the fractional energy density of DE $\Omega'_{D_{GH}}$, the pressure $p_{GH}$, the EoS parameter $\omega_{D_{GH}}$ and the deceleration parameter $q_{GH}$. \\
For the fractional energy density of DE, we use the general relation obtained in Eq. (\ref{eq:OmegaGH1}):
\begin{eqnarray}
\Omega_{D_{GH}}(x) &=& c^2 \left( 1 + \epsilon + \frac{\epsilon}{2h_{GH}^2(x)} \frac{d h_{GH}^2(x)}{dx} \right).\label{eq:OmegaGH2}
\end{eqnarray}
Using the expression of $h^2_{GH}$ derived in Eq. (\ref{st2}), we obtain the following expression for $\frac{d h_{GH}^2(x)}{dx} $:
\begin{eqnarray}
\frac{d h_{GH}^2(x)}{dx} &=& -3\Omega_{m0} e^{-3x} 
- 2\left[1 - \frac{2 \Omega_{m0}}{2 + c^2(\epsilon - 2)} \right] \left(1 - \alpha\right) e^{-2\left(1 - \alpha\right) x} \nonumber \\
&&+  \frac{3c^2 (\epsilon - 2)\Omega_{m0} e^{-3x}}{2 + c^2(\epsilon - 2)} .\label{renato1}
\end{eqnarray}
Therefore, the final expression of $\Omega_{D_{GH}}(x)$ is given by:
\begin{eqnarray}
\Omega_{D_{GH}}(x) &=& c^2 \left( 1 + \epsilon\right) +\frac{\epsilon c^2}{2}\cdot\left\{-3\Omega_{m0} e^{-3x} 
- 2\left[1 - \frac{2 \Omega_{m0}}{2 + c^2(\epsilon - 2)} \right] \left(1 - \alpha\right) e^{-2\left(1 - \alpha\right) x}\right. \nonumber \\
&&\left.\,\,\,\,\,\,\,\,\,\,\,\,\,\,\,\,\,\,\,\,\,\,\,\,\,\,\,\,\,\,\,\,\,\,\,\,\,\,\,\,\,\,\,\,\,\,\,+  \frac{3c^2 (\epsilon - 2)\Omega_{m0} e^{-3x}}{2 + c^2(\epsilon - 2)}  \right\}\times \nonumber \\
&&\left\{ \Omega_{m0} e^{-3x} 
+ \left[1 - \frac{2 \Omega_{m0}}{2 + c^2(\epsilon - 2)} \right] e^{-2\left(1 - \alpha\right)x} 
-  \frac{c^2 (\epsilon - 2)\Omega_{m0} e^{-3x}}{2 + c^2(\epsilon - 2)}  \right\}^{-1}.\label{eq:OmegaGH2new}
\end{eqnarray}
We now want to calculate the evolutionary form of the fractional energy density of DE $\Omega'_{D_{GH}}$.\\
If we put $h^2_{GH} \equiv A_1$, we have inside the brackets of Eq. (\ref{eq:OmegaGH2new}) a term which can be written as $\frac{A_1'}{A_1}$.
Therefore, we can write:
\begin{eqnarray}
\Omega_{D_{GH}}(x) &=& c^2 \left( 1 + \epsilon\right) +\frac{\epsilon c^2}{2}\left( \frac{A_1'}{A_1}  \right),
\end{eqnarray}
which implies:
\begin{eqnarray}
\Omega_{D_{GH}}'(x) &=& \frac{\epsilon c^2}{2}\left( \frac{A_1''A_1 - A_1'^2}{A_1^2}\right)\nonumber \\
&=&\frac{\epsilon c^2}{2}\left[ \frac{A_1''}{A_1}- \left(\frac{A_1'}{A_1}\right)^2\right],
\end{eqnarray}
where $A_1'' = \frac{d^2 h_{GH}^2(x)}{dx^2}$.\\
Using the general expression of $h^2_{GH}$ given in Eq. (\ref{st2}) or equivalently the expression of $\frac{d h_{GH}^2(x)}{dx}$ derived in Eq. (\ref{renato1}), we can derive the following expression for $\frac{d^2 h_{GH}^2(x)}{dx^2} $:
\begin{eqnarray}
    \frac{d^2 h_{GH}^2(x)}{dx^2} &=& 9\Omega_{m0} e^{-3x} 
+4\left[1 - \frac{2 \Omega_{m0}}{2 + c^2(\epsilon - 2)} \right] \left(1 - \alpha\right)^2 e^{-2\left(1 - \alpha\right) x} \nonumber \\
&&-  \frac{9c^2 (\epsilon - 2)\Omega_{m0} e^{-3x}}{2 + c^2(\epsilon - 2)}.
\end{eqnarray}
Then, the evolutionary form of the fractional energy density of DE is given by:
\begin{eqnarray}
\Omega_{D_{GH}}'(x) &=& \frac{\epsilon c^2}{2}\cdot\left\{  9\Omega_{m0} e^{-3x} 
+4\left[1 - \frac{2 \Omega_{m0}}{2 + c^2(\epsilon - 2)} \right] \left(1 - \alpha\right)^2 e^{-2\left(1 - \alpha\right) x}\right. \nonumber \\
&&\left.-  \frac{9c^2 (\epsilon - 2)\Omega_{m0} e^{-3x}}{2 + c^2(\epsilon - 2)}  \right\}\times\left\{   \Omega_{m0} e^{-3x}  
+ \left[1 - \frac{2 \Omega_{m0}}{2 + c^2(\epsilon - 2)} \right] e^{-2\left(1 - \alpha\right)x} \right.\nonumber \\
&&\left.
-  \frac{c^2 (\epsilon - 2) \Omega_{m0}e^{-3x} }{2 + c^2(\epsilon - 2)} \right\}^{-1}\nonumber \\
&&-\frac{\epsilon c^2}{2}\cdot\left\{  -3\Omega_{m0} e^{-3x} 
-2\left[1 - \frac{2 \Omega_{m0}}{2 + c^2(\epsilon - 2)} \right] \left(1 - \alpha\right) e^{-2\left(1 - \alpha\right) x}\right. \nonumber \\
&&\left.+  \frac{3c^2 (\epsilon - 2)\Omega_{m0} e^{-3x}}{2 + c^2(\epsilon - 2)}  \right\}^2\times\left\{   \Omega_{m0} e^{-3x} 
+ \left[1 - \frac{2 \Omega_{m0}}{2 + c^2(\epsilon - 2)} \right] e^{-2\left(1 - \alpha\right)x} \right.\nonumber \\
&&\left.
-  \frac{c^2 (\epsilon - 2) \Omega_{m0}e^{-3x}}{2 + c^2(\epsilon - 2)} \right\}^{-2}.
\end{eqnarray}


We now want to find the final expression of the pressure of DE $p_{D_{GH}}$ for this case.\\
The general expression of the pressure $p_{D_{GH}}$ is given by:
\begin{eqnarray}
p_{D_{GH}}&=&-\rho_{D_{GH}}-\frac{1}{3}\frac{d\rho_{D_{GH}}}{dx}\nonumber \\
&=& -\rho_{D_{GH}}-\frac{\rho'_{D_{GH}}}{3}.
\end{eqnarray}
Using the general expression fo $\rho_{D_{GH}}$ we derived in Eq. (\ref{density1}), we obtain the following expression for $\rho'_{D_{GH}} $:
\begin{equation}
\rho'_{D_{GH}}(x) =-2\left[1 - \frac{2 \Omega_{m0}}{2 + c^2(\epsilon - 2)} \right] \left(1 - \alpha\right) e^{-2\left(1 - \alpha\right) x} 
+  \frac{3c^2 (\epsilon - 2) \Omega_{m0} e^{-3x}}{2 + c^2(\epsilon - 2)}.\label{density1deri}
\end{equation}
Therefore, the final expression of the pressure is given by:
\begin{eqnarray}
p_{D_{GH}}(x)&=&\left[ \frac{2\left(1 - \alpha\right)}{3}-1  \right] \left[1 - \frac{2 \Omega_{m0}}{2 + c^2(\epsilon - 2)} \right] e^{-2\left(1 - \alpha\right)x}. \label{murmura}
\end{eqnarray}
We now want to obtain an expression for the EoS parameter $\omega_{D_{GH}}$.\\ The general definition of $\omega_{D_{GH}}$ is:
\begin{eqnarray}
\omega_{D_{GH}}(x) = -1 - \frac{\rho_{D_{GH}}'(x)}{3 \rho_{D_{GH}}(x)} \label{eosnongen}.
\end{eqnarray}
Using the expressions of $\rho_{D_{GH}}$ and $\rho'_{D_{GH}}$ obtained, respectively, in Eqs. (\ref{density1}) and (\ref{density1deri}), we can write:
\begin{eqnarray}
\omega_{D_{GH}}(x) 
&=&- 1 -\frac{1}{3}\cdot\frac{-2\left[1 - \frac{2 \Omega_{m0}}{2 + c^2(\epsilon - 2)} \right] \left(1 - \alpha\right) e^{-2\left(1 - \alpha\right) x} 
+  \frac{3c^2 (\epsilon - 2) \Omega_{m0} e^{-3x}}{2 + c^2(\epsilon - 2)}}{\left[1 - \frac{2 \Omega_{m0}}{2 + c^2(\epsilon - 2)} \right] e^{-2\left(1 - \alpha\right)x} 
-  \frac{c^2 (\epsilon - 2) \Omega_{m0} e^{-3x}}{2 + c^2(\epsilon - 2)}}\label{carolina60}.
\end{eqnarray}

\begin{figure}[htbp]
    \centering
    \begin{subfigure}{0.8\textwidth}
        \includegraphics[width=0.6\textwidth]{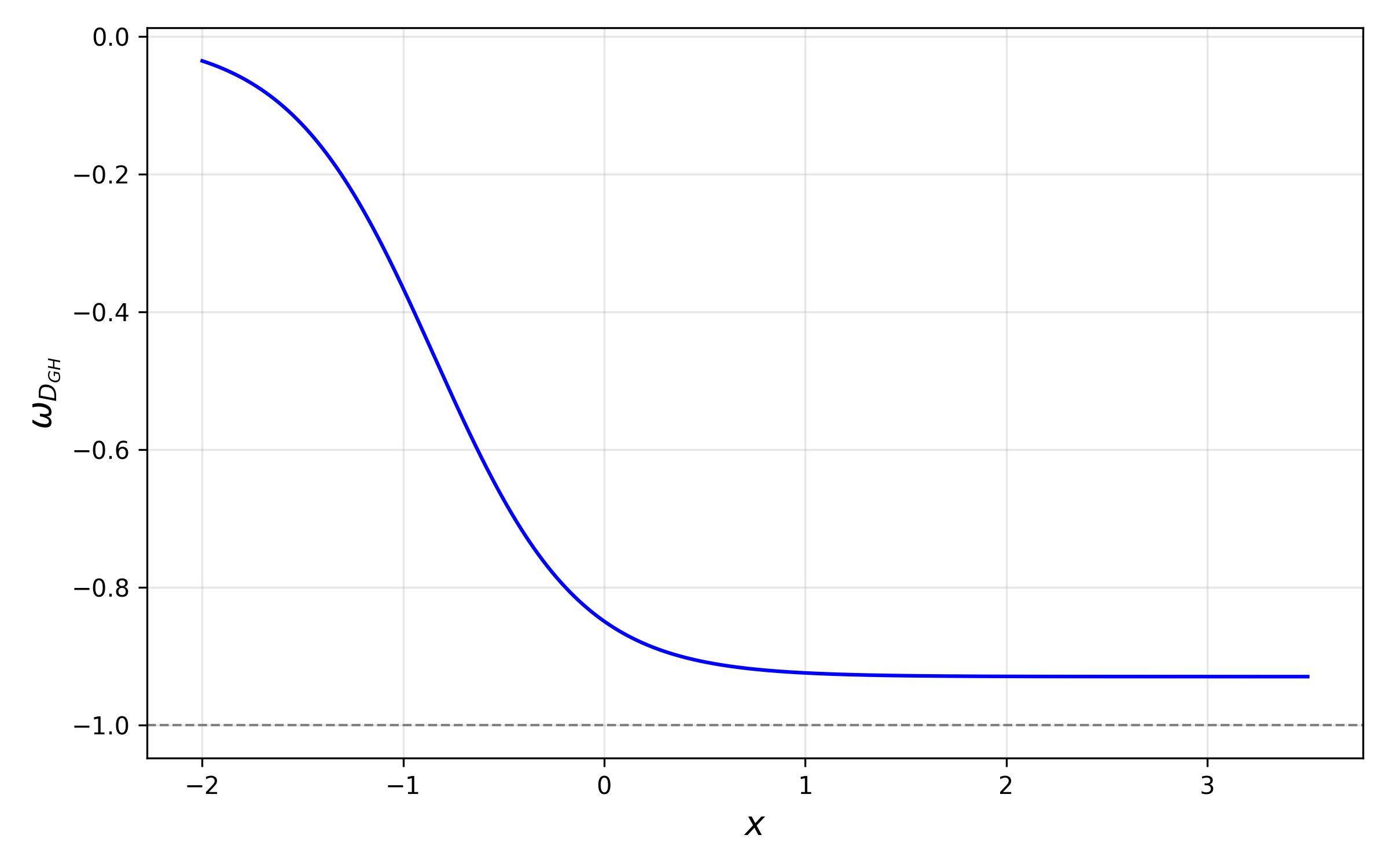}
        \caption{Plot of $\omega_{D_{GH}}(x)$ for $c^2=0.46$.}
        \label{EoS1}
    \end{subfigure}\\[0.5cm]
    \begin{subfigure}{0.8\textwidth}
        \includegraphics[width=0.6\textwidth]{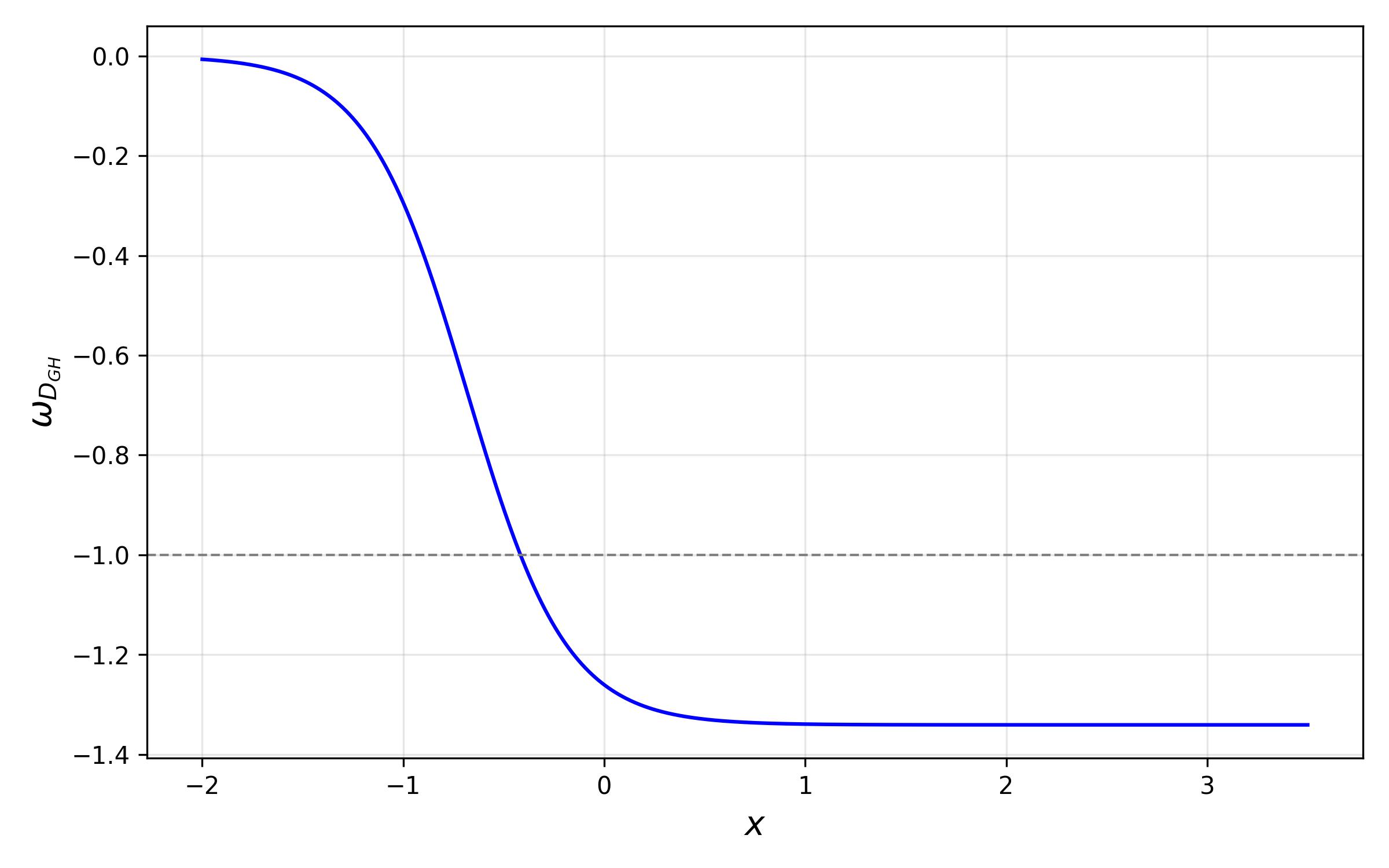}
        \caption{Plot of $\omega_{D_{GH}}(x)$ for $c=0.579$.}
        \label{EoS1-2}
    \end{subfigure}\\[0.5cm]
    \begin{subfigure}{0.8\textwidth}
        \includegraphics[width=0.6\textwidth]{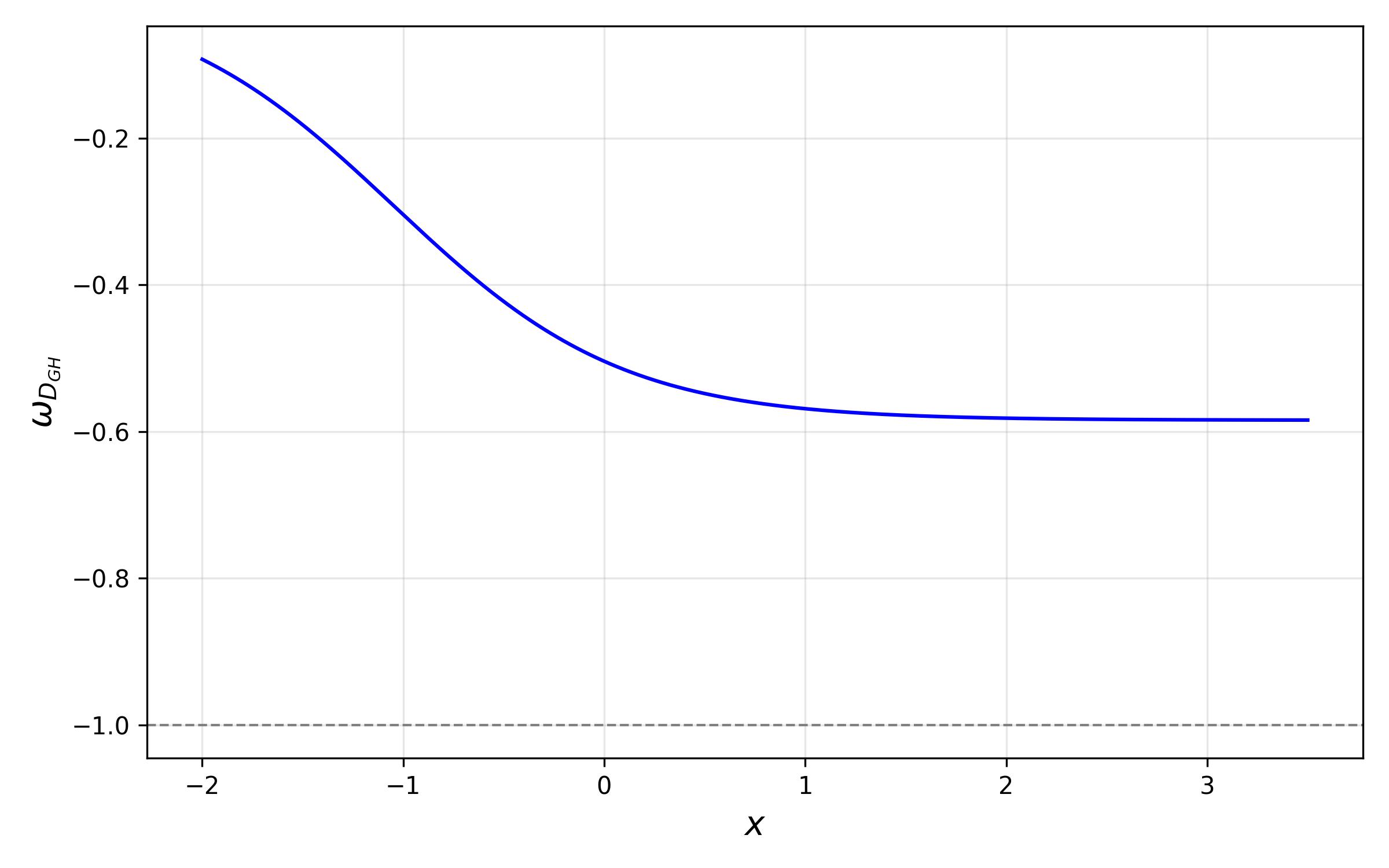}
        \caption{Plot of $\omega_{D_{GH}}(x)$ for $c=0.818$.}
        \label{EoS1-3}
    \end{subfigure}
    \caption{Comparison of the three plots of $\omega_{D_{GH}}(x)$.}
    \label{fig:EoS_all1}
\end{figure}

In Figs. (\ref{EoS1}), (\ref{EoS1-2}) and (\ref{EoS1-3})  we plot the expression of $\omega_{D_{GH}}(x) $ obtained in Eq. (\ref{carolina60}) for $c^2=0.46$, $c=0.579$ and $c=0.818$, respectively. The following values of the parameters involved have been chosen: $\Omega_{m0}=0.315$, $\epsilon =1.312$ along with the values of $c^2$ and $c$ previously indicated.

We obtain the present day values (i.e. the values corresponding to $x=0$) for the EoS for the three different cases: 
\begin{itemize}
    \item $\omega_{D_{GH},1}(0) \approx -0.849$
    \item $\omega_{D_{GH},2}(0) \approx -1.260$
    \item $\omega_{D_{GH},3}(0) \approx -0.504$
\end{itemize}

We now want to calculate the expression of the deceleration parameter $q_{GH}$.\\
The deceleration parameter $q$ quantifies the sign and magnitude of the cosmic acceleration, and is defined as:
\begin{eqnarray}
q &=& - \frac{\ddot{a}}{aH^2} = - \frac{a\ddot{a}}{\dot{a}^2}  \nonumber \\
&=&  - \frac{\dot{H}+H^2}{H^2} =-1 - \frac{1}{2 h^2} \frac{d h^2(x)}{dx}, 
\end{eqnarray}
where $a$ represents the scale factor, which is in general a function of the time.\\
For the model we are considering, we choose the following expression: 
\begin{eqnarray}
q_{GH}(x) &=&  -1 - \frac{1}{2 h_{GH}^2(x)} \frac{d h_{GH}^2(x)}{dx}. \label{deceleration}
\end{eqnarray}
Using the general expression of $h_{GH}^2$ given in Eq. (\ref{st2}) along with the expression of $\frac{d h_{GH}^2(x)}{dx}$ derived in Eq. (\ref{renato1}), we can write:
\begin{eqnarray}
q_{GH}(x) &=&  -1 + \frac{1}{2}\cdot \left\{ 3\Omega_{m0} e^{-3x} 
+ 2\left[1 + \frac{2 \Omega_{m0}}{2 + c^2(\epsilon - 2)} \right] \left(1 - \alpha\right) e^{-2\left(1 - \alpha\right) x}  \right.\nonumber \\
&&\,\,\,\,\,\,\,\,\,\,\,\,\,\,\,\,\,\,\,\,\,\,\,\,\,\,\,\,\,\,\,\,\,\,\,\,\,\left.-\frac{3c^2 (\epsilon - 2)\Omega_{m0} e^{-3x}}{2 + c^2(\epsilon - 2)}   \right\}\times \nonumber \\
&& \left\{ \Omega_{m0} e^{-3x} 
+ \left[1 - \frac{2 \Omega_{m0}}{2 + c^2(\epsilon - 2)} \right] e^{-2\left(1 - \alpha\right)x} 
-  \frac{c^2 (\epsilon - 2)\Omega_{m0} e^{-3x}}{2 + c^2(\epsilon - 2)} \right\}^{-1}\label{carolina70}.
\end{eqnarray}

\begin{figure}[htbp]
    \centering
    \begin{subfigure}{0.8\textwidth}
        \includegraphics[width=0.6\textwidth]{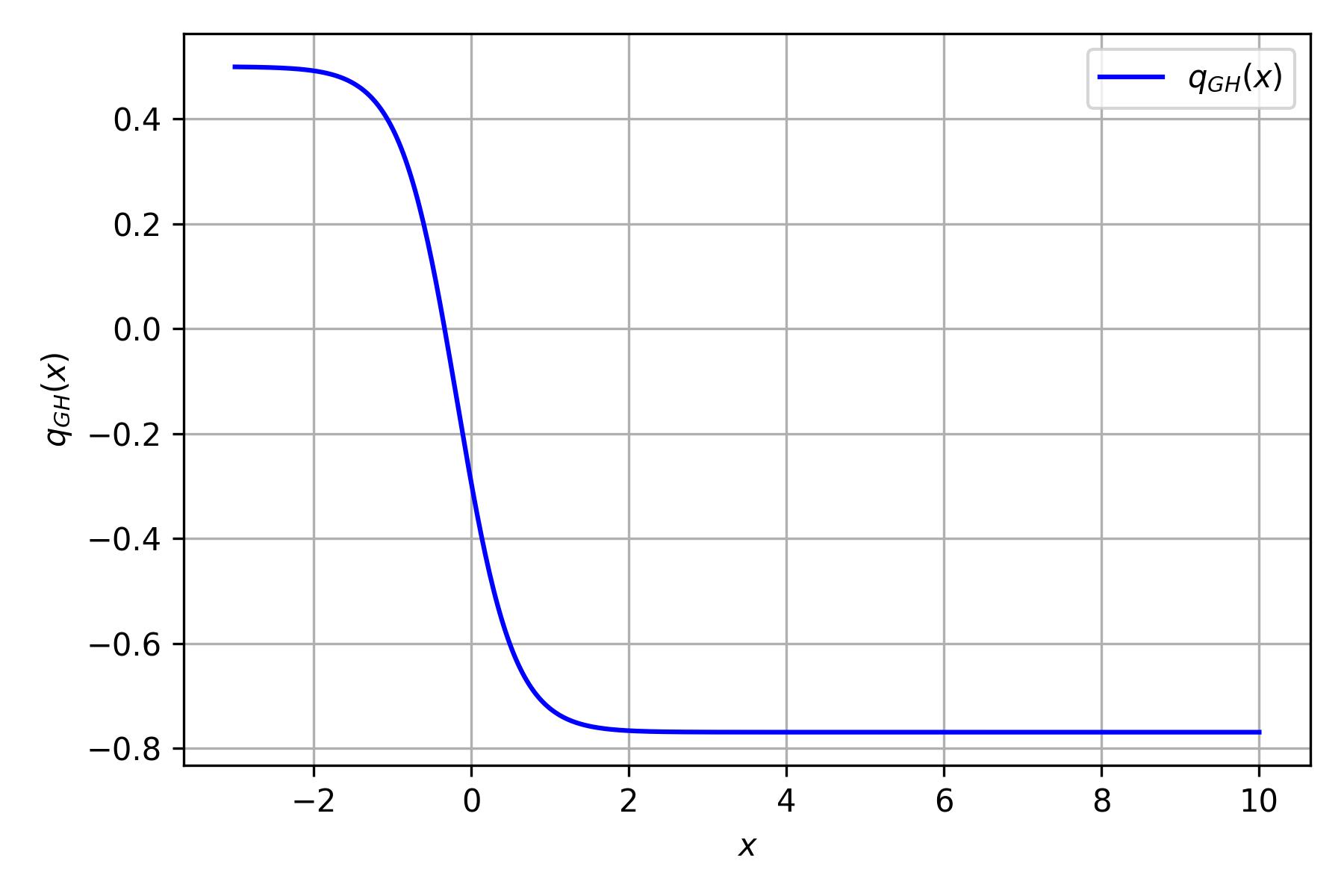}
        \caption{Plot of $q_{GH}(x)$ for $c^2=0.46$.}
        \label{q1}
    \end{subfigure}\\[0.5cm]
    \begin{subfigure}{0.8\textwidth}
        \includegraphics[width=0.6\textwidth]{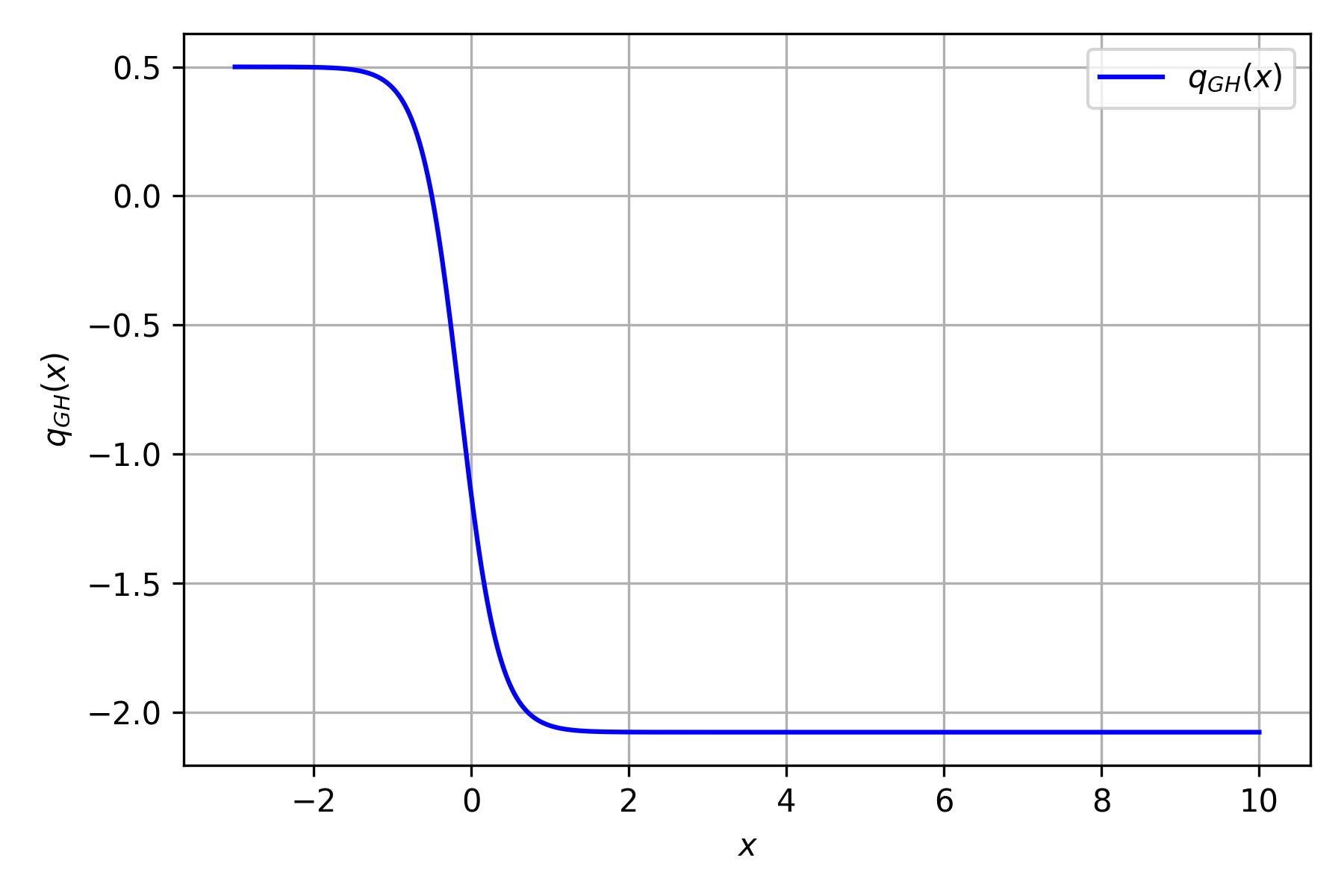}
        \caption{Plot of $q_{GH}(x)$ for $c=0.579$.}
        \label{q1-2}
    \end{subfigure}\\[0.5cm]
    \begin{subfigure}{0.8\textwidth}
        \includegraphics[width=0.6\textwidth]{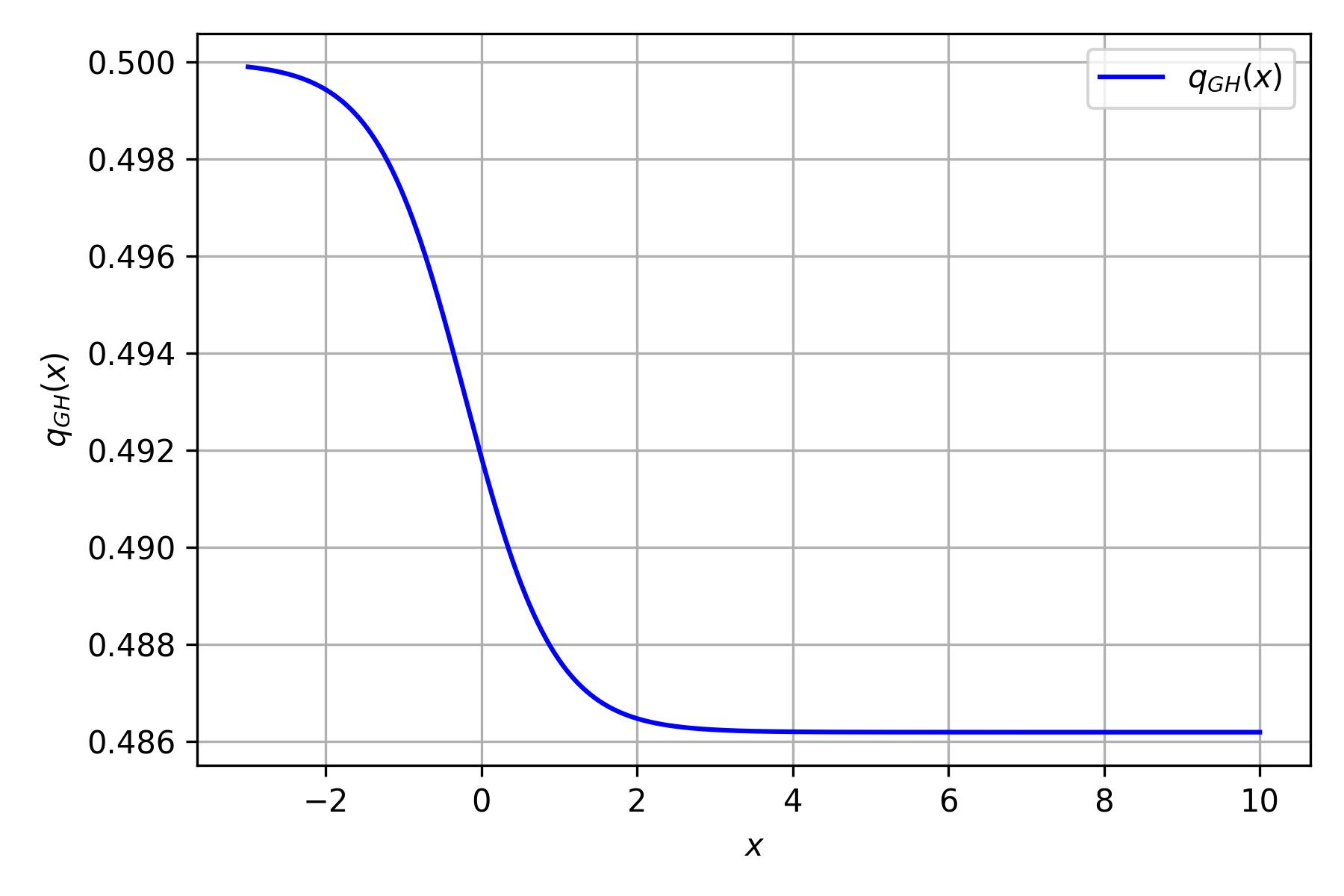}
        \caption{Plot of $q_{GH}(x)$ for $c=0.818$.}
        \label{q1-3}
    \end{subfigure}
    \caption{Comparison of the three cases for the deceleration parameter $q_{GH}(x)$.}
    \label{fig:q_all1}
\end{figure}

In Figs. (\ref{q1}), (\ref{q1-2}) and (\ref{q1-3}) we plot the expression of $q_{GH}(x) $ obtained in Eq. (\ref{carolina70}) for $c^2=0.46$, $c=0.579$ and $c=0.818$, respectively.

We obtain the present day values (i.e. the values corresponding to $x=0$) for $q_{GH}(x) $ for the three different cases: 
\begin{itemize}
    \item $q_{GH}(0) \approx -0.294 $
    \item $q_{GH}(0) \approx-1.159 $
    \item $q_{GH}(0) \approx 0.492 $
\end{itemize}
Therefore, the first two cases predict an accelerated expansion of the Universe as predicted by recent observations while the third one not.

There is also a general relation between $q_{GH}$ and $\Omega_{D_{GH}}$ given by:
\begin{equation}
q_{GH}(x)=\frac{1}{\epsilon}-\frac{\Omega_{D_{GH}}(x)}{c^2\epsilon}.
\end{equation}

In the limiting case of a Dark Dominated Universe, i.e. for $\Omega_m=\Omega_k=0$ and $\Omega_D=1$, we obtain the following expressions for the cosmological quantities we derived above:
\begin{eqnarray}
h_{GH,DD}^2(x) &=& e^{-2(1 - \alpha) x},\label{goddi1}\\
\rho_{D_{GH},DD}(x) &=&  e^{-2(1 - \alpha) x}, \\
p_{D_{GH},DD}(x)  &=& -\left(\frac{2\alpha+1}{3}  \right) e^{-2(1-\alpha)x}\\
\omega_{D_{GH},DD}&=&  -\frac{1}{3} - \frac{2}{3} \alpha =-\left(\frac{2\alpha+1}{3}\right) ,\\
q_{GH,DD} &=&  -\alpha.
\end{eqnarray}
We can clearly observe that $\omega_{D_{GH},DD}$ and $q_{GH,DD}$ are then two constants, and $\rho_{D_{GH},DD} = h^2_{GH,DD}$. \\
Passing from the variable $x$ to the time $t$, from the expression of $h^2_{GH,DD}$ we derive the following expression for the scale factor $a$ as a function of time:
\begin{eqnarray}
a_{GH,DD}(t) = \left[ (1 - \alpha)(H_0t + C_1) \right]^{\frac{1}{1 - \alpha}}, \label{scaleDD}
\end{eqnarray}
where $C_1$ is an integration constant.\\
Using the expression of the scale factor $a$ given in Eq- (\ref{scaleDD}), we obtain the following expression for the Hubble parameter as a function of the time:
\begin{eqnarray}
H_{GH,DD}(t) = \frac{H_0}{(1 - \alpha)(H_0t + C_1)}.
\end{eqnarray}
In order to avoid singularities, we must have $\alpha \neq 1$,  which implies $ c^2 \neq \frac{1}{1 + \epsilon}$. Moreover, we must have that $t\neq \frac{C_1}{H_0}$\\
As we found before, $\rho_{D_{GH},DD} = h^2_{GH,DD}$, therefore we can write:
\begin{eqnarray}
\rho_{D_{GH},DD}(t)  &=& \frac{H_0^2}{(1 - \alpha)^2(H_0t + C_1)^2}.
\end{eqnarray}
In the limiting case of $C_1=0$, we obtain:
\begin{eqnarray}
a_{GH,DD,lim}(t) &=& \left[ (1 - \alpha)H_0t \right]^{\frac{1}{1 - \alpha}},\\
H_{GH,DD,lim}(t) &=& \frac{1}{(1 - \alpha)}\cdot\frac{1}{t},\\
\rho_{D_{GH},DD,lim}(t)  &=&  \frac{1}{(1 - \alpha)^2}\cdot\frac{1}{t^2}.
\end{eqnarray}

\subsection{Case 2: $g\left(\frac{H^2}{R}\right)=1-\eta\left(1-\frac{H^2}{R}\right)$}
We now consider the second case of this paper, which corresponds to $g\left(\frac{H^2}{R}\right)=1-\eta\left(1-\frac{H^2}{R}\right)$. In this case, the energy density of DE can be written as:
\begin{eqnarray}
\rho_{GR}&=&3c^2M^{2}_{pl}\left[ 1-\eta\left(1-\frac{H^2}{R}\right) \right]R.
\label{eq:GRD}
\end{eqnarray}
We obtain the Friedmann equation given in Eq. (\ref{genfri}) can be rewritten as follows:
\begin{eqnarray}
H^2&=&\frac{1}{3M^{2}_{pl}}\left(\rho_{m}+\rho_{GR}\right)\nonumber\\
&=&H^2\Omega_{m}+c^2\left[1-\eta\left(1-\frac{H^2}{R}\right)\right]R\label{eq:FREGR}.
\end{eqnarray}
The fractional energy density of generalized Ricci dark energy is
\begin{eqnarray}
\Omega_{D_{GR}}&=&\frac{\rho_{GR}}{3M^{2}_{pl}H^2}\nonumber\\
&=&c^2\left[1-\eta\left(1-\frac{H^2}{R}\right)\right]\frac{R}{H^2}\nonumber\\
&=&c^2\left[(2-\eta)-\frac{(1-\eta)(1+z)}{2H_{GR}^2}\frac{d
H_{GR}^2}{dz}\right]\nonumber \\
&=&c^{2} \left[ (2-\eta) +  \frac{(1-\eta)}{2h_{GR}^{2}} \frac{d h_{GR}^{2}}{d x} \right].\label{eq:OmegaGR!}
\end{eqnarray}
Using the expression of $\Omega_{D_{GR}}$ obtained in Eq. (\ref{eq:OmegaGR!}), the Friedmann Eq. (\ref{genfri}) can be rewritten as the
differential equation of $H^2_{GR}(z)$ in the following way:
\begin{equation}
H^2_{GR}(z)\left\{1-c^2\left[(2-\eta)-(1-\eta)\frac{(1+z)}{2}\frac{d\ln
H^2_{GR}(z)}{dz}\right]\right\}=H^2_0\Omega_{m0}(1+z)^3,
\end{equation}
which has the following solution:
\begin{eqnarray}
h_{GR}^2(z) &=&\left[1+\frac{2\Omega_{m0}}{c^2(1+\eta)-2}\right](1+z)^{\lambda}-\frac{2\Omega_{m0}(1+z)^3}{c^2(1+\eta)-2} ,\label{hgr}
\end{eqnarray}
where $\lambda$ is defined as follows:
\begin{eqnarray}
\lambda = \frac{2\left[1 + c^2(\eta - 2)\right]}{c^2(\eta - 1)}.
\end{eqnarray}
$h_{GR}^2(z)$ can be also expressed in the following way:
\begin{eqnarray}
h^2_{GR}(z)&=& \Omega_{m0}(1+z)^3 +  \left[1 + \frac{2\Omega_{m0}}{c^2(1+\eta)-2} \right](1+z)^{\lambda}- \left[ \frac{c^2(1+\eta)}{c^2(1+\eta)-2} \right]\Omega_{m0}(1+z)^3.\label{st2-2}
\end{eqnarray}
The form obtained in Eq.~(\ref{st2-2}) is particularly convenient, as it separates $h_{GR}^2(z)$ into two distinct contributions: the term $\Omega_{m0}(1+z)^3$, which arises from dark matter (DM), and the remaining part, 
which is associated with dark energy (DE). 
Consequently, the energy density of DE, $\rho_{D_{GH}}(z)$, can be written as:
\begin{eqnarray}
\rho_{D_{GR}}(z)&=&  \left[1 + \frac{2\Omega_{m0}}{c^2(1+\eta)-2} \right](1+z)^{\lambda}- \left[\frac{c^2(1+\eta)}{c^2(1+\eta)-2} \right]\Omega_{m0}(1+z)^3 .
\end{eqnarray}
$h^2_{GR}(z)$ and $\rho_{D_{GR}}(z)$ can be written as functions of $x=\ln a$ in the following way:
\begin{eqnarray}
h^2_{GR}(x)&=& \Omega_{m0}e^{-3x} + \left[1 + \frac{2\Omega_{m0}}{c^2(1+\eta)-2} \right]e^{-\lambda x}- \left[\frac{c^2(1+\eta)}{c^2(1+\eta)-2} \right]\Omega_{m0}e^{-3x}, \label{hubblegr}\\
\rho_{D_{GR}}(x)&=& \left[1 + \frac{2\Omega_{m0}}{c^2(1+\eta)-2} \right]e^{-\lambda x}- \left[ \frac{c^2(1+\eta)}{c^2(1+\eta)-2} \right]\Omega_{m0}e^{-3x}. \label{endenGR}
\end{eqnarray}
We now want to calculate the expressions of the fractional energy density of DE $\Omega_{D_{GR}}$, the evolutionary form of the fractional energy density of DE $\Omega'_{D_{GR}}$, the pressure $p_{D_{GR}}$, the EoS parameter $\omega_{D_{GR}}$ and the deceleration parameter $q_{GR}$.\\
In order to obtain the final expression of the fractional energy density of DE for this case, we use the following relation:
\begin{eqnarray}
\Omega_{D_{GR}}(x)&=&c^{2} \left[ (2-\eta) +  \frac{(1-\eta)}{2h_{GR}^{2}(x)} \frac{d h_{GR}^{2}(x)}{d x} \right].\label{eq:OmegaGR}
\end{eqnarray}
Using the expression of $h_{GR}^{2}(x)$ obtained in Eq. (\ref{st2-2}), we obtain the following expression for $\frac{d h_{GR}^2(x)}{dx}$:
\begin{eqnarray}
\frac{d h_{GR}^2(x)}{dx}&=& -3\Omega_{m0}e^{-3x} - \left[1 + \dfrac{2\Omega_{m0}}{c^2(1+\eta)-2} \right]\lambda e^{-\lambda x}+ 3 \left[\dfrac{c^2(1+\eta)}{c^2(1+\eta)-2} \right] \Omega_{m0} e^{-3x}.\label{angie1}
\end{eqnarray}
Therefore, we obtain the following expression for $\Omega_{D_{GR}}(x)$:
\begin{eqnarray}
\Omega_{D_{GR}}(x)&=&c^{2}  (2-\eta) +\frac{c^2  (1-\eta) }{2}  \cdot  \left\{  -3\Omega_{m0}e^{-3x} - \left[1 + \dfrac{2\Omega_{m0}}{c^2(1+\eta)-2} \right]\lambda e^{-\lambda x}\right. \nonumber \\
&&\left.+ 3 \left[\dfrac{c^2(1+\eta)}{c^2(1+\eta)-2} \right] \Omega_{m0} e^{-3x} \right\}\times \left\{    \Omega_{m0}e^{-3x} + \left[1 + \frac{2\Omega_{m0}}{c^2(1+\eta)-2} \right]e^{-\lambda x}\right. \nonumber \\
&&\left.- \left[\frac{c^2(1+\eta)}{c^2(1+\eta)-2} \right]\Omega_{m0}e^{-3x} \right\}^{-1}.\label{eq:OmegaGRneww} 
\end{eqnarray}
We now want to calculate the evolutionary form of the fractional energy density of DE $\Omega'_{D_{GR}}(x)$.\\
If we put $h^2_{GR}(x) \equiv A_2$, we can write Eq. (\ref{eq:OmegaGRneww}) as:
\begin{eqnarray}
\Omega_{D_{GR}}(x)&=&c^{2} (2-\eta) +  \frac{c^2(1-\eta)}{2} \left( \frac{A_2'}{A_2}  \right). \label{renato-22}
\end{eqnarray}
Differentiating Eq. (\ref{renato-22}) with respect to $x$, we can write:
\begin{eqnarray}
\Omega'_{D_{GR}}(x)&=& \frac{c^2(1-\eta)}{2}\left[ \frac{A_2''}{A_2}-\left( \frac{A_2'}{A_2}  \right)^2 \right],\label{renato-2prime}
\end{eqnarray}
where $A_2''= \frac{d^2 h_{GR}^2(x)}{dx^2}$. \\
Using the general expression of $ h_{GR}^2(x)$ given in Eq. (\ref{st2-2}) or equivalently the expression of $\frac{d h_{GR}^2(x)}{dx}$ obtained in Eq. (\ref{angie1}), we can write:
\begin{eqnarray}
    \frac{d^2 h_{GR}^2(x)}{dx^2}  = 9\Omega_{m0}e^{-3x} + \left[1 + \dfrac{2\Omega_{m0}}{c^2(1+\eta)-2} \right]\lambda^2 e^{-\lambda x}- 9 \left[\dfrac{c^2(1+\eta)}{c^2(1+\eta)-2} \right] \Omega_{m0} e^{-3x}.\label{hubblegrprime}
\end{eqnarray}
Then, the evolutionary form of the fractional energy density of DE can be written as:
\begin{eqnarray}
\Omega'_{D_{GR}}(x)&=& \frac{c^2  (1-\eta) }{2}  \cdot  \left\{  9\Omega_{m0}e^{-3x} + \left[1 + \dfrac{2\Omega_{m0}}{c^2(1+\eta)-2} \right]\lambda^2 e^{-\lambda x}\right. \nonumber \\
&&\left.-9 \left[\dfrac{c^2(1+\eta)}{c^2(1+\eta)-2} \right] \Omega_{m0} e^{-3x} \right\}\times \left\{    \Omega_{m0}e^{-3x} + \left[1 + \frac{2\Omega_{m0}}{c^2(1+\eta)-2} \right]e^{-\lambda x}\right. \nonumber \\
&&\left.- \left[\frac{c^2(1+\eta)}{c^2(1+\eta)-2} \right]\Omega_{m0}e^{-3x} \right\}^{-1} \nonumber \\
&&-\frac{c^2  (1-\eta) }{2}  \cdot  \left\{  -3\Omega_{m0}e^{-3x} - \left[1 + \dfrac{2\Omega_{m0}}{c^2(1+\eta)-2} \right]\lambda e^{-\lambda x}\right. \nonumber \\
&&\left.+ 3 \left[\dfrac{c^2(1+\eta)}{c^2(1+\eta)-2} \right] \Omega_{m0} e^{-3x} \right\}^2\times \left\{    \Omega_{m0}e^{-3x} + \left[1 + \frac{2\Omega_{m0}}{c^2(1+\eta)-2} \right]e^{-\lambda x}\right. \nonumber \\
&&\left.- \left[\frac{c^2(1+\eta)}{c^2(1+\eta)-2} \right]\Omega_{m0}e^{-3x} \right\}^{-2}.
\end{eqnarray}

We now want to obtain the final expression of the pressure $p_{D_{GR}}(x)$.\\
The general expression of $p_{D_{GR}}$ is given by:
\begin{eqnarray}
\label{eq:conserve}
p_{D_{GR}}(x)&=& -\rho_{D_{GR}}(x)-\frac{\rho'_{D_{GR}}(x)}{3}.
\end{eqnarray}
Differentiating with respect to $x$ the expression of $\rho_{D_{GR}}(x)$ obtained in Eq. (\ref{endenGR}), we obtain:
\begin{eqnarray}
\rho'_{D_{GR}}(x)=-\left[1 + \dfrac{2\Omega_{m0}}{c^2(1+\eta)-2} \right]\lambda e^{-\lambda x}+ 3 \left[ \dfrac{c^2(1+\eta)}{c^2(1+\eta)-2} \right] \Omega_{m0} e^{-3x}. \label{murmuraGR}
\end{eqnarray}
Therefore, the final expression of $p_{D_{GR}}(x)$ is given by:
\begin{eqnarray}
p_{D_{GR}}(x)
&=& \left(\frac{\lambda}{3} -1   \right)  \left[1 + \frac{2\Omega_{m0}}{c^2(1+\eta)-2} \right]e^{-\lambda x}.  \label{murmuraGRp}
\end{eqnarray}

We now want to derive the final expression of the EoS parameter for this case.\\
The general definition of $\omega_{D_{GR}}(x)$ is:
\begin{eqnarray}
\omega_{D_{GR}}(x) = -1 - \frac{\rho'_{D_{GR}}(x)}{3 \rho_{D_{GR}}(x)} \label{eosnongen}.
\end{eqnarray}

Therefore, using the expressions of $\rho_{D_{GR}}(x)$ and $\rho'_{D_{GR}}(x)$ we derived in Eqs. (\ref{endenGR}) and (\ref{murmuraGR}), we obtain the following expression for $\omega_{D_{GR}}(x)$:
\begin{eqnarray}
\omega_{D_{GR}}(x) = -1 + \frac{1}{3} \cdot \frac{
 \left[1 + \dfrac{2\Omega_{m0}}{c^2(1+\eta)-2} \right]\lambda e^{-\lambda x}- 3 \left[\dfrac{c^2(1+\eta)}{c^2(1+\eta)-2} \right] \Omega_{m0} e^{-3x}}{\left[1 + \dfrac{2\Omega_{m0}}{c^2(1+\eta)-2} \right] e^{-\lambda x} - \left[\dfrac{c^2(1+\eta)}{c^2(1+\eta)-2} \right] \Omega_{m0} e^{-3x} }.
\end{eqnarray}
In order to find the final expression of the deceleration parameter $q_{GR}(x)$, we use the general expression:
\begin{eqnarray}
q_{GR}(x) &=&  -1 - \frac{1}{2 h_{GR}^2(x)} \frac{d h_{GR}^2(x)}{dx}. \label{deceleration}
\end{eqnarray}
Using the expression of $ \frac{d h_{GR}^2(x)}{dx}$ we derived in Eq. (\ref{angie1}) along with the expression of $h_{GR}^2(x)$ obtained in Eq. (\ref{hubblegr}), we can write:
\begin{eqnarray}
q_{GR}(x) &=&  -1 + \frac{1}{2}\cdot \left\{ 3\Omega_{m0}e^{-3x} + \left[1 + \dfrac{2\Omega_{m0}}{c^2(1+\eta)-2} \right]\lambda e^{-\lambda x}- 3 \left[\dfrac{c^2(1+\eta)}{c^2(1+\eta)-2} \right] \Omega_{m0} e^{-3x}  \right\} \times \nonumber \\
&& \left\{  \Omega_{m0}e^{-3x} + \left[1 + \frac{2\Omega_{m0}}{c^2(1+\eta)-2} \right]e^{-\lambda x}- \left[\frac{c^2(1+\eta)}{c^2(1+\eta)-2} \right]\Omega_{m0}e^{-3x}  \right\}^{-1}. \label{deceleration}
\end{eqnarray}
A relation between $q_{GR}(x)$ and $\Omega_{GR}(x)$ is given by:
\begin{equation}
q_{GR}(x)=\frac{1}{1-\eta}-\frac{\Omega_{D_{GR}}(x)}{(1-\eta)c^2}.
\end{equation}

One can immediately find out that the results we obtained for these two cases are equivalent when:
\begin{equation}
\epsilon=1-\eta.
\end{equation}

In the limiting case of a Dark Dominated Universe, i.e. $\Omega_m=\Omega_k=0$ and $\Omega_D=1$, we obtain the following expressions for the quantities we derived:
\begin{eqnarray}
h_{GR,DD}^2(x) &=& e^{-\lambda x}, \\
\rho_{D_{GR},DD}(x) &=& e^{-\lambda x}, \\
p_{D_{GR},DD}(x) &=&  \left(\frac{\lambda -3}{3}  \right) e^{-\lambda x},\\
\omega_{D_{GR},DD} &=& -1 + \frac{\lambda}{3} = \frac{\lambda -3}{3},\\
q_{GR,DD} &=& -1 + \frac{\lambda}{2} = \frac{\lambda-2}{2}.
\end{eqnarray}

We can clearly observe that in this case $\omega_{D_{GR},DD}$ and $q_{GR,DD}$ are constants depending on the parameters of the model considered.

Passing from the variable $x$ to the time $t$, from the expression of  $h^2_{GR,DD}$ we derive the following expression for the scale factor:
\begin{eqnarray}
a_{GR,DD}(t) = \left[ \frac{\lambda}{2}\left(H_0t + C_2\right) \right]^{\frac{2}{\lambda}}.
\end{eqnarray}

Then, the expression of the Hubble parameter as function of the time is given by:
\begin{eqnarray}
H_{GR,DD}(t) =\frac{\dot{a}(t)}{a(t)}=\frac{2H_0}{\lambda (H_0t+C_2)}.
\end{eqnarray}
Therefore, the energy density of the DE is given by:
\begin{eqnarray}
\rho_{D_{GR},DD}(t) =H^2_{2,DD}(t) =\frac{4H_0^2}{\lambda^2 (H_0t+C_2)^2}.
\end{eqnarray}
In the limiting case corresponding to $C_2=0$, we derive the following expressions for the quantities derived above:
\begin{eqnarray}
a_{GR,DD,lim}(t) &=& \left( \frac{\lambda H_0 t}{2} \right)^{\frac{2}{\lambda}},\\
H_{GR,DD,lim}(t) &=&\frac{2}{\lambda t},\\
\rho_{D_{GR},DD,lim}(t) &=& \frac{4}{\lambda^2  t^2}.
\end{eqnarray}

\section{Equations with Curvature}
We now want to extend the calculations we did in the previous subsection considering the curvature term.\\
\subsection{Case 1: $f\left(\frac{R}{H^2}\right)=1-\epsilon\left(1-\frac{R}{H^2}\right)$}
We start considering the first model we are studying.\\
We begin by deriving the main expressions for the Hubble parameter 
and the dark energy density as functions of redshift $z$ , and subsequently in terms of the variable $x$.\\
In this case, the Friedmann equation given in Eq. (\ref{genfri}) can be rewritten as follows:
\begin{eqnarray}
H^2&=&\frac{1}{3M^{2}_{pl}}\left(\rho_{m}+\rho_{GH}\right)-\frac{k}{a^2}\nonumber\\
&=&H^2\Omega_{m}- H^2\Omega_k+c^2\left[1-\epsilon\left(1-\frac{R}{H^2}\right)\right]H^2\label{eq:FREGH}.
\end{eqnarray}
 The fractional energy density of the generalized holographic dark energy is given by:
\begin{eqnarray}
\Omega_{D_{GH},k}&=&\frac{\rho_{D_{GH},k}}{3M^{2}_{pl}H^2}\nonumber\\
&=&c^2\left[1-\epsilon\left(1-\frac{R}{H^2}\right)\right]\nonumber\\
&=& c^2 \left[ 1 + \epsilon + \frac{\epsilon}{2H^2_{D_{GH},k}} \frac{d H^2_{D_{GH},k}}{dx}+ \epsilon H_0^2\Omega_{k0} (1+z)^2 \right]\nonumber \\
&=& c^2 \left[ 1 + \epsilon + \frac{\epsilon}{2h^2_{D_{GH},k}} \frac{d h^2_{D_{GH},k}}{dx}+ \epsilon H_0^2\Omega_{k0} e^{-2x} \right].\label{eq:OmegaGHcurv}
\end{eqnarray}
Using the expression of $\Omega_{D_{GH}}$ given in Eq. (\ref{eq:OmegaGHcurv}), the Friedmann equation obtained in Eq. (\ref{eq:FREGH}) can be rewritten as a differential equation of $H^2(z)$ in the following way:
\begin{eqnarray}
H_{GH,k}^2(z)\left\{1-c^2\left[1+\epsilon-\frac{\epsilon(1+z)}{2}\frac{d \ln
H_{GH,k}^2(z)}{dz}\right]\right\}&=&H^2_0\Omega_{m0}(1+z)^3\nonumber \\
&&-(1-\epsilon c^2)H^2_0\Omega_{k0}(1+z)^2, \label{cledison1}
\end{eqnarray}
which is equivalent to:
\begin{eqnarray}
H_{GH,k}^2(z)\left\{1-c^2\left[1+\epsilon-\frac{\epsilon(1+z)}{2H_{GH,k}^2(z)}\frac{d 
H_{GH,k}^2(z)}{dz}\right]\right\}&=&H^2_0\Omega_{m0}(1+z)^3\nonumber \\
&&-(1-\epsilon c^2)H^2_0\Omega_{k0}(1+z)^2. \label{cledison2}
\end{eqnarray}
The general solution of Eq. (\ref{cledison1}) or equivalently of Eq. (\ref{cledison2}) is given by:
\begin{eqnarray}
h^2_{D_{GH},k}(z) &=&
\frac{3\Omega_{m0} (1+z)^3}{2 + c^2(\epsilon - 2)}
-
\left(\frac{1-\epsilon c^2}{1-c^2}\right) \Omega_{k0}(1+z)^2
\nonumber \\
&&+
\left[  1 - 
\frac{2\Omega_{m0}}{2 + c^2(\epsilon - 2)} 
+
\left(\frac{1-\epsilon c^2}{1-c^2}\right) \Omega_{k0} \right] \, (1+z)^{2\left(1-\alpha \right)}, \label{picu1}
\end{eqnarray}
where:
\begin{eqnarray}
\alpha = \frac{1}{\epsilon} \left( \frac{1}{c^2} - 1 \right).
\end{eqnarray}
We can also write $h^2_{D_{GH},k}(z)$ in the following way:
\begin{eqnarray}
h^2_{D_{GH},k}(z) &=&\Omega_{m0} (1+z)^3
\nonumber \\
&& -\Omega_{k0} (1+z)^2-
\left[  \frac{ c^2(\epsilon - 2)}{2 + c^2(\epsilon - 2)}\right]\Omega_{m0} (1+z)^3
+\left[\frac{c^2(\epsilon - 1)}{1-c^2}\right]\Omega_{k0}(1+z)^2\nonumber \\
&&+\left[  1 - \frac{2\Omega_{m0}}{2 + c^2(\epsilon - 2)} 
+\left(\frac{1-\epsilon c^2}{1-c^2}\right) \Omega_{k0}  \right] \, (1+z)^{2\left(1-\alpha \right)}.\label{cledison3}
\end{eqnarray}

The form obtained in Eq.~(\ref{cledison3}) is particularly useful, as it allows us to express $ h_{GH}^2(z) $ as the sum of three distinct contributions: one arising from dark matter, one arising from curvature and the other attributed to dark energy. The term $ \Omega_{m0} (1+z)^3 $ accounts for the dark matter component while the term $ \Omega_{k0} (1+z)^2 $ accounts for the curvature contribution. Consequently, we can derive the following expression for $ \rho_{D_{GH},k}(z) $:
\begin{eqnarray}
\rho_{D_{GH},k}(z)  &=&-
\left[  \frac{ c^2(\epsilon - 2)}{2 + c^2(\epsilon - 2)}\right]\Omega_{m0} (1+z)^3
+ \left[\frac{c^2(\epsilon - 1)}{1-c^2}\right]\Omega_{k0}(1+z)^2\nonumber \\
&&+\left[  1 - \frac{2\Omega_{m0}}{2 + c^2(\epsilon - 2)} 
+\left(\frac{1-\epsilon c^2}{1-c^2}\right) \Omega_{k0}   \right] \, (1+z)^{2\left(1-\alpha \right)}.
\end{eqnarray}
We can now write $h^2_{GH,k}$ and $\rho_{D_{GH},k}$ as functions of $x$:
\begin{eqnarray}
h^2_{GH,k}(x) &=&\Omega_{m0} e^{-3x} -\Omega_{k0} e^{-2x} -
\left[  \frac{ c^2(\epsilon - 2)}{2 + c^2(\epsilon - 2)}\right]\Omega_{m0} e^{-3x} 
+\left[\frac{c^2(\epsilon - 1)}{1-c^2}\right]\Omega_{k0}e^{-2x} \nonumber \\
&&
+\left[  1 - \frac{2\Omega_{m0}}{2 + c^2(\epsilon - 2)} 
+\left(\frac{1-\epsilon c^2}{1-c^2}\right) \Omega_{k0}  \right] \, e^{-2\left(1-\alpha \right)} ,\label{cledison4}\\
\rho_{D_{GH},k}(x)  &=&-
\left[  \frac{ c^2(\epsilon - 2)}{2 + c^2(\epsilon - 2)}\right]\Omega_{m0} e^{-3x}
+ \left[\frac{c^2(\epsilon - 1)}{1-c^2}\right]\Omega_{k0}e^{-2x}\nonumber \\
&&+\left[  1 - \frac{2\Omega_{m0}}{2 + c^2(\epsilon - 2)} 
+\left(\frac{1-\epsilon c^2}{1-c^2}\right) \Omega_{k0}  \right] \,e^{-2\left(1-\alpha \right)x}. \label{cledi1}
\end{eqnarray}
We now want to calculate the final expressions of other important cosmological quantities, like the fractional energy density of dark energy $\Omega_{D_{GH},k}$, the evolutionary form of the fractional energy density of DE $\Omega'_{D_{GH},k}$, the pressure $p_{GH,k}$, the EoS parameter $\omega_{D_{GH},k}$ and the deceleration parameter $q_{GH,k}$. \\
For the fractional energy density of DE, we use the general relation obtained in Eq. (\ref{eq:OmegaGHcurv})
\begin{eqnarray}
\Omega_{D_{GH},k}(x)
&=& c^2 \left[ 1 + \epsilon + \frac{\epsilon}{2h^2_{GH,k}} \frac{d h^2_{GH,k}}{dx}+ \epsilon H_0^2\Omega_{k0} e^{-2x} \right].\label{eq:OmegaGH}
\end{eqnarray}
Using the expression of $h^2_{GH,k}$ obtained in Eq. (\ref{cledison4}), we can write:
\begin{eqnarray}
    \frac{d h^2_{GH,k}(x)}{dx} &=& -3\Omega_{m0} e^{-3x} +2\Omega_{k0} e^{-2x} +3
\left[  \frac{ c^2(\epsilon - 2)}{2 + c^2(\epsilon - 2)}\right]\Omega_{m0} e^{-3x} 
-2\left[\frac{c^2(\epsilon - 1)}{1-c^2}\right]\Omega_{k0}e^{-2x} \nonumber \\
&&
-2\left(1-\alpha \right)\left[  1 - \frac{2\Omega_{m0}}{2 + c^2(\epsilon - 2)} 
+\left(\frac{1-\epsilon c^2}{1-c^2}\right) \Omega_{k0}  \right] \, e^{-2\left(1-\alpha \right)x}.
\end{eqnarray}
Therefore, we obtain the following expression for $\Omega_{D_{GH},k}(x)$:
\begin{eqnarray}
\Omega_{D_{GH},k}(x)
&=& c^2\left[1 + \epsilon + \epsilon H_0^2\Omega_{k0} e^{-2x}\right] \nonumber \\
&&-\frac{\epsilon c^2}{2}\cdot \left\{3\Omega_{m0} e^{-3x} -2\Omega_{k0} e^{-2x} -3
\left[  \frac{ c^2(\epsilon - 2)}{2 + c^2(\epsilon - 2)}\right]\Omega_{m0} e^{-3x} 
+2\left[\frac{c^2(\epsilon - 1)}{1-c^2}\right]\Omega_{k0}e^{-2x}\right. \nonumber \\
&&\left.
+2\left(1-\alpha \right)\left[  1 - \frac{2\Omega_{m0}}{2 + c^2(\epsilon - 2)} 
+\left(\frac{1-\epsilon c^2}{1-c^2}\right) \Omega_{k0}  \right] \, e^{-2\left(1-\alpha \right)x}   \right\}\times \left\{ \Omega_{m0} e^{-3x} -\Omega_{k0} e^{-2x}
\right. \nonumber \\
&&\left. -
\left[  \frac{ c^2(\epsilon - 2)}{2 + c^2(\epsilon - 2)}\right]\Omega_{m0} e^{-3x} +\left[\frac{c^2(\epsilon - 1)}{1-c^2}\right]\Omega_{k0}e^{-2x} 
\nonumber \right.\\
&&\left.+\left[  1 - \frac{2\Omega_{m0}}{2 + c^2(\epsilon - 2)} 
+\left(\frac{1-\epsilon c^2}{1-c^2}\right) \Omega_{k0}  \right] \, e^{-2\left(1-\alpha \right)x} \right\}^{-1}.
\end{eqnarray}

We now want to calculate the evolutionary form of the fractional energy density of DE $\Omega'_{D_{GH},k}(x)$.\\
If we put $h^2_{GH,k} \equiv A_3$, we can write $\Omega_{D_{GH},k}(x) $ in the following way:
\begin{eqnarray}
\Omega_{D_{GH},k}(x) &=&c^2\left[1 + \epsilon + \epsilon H_0^2\Omega_{k0} e^{-2x}\right]  +\frac{\epsilon c^2}{2}\left( \frac{A_3'}{A_3}  \right),
\end{eqnarray}
which implies:
\begin{eqnarray}
\Omega_{D_{GH},k}'(x) &=&-2\epsilon c^2H_0^2\Omega_{k0} e^{-2x}+\frac{\epsilon c^2}{2}\left[ \frac{A_3''}{A_3}- \left(\frac{A_3'}{A_3}\right)^2\right],
\end{eqnarray}
where $A_3'' = \frac{d^2 h_{GH}^2(x)}{dx^2}$.\\
We obtain then:
\begin{eqnarray}
\Omega'_{D_{GH},k}(x)&=&
\frac{\epsilon c^2}{2}\cdot \left\{9\Omega_{m0} e^{-3x} -4\Omega_{k0} e^{-2x} -9
\left[  \frac{ c^2(\epsilon - 2)}{2 + c^2(\epsilon - 2)}\right]\Omega_{m0} e^{-3x} 
+4\left[\frac{c^2(\epsilon - 1)}{1-c^2}\right]\Omega_{k0}e^{-2x}\right. \nonumber \\
&&\left.
+4\left(1-\alpha \right)^2\left[  1 - \frac{2\Omega_{m0}}{2 + c^2(\epsilon - 2)} 
+\left(\frac{1-\epsilon c^2}{1-c^2}\right) \Omega_{k0}  \right] \, e^{-2\left(1-\alpha \right)x}   \right\}\times \left\{ \Omega_{m0} e^{-3x} -\Omega_{k0} e^{-2x} 
\right. \nonumber \\
&&\left.-
\left[  \frac{ c^2(\epsilon - 2)}{2 + c^2(\epsilon - 2)}\right]\Omega_{m0} e^{-3x} +\left[\frac{c^2(\epsilon - 1)}{1-c^2}\right]\Omega_{k0}e^{-2x} \right.
\nonumber\\
&&\left.+\left[  1 - \frac{2\Omega_{m0}}{2 + c^2(\epsilon - 2)} 
+\left(\frac{1-\epsilon c^2}{1-c^2}\right) \Omega_{k0}  \right]  e^{-2\left(1-\alpha \right)x} \right\}^{-1}\nonumber \\
&&+\frac{\epsilon c^2}{2}\cdot \left\{ -3\Omega_{m0} e^{-3x} +2\Omega_{k0} e^{-2x} +3
\left[  \frac{ c^2(\epsilon - 2)}{2 + c^2(\epsilon - 2)}\right]\Omega_{m0} e^{-3x} 
-2\left[\frac{c^2(\epsilon - 1)}{1-c^2}\right]\Omega_{k0}e^{-2x}\right. \nonumber \\
&&\left.
-2\left(1-\alpha \right)\left[  1 - \frac{2\Omega_{m0}}{2 + c^2(\epsilon - 2)} 
+\left(\frac{1-\epsilon c^2}{1-c^2}\right) \Omega_{k0}  \right] \, e^{-2\left(1-\alpha \right)x}\right\}^2\times \left\{ \Omega_{m0} e^{-3x} -\Omega_{k0} e^{-2x}
\right. \nonumber \\
&&\left. -
\left[  \frac{ c^2(\epsilon - 2)}{2 + c^2(\epsilon - 2)}\right]\Omega_{m0} e^{-3x} +\left[\frac{c^2(\epsilon - 1)}{1-c^2}\right]\Omega_{k0}e^{-2x} \right.\nonumber \\
&&\left. 
+\left[  1 - \frac{2\Omega_{m0}}{2 + c^2(\epsilon - 2)} 
+\left(\frac{1-\epsilon c^2}{1-c^2}\right) \Omega_{k0}  \right] \, e^{-2\left(1-\alpha \right)x} \right\}^{-2}-2\epsilon c^2H_0^2\Omega_{k0} e^{-2x}.
\end{eqnarray}
We now want to obtain the final expression of the pressure $ p_{D_{GH},k} $.\\
The general expression of the pressure $ p_{D_{GH},k} $ is given by:
\begin{eqnarray}
\label{eq:conserve}
p_{D_{GH},k}(x)  &=& -\rho_{D_{GH},k}(x) - \frac{\rho'_{D_{GH},k}(x)}{3}.
\end{eqnarray}
 Using the general expression for $ \rho_{D_{GH}.k} $ derived in Eq.~(\ref{cledi1}), we obtain the following result:
\begin{eqnarray}
    \rho'_{D_{GH},k}(x)  &=&3
\left[  \frac{ c^2(\epsilon - 2)}{2 + c^2(\epsilon - 2)}\right]\Omega_{m0} e^{-3x}
-2 \left[\frac{c^2(\epsilon - 1)}{1-c^2}\right]\Omega_{k0}e^{-2x}\nonumber \\
&&-2(1-\alpha)\left[  1 - \frac{2\Omega_{m0}}{2 + c^2(\epsilon - 2)} 
+\left(\frac{1-\epsilon c^2}{1-c^2}\right) \Omega_{k0}  \right] \,e^{-2(1-\alpha)x}\label{cledi2}.
\end{eqnarray}
Therefore, we obtain the following expression for $p_{D_{GH},k}(x) $:
\begin{eqnarray}
p_{D_{GH},k}(x) &=&
- \frac{1}{3}\left[\frac{c^2(\epsilon - 1)}{1-c^2}\right]\Omega_{k0}e^{-2x}\nonumber \\
&&-\frac{1+2\alpha}{3}\left[  1 - \frac{2\Omega_{m0}}{2 + c^2(\epsilon - 2)} 
+\left(\frac{1-\epsilon c^2}{1-c^2}\right) \Omega_{k0}(1+z)^2  \right] \,e^{-2\left(1-\alpha \right)x}. \label{carolina50}
\end{eqnarray}
We now want to obtain the EoS parameter $\omega_{D_{GH},k}$. The general definition is given by:
\begin{eqnarray}
\omega_{D_{GH},k}(x) = -1 - \frac{\rho_{D_{GH},k}'(x)}{3 \rho_{D_{GH},k}(x)} \label{eosnongen}.
\end{eqnarray}
Using the expressions of $\rho_{D_{GH},k}(x)$ and $\rho'_{D_{GH},k}(x)$ obtained in Eqs. (\ref{cledi1}) and (\ref{cledi2}), we obtain the following expression for $\omega_{D_{GH},k}(x)$: 
\begin{eqnarray}
\omega_{D_{GH},k}(x) &=& -1 \nonumber \\
&&-\frac{1}{3}\cdot \left\{ 3
\left[  \frac{ c^2(\epsilon - 2)}{2 + c^2(\epsilon - 2)}\right]\Omega_{m0} e^{-3x}
-2 \left[\frac{c^2(\epsilon - 1)}{1-c^2}\right]\Omega_{k0}e^{-2x}\right.\nonumber \\
&&\left.-2(1-\alpha)\left[  1 - \frac{2\Omega_{m0}}{2 + c^2(\epsilon - 2)} 
+\left(\frac{1-\epsilon c^2}{1-c^2}\right) \Omega_{k0}  \right] \,e^{-2(1-\alpha)x}  \right\} \times\nonumber \\ 
&&\left\{-
\left[  \frac{ c^2(\epsilon - 2)}{2 + c^2(\epsilon - 2)}\right]\Omega_{m0} e^{-3x}+ \left[\frac{c^2(\epsilon - 1)}{1-c^2}\right]\Omega_{k0}e^{-2x}
\right.\nonumber \\
&&\left.+\left[  1 - \frac{2\Omega_{m0}}{2 + c^2(\epsilon - 2)} 
+\left(\frac{1-\epsilon c^2}{1-c^2}\right) \Omega_{k0}  \right] \,e^{-2(1-\alpha)x} \right\}^{-1}. \label{carolina61}
\end{eqnarray}

\begin{figure}[htbp]
    \centering
    \begin{subfigure}{0.8\textwidth}
        \includegraphics[width=0.6\textwidth]{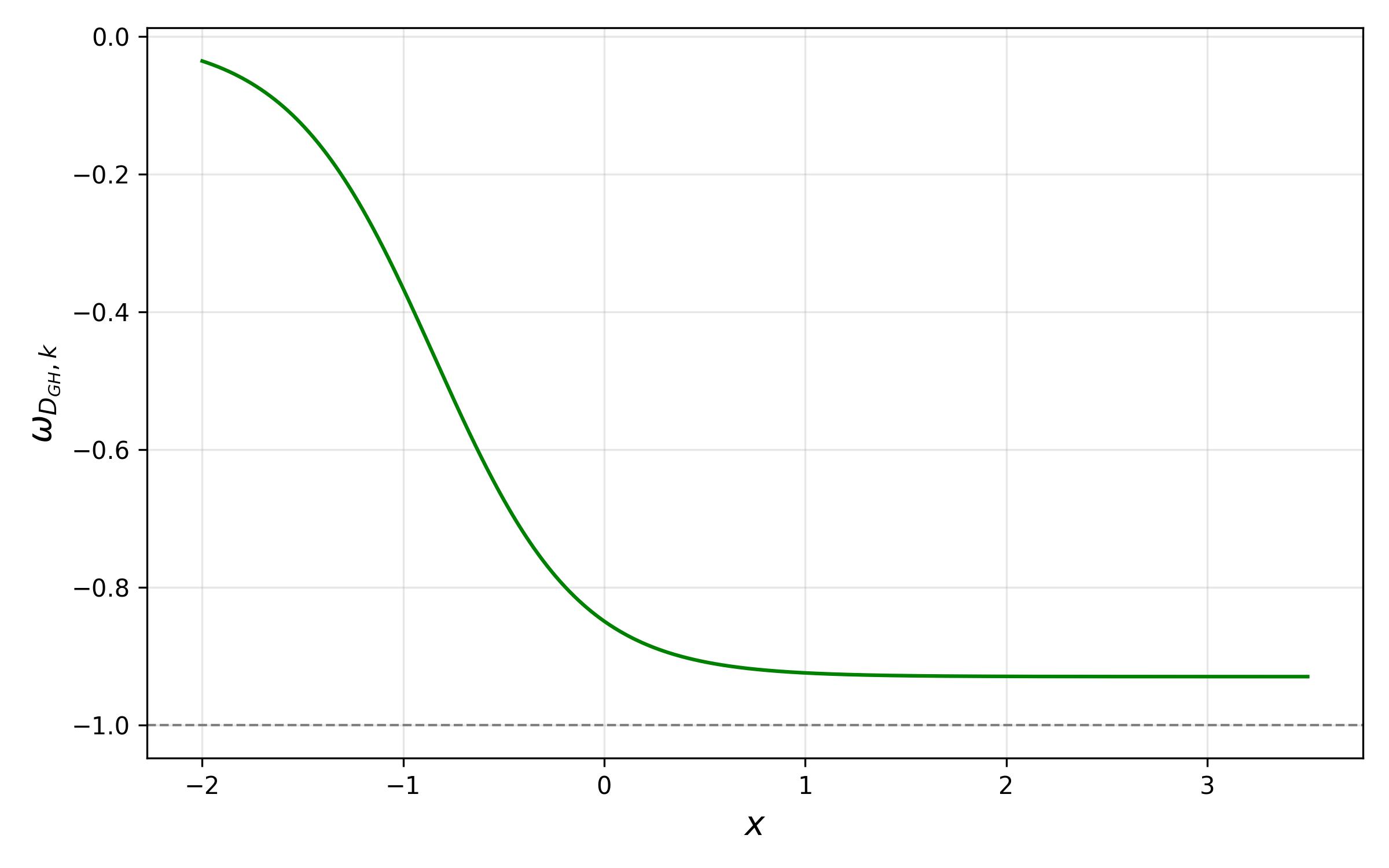}
        \caption{Plot of $\omega_{D_{GH},k}(x)$ for $c^2=0.46$.}
        \label{EoS2}
    \end{subfigure}\\[0.5cm]
    \begin{subfigure}{0.8\textwidth}
        \includegraphics[width=0.6\textwidth]{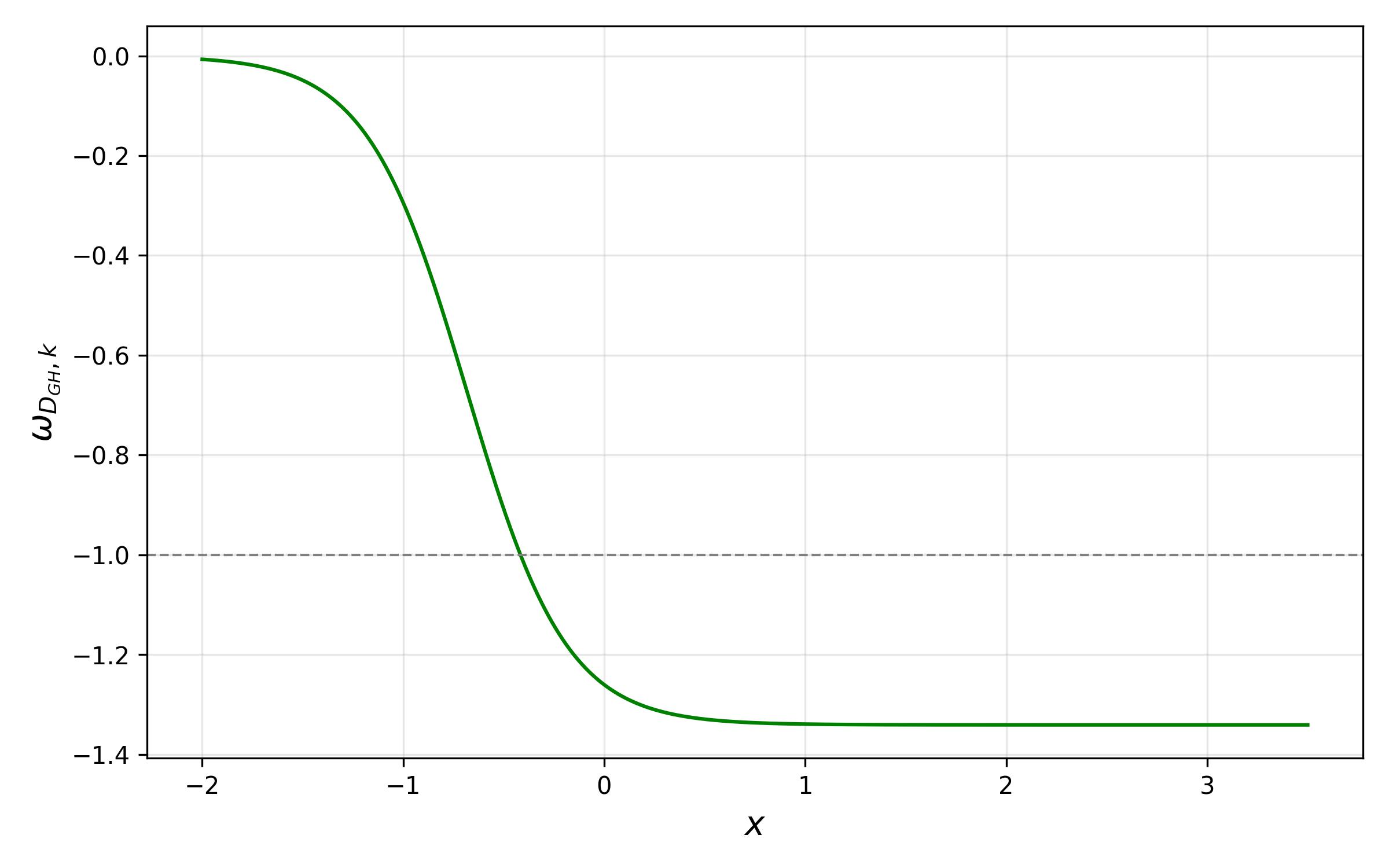}
        \caption{Plot of $\omega_{D_{GH},k}(x)$ for $c=0.579$.}
        \label{EoS2-2}
    \end{subfigure}\\[0.5cm]
    \begin{subfigure}{0.8\textwidth}
        \includegraphics[width=0.6\textwidth]{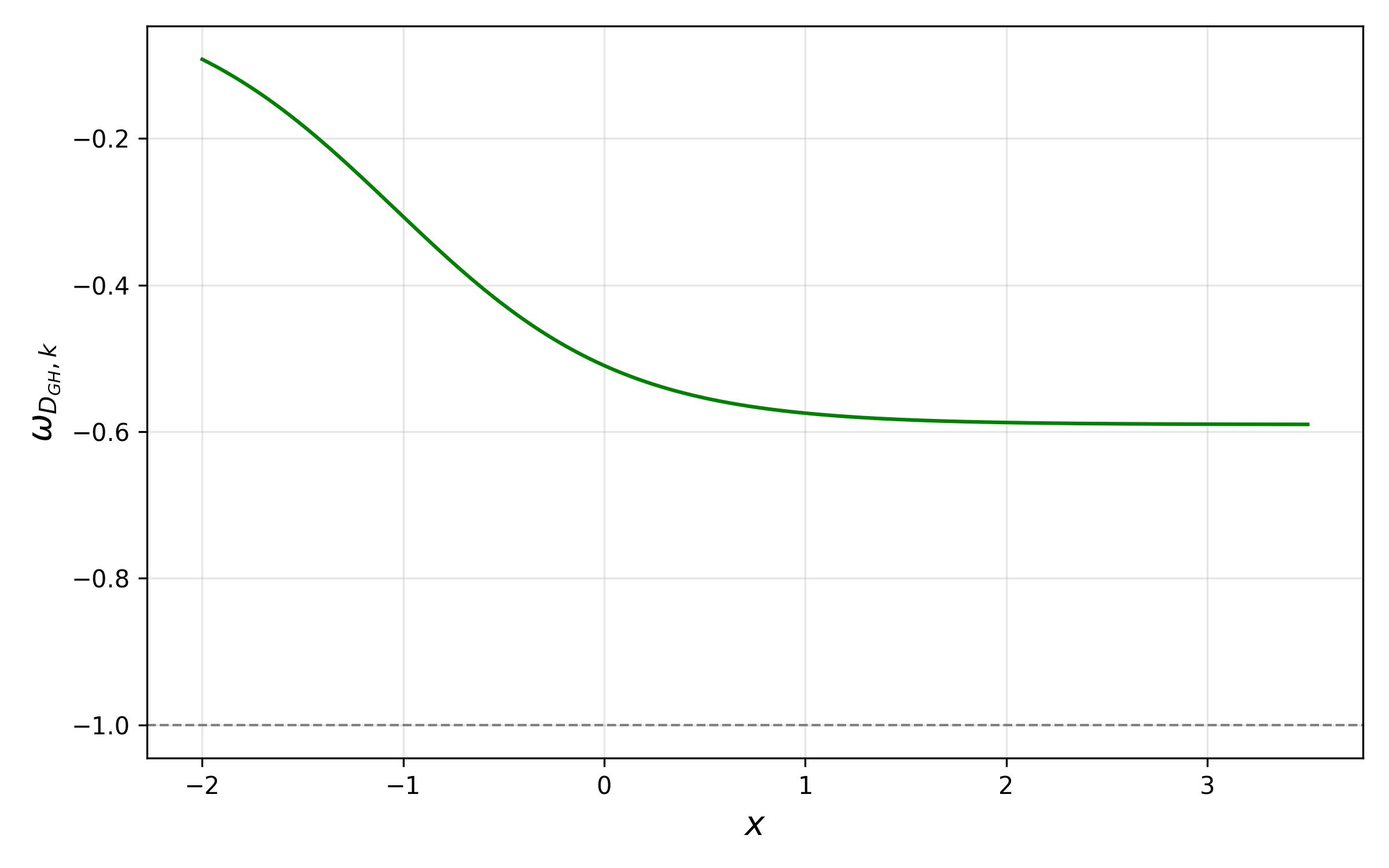}
        \caption{Plot of $\omega_{D_{GH},k}(x)$ for $c=0.815$.}
        \label{EoS2-3}
    \end{subfigure}
    \caption{Comparison of the three plots of $\omega_{D_{GH},k}(x)$.}
    \label{fig:EoS_all2}
\end{figure}

In Figs. (\ref{EoS2}), (\ref{EoS2-2}) and (\ref{EoS2-3})  we plot the expression of $\omega_{D_{GH},k}(x) $ obtained in Eq. (\ref{carolina61}) for $c^2=0.46$, $c=0.579$ and $c=0.815$, respectively. In order to obtain the Figures for the curvature cases, we have considered $\Omega_{k0}=0.001$

We obtain the present day values (i.e. the values corresponding to $x=0$) for the EoS for the three different cases: 
\begin{itemize}
    \item $\omega_{D_{GH},k,1}(0) \approx -0.849$
    \item $\omega_{D_{GH},k,2}(0) \approx -1.260$
    \item $\omega_{D_{GH},k,3}(0) \approx -0.509$
\end{itemize}

We now want to calculate the expression of the deceleration parameter $q_{GH,k}(x)$.\\
For the model we are considering, we choose the following expression in order to derive $q_{GH,k}(x)$: 
\begin{eqnarray}
q_{GH,k}(x) &=&  -1 - \frac{1}{2 h_{GH,k}^2(x)} \frac{d h_{GH,k}^2(x)}{dx}.
\end{eqnarray}
Using the general expression of $h_{GH,k}^2$ given in Eq. (\ref{cledison4}) along with the expression of $\frac{d h_{GH,k}^2(x)}{dx}$ derived in Eq. (\ref{renato1}), we can write:
\begin{eqnarray}
q_{GH,k}(x) &=&  -1 + \frac{1}{2} \cdot \left\{  3\Omega_{m0} e^{-3x} -2\Omega_{k0} e^{-2x} -3
\left[  \frac{ c^2(\epsilon - 2)}{2 + c^2(\epsilon - 2)}\right]\Omega_{m0} e^{-3x} 
+2\left[\frac{c^2(\epsilon - 1)}{1-c^2}\right]\Omega_{k0}e^{-2x}\right. \nonumber \\
&&\left.
-2(1-\alpha)\left[  1 - \frac{2\Omega_{m0}}{2 + c^2(\epsilon - 2)} 
+\left(\frac{1-\epsilon c^2}{1-c^2}\right) \Omega_{k0}  \right] \, e^{-2(1-\alpha)x}  \right\}\times \left\{ \Omega_{m0} e^{-3x} -\Omega_{k0} e^{-2x}
\right. \nonumber \\
&&\left. -
\left[  \frac{ c^2(\epsilon - 2)}{2 + c^2(\epsilon - 2)}\right]\Omega_{m0} e^{-3x} +\left[\frac{c^2(\epsilon - 1)}{1-c^2}\right]\Omega_{k0}e^{-2x}\right. \nonumber \\
&&\left.
+\left[  1 - \frac{2\Omega_{m0}}{2 + c^2(\epsilon - 2)} 
+\left(\frac{1-\epsilon c^2}{1-c^2}\right) \Omega_{k0}  \right] \, e^{-2(1-\alpha)x}  \right\}^{-1}.\label{carolina71}
\end{eqnarray}

\begin{figure}[htbp]
    \centering
    \begin{subfigure}{0.8\textwidth}
        \includegraphics[width=0.6\textwidth]{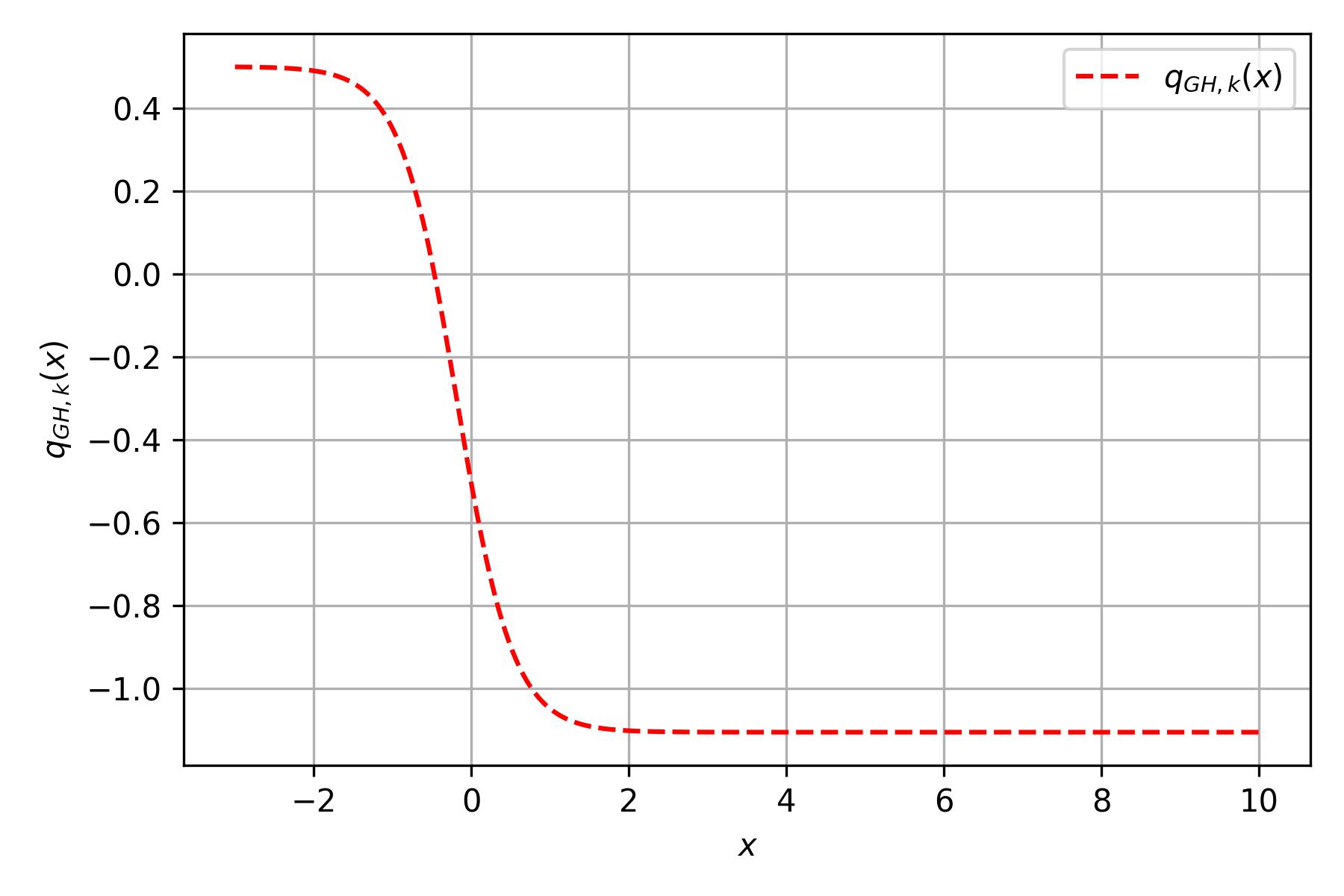}
        \caption{Plot of $q_{GH,k}(x)$ for $c^2=0.46$.}
        \label{q2}
    \end{subfigure}\\[0.5cm]
    \begin{subfigure}{0.8\textwidth}
        \includegraphics[width=0.6\textwidth]{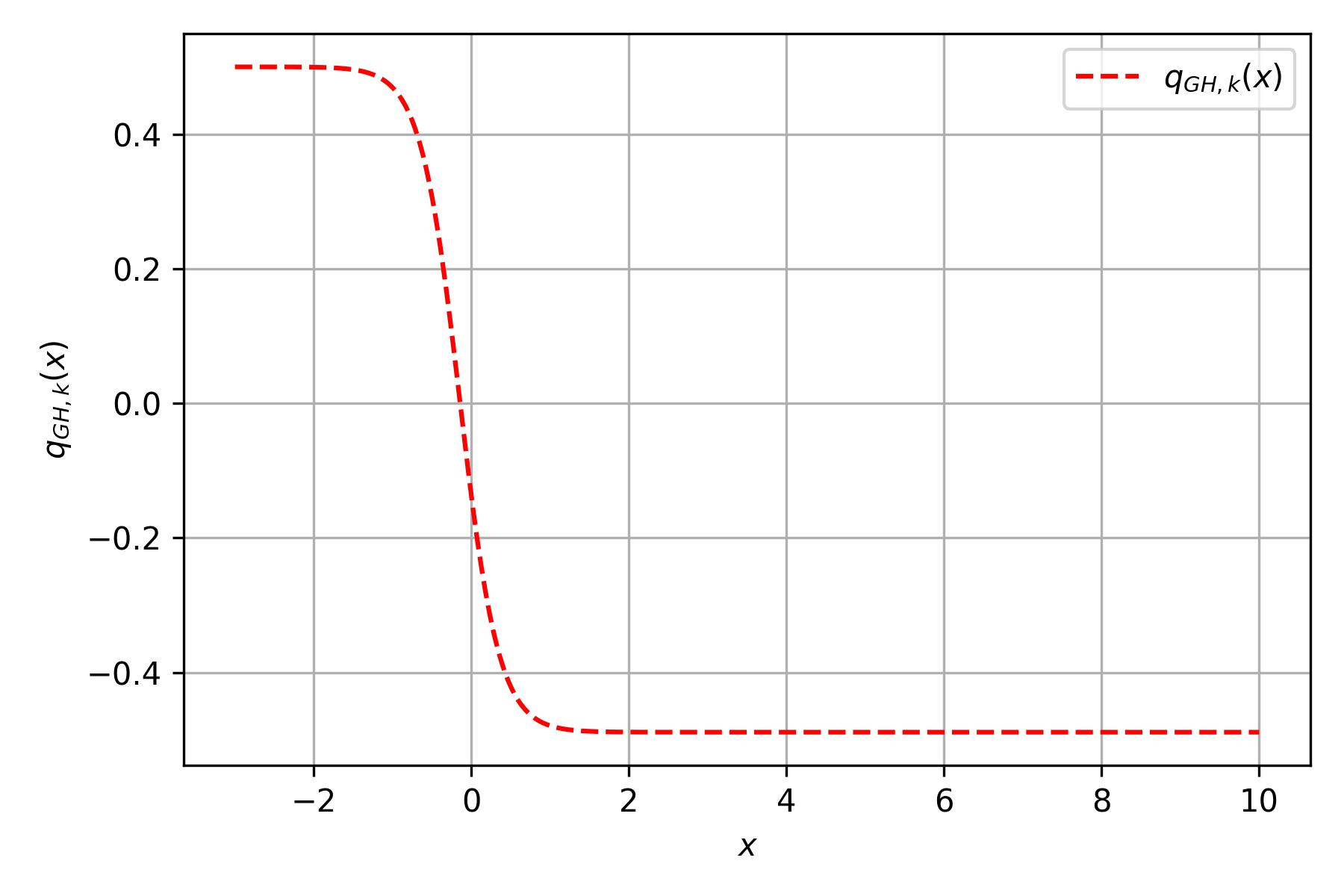}
        \caption{Plot of $q_{GH,k}(x)$ for $c=0.579$.}
        \label{q2-2}
    \end{subfigure}\\[0.5cm]
    \begin{subfigure}{0.8\textwidth}
        \includegraphics[width=0.6\textwidth]{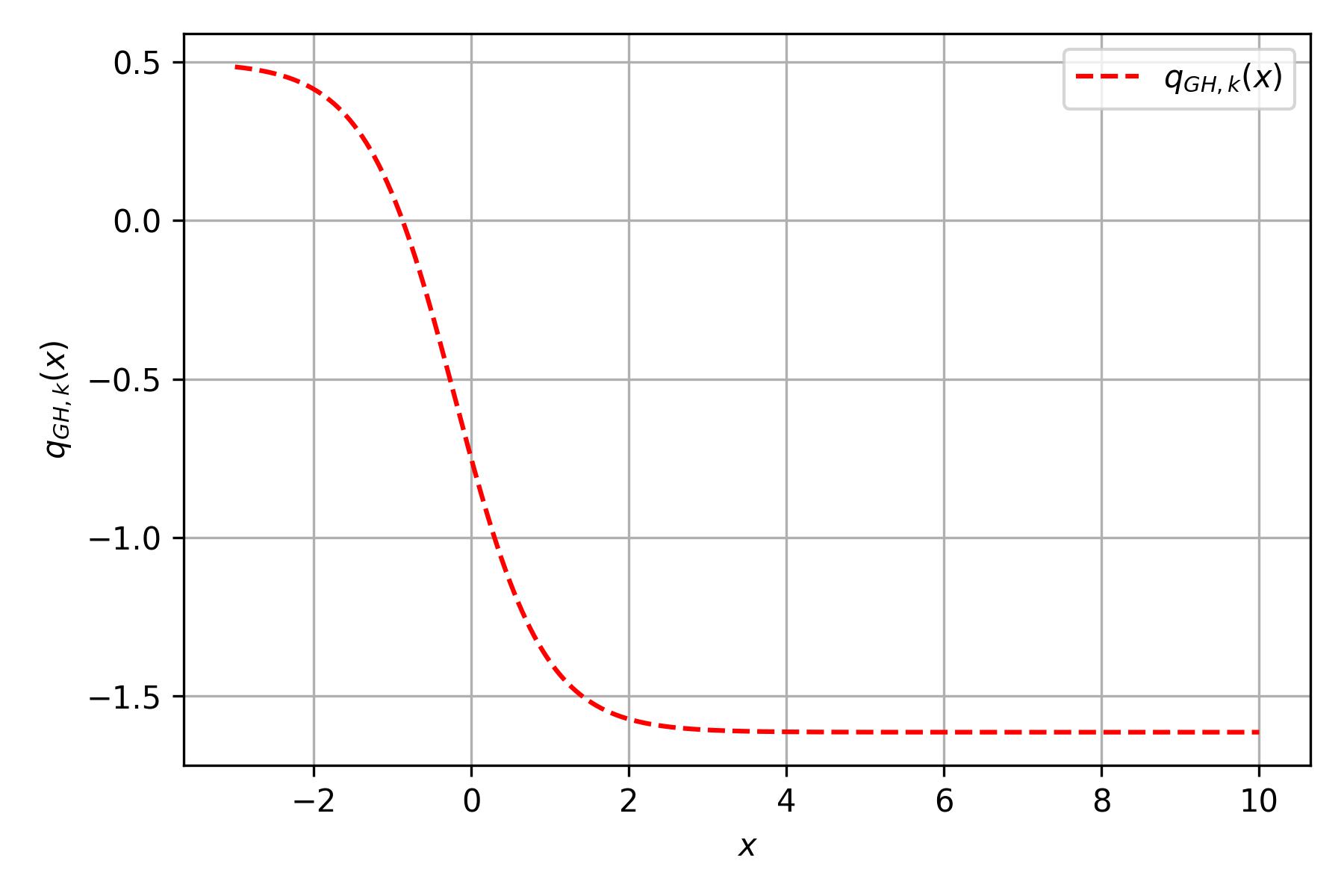}
        \caption{Plot of $q_{GH,k}(x)$ for $c=0.815$.}
        \label{q2-3}
    \end{subfigure}
    \caption{Comparison of the three cases for the deceleration parameter $q_{GH,k}(x)$.}
    \label{fig:q_all2}
\end{figure}

In Figs. (\ref{q2}), (\ref{q2-2}) and (\ref{q2-3}) we plot the expression of $q_{GH,k}(x) $ obtained in Eq. (\ref{carolina71}) for $c^2=0.46$, $c=0.579$ and $c=0.815$, respectively.

We obtain the present day values (i.e. the values corresponding to $x=0$) for $q_{GH,k}(x) $ for the three different cases: 
\begin{itemize}
    \item $q_{GH,k}(0) \approx -0.505$
    \item $q_{GH,k}(0) \approx-0.137$
    \item $q_{GH,k}(0) \approx-0.752$
\end{itemize}
All the three cases indicate an accelerated expansions of the Universe as predicted by recent observations.

Moreover, the following relation between $q_{GH,k}(x)$ and $\Omega_{D_{GH},k}(x)$ is valid:
\begin{equation}
q_{GH,k}(x) =\frac{1}{\epsilon} + H_0^2 \Omega_{k0} e^{-2x}- \frac{\Omega_{D_{GH},k}(x)}{\epsilon c^2} .
\end{equation}

\subsection{Case 2: $g\left(\frac{H^2}{R}\right)=1-\eta\left(1-\frac{H^2}{R}\right)$}
We now make the same calculations considering the curvature term for the second model of this paper.\\
We begin by deriving the main expressions for the Hubble parameter 
$H$ and the dark energy density $\rho_D$ as functions of redshift $z$ , and subsequently in terms of the variable $x$.\\
The fractional energy density of generalized Ricci dark energy is
\begin{eqnarray}
\Omega_{D_{GR},k}&=&c^{2} \left[ (2-\eta) + H_0^2(1-\eta)\Omega_{k0}(1+z)^2+ \frac{(1-\eta)}{2h_{GR,k}^{2}} \frac{d h_{GR,k}^{2}}{d x} \right].\label{cledison5}
\end{eqnarray}
Using the expression of $\Omega_{D_{GR},k}$ given in Eq. (\ref{cledison5}), the Friedmann equation obtained in Eq. (\ref{genfri}) can be rewritten as a differential equation of $H^2_{GR,k}(z)$ in the following way:
\begin{eqnarray}
\frac{H^2_{GR,k}(z)}{H_0^2}\left[\frac{c^2(1-\eta)(1+z)}{2}\frac{d\ln
H^2_{GR,k}(z)}{dz}\right. \nonumber \\
\left.+1-c^2(2-\eta)\right]&=&\Omega_{m0}(1+z)^3\nonumber \\
&&-\left[1- c^2(1-\eta)\right]\Omega_{k0}(1+z)^2,
\end{eqnarray}
which is equivalent to:
\begin{eqnarray}
\frac{H^2_{GR,k}(z)}{H_0^2}\left[\frac{c^2(1-\eta)(1+z)}{2H^2_{GR,k}(z)}\frac{d
H^2_{GR,k}(z)}{dz}\right. \nonumber \\
\left.+1-c^2(2-\eta)\right]&=&\Omega_{m0}(1+z)^3-\left[1- c^2(1-\eta)\right]\Omega_{k0}(1+z)^2 .\label{borgai}
\end{eqnarray}
Eq. (\ref{borgai}) has the following general solution:
\begin{eqnarray}
h^2_{GR,k}(z) &=&  \Omega_{m0}(1+z)^{3}  -\Omega_{k0}(1+z)^2 \nonumber\\
&& + \left[ \frac{c^2 (1 + \eta)}{2 - c^2 (1 + \eta)} \right]\Omega_{m0}(1+z)^{3}\nonumber \\
 && + \left[ \frac{1 + c^2 (1-\eta)}{1 - c^2} + 1 \right]\Omega_{k0}(1+z)^2 \nonumber\\
 &&+ \left \{1 + \frac{2\Omega_{m0}}{
c^2 (1 + \eta)-2} - \frac{ \left[ 1 + c^2 (1-\eta) \right]\Omega_{k0}}{1 - c^2}
\right\}
(1+z)^{\lambda  },\label{cledison6}
\end{eqnarray}
where $\lambda$ is defined as follows:
\begin{eqnarray}
\lambda = \frac{2\left[1 + c^2(\eta - 2)\right]}{c^2(\eta - 1)}.
\end{eqnarray}
The form obtained in Eq.~(\ref{cledison6}) is particularly convenient, as it allows 
$ h_{GR,k}^2(z) $ to be written as the sum of three separate contributions: 
one associated with dark matter, one with spatial curvature, and one with dark energy. 
Specifically, the term $ \Omega_{m0} (1+z)^3 $ represents the dark matter component,  while $ \Omega_{k0} (1+z)^2 $ corresponds to the curvature contribution.  Based on this decomposition, we can write the following expression for  $ \rho_{D_{GH},k}(z) $:

\begin{eqnarray}
\rho_{D_{GH},k}(z)= 
&&  \left[ \frac{c^2 (1 + \eta)}{2 - c^2 (1 + \eta)} \right]\Omega_{m0}(1+z)^{3}\nonumber \\
 && + \left[ \frac{1 + c^2 (1-\eta)}{1 - c^2} + 1 \right]\Omega_{k0}(1+z)^2 \nonumber\\
 &&+ \left \{1 + \frac{2\Omega_{m0}}{
c^2 (1 + \eta)-2} - \frac{ \left[ 1 + c^2 (1-\eta) \right]\Omega_{k0}}{1 - c^2}
\right\}
(1+z)^{\lambda  }.
\end{eqnarray}
We can now write $h^2_{GH,k}$ and $\rho_{D_{GH},k}$ as functions of the variable $x=\ln a$:
\begin{eqnarray}
h^2_{GR,k}(x) &=&  \Omega_{m0}e^{-3x}  -\Omega_{k0}e^{-2x}  \nonumber\\
&& + \left[ \frac{c^2 (1 + \eta)}{2 - c^2 (1 + \eta)} \right]\Omega_{m0}e^{-3x} \nonumber \\
 && + \left[ \frac{1 + c^2 (1-\eta)}{1 - c^2} + 1 \right]\Omega_{k0}e^{-2x}  \nonumber\\
 &&+ \left \{1 + \frac{2\Omega_{m0}}{
c^2 (1 + \eta)-2} - \frac{ \left[ 1 + c^2 (1-\eta) \right]\Omega_{k0}}{1 - c^2}
\right\}e^{-\lambda x}, \label{cledison7} \\
\rho_{D_{GR},k}(x)
&=& \left[ \frac{c^2 (1 + \eta)}{2 - c^2 (1 + \eta)} \right]\Omega_{m0}e^{-3x} \nonumber \\
 && + \left[ \frac{1 + c^2 (1-\eta)}{1 - c^2} + 1 \right]\Omega_{k0}e^{-2x}  \nonumber\\
 &&+ \left \{1 + \frac{2\Omega_{m0}}{
c^2 (1 + \eta)-2} - \frac{ \left[ 1 + c^2 (1-\eta) \right]\Omega_{k0}}{1 - c^2}
\right\}e^{-\lambda x} .\label{cledison12}
\end{eqnarray}

We now want to calculate the expressions of the fractional energy density of DE $\Omega_{D_{GR},k}$, the evolutionary form of the fractional energy density of DE $\Omega'_{D_{GR},k}$, the pressure $p_{D_{GR},k}$, the EoS parameter $\omega_{D_{GR},k}$ and the deceleration parameter $q_{GR,k}$ for this case.\\
For the fractional energy density we start from Eq. (\ref{cledison5}).\\
Using the expression of $h_{GR,k}^{2}(x)$ obtained in Eq. (\ref{cledison7}), we obtain:
\begin{eqnarray}
    \frac{dh^2_{GR,k}(x)}{dx} &=&  -3\Omega_{m0}e^{-3x}  +2\Omega_{k0}e^{-2x}  \nonumber\\
&& -3 \left[ \frac{c^2 (1 + \eta)}{2 - c^2 (1 + \eta)} \right]\Omega_{m0}e^{-3x} \nonumber \\
 && -2 \left[ \frac{1 + c^2 (1-\eta)}{1 - c^2} + 1 \right]\Omega_{k0}e^{-2x}  \nonumber\\
 &&-\lambda \left \{1 + \frac{2\Omega_{m0}}{
c^2 (1 + \eta)-2} - \frac{ \left[ 1 + c^2 (1-\eta) \right]\Omega_{k0}}{1 - c^2}
\right\}e^{-\lambda x}. \label{cledison10}
\end{eqnarray}
Therefore, the expression of the fractional energy density of DE is given by:
\begin{eqnarray}
\Omega_{D_{GR},k}(x)&=&c^2 \left[ (2-\eta) + H_0^2(1-\eta)\Omega_{k0}   \right]\nonumber \\
&&-\frac{c^2(1-\eta)}{2} \cdot \left\{ 3\Omega_{m0}e^{-3x}  -2\Omega_{k0}e^{-2x} \right. \nonumber\\
&&\left. +3 \left[ \frac{c^2 (1 + \eta)}{2 - c^2 (1 + \eta)} \right]\Omega_{m0}e^{-3x} \right.\nonumber \\
 &&\left. +2 \left[ \frac{1 + c^2 (1-\eta)}{1 - c^2} + 1 \right]\Omega_{k0}e^{-2x} \right. \nonumber\\
 &&\left.+\lambda \left \{1 + \frac{2\Omega_{m0}}{
c^2 (1 + \eta)-2} - \frac{ \left[ 1 + c^2 (1-\eta) \right]\Omega_{k0}}{1 - c^2}
\right\}e^{-\lambda x}  \right\} \times \nonumber \\
&&\left\{ \Omega_{m0}e^{-3x}  -\Omega_{k0}e^{-2x}\right.  \nonumber\\
&&\left. + \left[ \frac{c^2 (1 + \eta)}{2 - c^2 (1 + \eta)} \right]\Omega_{m0}e^{-3x} \right.\nonumber \\
 &&\left. + \left[ \frac{1 + c^2 (1-\eta)}{1 - c^2} + 1 \right]\Omega_{k0}e^{-2x} \right. \nonumber\\
 &&\left.+ \left \{1 + \frac{2\Omega_{m0}}{
c^2 (1 + \eta)-2} - \frac{ \left[ 1 + c^2 (1-\eta) \right]\Omega_{k0}}{1 - c^2}
\right\}e^{-\lambda x}  \right\}^{-1}. \label{cledison11}
\end{eqnarray}
We now want to calculate the evolutionary form of the fractional energy density of DE.\\
If we put $h^2_{GR,k}(x) \equiv A_4$, we have that Eq. (\ref{cledison11}) can be written as
\begin{eqnarray}
\Omega_{D_{GR},k}(x)&=&c^2 \left[ (2-\eta) + H_0^2(1-\eta)\Omega_{k0}   \right] +  \frac{c^2(1-\eta)}{2} \left( \frac{A_4'}{A_4}  \right). \label{renato-2!!!}
\end{eqnarray}
Differentiating Eq. (\ref{renato-2!!!}) with respect to $x$, we can write:
\begin{eqnarray}
\Omega'_{D_{GR},k}(x)&=& \frac{c^2(1-\eta)}{2}\left[ \frac{A_4''}{A_4}-\left( \frac{A_4'}{A_4}  \right)^2 \right],\label{renato-2prime}
\end{eqnarray}
where $A_4''= \frac{d^2 h_{GR}^2(x)}{dx^2}$. \\
Using the general expression of $ h_{GR,k}^2(x)$ given in Eq. (\ref{cledison7}) or equivalently the expression of $\frac{d h_{GR,k}^2(x)}{dx}$ obtained in Eq. (\ref{cledison10}), we can write:

\begin{eqnarray}
    \frac{d^2h^2_{GR,k}(x)}{dx^2} &=&  9\Omega_{m0}e^{-3x}  -4\Omega_{k0}e^{-2x}  \nonumber\\
&& +9 \left[ \frac{c^2 (1 + \eta)}{2 - c^2 (1 + \eta)} \right]\Omega_{m0}e^{-3x} \nonumber \\
 && +4 \left[ \frac{1 + c^2 (1-\eta)}{1 - c^2} + 1 \right]\Omega_{k0}e^{-2x}  \nonumber\\
 &&+\lambda^2 \left \{1 + \frac{2\Omega_{m0}}{
c^2 (1 + \eta)-2} - \frac{ \left[ 1 + c^2 (1-\eta) \right]\Omega_{k0}}{1 - c^2}
\right\}e^{-\lambda x}.
\end{eqnarray}
Therefore, we obtain that evolutionary form of the fractional energy density can be written as:
\begin{eqnarray}
\Omega_{D_{GR},k}'(x)&=&
\frac{c^2(1-\eta)}{2} \cdot \left\{  9\Omega_{m0}e^{-3x}  -4\Omega_{k0}e^{-2x} \right. \nonumber\\
&&\left. +9 \left[ \frac{c^2 (1 + \eta)}{2 - c^2 (1 + \eta)} \right]\Omega_{m0}e^{-3x}\right. \nonumber \\
 &&\left. +4 \left[ \frac{1 + c^2 (1-\eta)}{1 - c^2} + 1 \right]\Omega_{k0}e^{-2x} \right. \nonumber\\
 &&\left.+\lambda^2 \left \{1 + \frac{2\Omega_{m0}}{
c^2 (1 + \eta)-2} - \frac{ \left[ 1 + c^2 (1-\eta) \right]\Omega_{k0}}{1 - c^2}
\right\}e^{-\lambda x}    \right\} \times \nonumber \\
&&\left\{ \Omega_{m0}e^{-3x}  -\Omega_{k0}e^{-2x}\right.  \nonumber\\
&&\left. + \left[ \frac{c^2 (1 + \eta)}{2 - c^2 (1 + \eta)} \right]\Omega_{m0}e^{-3x} \right.\nonumber \\
 &&\left. + \left[ \frac{1 + c^2 (1-\eta)}{1 - c^2} + 1 \right]\Omega_{k0}e^{-2x} \right. \nonumber\\
 &&\left.+ \left \{1 + \frac{2\Omega_{m0}}{
c^2 (1 + \eta)-2} - \frac{ \left[ 1 + c^2 (1-\eta) \right]\Omega_{k0}}{1 - c^2}
\right\}e^{-\lambda x}  \right\}^{-1}\nonumber \\
&&-\frac{c^2(1-\eta)}{2} \cdot \left\{ 3\Omega_{m0}e^{-3x}  -2\Omega_{k0}e^{-2x} \right. \nonumber\\
&&\left. +3 \left[ \frac{c^2 (1 + \eta)}{2 - c^2 (1 + \eta)} \right]\Omega_{m0}e^{-3x} \right.\nonumber \\
 &&\left. +2 \left[ \frac{1 + c^2 (1-\eta)}{1 - c^2} + 1 \right]\Omega_{k0}e^{-2x} \right. \nonumber\\
 &&\left.+\lambda \left \{1 + \frac{2\Omega_{m0}}{
c^2 (1 + \eta)-2} - \frac{ \left[ 1 + c^2 (1-\eta) \right]\Omega_{k0}}{1 - c^2}
\right\}e^{-\lambda x}  \right\}^2 \times \nonumber \\
&&\left\{ \Omega_{m0}e^{-3x}  -\Omega_{k0}e^{-2x}\right.  \nonumber\\
&&\left. + \left[ \frac{c^2 (1 + \eta)}{2 - c^2 (1 + \eta)} \right]\Omega_{m0}e^{-3x} \right.\nonumber \\
 &&\left. + \left[ \frac{1 + c^2 (1-\eta)}{1 - c^2} + 1 \right]\Omega_{k0}e^{-2x} \right. \nonumber\\
 &&\left.+ \left \{1 + \frac{2\Omega_{m0}}{
c^2 (1 + \eta)-2} - \frac{ \left[ 1 + c^2 (1-\eta) \right]\Omega_{k0}}{1 - c^2}
\right\}e^{-\lambda x}  \right\}^{-2}.
\end{eqnarray}

We now want to find the final expression for the pressure $p_{D_{GR},k}(x) $. \\
The general expression of the pressure $ p_{D_{GR},k}(x)  $ is given by:
\begin{eqnarray}
\label{eq:conserve}
p_{D_{GR},k}(x) 
           &=& -\rho_{D_{GR},k}(x)  - \frac{\rho'_{D_{GR},k}(x) }{3}.
\end{eqnarray}
 Using the general expression for $ \rho_{D_{GR},k}(x) $ previously derived in Eq.~(\ref{cledison12}), we obtain the following result for $ \rho'_{D_{GR},k}(x) $:
\begin{eqnarray}
\rho_{D_{GR},k}'(x) 
&=& -3 \left[ \frac{c^2 (1 + \eta)}{2 - c^2 (1 + \eta)} \right]\Omega_{m0}e^{-3x} \nonumber \\
 && -2 \left[ \frac{1 + c^2 (1-\eta)}{1 - c^2} + 1 \right]\Omega_{k0}e^{-2x}  \nonumber\\
 &&-\lambda \left \{1 + \frac{2\Omega_{m0}}{
c^2 (1 + \eta)-2} - \frac{ \left[ 1 + c^2 (1-\eta) \right]\Omega_{k0}}{1 - c^2}
\right\}e^{-\lambda x} .\label{cledison13}
\end{eqnarray}
Therefore, we obtain the following expression for the pressure:
\begin{eqnarray}
p_{D_{GR},k}(x)   &=& 
 -\frac{1}{3} \left[ \frac{1 + c^2 (1-\eta)}{1 - c^2} + 1 \right]\Omega_{k0}e^{-2x}  \nonumber\\
 &&- \frac{2\lambda}{3}\left \{1 + \frac{2\Omega_{m0}}{
c^2 (1 + \eta)-2} - \frac{ \left[ 1 + c^2 (1-\eta) \right]\Omega_{k0}}{1 - c^2}
\right\}e^{-\lambda x} .
\end{eqnarray}

We now want to obtain the EoS parameter $\omega_{D_{GH},k}$. \\
The general relation for this case is given by:
\begin{eqnarray}
\omega_{D_{GR},k}(x)  = -1 - \frac{\rho_{D_{GR},k}'(x) }{3 \rho_{D_{GR},k}(x) } \label{eosnongen}.
\end{eqnarray}
Using the expressions of $\rho_{D_{GR},k}$ and $\rho'_{D_{GR},k}$ obtained in Eqs. (\ref{cledison12}) and (\ref{cledison13}), we obtain the following expression for $\omega_{D_{GR},k} (x) $:
\begin{eqnarray}
\omega_{D_{GR},k} (x) 
&=&- 1 +\frac{1}{3}\cdot \left\{  3 \left[ \frac{c^2 (1 + \eta)}{2 - c^2 (1 + \eta)} \right]\Omega_{m0}e^{-3x}\right. \nonumber \\
 && \left.+2 \left[ \frac{1 + c^2 (1-\eta)}{1 - c^2} + 1 \right]\Omega_{k0}e^{-2x}\right.  \nonumber\\
 &&\left.+\left \{1 + \frac{2\Omega_{m0}}{
c^2 (1 + \eta)-2} - \frac{ \left[ 1 + c^2 (1-\eta) \right]\Omega_{k0}}{1 - c^2}\right\}\lambda e^{-\lambda x}  \right\}\times \nonumber \\
&&\left\{ \left[ \frac{c^2 (1 + \eta)}{2 - c^2 (1 + \eta)} \right]\Omega_{m0}e^{-3x}\right. \nonumber \\
 &&\left. + \left[ \frac{1 + c^2 (1-\eta)}{1 - c^2} + 1 \right]\Omega_{k0}e^{-2x}\right.  \nonumber\\
 &&\left.+ \left \{1 + \frac{2\Omega_{m0}}{
c^2 (1 + \eta)-2} - \frac{ \left[ 1 + c^2 (1-\eta) \right]\Omega_{k0}}{1 - c^2}
\right\}e^{-\lambda x}    \right\}^{-1}.
\end{eqnarray}

We now want to calculate the expression of the deceleration parameter $q_{GR,k}(x) $.\\
For the model we are considering, we choose the following expression: 
\begin{eqnarray}
q_{GR,k}(x)  &=&  -1 - \frac{1}{2 h_{GR,k}^2(x) } \frac{d h_{GR,k}^2(x)}{dx}. \label{deceleration}
\end{eqnarray}
Using the general expression of $h_{GR,k}^2$ given in Eq. (\ref{cledison7}) along with the expression of $\frac{d h_{GR,k}^2(x)}{dx}$ derived in Eq. (\ref{cledison10}), we can write:
\begin{eqnarray}
q_{GR,k} (x) &=&  -1 + \frac{1}{2}\cdot \left\{3\Omega_{m0}e^{-3x} -2\Omega_{k0}e^{-2x}\right.  \nonumber\\
&& \left.+3 \left[ \frac{c^2 (1 + \eta)}{2 - c^2 (1 + \eta)} \right]\Omega_{m0}e^{-3x}\right. \nonumber \\
 &&\left.+2 \left[ \frac{1 + c^2 (1-\eta)}{1 - c^2} + 1 \right]\Omega_{k0}e^{-2x} \right. \nonumber\\
 &&\left. \left \{1 + \frac{2\Omega_{m0}}{
c^2 (1 + \eta)-2} - \frac{ \left[ 1 + c^2 (1-\eta) \right]\Omega_{k0}}{1 - c^2}
\right\}\lambda e^{-\lambda x}   \right\} \times \nonumber \\
&& \left\{  \Omega_{m0}e^{-3x}  -\Omega_{k0}e^{-2x} \right.  \nonumber\\
&&\left. + \left[ \frac{c^2 (1 + \eta)}{2 - c^2 (1 + \eta)} \right]\Omega_{m0}e^{-3x}\right.  \nonumber \\
 && \left.+ \left[ \frac{1 + c^2 (1-\eta)}{1 - c^2} + 1 \right]\Omega_{k0}e^{-2x} \right.  \nonumber\\
 &&\left.+ \left \{1 + \frac{2\Omega_{m0}}{
c^2 (1 + \eta)-2} - \frac{ \left[ 1 + c^2 (1-\eta) \right]\Omega_{k0}}{1 - c^2}
\right\}e^{-\lambda x}    \right\}^{-1}.
\end{eqnarray}

Moreover, we have that the following relation is valid:
\begin{equation}
q_{GR,k}(x) =\frac{1}{1-\eta} + H_0^2 \Omega_{k0} e^{-2x}- \frac{\Omega_{D_{GH},k}(x)}{c^2(1-\eta)}. 
\end{equation}

\section{Interacting Case}
We now extend our work considering the presence of interaction between Dark Sectors.\\
The interaction between dark sectors refers to the possibility that dark matter and dark energy are not entirely independent, but may exchange energy or momentum. Such interactions are motivated by attempts to address the coincidence problem, which questions why the energy densities of dark matter and dark energy are of the same order today despite evolving differently over cosmic time. Models with interacting dark sectors modify the standard cosmological evolution and can lead to distinct observational signatures, such as deviations in the expansion history, structure growth, or cosmic microwave background anisotropies. These scenarios are actively explored as alternatives or extensions to the standard $\Lambda$CDM model.\\
When an interaction between the two Dark Sectors is present, the energy densities of dark energy (DE) and dark matter (DM) obey modified conservation laws. These equations can be written as
\begin{align}
\dot{\rho}_{D} + 3H \rho_{D} (1+\omega_{D}) &= -Q, \label{eq:DE_cons}\\
\dot{\rho}_{m} + 3H \rho_{m} &= Q, \label{eq:DM_cons}
\end{align}
where the quantity $Q$ encodes the energy exchange between the two components. In general, $Q$ may depend on several cosmological parameters, such as the Hubble parameter $H$, the deceleration parameter $q$, and the energy densities $\rho_{m}$ and $\rho_{D}$, i.e. $Q=Q(\rho_{m},\rho_{D},H,q)$. Various forms for this interaction have been investigated in the literature. In the present work, we adopt the following phenomenological choice:
\begin{equation}
Q = 3 d^{2} H \rho_{m},
\end{equation}
where $d^{2}$ is a dimensionless coupling constant that measures the strength of the interaction between DE and DM,   often referred to in the literature as transfer strength or interaction parameter~\cite{ref144d,ref145d,ref146d}. 
Analyses based on different cosmological datasets - including the Gold SNe~Ia sample, the CMB observations from WMAP, and BAO measurements from the Sloan Digital Sky Survey (SDSS) - indicate that this parameter  should take a small positive value. Such a result is consistent both with the requirements imposed by the  cosmic coincidence problem and with the constraints derived from the second law of thermodynamics~\cite{ref147d}. 
Furthermore, independent studies of CMB anisotropies and galaxy clusters point to the range  $0 < d^{2}< 0.025$~\cite{ref148d}. More generally, the interaction parameter is considered within the  interval $[0,1]$, with the special case $d^{2}=0$ corresponding to the standard non-interacting FLRW  scenario. It is also worth noting that several other possible interaction terms between  DM and DE have been explored in the literature.

Eqs. (\ref{eq:DE_cons}) and (\ref{eq:DM_cons}) can be also written as:
\begin{align}
\rho'_{D} + 3 \rho_{D} (1+\omega_{D}) &= -Q, \label{eq:DE_consprime}\\
\rho'_{m} + 3 \rho_{m} &= Q. \label{eq:DM_consprime}
\end{align}

\subsection{Case 1: $f\left(\frac{R}{H^2}\right)=1-\epsilon\left(1-\frac{R}{H^2}\right)$}
We start by studying the first model of this paper. \\
The fractional energy density is given by:
\begin{eqnarray}
\Omega_{D_{GH},I} &=& c^2 \left( 1 + \epsilon + \frac{\epsilon}{2h_{GH,I}^2} \frac{d h_{GH,I}^2}{dx} \right).\label{carolina4}
\end{eqnarray}

For the interacting case, we have:
\begin{eqnarray}
    \rho_{m,I}&=& \rho_{m0}e^{-3(1-d^2)x}\nonumber \\
    &=& \Omega_{m0}e^{-3(1-d^2)x},
\end{eqnarray}
which leads to
\begin{eqnarray}
    \rho'_{m,I} &=& -3(1-d^2)\rho_{m0}e^{-3(1-d^2)x}\nonumber \\
    &=&-3(1-d^2)\Omega_{m0}e^{-3(1-d^2)x}, \label{rhoI}
\end{eqnarray}
where we used the fact that $\rho_{m0}=\Omega_{m0}$ since we are using normalized quantities.

In this case, we obtain that the Friedmann equation given in Eq. (\ref{genfri}) can be rewritten in the following way:
\begin{equation}
H_{GH,I}^2(z)\left\{1-c^2\left[1+\epsilon-\frac{\epsilon(1+z)}{2}\frac{d \ln
H_{GH,I}^2(z)}{dz}\right]\right\}=H^2_0\Omega_{m0}(1+z)^{3(1-d^2)},
\end{equation}
which is equivalent to:
\begin{equation}
H_{GH,I}^2(z)\left\{1-c^2\left[1+\epsilon-\frac{\epsilon(1+z)}{2H_{GH,I}^2(z)}\frac{d 
H_{GH,I}^2(z)}{dz}\right]\right\}=H^2_0\Omega_{m0}(1+z)^{3(1-d^2)}.\label{carolina1}
\end{equation}

The general solution of Eq. (\ref{carolina1}) is given by:
\begin{eqnarray}
h_{GH,I}^2(z) &=& \frac{2  \Omega_{m0}(1+z)^{3(1 - d^2)}}{2 \left[1 - c^2 (1+\epsilon)\right] + 3 c^2 \epsilon (1 - d^2)} \, \nonumber \\
&& + \left\{1 - \frac{2 \Omega_{m0}}{2 \left[1 - c^2 (1+\epsilon)\right] + 3 c^2 \epsilon (1 - d^2)} \right\}\, (1+z)^{2(1-\alpha)}.
\end{eqnarray}
which can be written in the equivalent form:
\begin{eqnarray}
h_{GH,I}^2(z) &=&\Omega_{m0}(1+z)^{3(1 - d^2)} \nonumber  \\
&& +\left\{\frac{2  }{2 \left[1 - c^2 (1+\epsilon)\right] + 3 c^2 \epsilon (1 - d^2)}-1\right\}\Omega_{m0}(1+z)^{3(1 - d^2)}  \nonumber  \\
&& + \left\{1 - \frac{2 \Omega_{m0}}{2 \left[1 - c^2 (1+\epsilon)\right] + 3 c^2 \epsilon (1 - d^2)} \right\}\, (1+z)^{2(1-\alpha)}. \label{picu2}
\end{eqnarray}
The form obtained in Eq.~(\ref{picu2}) is particularly useful because it allows  $ h_{GH,I}^2(z) $ to be written as the sum of two distinct contributions:  one from the matter component and one from Dark Energy. \\
The term $ \Omega_{m0} (1+z)^{3(1 - d^2)} $ represents the Dark Matter component; consequently, the following expression for $ \rho_{D_{GH},I}(z) $ can be obtained:
\begin{eqnarray}
\rho_{D_{GH},I}(z) &=& \left\{\frac{2  }{2 \left[1 - c^2 (1+\epsilon)\right] + 3 c^2 \epsilon (1 - d^2)}-1\right\}\Omega_{m0}(1+z)^{3(1 - d^2)} \nonumber   \\
&& + \left\{1 - \frac{2 \Omega_{m0}}{2 \left[1 - c^2 (1+\epsilon)\right] + 3 c^2 \epsilon (1 - d^2)} \right\}\, (1+z)^{2(1-\alpha)}.
\end{eqnarray}

We can now write $h_{GH,I}^2$ and $\rho_{D_{GH},I}$ as functions of $x$:
\begin{eqnarray}
h_{GH,I}^2(x) &=&\Omega_{m0}e^{-3(1 - d^2)x} \nonumber  \\
&& +\left\{\frac{2  }{2 \left[1 - c^2 (1+\epsilon)\right] + 3 c^2 \epsilon (1 - d^2)}-1\right\}\Omega_{m0}e^{-3(1 - d^2)x} \, \nonumber   \\
&& + \left\{1 - \frac{2 \Omega_{m0}}{2 \left[1 - c^2 (1+\epsilon)\right] + 3 c^2 \epsilon (1 - d^2)} \right\}\, e^{-2(1-\alpha)x} ,\label{carolina2}\\
\rho_{D_{GH}.I}(x) &=&\left\{\frac{2  }{2 \left[1 - c^2 (1+\epsilon)\right] + 3 c^2 \epsilon (1 - d^2)}-1\right\}\Omega_{m0}e^{-3(1 - d^2)x} \, \nonumber   \\
&& + \left\{1 - \frac{2 \Omega_{m0}}{2 \left[1 - c^2 (1+\epsilon)\right] + 3 c^2 \epsilon (1 - d^2)} \right\}\, e^{-2(1-\alpha)x}.\label{carolina3}
\end{eqnarray}

For the fractional energy density of DE, we use the general relation obtained in Eq. (\ref{carolina4}). \\
Using the expression of $h^2_{GH,I}(x)$ derived in Eq. (\ref{carolina2}), we can write:
\begin{eqnarray}
\frac{d h_{GH,I}^2(x)}{dx}&=&-3(1 - d^2)\Omega_{m0}e^{-3(1 - d^2)x} \nonumber  \\
&& -3(1 - d^2)\left\{\frac{2  }{2 \left[1 - c^2 (1+\epsilon)\right] + 3 c^2 \epsilon (1 - d^2)}-1\right\}\Omega_{m0}e^{-3(1 - d^2)x} \, \nonumber   \\
&& -2(1-\alpha)\left\{1 - \frac{2 \Omega_{m0}}{2 \left[1 - c^2 (1+\epsilon)\right] + 3 c^2 \epsilon (1 - d^2)} \right\}\, e^{-2(1-\alpha)x}. \label{carolina5}
\end{eqnarray}
Therefore, the fractional energy density of DE can be written as:
\begin{eqnarray}
\Omega_{D_{GH},I}(x) &=& c^2(1+\epsilon )  + \left(\frac{\epsilon c^2}{2 }\right)\cdot\left\{ -3(1 - d^2)\Omega_{m0}e^{-3(1 - d^2)x} \nonumber \right.  \\
&& \left. -3(1 - d^2)\left\{\frac{2  }{2 \left[1 - c^2 (1+\epsilon)\right] + 3 c^2 \epsilon (1 - d^2)}-1\right\}\Omega_{m0}e^{-3(1 - d^2)x} \, \right.\nonumber   \\
&&\left. -2(1-\alpha)\left\{1 - \frac{2 \Omega_{m0}}{2 \left[1 - c^2 (1+\epsilon)\right] + 3 c^2 \epsilon (1 - d^2)} \right\}\, e^{-2(1-\alpha)x}   \right\}         \times \left\{\Omega_{m0}e^{-3(1 - d^2)x} \nonumber \right. \\
&&\left. +\left\{\frac{2  }{2 \left[1 - c^2 (1+\epsilon)\right] + 3 c^2 \epsilon (1 - d^2)}-1\right\}\Omega_{m0}e^{-3(1 - d^2)x} \, \nonumber \right.  \\
&& \left.+ \left\{1 - \frac{2 \Omega_{m0}}{2 \left[1 - c^2 (1+\epsilon)\right] + 3 c^2 \epsilon (1 - d^2)} \right\}\, e^{-2(1-\alpha)x}    \right\}^{-1}.
\end{eqnarray}

We now want to calculate the evolutionary form of the fractional energy density of DE.\\
If we put $h^2_{GH,I} \equiv A_5$, the fractional energy density of DE can be written as:
\begin{eqnarray}
\Omega_{D_{GH},I}(x) &=& c^2 \left( 1 + \epsilon\right) +\frac{\epsilon c^2}{2}\left( \frac{A'_5}{A_5}  \right),
\end{eqnarray}
which implies:
\begin{eqnarray}
\Omega_{D_{GH},I}'(x) 
&=&\frac{\epsilon c^2}{2}\left[ \frac{A_5''}{A_5}- \left(\frac{A'_5}{A_5}\right)^2\right],
\end{eqnarray}
where $A_5'' = \frac{d^2 h_{GH,I}^2(x)}{dx^2}$. \\
Using the general expression of $h^2_{GH.I}$ given in Eq. (\ref{carolina2}) or equivalently the expression of $\frac{d h_{GH,I}^2(x)}{dx}$ derived in Eq. (\ref{carolina5}), we obtain:
\begin{eqnarray}
    \frac{d^2 h_{GH,I}^2(x)}{dx^2} &=& 9(1 - d^2)^2\Omega_{m0}e^{-3(1 - d^2)x} \nonumber  \\
&& +9(1 - d^2)^2\left\{\frac{2  }{2 \left[1 - c^2 (1+\epsilon)\right] + 3 c^2 \epsilon (1 - d^2)}-1\right\}\Omega_{m0}e^{-3(1 - d^2)x} \, \nonumber   \\
&& +4(1-\alpha)^2\left\{1 - \frac{2 \Omega_{m0}}{2 \left[1 - c^2 (1+\epsilon)\right] + 3 c^2 \epsilon (1 - d^2)} \right\}\, e^{-2(1-\alpha)x}.
\end{eqnarray}
Then, the evolutionary form of the fractional energy density of DE can be written as:
\begin{eqnarray}
\Omega_{D_{GH},I}'(x) &=&  \left(\frac{\epsilon c^2}{2 }\right)\cdot\left\{ 9(1 - d^2)^2\Omega_{m0}e^{-3(1 - d^2)x} \nonumber \right.  \\
&& \left. +9(1 - d^2)^2\left\{\frac{2  }{2 \left[1 - c^2 (1+\epsilon)\right] + 3 c^2 \epsilon (1 - d^2)}-1\right\}\Omega_{m0}e^{-3(1 - d^2)x} \, \right.\nonumber   \\
&&\left. +4(1-\alpha)^2\left\{1 - \frac{2 \Omega_{m0}}{2 \left[1 - c^2 (1+\epsilon)\right] + 3 c^2 \epsilon (1 - d^2)} \right\}\, e^{-2(1-\alpha)x}   \right\}       \times \left\{\Omega_{m0}e^{-3(1 - d^2)x} \nonumber \right. \\
&&\left. +\left\{\frac{2  }{2 \left[1 - c^2 (1+\epsilon)\right] + 3 c^2 \epsilon (1 - d^2)}-1\right\}\Omega_{m0}e^{-3(1 - d^2)x} \, \nonumber \right.  \\
&& \left.+ \left\{1 - \frac{2 \Omega_{m0}}{2 \left[1 - c^2 (1+\epsilon)\right] + 3 c^2 \epsilon (1 - d^2)} \right\}\, e^{-2(1-\alpha)x}    \right\}^{-1}\nonumber \\
&&- \left(\frac{\epsilon c^2}{2 }\right)\cdot\left\{ -3(1 - d^2)\Omega_{m0}e^{-3(1 - d^2)x} \nonumber \right.  \\
&& \left. -3(1 - d^2)\left\{\frac{2  }{2 \left[1 - c^2 (1+\epsilon)\right] + 3 c^2 \epsilon (1 - d^2)}-1\right\}\Omega_{m0}e^{-3(1 - d^2)x} \, \right.\nonumber   \\
&&\left. -2(1-\alpha)\left\{1 - \frac{2 \Omega_{m0}}{2 \left[1 - c^2 (1+\epsilon)\right] + 3 c^2 \epsilon (1 - d^2)} \right\}\, e^{-2(1-\alpha)x}   \right\}^2       \times \left\{\Omega_{m0}e^{-3(1 - d^2)x} \nonumber \right. \\
&&\left. +\left\{\frac{2  }{2 \left[1 - c^2 (1+\epsilon)\right] + 3 c^2 \epsilon (1 - d^2)}-1\right\}\Omega_{m0}e^{-3(1 - d^2)x} \, \nonumber \right.  \\
&& \left.+ \left\{1 - \frac{2 \Omega_{m0}}{2 \left[1 - c^2 (1+\epsilon)\right] + 3 c^2 \epsilon (1 - d^2)} \right\}\, e^{-2(1-\alpha)x}    \right\}^{-2}.
\end{eqnarray}

We now want to derive the final expression of the pressure for this case.\\
The general expression of pressure $p_{D_{GH},I}$ is given by:
\begin{eqnarray}
\label{eq:conserve}
p_{D_{GH},I}(x)&=& -\rho_{D_{GH},I}(x)-\frac{\rho'_{D_{GH},I}(x)}{3} - \frac{Q}{3H}.
\end{eqnarray}
The term $\frac{Q}{3H}$ is given by:
\begin{eqnarray}
    \frac{Q}{3H} = d^2\rho_{m,I} = d^2\Omega_{m0}e^{-3(1-d^2)x}.\label{q/3}
\end{eqnarray}
Using the general expression fo $\rho_{D_{GH},I}$ we derived in Eq. (\ref{carolina3}), we obtain the following expression for $\rho'_{D_{GH},I} $:
\begin{eqnarray}
\rho'_{D_{GH},I}(x) &=&-3\left\{\frac{2  }{2 \left[1 - c^2 (1+\epsilon)\right] + 3 c^2 \epsilon (1 - d^2)}-1\right\}(1-d^2)\Omega_{m0}e^{-3(1 - d^2)x} \, \nonumber   \\
&& -2 \left\{1 - \frac{2 \Omega_{m0}}{2 \left[1 - c^2 (1+\epsilon)\right] + 3 c^2 \epsilon (1 - d^2)} \right\}(1-\alpha) e^{-2(1-\alpha)x}. \label{carolina6}
\end{eqnarray}
Therefore, we obtain the following relation for the pressure:
\begin{eqnarray}
p_{D_{GH},I}(x) &=& - d^2 \left\{ \frac{2}{2[1 - c^2(1+\epsilon)] + 3 c^2 \epsilon (1-d^2)} \right\} \Omega_{m0} \, e^{-3(1-d^2)x} \, \nonumber \\ && - \frac{1}{3} \left\{ 1 - \frac{2 \Omega_{m0}}{2[1 - c^2(1+\epsilon)] + 3 c^2 \epsilon (1-d^2)} \right\} (1 + 2\alpha) \, e^{-2(1-\alpha)x}.\label{carolina51}
\end{eqnarray}
Therefore, we observe a difference compared to the non-interacting case: in the interacting scenario, an additional term proportional to the dark matter component appears in the equations.\\
We now want to obtain the final expression EoS parameter $\omega_{D_{GH},I}$. The general definition is given by:
\begin{eqnarray}
\omega_{D_{GH},I}(x) = -1 - \frac{\rho_{D_{GH},I}'(x)}{3 \rho_{D_{GH},I}(x)} - \frac{Q}{3H\rho_{D_{GH},I}(x)} \label{eosnongen}.
\end{eqnarray}
The term $\frac{Q}{3H\rho_{D_{GH},I}(x)}$ is given by:
\begin{eqnarray}
\frac{Q}{3H\rho_{D_{GH},I}(x)} &=& d^2\left(\frac{ \rho_{m,I}}{\rho_{D_{GH},I}(x)}\right)\nonumber \\
&=&d^2  \Omega_{m0}e^{-3(1-d^2)x}\times \nonumber \\
&&\left\{ \left\{\frac{2  }{2 \left[1 - c^2 (1+\epsilon)\right] + 3 c^2 \epsilon (1 - d^2)}-1\right\}\Omega_{m0}e^{-3(1 - d^2)}\right. \nonumber   \\
&&\left. + \left\{1 - \frac{2 \Omega_{m0}}{2 \left[1 - c^2 (1+\epsilon)\right] + 3 c^2 \epsilon (1 - d^2)} \right\}\, e^{-2(1-\alpha)}\right\}^{-1}. \label{sergio1}
\end{eqnarray}
Using the expressions of $\rho_{D_{GH},I}$ and $\rho'_{D_{GH},I}$ obtained in Eqs. (\ref{carolina3}) and (\ref{carolina6}) along with the result of Eq. (\ref{sergio1}),  we obtain the following result:
\begin{eqnarray}
\omega_{D_{GH},I} (x)
&=&- 1 +\frac{1}{3}\left\{3\left\{\frac{2  }{2 \left[1 - c^2 (1+\epsilon)\right] + 3 c^2 \epsilon (1 - d^2)}-1\right\}(1-d^2)\Omega_{m0}e^{-3(1 - d^2)x} \, \right.\nonumber   \\
&& \left.+2 \left\{1 - \frac{2 \Omega_{m0}}{2 \left[1 - c^2 (1+\epsilon)\right] + 3 c^2 \epsilon (1 - d^2)} \right\}(1-\alpha) e^{-2(1-\alpha)x}\right\}\times \nonumber \\
&&\left\{ \left\{\frac{2  }{2 \left[1 - c^2 (1+\epsilon)\right] + 3 c^2 \epsilon (1 - d^2)}-1\right\}\Omega_{m0}e^{-3(1 - d^2)x} \,\right. \nonumber   \\
&& \left.+ \left\{1 - \frac{2 \Omega_{m0}}{2 \left[1 - c^2 (1+\epsilon)\right] + 3 c^2 \epsilon (1 - d^2)} \right\}\, e^{-2(1-\alpha)x}  \right\}^{-1}\nonumber \\
&&-d^2  \Omega_{m0}e^{-3(1-d^2)x}\times \nonumber \\
&&\left\{ \left\{\frac{2  }{2 \left[1 - c^2 (1+\epsilon)\right] + 3 c^2 \epsilon (1 - d^2)}-1\right\}\Omega_{m0}(1+z)^{3(1 - d^2)}\right. \nonumber   \\
&&\left. + \left\{1 - \frac{2 \Omega_{m0}}{2 \left[1 - c^2 (1+\epsilon)\right] + 3 c^2 \epsilon (1 - d^2)} \right\}\, (1+z)^{2(1-\alpha)}\right\}^{-1}. \label{carolina62}
\end{eqnarray}

\begin{figure}[htbp]
    \centering
    \begin{subfigure}{0.8\textwidth}
        \includegraphics[width=0.6\textwidth]{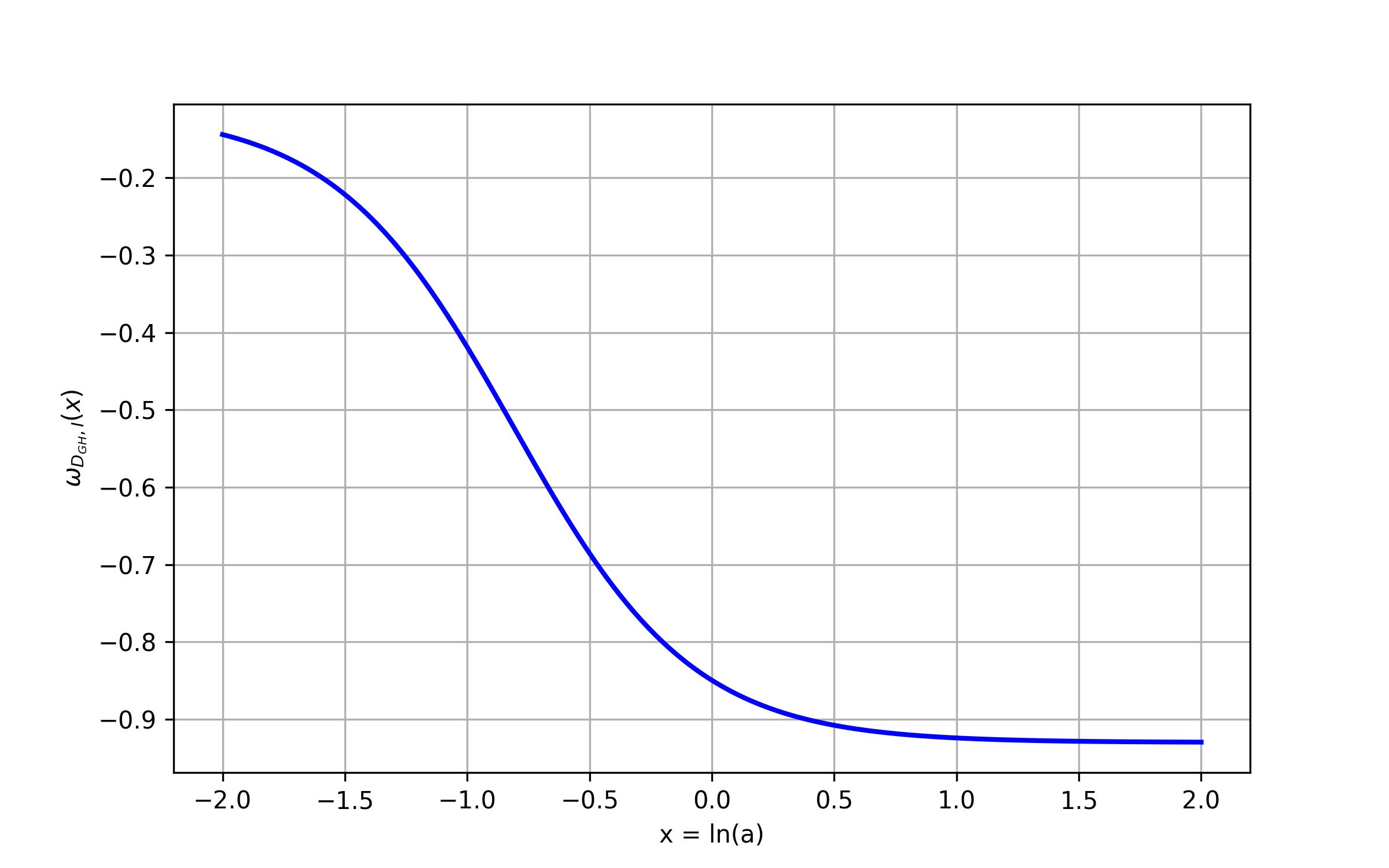}
        \caption{Plot of $\omega_{D_{GH},I}(x)$ for $c^2=0.46$.}
        \label{EoS3}
    \end{subfigure}\\[0.5cm]
    \begin{subfigure}{0.8\textwidth}
        \includegraphics[width=0.6\textwidth]{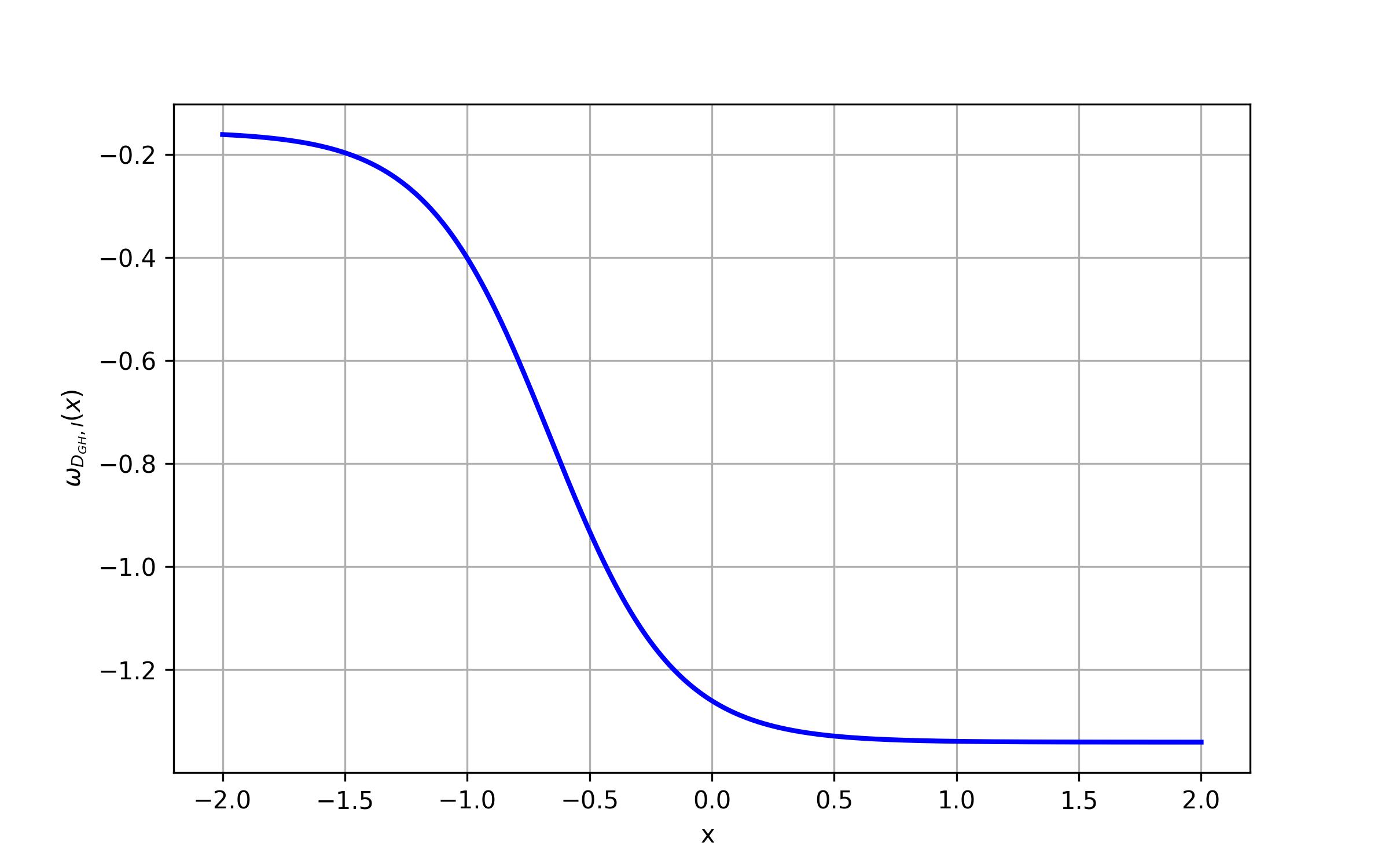}
        \caption{Plot of $\omega_{D_{GH},I}(x)$ for $c=0.579$.}
        \label{EoS3-2}
    \end{subfigure}\\[0.5cm]
    \begin{subfigure}{0.8\textwidth}
        \includegraphics[width=0.6\textwidth]{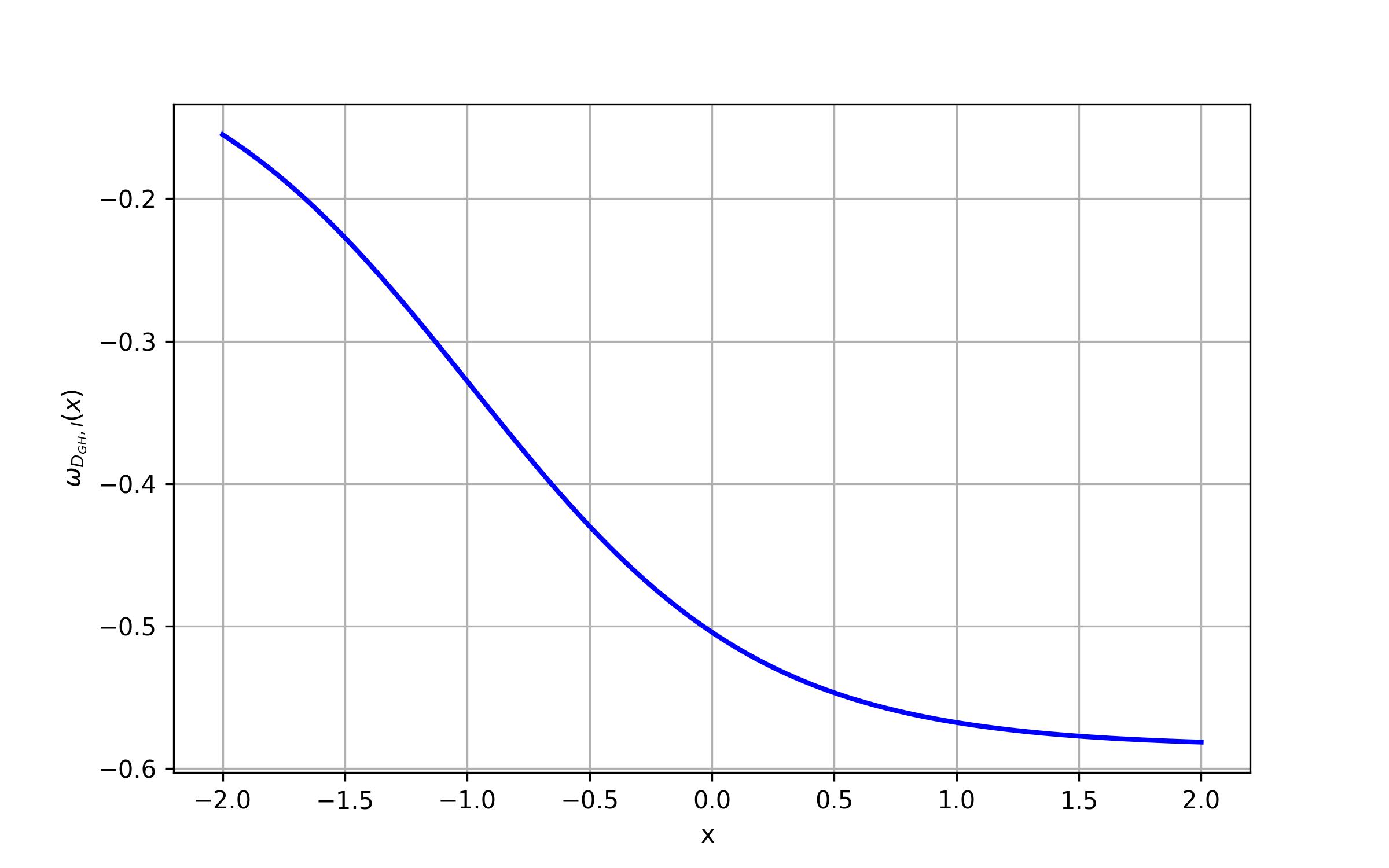}
        \caption{Plot of $\omega_{D_{GH},I}(x)$ for $c=0.818$.}
        \label{EoS3-3}
    \end{subfigure}
    \caption{Comparison of the three plots of $\omega_{D_{GH},I}(x)$.}
    \label{fig:EoS_all3}
\end{figure}

In Figs. (\ref{EoS3}), (\ref{EoS3-2}) and (\ref{EoS3-3})  we plot the expression of $\omega_{D_{GH},I}(x) $ obtained in Eq. (\ref{carolina62}) for $c^2=0.46$, $c=0.579$ and $c=0.818$, respectively. We have considered a value of $d^2=0.02$ in order to obtain the Figures for the interacting cases. 

We obtain the present day values (i.e. the values corresponding to $x=0$) for the EoS for the three different cases: 
\begin{itemize}
    \item $\omega_{D_{GH},I,1}(0) \approx -0.849$
    \item $\omega_{D_{GH},I,2}(0) \approx -1.261$
    \item $\omega_{D_{GH},I,3}(0) \approx -0.504$
\end{itemize}


We now want to calculate the expression of the deceleration parameter $q_{GH,I}(x)$.\\
For the model we are considering, we choose the following expression: 
\begin{eqnarray}
q_{GH,I}(x) &=&  -1 - \frac{1}{2 h_{GH,I}^2(x)} \frac{d h_{GH,I}^2(x)}{dx}. \label{deceleration}
\end{eqnarray}
Using the general expression of $h_{GH,I}^2$ given in Eq. (\ref{carolina2}) along with the expression of $\frac{d h_{GH,I}^2(x)}{dx}$ derived in Eq. (\ref{carolina5}), we can write:
\begin{eqnarray}
q_{GH,I} (x)&=&  -1+\frac{1}{2}\left\{ 3(1 - d^2)\Omega_{m0}e^{-3(1 - d^2)x}\right. \nonumber  \\
&& \left.+3(1 - d^2)\left\{\frac{2  }{2 \left[1 - c^2 (1+\epsilon)\right] + 3 c^2 \epsilon (1 - d^2)}-1\right\}\Omega_{m0}e^{-3(1 - d^2)x} \, \right.\nonumber   \\
&& \left.+2(1-\alpha)\left\{1 - \frac{2 \Omega_{m0}}{2 \left[1 - c^2 (1+\epsilon)\right] + 3 c^2 \epsilon (1 - d^2)} \right\}\, e^{-2(1-\alpha)x}  \right\} \times \nonumber \\
&&\left\{ \Omega_{m0}e^{-3(1 - d^2)x} \right.\nonumber  \\
&&\left. +\left\{\frac{2  }{2 \left[1 - c^2 (1+\epsilon)\right] + 3 c^2 \epsilon (1 - d^2)}-1\right\}\Omega_{m0}e^{-3(1 - d^2)x} \, \right.\nonumber   \\
&& \left.+ \left\{1 - \frac{2 \Omega_{m0}}{2 \left[1 - c^2 (1+\epsilon)\right] + 3 c^2 \epsilon (1 - d^2)} \right\}\, e^{-2(1-\alpha)x}   \right\}^{-1}\label{carolina72}.
\end{eqnarray} 

\begin{figure}[htbp]
    \centering
    \begin{subfigure}{0.8\textwidth}
        \includegraphics[width=0.6\textwidth]{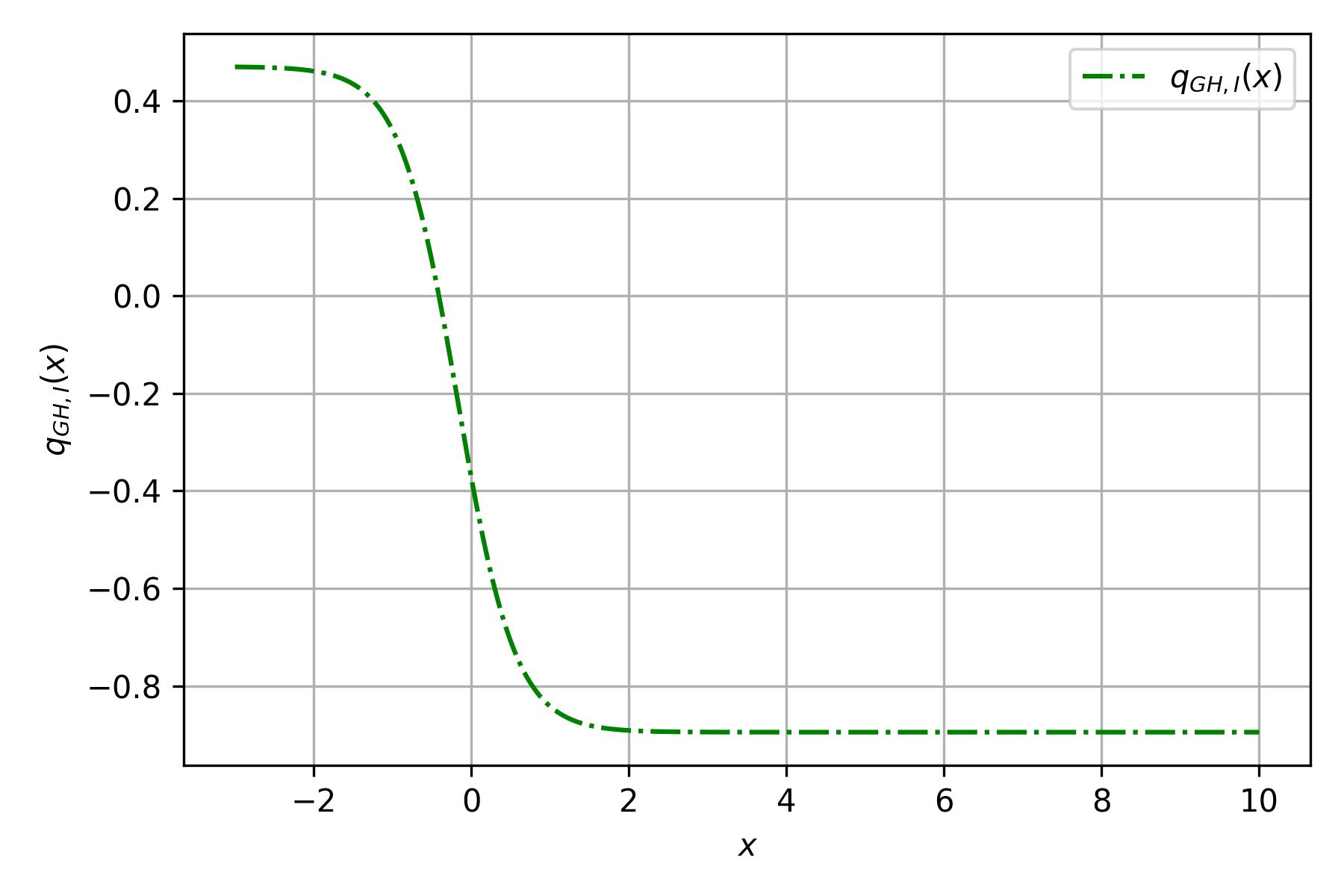}
        \caption{Plot of $q_{GH,I}(x)$ for $c^2=0.46$.}
        \label{q3}
    \end{subfigure}\\[0.5cm]
    \begin{subfigure}{0.8\textwidth}
        \includegraphics[width=0.6\textwidth]{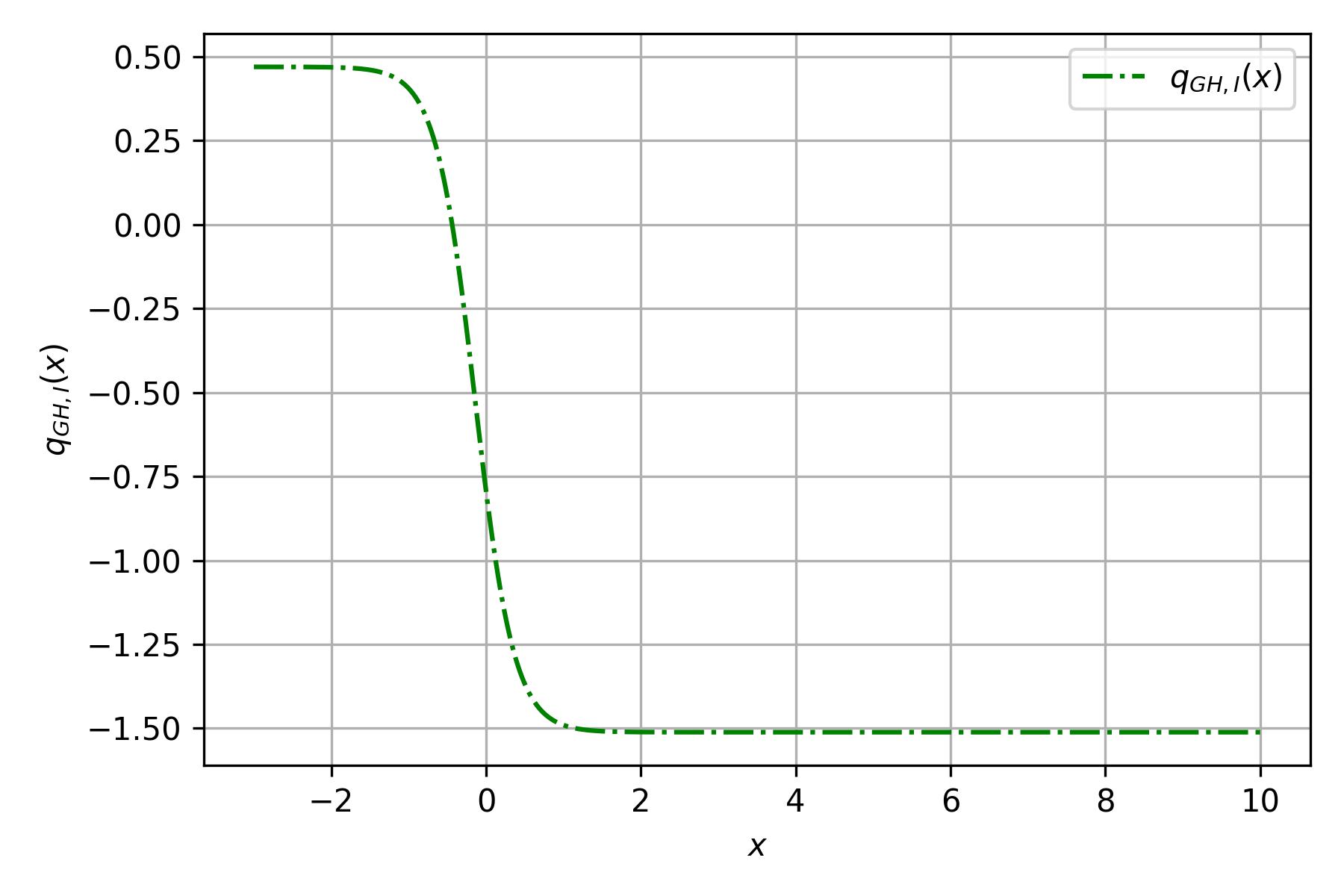}
        \caption{Plot of $q_{GH,I}(x)$ for $c=0.579$.}
        \label{q3-2}
    \end{subfigure}\\[0.5cm]
    \begin{subfigure}{0.8\textwidth}
        \includegraphics[width=0.6\textwidth]{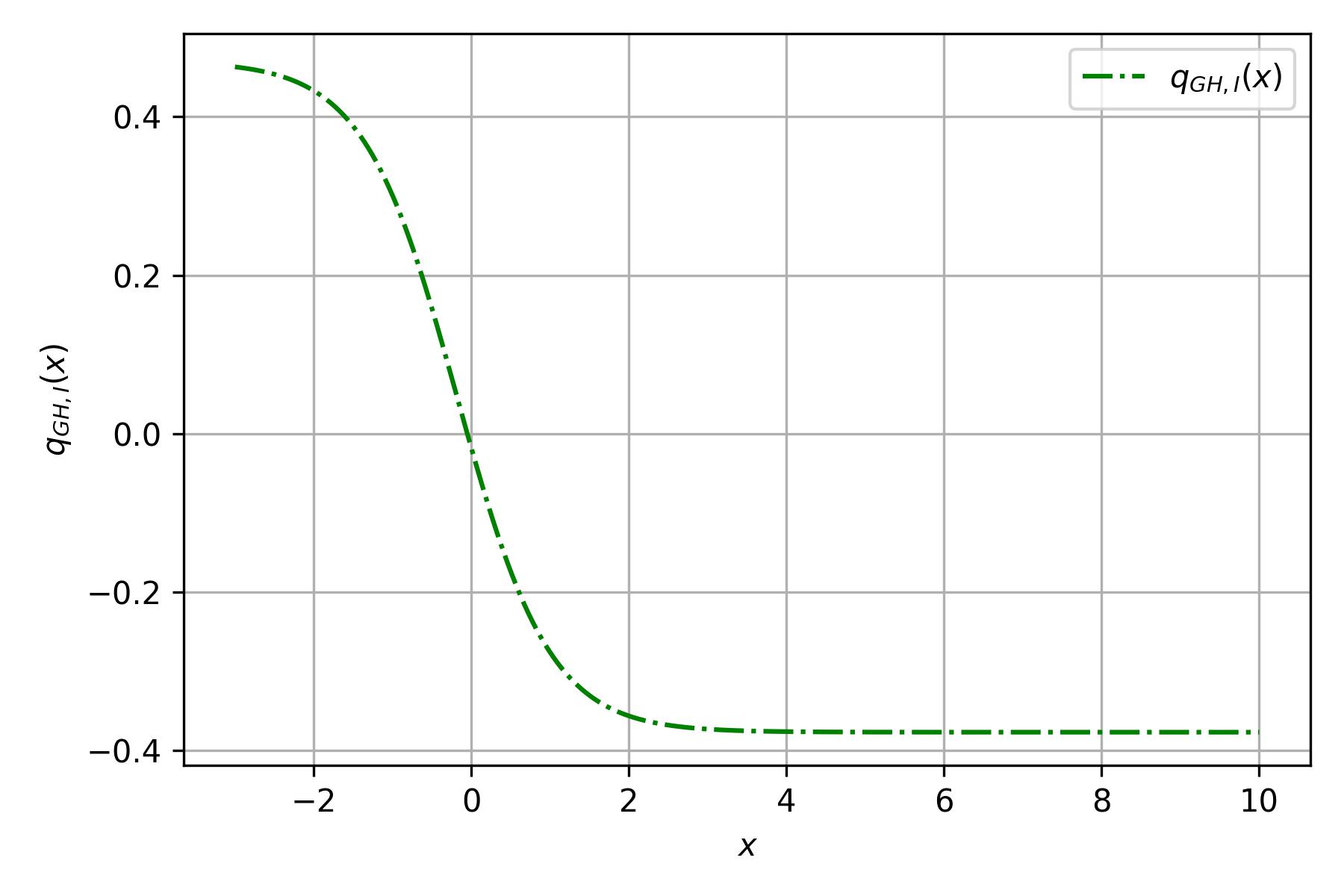}
        \caption{Plot of $q_{GH,I}(x)$ for $c=0.818$.}
        \label{q3-3}
    \end{subfigure}
    \caption{Comparison of the three cases for the deceleration parameter $q_{GH,I}(x)$.}
    \label{fig:q_all3}
\end{figure}

In Figs. (\ref{q3}), (\ref{q3-2}) and (\ref{q3-3}) we plot the expression of $q_{GH,I}(x) $ obtained in Eq. (\ref{carolina72}) for $c^2=0.46$, $c=0.579$ and $c=0.818$, respectively.

We obtain the present day values (i.e. the values corresponding to $x=0$) for $q_{GH,I}$ for the three different cases: 
\begin{itemize}
    \item $q_{GH,I}(0) \approx -0.373 $
    \item $q_{GH,I}(0) \approx-0.795$
    \item $q_{GH,I}(0) \approx -0.018$
\end{itemize}
Therefore, all the three cases predict an accelerated expansion of the Universe as predicted by recent observations. 

\subsection{Case 2: $g\left(\frac{H^2}{R}\right)=1-\eta\left(1-\frac{H^2}{R}\right)$}
We now consider the second case of this work. \\
Also in this case, we have:
\begin{eqnarray}
    \rho_{m,I}&=& \rho_{m0}e^{-3(1-d^2)x},\\
    \rho'_{m,I} &=& -3(1-d^2)\rho_{m0}e^{-3(1-d^2)x}.
\end{eqnarray}
The fractional energy density of DE is given by:
\begin{eqnarray}
\Omega_{D_{GR},I}&=&c^{2} \left[ (2-\eta) +  \frac{(1-\eta)}{2h_{GR,I}^{2}} \frac{d h_{GR,I}^{2}}{d x} \right].\label{carolina11}
\end{eqnarray}
Using the expression of $\Omega_{D_{GR}}$, the Friedmann Eq. (\ref{genfri}) can be rewritten as the differential equation of $H(z)$ in the following way:
\begin{equation}
H^2_{GR,I}(z)\left\{1-c^2\left[(2-\eta)-\frac{(1-\eta)(1+z)}{2}\frac{d\ln
H^2_{GR,I}(z)}{dz}\right]\right\}=H^2_0\Omega_{m0}(1+z)^{3(1-d^2)},
\end{equation}
which is equivalent to:
\begin{equation}
H^2_{GR,I}(z)\left\{1-c^2\left[(2-\eta)-\frac{(1-\eta)(1+z)}{2H^2_{GR,I}}\frac{d
H^2_{GR,I}(z)}{dz}\right]\right\}=H^2_0\Omega_{m0}(1+z)^{3(1-d^2)}. \label{carolina7}
\end{equation}
The solution of Eq. (\ref{carolina7}) is given by:
\begin{eqnarray}
h^2_{GR,I}(z) &=&   
\frac{2 \, \Omega_{m0}(1+z)^{3(1-d^2)} }{2 \left[ 1 - c^2 (2-\eta) \right] + 3 c^2 (1-\eta)(1 - d^2)} \nonumber  \\
&& +  \left\{ 1 - \frac{2 \, \Omega_{m0}}{2 \left[ 1 - c^2 (2-\eta) \right] + 3 c^2 (1-\eta)(1 - d^2)}  \right\} (1+z)^{\lambda},
\end{eqnarray}
  which can be written as:
\begin{eqnarray}
h^2_{GR,I}(z) &=&   \Omega_{m0}(1+z)^{3(1-d^2)}\nonumber \\
&&
+\left\{\frac{2 \,}{2 \left[ 1 - c^2 (2-\eta) \right] + 3 c^2 (1-\eta)(1 - d^2)}-1\right\} \Omega_{m0}(1+z)^{3(1-d^2)}  \nonumber  \\
&& +  \left\{ 1 - \frac{2 \, \Omega_{m0}}{2 \left[ 1 - c^2 (2-\eta) \right] + 3 c^2 (1-\eta)(1 - d^2)} \right\} (1+z)^{\lambda}.\label{carolina8}
\end{eqnarray}
The expression obtained in Eq.~(\ref{carolina8}) is particularly useful because it allows  $ h_{GR,I}^2(z) $ to be written as the sum of two distinct contributions:  one from the matter component and one from dark energy.  The term $ \Omega_{m0} (1+z)^{3(1-d^2)} $ represents the dark matter component; consequently, the following expression for $ \rho_{D_{GR},I}(z) $ can be easily obtained:
\begin{eqnarray}
\rho_{D_{GR},I}(z)&=&
\left\{\frac{2 \,}{2 \left[ 1 - c^2 (2-\eta) \right] + 3 c^2 (1-\eta)(1 - d^2)}-1\right\} \Omega_{m0}(1+z)^{3(1-d^2)}  \nonumber  \\
&& +  \left\{ 1 - \frac{2 \, \Omega_{m0}}{2 \left[ 1 - c^2 (2-\eta) \right] + 3 c^2 (1-\eta)(1 - d^2)} \right\} (1+z)^{\lambda} .
\end{eqnarray}
We can now write $h^2_{GR,I}(z)$ and $\rho_{D_{GR},I}(z)$ as functions of the parameter $x=\ln a$ as follows:
\begin{eqnarray}
h^2_{GR,I}(x) &=&   \Omega_{m0}e^{-3(1-d^2)x}\nonumber \\
&&
+\left\{\frac{2 \,}{2 \left[ 1 - c^2 (2-\eta) \right] + 3 c^2 (1-\eta)(1 - d^2)}-1\right\} \Omega_{m0}3^{-3(1-d^2)x}  \nonumber  \\
&& +  \left\{ 1 - \frac{2 \, \Omega_{m0}}{2 \left[ 1 - c^2 (2-\eta) \right] + 3 c^2 (1-\eta)(1 - d^2)} \right\} e^{-\lambda x}, \label{carolina9}\\
\rho_{D_{GR},I}(x)&=&
\left\{\frac{2 \,}{2 \left[ 1 - c^2 (2-\eta) \right] + 3 c^2 (1-\eta)(1 - d^2)}-1\right\} \Omega_{m0}e^{-3(1-d^2)x}  \nonumber  \\
&& +  \left\{ 1 - \frac{2 \, \Omega_{m0}}{2 \left[ 1 - c^2 (2-\eta) \right] + 3 c^2 (1-\eta)(1 - d^2)} \right\} e^{-\lambda x}. \label{carolina10}
\end{eqnarray}

We now want to calculate the expressions of the fractional energy density of DE $\Omega_{D_{GR},I}$, the evolutionary form of the fractional energy density of DE $\Omega'_{D_{GR},I}$, the pressure $p_{D_{GR},I}$, the EoS parameter $\omega_{D_{GR},I}$ and the deceleration parameter $q_{GR,I}$ for this case.\\
The general expression of the fractional energy density of DE is given in Eq. (\ref{carolina11}).\\
Using the expression of $h_{GR,I}^{2}$ obtained in Eq. (\ref{carolina9}), we obtain:
\begin{eqnarray}
\frac{d h_{GR,I}^2(x)}{dx}&=&  -3(1-d^2)\Omega_{m0}e^{-3(1-d^2)x}\nonumber \\
&&
-3(1-d^2)\left\{\frac{2 \,}{2 \left[ 1 - c^2 (2-\eta) \right] + 3 c^2 (1-\eta)(1 - d^2)}-1\right\} \Omega_{m0}3^{-3(1-d^2)x}  \nonumber  \\
&& -\lambda \left\{ 1 - \frac{2 \, \Omega_{m0}}{2 \left[ 1 - c^2 (2-\eta) \right] + 3 c^2 (1-\eta)(1 - d^2)} \right\} e^{-\lambda x} .\label{carolina13}
\end{eqnarray}
Therefore, we obtain the following equation for $\Omega_{D_{GR},I}(x)$:
\begin{eqnarray}
\Omega_{D_{GR},I}(x)&=&c^{2}  (2-\eta) -\frac{c^2  (1-\eta) }{2}  \cdot  \left\{   3(1-d^2)\Omega_{m0}e^{-3(1-d^2)x}\right.\nonumber \\
&&\left.
+3(1-d^2)\left\{\frac{2 \,}{2 \left[ 1 - c^2 (2-\eta) \right] + 3 c^2 (1-\eta)(1 - d^2)}-1\right\} \Omega_{m0}3^{-3(1-d^2)x}  \right.\nonumber  \\
&& \left.+\lambda \left\{ 1 - \frac{2 \, \Omega_{m0}}{2 \left[ 1 - c^2 (2-\eta) \right] + 3 c^2 (1-\eta)(1 - d^2)} \right\} e^{-\lambda x}     \right\}\times\nonumber \\
&&\left\{ \Omega_{m0}e^{-3(1-d^2)x}+ \right.\nonumber \\
&&\left.\left\{\frac{2 \,}{2 \left[ 1 - c^2 (2-\eta) \right] + 3 c^2 (1-\eta)(1 - d^2)}-1\right\} \Omega_{m0}e^{-3(1-d^2)x}\right.  \nonumber  \\
&&\left. +  \left\{ 1 - \frac{2 \, \Omega_{m0}}{2 \left[ 1 - c^2 (2-\eta) \right] + 3 c^2 (1-\eta)(1 - d^2)} \right\} e^{-\lambda x}    \right\}^{-1}.\label{eq:OmegaGRnew}
\end{eqnarray}
We now want to calculate the evolutionary form of the fractional energy density of DE.\\
If we put $h^2_{GR,I} \equiv A_6$, we can rewrite $\Omega_{D_{GR,I}}$ in the following way:
\begin{eqnarray}
\Omega_{D_{GR,I}}&=&c^{2} (2-\eta) +  \frac{c^2(1-\eta)}{2} \left( \frac{A_6'}{A_6}  \right). \label{carolina12}
\end{eqnarray}
Differentiating Eq. (\ref{carolina12}) with respect to $x$, we obtain:
\begin{eqnarray}
\Omega_{D_{GR},I}'&=& \frac{c^2(1-\eta)}{2}\left[ \frac{A_6''}{A_6}-\left( \frac{A_6'}{A_6}  \right)^2 \right],\label{renato-2prime}
\end{eqnarray}
where $A_6''= \frac{d^2 h_{GR,I}^2(x)}{dx^2}$. \\
Using the general expression of $ h_{GR,I}^2(x)$ given in Eq. (\ref{carolina9}) or equivalently the expression of $\frac{d h_{GR,I}^2(x)}{dx}$ obtained in Eq. (\ref{carolina13}), we can write:
\begin{eqnarray}
    \frac{d^2 h_{GR,I}^2(x)}{dx^2}  &=& 9(1-d^2)^2\Omega_{m0}e^{-3(1-d^2)x}\nonumber \\
&&
+9(1-d^2)^2\left\{\frac{2 \,}{2 \left[ 1 - c^2 (2-\eta) \right] + 3 c^2 (1-\eta)(1 - d^2)}-1\right\} \Omega_{m0}3^{-3(1-d^2)x}  \nonumber  \\
&& +\lambda^2 \left\{ 1 - \frac{2 \, \Omega_{m0}}{2 \left[ 1 - c^2 (2-\eta) \right] + 3 c^2 (1-\eta)(1 - d^2)} \right\} e^{-\lambda x}.
\end{eqnarray}
Then, the final expression of $\Omega'_{D_{GR},I}(x)$ is given by:
\begin{eqnarray}
\Omega'_{D_{GR},I}(x)&=& \frac{c^2  (1-\eta) }{2} \times \left\{  9(1-d^2)^2\Omega_{m0}e^{-3(1-d^2)x}\right.\nonumber \\
&&\left.
+9(1-d^2)^2\left\{\frac{2 \,}{2 \left[ 1 - c^2 (2-\eta) \right] + 3 c^2 (1-\eta)(1 - d^2)}-1\right\} \Omega_{m0}3^{-3(1-d^2)x}  \right.\nonumber  \\
&& \left.+\lambda^2 \left\{ 1 - \frac{2 \, \Omega_{m0}}{2 \left[ 1 - c^2 (2-\eta) \right] + 3 c^2 (1-\eta)(1 - d^2)} \right\} e^{-\lambda x} \right\} \times \nonumber \\
&& \left\{  \Omega_{m0}e^{-3(1-d^2)x}\right.\nonumber \\
&&\left.
+\left\{\frac{2 \,}{2 \left[ 1 - c^2 (2-\eta) \right] + 3 c^2 (1-\eta)(1 - d^2)}-1\right\} \Omega_{m0}3^{-3(1-d^2)x} \right. \nonumber  \\
&& \left.+  \left\{ 1 - \frac{2 \, \Omega_{m0}}{2 \left[ 1 - c^2 (2-\eta) \right] + 3 c^2 (1-\eta)(1 - d^2)} \right\} e^{-\lambda x}  \right\}^{-1}\nonumber \\
&&-\frac{c^2(1-\eta)}{2} \left\{  3(1-d^2)\Omega_{m0}e^{-3(1-d^2)x}\right.\nonumber \\
&&\left.
+3(1-d^2)\left\{\frac{2 \,}{2 \left[ 1 - c^2 (2-\eta) \right] + 3 c^2 (1-\eta)(1 - d^2)}-1\right\} \Omega_{m0}3^{-3(1-d^2)x}  \right.\nonumber  \\
&&\left. +\lambda \left\{ 1 - \frac{2 \, \Omega_{m0}}{2 \left[ 1 - c^2 (2-\eta) \right] + 3 c^2 (1-\eta)(1 - d^2)} \right\} e^{-\lambda x}    \right\}^2 \times \nonumber \\
&&\left\{  \Omega_{m0}e^{-3(1-d^2)x}\right.\nonumber \\
&&\left.
+\left\{\frac{2 \,}{2 \left[ 1 - c^2 (2-\eta) \right] + 3 c^2 (1-\eta)(1 - d^2)}-1\right\} \Omega_{m0}3^{-3(1-d^2)x} \right. \nonumber  \\
&& \left.+  \left\{ 1 - \frac{2 \, \Omega_{m0}}{2 \left[ 1 - c^2 (2-\eta) \right] + 3 c^2 (1-\eta)(1 - d^2)} \right\} e^{-\lambda x}  \right\}^{-2}.
\end{eqnarray}

We now want to obtain the final expression of the pressure $p_{D_{GR},I}(x)$. \\The general expression of $p_{D_{GR},I}$ is given by:
\begin{eqnarray}
\label{eq:conserve}
p_{D_{GR},I}(x)
&=& -\rho_{D_{GR},I}(x)-\frac{\rho'_{D_{GR},I}(x)}{3}-\frac{Q}{3H}.
\end{eqnarray}
In this case, we have that $\rho'_{D_{GR},I}$ is given by:
\begin{eqnarray}
\rho'_{D_{GR},I}(x)&=& -3\left(1-d^2\right)\left\{\frac{2 \,}{2 \left[ 1 - c^2 (2-\eta) \right] + 3 c^2 (1-\eta)(1 - d^2)}-1\right\} \Omega_{m0}e^{-3(1-d^2)x}  \nonumber  \\
&& -\lambda  \left\{ 1 - \frac{2 \, \Omega_{m0}}{2 \left[ 1 - c^2 (2-\eta) \right] + 3 c^2 (1-\eta)(1 - d^2)} \right\} e^{-\lambda x}. \label{carolina15}
\end{eqnarray}
Therefore, we obtain:
\begin{eqnarray}
p_{D_{GR},I}(x)&=& -\frac{2 d^2\Omega_{m0}e^{-3(1-d^2)x}}{2 \left[ 1 - c^2 (2-\eta) \right] + 3 c^2 (1-\eta)(1 - d^2)}   \nonumber  \\
&& +\left(\frac{\lambda -3}{3} \right) \left\{ 1 - \frac{2 \, \Omega_{m0}}{2 \left[ 1 - c^2 (2-\eta) \right] + 3 c^2 (1-\eta)(1 - d^2)} \right\} e^{-\lambda x}.
\end{eqnarray}

We now want to calculate the final expression for the EoS parameter for this case. \\
The general expression for this case is given by:
\begin{eqnarray}
\omega_{D_{GR},I}(x) = -1 - \frac{\rho_{D_{GR},I}'(x)}{3 \rho_{D_{GR},I}(x)}- \frac{Q}{3H\rho_{D_{GR},I}(x)}.
\end{eqnarray}
Using the expressions of $\rho_{D_{GR},I}$ and $\rho_{D_{GR},I}'$ we derived in Eqs. (\ref{carolina10}) and (\ref{carolina15}), we can write:
\begin{eqnarray}
\omega_{D_{GR},I}(x) &=& -1 + \frac{1}{3}\cdot \left\{ 3\left(1-d^2\right)\left\{\frac{2 \,}{2 \left[ 1 - c^2 (2-\eta) \right] + 3 c^2 (1-\eta)(1 - d^2)}-1\right\} \Omega_{m0}e^{-3(1-d^2)x}  \nonumber \right. \\
&&\left. +\lambda  \left\{ 1 - \frac{2 \, \Omega_{m0}}{2 \left[ 1 - c^2 (2-\eta) \right] + 3 c^2 (1-\eta)(1 - d^2)} \right\} e^{-\lambda x}  \right\} \nonumber \\
&&\times \left\{ \left\{\frac{2 \,}{2 \left[ 1 - c^2 (2-\eta) \right] + 3 c^2 (1-\eta)(1 - d^2)}-1\right\} \Omega_{m0}e^{-3(1-d^2)x}  \nonumber\right.  \\
&&\left. +  \left\{ 1 - \frac{2 \, \Omega_{m0}}{2 \left[ 1 - c^2 (2-\eta) \right] + 3 c^2 (1-\eta)(1 - d^2)} \right\} e^{-\lambda x}   \right\}^{-1}\nonumber \\
&&-d^2  \Omega_{m0}e^{-3(1-d^2)x}\times \nonumber \\
&& \left\{ \left\{\frac{2 \,}{2 \left[ 1 - c^2 (2-\eta) \right] + 3 c^2 (1-\eta)(1 - d^2)}-1\right\} \Omega_{m0}e^{-3(1-d^2)x}  \nonumber\right.  \\
&&\left. +  \left\{ 1 - \frac{2 \, \Omega_{m0}}{2 \left[ 1 - c^2 (2-\eta) \right] + 3 c^2 (1-\eta)(1 - d^2)} \right\} e^{-\lambda x}   \right\}^{-1}.
\end{eqnarray}

In order to find the final expression of $q_{GR,I}(x)$, we use the general expression:
\begin{eqnarray}
q_{GR,I}(x) &=&  -1 - \frac{1}{2 h_{GR,I}^2} \frac{d h_{GR,I}^2(x)}{dx}. \label{deceleration}
\end{eqnarray}
Using the expression of $ \frac{d h_{GR,I}^2(x)}{dx}$ we derived in Eq. (\ref{carolina13}) along with the expression of $h_{GR,I}^2$ obtained in Eq. (\ref{carolina9}), we can write:
\begin{eqnarray}
q_{GR,I}(x) &=&  -1 +\frac{1}{2} \left\{   3(1-d^2)\Omega_{m0}e^{-3(1-d^2)x}\right.\nonumber \\
&&\left.
+3(1-d^2)\left\{\frac{2 \,}{2 \left[ 1 - c^2 (2-\eta) \right] + 3 c^2 (1-\eta)(1 - d^2)}-1\right\} \Omega_{m0}3^{-3(1-d^2)x}  \right.\nonumber  \\
&&\left. +\lambda \left\{ 1 - \frac{2 \, \Omega_{m0}}{2 \left[ 1 - c^2 (2-\eta) \right] + 3 c^2 (1-\eta)(1 - d^2)} \right\} e^{-\lambda x}  \right\} \times\nonumber \\
&&\left\{  \Omega_{m0}e^{-3(1-d^2)x}\right.\nonumber \\
&&
\left.+\left\{\frac{2 \,}{2 \left[ 1 - c^2 (2-\eta) \right] + 3 c^2 (1-\eta)(1 - d^2)}-1\right\} \Omega_{m0}3^{-3(1-d^2)x} \right. \nonumber  \\
&& \left.+  \left\{ 1 - \frac{2 \, \Omega_{m0}}{2 \left[ 1 - c^2 (2-\eta) \right] + 3 c^2 (1-\eta)(1 - d^2)} \right\} e^{-\lambda x}  \right\}^{-1}.
\end{eqnarray}

\section{Case with interaction and curvature}
We now consider the first model in the case both spatial curvature and interaction are present.
\subsection{Case 1: $f\left(\frac{R}{H^2}\right)=1-\epsilon\left(1-\frac{R}{H^2}\right)$}
In this case, we have that the Friedmann equation is given in Eq. (\ref{genfri}) and it can be rewritten in the following way:
\begin{eqnarray}
H^2&=&\frac{1}{3M^{2}_{pl}}\left(\rho_{m}+\rho_{GH}\right)-\frac{k}{a^2}.
\end{eqnarray}
Instead, the dimensionless
energy density of the generalized holographic DE is given by:
\begin{eqnarray}
\Omega_{D_{GH},I,k}&=&\frac{\rho_{D_{GH},I,k}}{3M^{2}_{pl}H^2}\nonumber\\
&=&c^2\left[1-\epsilon\left(1-\frac{R}{H^2}\right)\right]\nonumber\\
&=&c^2\left[1+\epsilon-\frac{\epsilon(1+z)}{2}\frac{d \ln
H^2_{{GH},k}}{dz} + \epsilon H_0^2\Omega_{k0} (1+z)^2\right]\nonumber \\
&=& c^2 \left[ 1 + \epsilon + \frac{\epsilon}{2H^2_{{GH},I,k}} \frac{d H^2_{{GH},I,k}}{dx}+ \epsilon H_0^2\Omega_{k0} (1+z)^2 \right]\nonumber \\
&=& c^2 \left[ 1 + \epsilon + \frac{\epsilon}{2h^2_{{GH},I,k}} \frac{d h^2_{{GH},I,k}}{dx}+ \epsilon H_0^2\Omega_{k0} (1+z)^2 \right].\label{carolina22}
\end{eqnarray}
Therefore, the Friedmann equation can we written as:
\begin{eqnarray}
H_{GH,I,k}^2(z)\left\{1-c^2\left[1+\epsilon-\frac{\epsilon(1+z)}{2H_{GH,I,k}^2(z)}\frac{d 
H_{GH,I,k}^2(z)}{dz}\right]\right\}&=&H^2_0\Omega_{m0}(1+z)^{3(1-d^2)}\nonumber \\
&&-(1-\epsilon c^2)H^2_0\Omega_{k0}(1+z)^2,
\end{eqnarray}
which has the following general solution:
\begin{eqnarray}
h_{GH,I,k}^2(z) &=& \frac{2  \Omega_{m0}(1+z)^{3(1 - d^2)}}{  2 \left[1 - c^2 (1 + \epsilon)\right] + 3(1 -  d^2)c^2 \epsilon } \nonumber  \\
&&-\left( \frac{ 1 - \epsilon c^2 }{  1-c^2 }\right)\Omega_{k0}(1+z)^2  \nonumber \\
&& + \left\{ 1 - \frac{2  \Omega_{m0}}{  2 \left[1 - c^2 (1 + \epsilon)\right] + 3(1 -  d^2)c^2 \epsilon } \right.\nonumber \\
&&\left.+ \frac{2 (1 - \epsilon c^2)  \Omega_{k0}}{2 \left[1 - c^2 (1 + \epsilon)\right] + 2c^2 \epsilon} \right\}(1+z)^{2(1-\alpha)}.\label{lucia1}
\end{eqnarray}

We can write Eq. (\ref{lucia1}) as follows:
\begin{eqnarray}
h_{GH,I,k}^2(z) &=& \Omega_{m0}(1+z)^{3(1 - d^2)} - \Omega_{k0}(1+z)^2\nonumber  \\
&&\left\{\frac{2 }{  2 \left[1 - c^2 (1 + \epsilon)\right] + 3(1 -  d^2)c^2 \epsilon }-1\right\} \Omega_{m0}(1+z)^{3(1 - d^2)} \nonumber  \\
&&+\left[\frac{c^2(\epsilon - 1)}{1-c^2}\right]\Omega_{k0}(1+z)^2  \nonumber \\
&& + \left\{ 1 - \frac{2  \Omega_{m0}}{  2 \left[1 - c^2 (1 + \epsilon)\right] + 3(1 -  d^2)c^2 \epsilon } + \left( \frac{ 1 - \epsilon c^2 }{  1-c^2 }\right)\Omega_{k0} \right\}(1+z)^{2(1-\alpha)}.\label{lucia2}
\end{eqnarray}
The form obtained in Eq.~(\ref{lucia2}) is particularly useful because it allows  $ h_{GH,I,k}^2(z) $ to be written as the sum of three distinct contributions:  one from matter, one from curvature, and one from dark energy. 
The term $ \Omega_{m0} (1+z)^{3(1 - d^2)} $ represents the dark matter component,  while $ \Omega_{k0} (1+z)^2 $ corresponds to the curvature contribution. 
Consequently, the following expression for $ \rho_{D_{GH},k}(z) $ can be obtained:
\begin{eqnarray}
\rho_{D_{GH},I,k}(z) &=&\left\{\frac{2 }{  2 \left[1 - c^2 (1 + \epsilon)\right] + 3(1 -  d^2)c^2 \epsilon }-1\right\} \Omega_{m0}(1+z)^{3(1 - d^2)} \nonumber  \\
&&+\left[\frac{c^2(\epsilon - 1)}{1-c^2}\right]\Omega_{k0}(1+z)^2  \nonumber \\
&& + \left\{ 1 - \frac{2  \Omega_{m0}}{  2 \left[1 - c^2 (1 + \epsilon)\right] + 3(1 -  d^2)c^2 \epsilon } + \left( \frac{ 1 - \epsilon c^2 }{  1-c^2 }\right)\Omega_{k0} \right\}(1+z)^{2(1-\alpha)}.
\end{eqnarray}

We can now write $h_{GH,I,k}^2$ and $\rho_{D_{GH},I,k}$ as functions of the parameter $x$:
\begin{eqnarray}
h_{GH,I,k}^2(x) &=& \Omega_{m0}e^{-3(1 - d^2)x} - \Omega_{k0}e^{-2x}\nonumber  \\
&&+\left\{\frac{2 }{  2 \left[1 - c^2 (1 + \epsilon)\right] + 3(1 -  d^2)c^2 \epsilon }-1\right\} \Omega_{m0}e^{-3(1 - d^2)x} \nonumber  \\
&&+\left[\frac{c^2(\epsilon - 1)}{1-c^2}\right]\Omega_{k0}e^{-2x}  \nonumber \\
&& + \left\{ 1 - \frac{2  \Omega_{m0}}{  2 \left[1 - c^2 (1 + \epsilon)\right] + 3(1 -  d^2)c^2 \epsilon } + \left( \frac{ 1 - \epsilon c^2 }{  1-c^2 }\right)\Omega_{k0} \right\}e^{-2(1-\alpha)x} ,\label{carolina20}\\
\rho_{D_{GH},I,k}(x) &=&\left\{\frac{2 }{  2 \left[1 - c^2 (1 + \epsilon)\right] + 3(1 -  d^2)c^2 \epsilon }-1\right\} \Omega_{m0}e^{-3(1 - d^2)x} \nonumber  \\
&&+\left[\frac{c^2(\epsilon - 1)}{1-c^2}\right]\Omega_{k0}e^{-2x}  \nonumber \\
&& + \left\{ 1 - \frac{2  \Omega_{m0}}{  2 \left[1 - c^2 (1 + \epsilon)\right] + 3(1 -  d^2)c^2 \epsilon } + \left( \frac{ 1 - \epsilon c^2 }{  1-c^2 }\right)\Omega_{k0} \right\}e^{-2(1-\alpha)x}. \label{carolina21}
\end{eqnarray}

We now want to calculate the final expressions of other important cosmological quantities, like the fractional energy density of dark energy $\Omega_{D_{GH},I,k}$, the evolutionary form of the fractional energy density of DE $\Omega'_{D_{GH},I,k}$, the pressure $p_{GH},I,k$, the EoS parameter $\omega_{D_{GH},I,k}$ and the deceleration parameter $q_{GH,I,k}$. \\
For the fractional energy density of DE, we use the general relation obtained in Eq. (\ref{carolina22}):
\begin{eqnarray}
\Omega_{D_{GH},I,k}(x)&=& c^2 \left[ 1 + \epsilon + \frac{\epsilon}{2h^2_{D_{GH},I,k}} \frac{d h^2_{D_{GH},I,k}}{dx}+ \epsilon H_0^2\Omega_{k0} e^{-2x} \right].\label{}
\end{eqnarray}
Using the expression of $h^2_{GH,I,k}$ obtained in Eq. (\ref{carolina20}), we can write:
\begin{eqnarray}
    \frac{d h^2_{GH,I,k}}{dx} &=&  -3(1 - d^2)\Omega_{m0}e^{-3(1 - d^2)x} +2 \Omega_{k0}e^{-2x}\nonumber  \\
&&-3(1 - d^2)\left\{\frac{2 }{  2 \left[1 - c^2 (1 + \epsilon)\right] + 3(1 -  d^2)c^2 \epsilon }-1\right\} \Omega_{m0}e^{-3(1 - d^2)x} \nonumber  \\
&&-2\left[\frac{c^2(\epsilon - 1)}{1-c^2}\right]\Omega_{k0}e^{-2x}  \nonumber \\
&& -2(1-\alpha) \left\{ 1 - \frac{2  \Omega_{m0}}{  2 \left[1 - c^2 (1 + \epsilon)\right] + 3(1 -  d^2)c^2 \epsilon } \right.\nonumber \\
&&\left.+ \left( \frac{ 1 - \epsilon c^2 }{  1-c^2 }\right)\Omega_{k0} \right\}e^{-2(1-\alpha)x}.\label{carolina30}
\end{eqnarray}
Therefore, $\Omega_{D_{GH},Ik}$ is given by the following expression:
\begin{eqnarray}
\Omega_{D_{GH},I,k}(x)
&=& c^2 \left[ 1 + \epsilon + \epsilon H_0^2\Omega_{k0}e^{-2x} \right]\nonumber \\
&&- \frac{c^2\epsilon}{2} \cdot \left\{  3(1 - d^2)\Omega_{m0}e^{-3(1 - d^2)x} +2 \Omega_{k0}e^{-2x}\right.\nonumber  \\
&&\left.+3(1 - d^2)\left\{\frac{2 }{  2 \left[1 - c^2 (1 + \epsilon)\right] + 3(1 -  d^2)c^2 \epsilon }-1\right\} \Omega_{m0}e^{-3(1 - d^2)x}\right. \nonumber  \\
&&\left.+2\left[\frac{c^2(\epsilon - 1)}{1-c^2}\right]\Omega_{k0}e^{-2x}+2(1-\alpha)\times \right.\nonumber \\
&&\left.  \left\{ 1 - \frac{2  \Omega_{m0}}{  2 \left[1 - c^2 (1 + \epsilon)\right] + 3(1 -  d^2)c^2 \epsilon } + \left( \frac{ 1 - \epsilon c^2 }{  1-c^2 }\right)\Omega_{k0} \right\}e^{-2(1-\alpha)x} \right\}\times \nonumber \\
&&\left\{  \Omega_{m0}e^{-3(1 - d^2)x} - \Omega_{k0}e^{-2x}\right.\nonumber  \\
&&\left.+\left\{\frac{2 }{  2 \left[1 - c^2 (1 + \epsilon)\right] + 3(1 -  d^2)c^2 \epsilon }-1\right\} \Omega_{m0}e^{-3(1 - d^2)x} \right.\nonumber  \\
&&\left.+\left[\frac{c^2(\epsilon - 1)}{1-c^2}\right]\Omega_{k0}e^{-2x} \right. \nonumber \\
&&\left. + \left\{ 1 - \frac{2  \Omega_{m0}}{  2 \left[1 - c^2 (1 + \epsilon)\right] + 3(1 -  d^2)c^2 \epsilon } + \left( \frac{ 1 - \epsilon c^2 }{  1-c^2 }\right)\Omega_{k0} \right\}e^{-2(1-\alpha)x}   \right\}^{-1}.
\end{eqnarray}

We now want to calculate the evolutionary form of the fractional energy density of DE.\\
If we put $h^2_{GH,I,k} \equiv A_7$, we can write:
\begin{eqnarray}
\Omega_{D_{GH},I,k} &=&  \left[ 1 + \epsilon + \epsilon H_0^2\Omega_{k0} e^{-2x} \right]+ \frac{c^2\epsilon}{2} \cdot\left( \frac{A_7'}{A_7}\right),
\end{eqnarray}
which implies:
\begin{eqnarray}
\Omega_{D_{GH},I,k}' &=& \frac{c^2\epsilon}{2} \cdot\left[\frac{A_7''}{A_7} -\left( \frac{A_7'}{A_7}\right)^2\right]-2\epsilon H_0^2\Omega_{k0} e^{-2x},
\end{eqnarray}
where $A_7'' = \frac{d^2 h_{GH,I,k}^2(x)}{dx^2}$.\\
Using the expression of $h_{GH,I,k}^2(x)$ given in Eq. (\ref{carolina20}) or equivalently the expression of $\frac{d h_{GH,I,k}^2(x)}{dx}$ obtained in Eq. (\ref{carolina30}), we can write:
\begin{eqnarray}
    \frac{d^2 h^2_{GH,I,k}}{dx^2} &=&  9(1 - d^2)^2\Omega_{m0}e^{-3(1 - d^2)x} -4 \Omega_{k0}e^{-2x}\nonumber  \\
&&+9(1 - d^2)^2\left\{\frac{2 }{  2 \left[1 - c^2 (1 + \epsilon)\right] + 3(1 -  d^2)c^2 \epsilon }-1\right\} \Omega_{m0}e^{-3(1 - d^2)x} \nonumber  \\
&&+4\left[\frac{c^2(\epsilon - 1)}{1-c^2}\right]\Omega_{k0}e^{-2x}  \nonumber \\
&& +4(1-\alpha)^2 \left\{ 1 - \frac{2  \Omega_{m0}}{  2 \left[1 - c^2 (1 + \epsilon)\right] + 3(1 -  d^2)c^2 \epsilon }\right.\nonumber \\
&&\left.+ \left( \frac{ 1 - \epsilon c^2 }{  1-c^2 }\right)\Omega_{k0} \right\}e^{-2(1-\alpha)x}\label{}.
\end{eqnarray}
Therefore, the final expression of $\Omega'_{D_{GH},I,k}(x)$ is given by:
\begin{eqnarray}
\Omega'_{D_{GH},I,k}(x)&=& -2\epsilon H_0^2\Omega_{k0} e^{-2x}+\frac{c^2\epsilon}{2} \times\nonumber \\
&& \left\{9(1 - d^2)^2\Omega_{m0}e^{-3(1 - d^2)x} -4 \Omega_{k0}e^{-2x}\right.\nonumber  \\
&&\left.+9(1 - d^2)^2\left\{\frac{2 }{  2 \left[1 - c^2 (1 + \epsilon)\right] + 3(1 -  d^2)c^2 \epsilon }-1\right\} \Omega_{m0}e^{-3(1 - d^2)x} \right.\nonumber  \\
&&\left.+4\left[\frac{c^2(\epsilon - 1)}{1-c^2}\right]\Omega_{k0}e^{-2x} \right. \nonumber \\
&& \left.+4(1-\alpha)^2 \left\{ 1 - \frac{2  \Omega_{m0}}{  2 \left[1 - c^2 (1 + \epsilon)\right] + 3(1 -  d^2)c^2 \epsilon } + \left( \frac{ 1 - \epsilon c^2 }{  1-c^2 }\right)\Omega_{k0} \right\}e^{-2(1-\alpha)x}\right\}\times \nonumber \\
&&\left\{  \Omega_{m0}e^{-3(1 - d^2)x} - \Omega_{k0}e^{-2x}\right.\nonumber  \\
&&\left.+\left\{\frac{2 }{  2 \left[1 - c^2 (1 + \epsilon)\right] + 3(1 -  d^2)c^2 \epsilon }-1\right\} \Omega_{m0}e^{-3(1 - d^2)x} \right.\nonumber  \\
&&\left.+\left[\frac{c^2(\epsilon - 1)}{1-c^2}\right]\Omega_{k0}e^{-2x} \right. \nonumber \\
&&\left. + \left\{ 1 - \frac{2  \Omega_{m0}}{  2 \left[1 - c^2 (1 + \epsilon)\right] + 3(1 -  d^2)c^2 \epsilon } + \left( \frac{ 1 - \epsilon c^2 }{  1-c^2 }\right)\Omega_{k0} \right\}e^{-2(1-\alpha)x}   \right\}^{-1}\nonumber \\ 
&&- \frac{c^2\epsilon}{2} \cdot \left\{  3(1 - d^2)\Omega_{m0}e^{-3(1 - d^2)x} -2 \Omega_{k0}e^{-2x}\right.\nonumber  \\
&&\left.+3(1 - d^2)\left\{\frac{2 }{  2 \left[1 - c^2 (1 + \epsilon)\right] + 3(1 -  d^2)c^2 \epsilon }-1\right\} \Omega_{m0}e^{-3(1 - d^2)x}\right. \nonumber  \\
&&\left.+2\left[\frac{c^2(\epsilon - 1)}{1-c^2}\right]\Omega_{k0}e^{-2x}+2(1-\alpha)\times \right.\nonumber \\
&&\left.  \left\{ 1 - \frac{2  \Omega_{m0}}{  2 \left[1 - c^2 (1 + \epsilon)\right] + 3(1 -  d^2)c^2 \epsilon } + \left( \frac{ 1 - \epsilon c^2 }{  1-c^2 }\right)\Omega_{k0} \right\}e^{-2(1-\alpha)x} \right\}^2\times \nonumber \\
&&\left\{  \Omega_{m0}e^{-3(1 - d^2)x} - \Omega_{k0}e^{-2x}\right.\nonumber  \\
&&\left.+\left\{\frac{2 }{  2 \left[1 - c^2 (1 + \epsilon)\right] + 3(1 -  d^2)c^2 \epsilon }-1\right\} \Omega_{m0}e^{-3(1 - d^2)x} \right.\nonumber  \\
&&\left.+\left[\frac{c^2(\epsilon - 1)}{1-c^2}\right]\Omega_{k0}e^{-2x} \right. \nonumber \\
&&\left. + \left\{ 1 - \frac{2  \Omega_{m0}}{  2 \left[1 - c^2 (1 + \epsilon)\right] + 3(1 -  d^2)c^2 \epsilon } + \left( \frac{ 1 - \epsilon c^2 }{  1-c^2 }\right)\Omega_{k0} \right\}e^{-2(1-\alpha)x}   \right\}^{-2}.
\end{eqnarray}
We now want to find the final expression of the pressure of DE for this case. \\
The general expression of the pressure $ p_{D_{GH},I,k} $ is given by:
\begin{eqnarray}
\label{eq:conserve}
p_{D_{GH},I,k}(x)&=& -\rho_{D_{GH},I,k}(x) - \frac{\rho'_{D_{GH},I,k}(x)}{3}-\frac{Q}{3H}.
\end{eqnarray}
Also in this case, we have that:
\begin{eqnarray}
    \frac{Q}{3H}=d^2\rho_{m,I}= d^2\Omega_{m0}e^{-3(1-d^2)x}.
\end{eqnarray}
Using the general expression for $ \rho_{D_{GH},I,k} $ previously derived in Eq.~(\ref{carolina21}) with the expression of $\frac{Q}{3H}$, we obtain the following result for $ \rho'_{D_{GH},I,k} $:
\begin{eqnarray}
    \rho'_{D_{GH},I,k} (x) &=& -3(1 - d^2)\left\{\frac{2 }{  2 \left[1 - c^2 (1 + \epsilon)\right] + 3(1 -  d^2)c^2 \epsilon }-1\right\} \Omega_{m0}e^{-3(1 - d^2)x} \nonumber  \\
&&-2\left[\frac{c^2(\epsilon - 1)}{1-c^2}\right]\Omega_{k0}e^{-2x}  \nonumber \\
&& -2(1-\alpha) \left\{ 1 - \frac{2  \Omega_{m0}}{  2 \left[1 - c^2 (1 + \epsilon)\right] + 3(1 -  d^2)c^2 \epsilon } \right.\nonumber \\
&&\left.+ \left( \frac{ 1 - \epsilon c^2 }{  1-c^2 }\right)\Omega_{k0} \right\}e^{-2(1-\alpha)x}.\label{carolina31}
\end{eqnarray}
Therefore, we obtain the following expression for $p_{D_{GH},I,k}(x)$:
\begin{eqnarray}
p_{D_{GH},I,k}(x) &=&-\frac{2d^2\Omega_{m0}e^{-3(1 - d^2)x} }{  2 \left[1 - c^2 (1 + \epsilon)\right] + 3(1 -  d^2)c^2 \epsilon } \nonumber  \\
&&-\frac{1}{3}\left[\frac{c^2(\epsilon - 1)}{1-c^2}\right]\Omega_{k0}e^{-2x}  \nonumber \\
&& -\frac{1+2\alpha}{3} \left\{ 1 - \frac{2  \Omega_{m0}}{  2 \left[1 - c^2 (1 + \epsilon)\right] + 3(1 -  d^2)c^2 \epsilon } \right.\nonumber \\
&&\left.+ \left( \frac{ 1 - \epsilon c^2 }{  1-c^2 }\right)\Omega_{k0} \right\}e^{-2(1-\alpha)x}.  \label{carolina52}
\end{eqnarray}
We now want to obtain the EoS parameter $\omega_{D_{GH},I,k}$.\\
The general definition is given by:
\begin{eqnarray}
\omega_{D_{GH},I,k}(x) = -1 - \frac{\rho_{D_{GH},I,k}'(x)}{3 \rho_{D_{GH},I,k}(x)}- \frac{Q}{3H\rho_{D_{GH},I,k}(x)} \label{eosnongen}.
\end{eqnarray}
The term $\frac{Q}{3H\rho_{D_{GH},I}(x)}$ is given by:
\begin{eqnarray}
    \frac{Q}{3H\rho_{D_{GH},I,k}(x)}&=&d^2\left( \frac{\rho_{m,I}}{\rho_{D_{GH},I,k}}  \right)\nonumber \\
    &=& d^2  \Omega_{m0}e^{-3(1-d^2)x} \times \nonumber \\
    &&\left\{\left\{\frac{2 }{  2 \left[1 - c^2 (1 + \epsilon)\right] + 3(1 -  d^2)c^2 \epsilon }-1\right\} \Omega_{m0}e^{-3(1 - d^2)x} \right.\nonumber  \\
&&\left.+\left[\frac{c^2(\epsilon - 1)}{1-c^2}\right]\Omega_{k0}e^{-2x} \right. \nonumber \\
&&\left. + \left\{ 1 - \frac{2  \Omega_{m0}}{  2 \left[1 - c^2 (1 + \epsilon)\right] + 3(1 -  d^2)c^2 \epsilon } \right.\right.\nonumber \\
&&\left.\left.+ \left( \frac{ 1 - \epsilon c^2 }{  1-c^2 }\right)\Omega_{k0} \right\}e^{-2(1-\alpha)x}\right\}^{-1}.\label{sergio222}
\end{eqnarray}

Using the expressions of $\rho_{D_{GH},I,k}$ and $\rho'_{D_{GH},I,k}$ obtained in Eqs. (\ref{carolina21}) and (\ref{carolina31}) along with the result of Eq. (\ref{sergio222}), we can write:
\begin{eqnarray}
\omega_{D_{GH},I,k}(x) &=& -1 +\frac{1}{3}\cdot \left\{ 3(1 - d^2)\left\{\frac{2 }{  2 \left[1 - c^2 (1 + \epsilon)\right] + 3(1 -  d^2)c^2 \epsilon }-1\right\} \Omega_{m0}e^{-3(1 - d^2)x} \right.\nonumber  \\
&&\left.+2\left[\frac{c^2(\epsilon - 1)}{1-c^2}\right]\Omega_{k0}e^{-2x}  \right.\nonumber \\
&&\left. +2(1-\alpha) \left\{ 1 - \frac{2  \Omega_{m0}}{  2 \left[1 - c^2 (1 + \epsilon)\right] + 3(1 -  d^2)c^2 \epsilon } + \left( \frac{ 1 - \epsilon c^2 }{  1-c^2 }\right)\Omega_{k0} \right\}e^{-2(1-\alpha)x}  \right\}\times \nonumber\\
&&\left\{ \left\{\frac{2 }{  2 \left[1 - c^2 (1 + \epsilon)\right] + 3(1 -  d^2)c^2 \epsilon }-1\right\} \Omega_{m0}e^{-3(1 - d^2)x} \right.\nonumber  \\
&&\left.+\left[\frac{c^2(\epsilon - 1)}{1-c^2}\right]\Omega_{k0}e^{-2x} \right. \nonumber \\
&&\left. + \left\{ 1 - \frac{2  \Omega_{m0}}{  2 \left[1 - c^2 (1 + \epsilon)\right] + 3(1 -  d^2)c^2 \epsilon } + \left( \frac{ 1 - \epsilon c^2 }{  1-c^2 }\right)\Omega_{k0} \right\}e^{-2(1-\alpha)x}  \right\}^{-1}\nonumber \\
&&-d^2  \Omega_{m0}e^{-3(1-d^2)x} \times \nonumber \\
    &&\left\{\left\{\frac{2 }{  2 \left[1 - c^2 (1 + \epsilon)\right] + 3(1 -  d^2)c^2 \epsilon }-1\right\} \Omega_{m0}e^{-3(1 - d^2)x} \right.\nonumber  \\
&&\left.+\left[\frac{c^2(\epsilon - 1)}{1-c^2}\right]\Omega_{k0}e^{-2x} \right. \nonumber \\
&&\left. + \left\{ 1 - \frac{2  \Omega_{m0}}{  2 \left[1 - c^2 (1 + \epsilon)\right] + 3(1 -  d^2)c^2 \epsilon } \right.\right.\nonumber \\
&&\left.\left.+ \left( \frac{ 1 - \epsilon c^2 }{  1-c^2 }\right)\Omega_{k0} \right\}e^{-2(1-\alpha)x}\right\}^{-1}. \label{carolina63}
\end{eqnarray}

\begin{figure}[htbp]
    \centering
    \begin{subfigure}{0.8\textwidth}
        \includegraphics[width=0.6\textwidth]{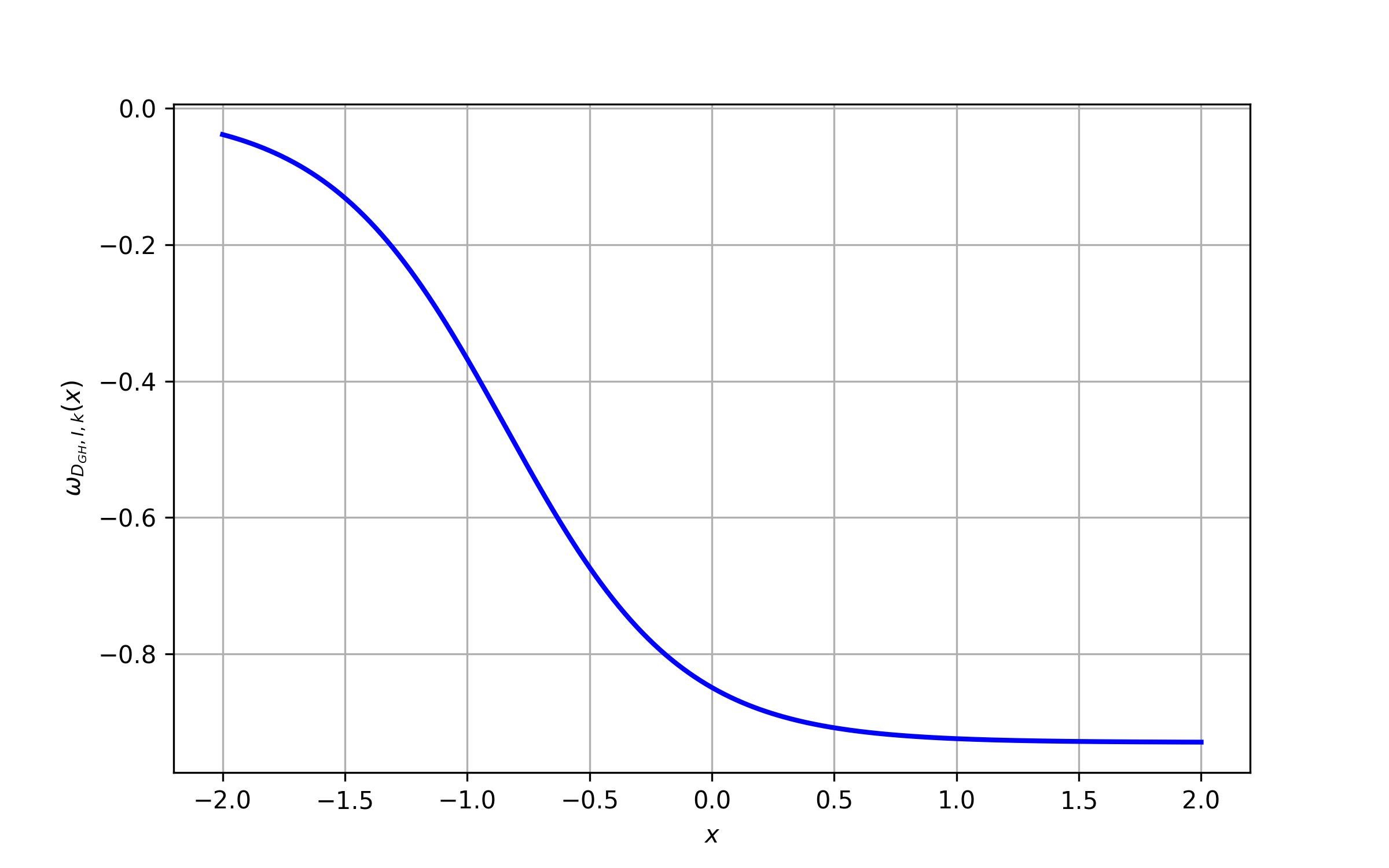}
        \caption{Plot of $\omega_{D_{GH},I,k}(x)$ for $c^2=0.46$.}
        \label{EoS4}
    \end{subfigure}\\[0.5cm]
    \begin{subfigure}{0.8\textwidth}
        \includegraphics[width=0.6\textwidth]{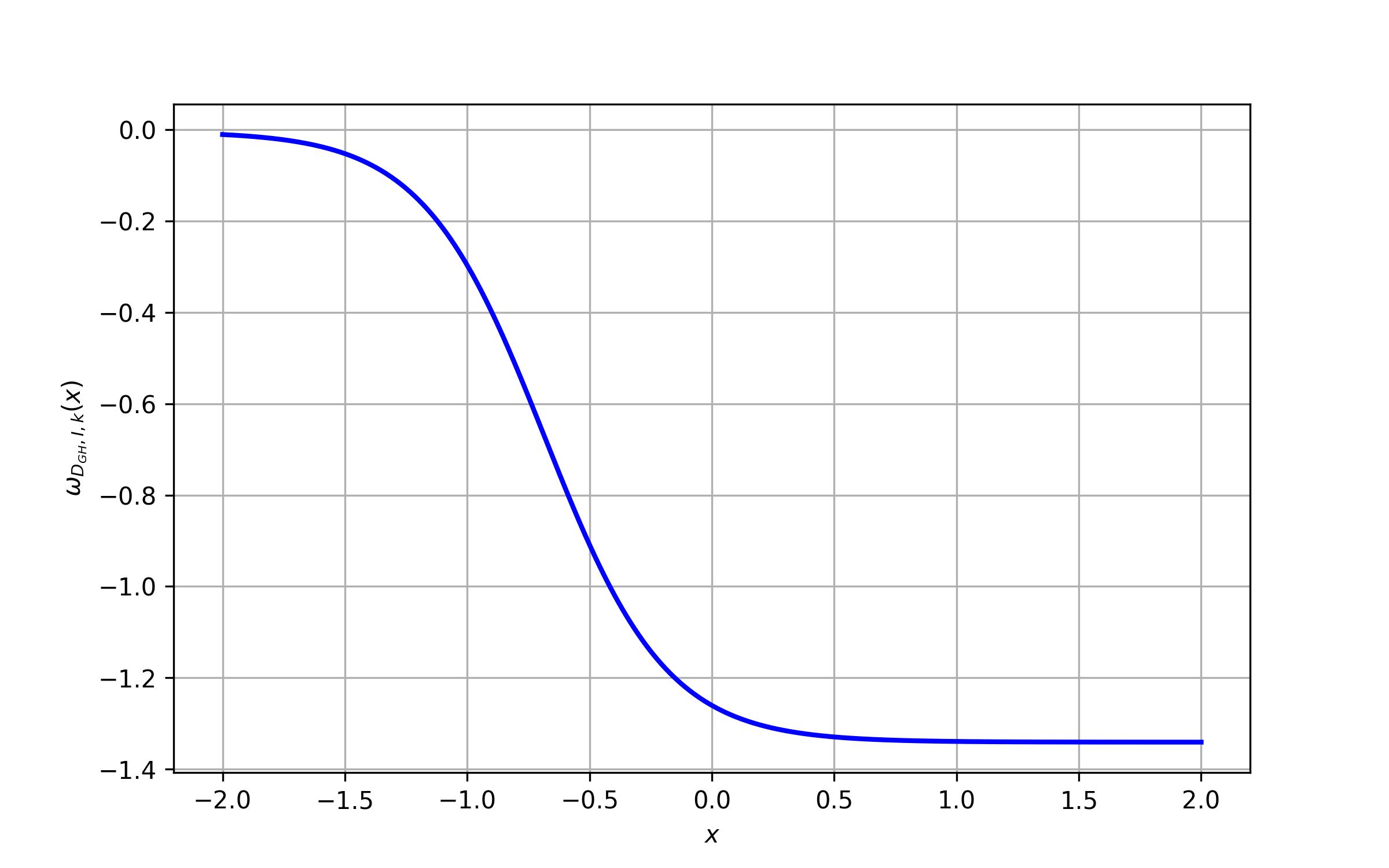}
        \caption{Plot of $\omega_{D_{GH},I,k}(x)$ for $c=0.579$.}
        \label{EoS4-2}
    \end{subfigure}\\[0.5cm]
    \begin{subfigure}{0.8\textwidth}
        \includegraphics[width=0.6\textwidth]{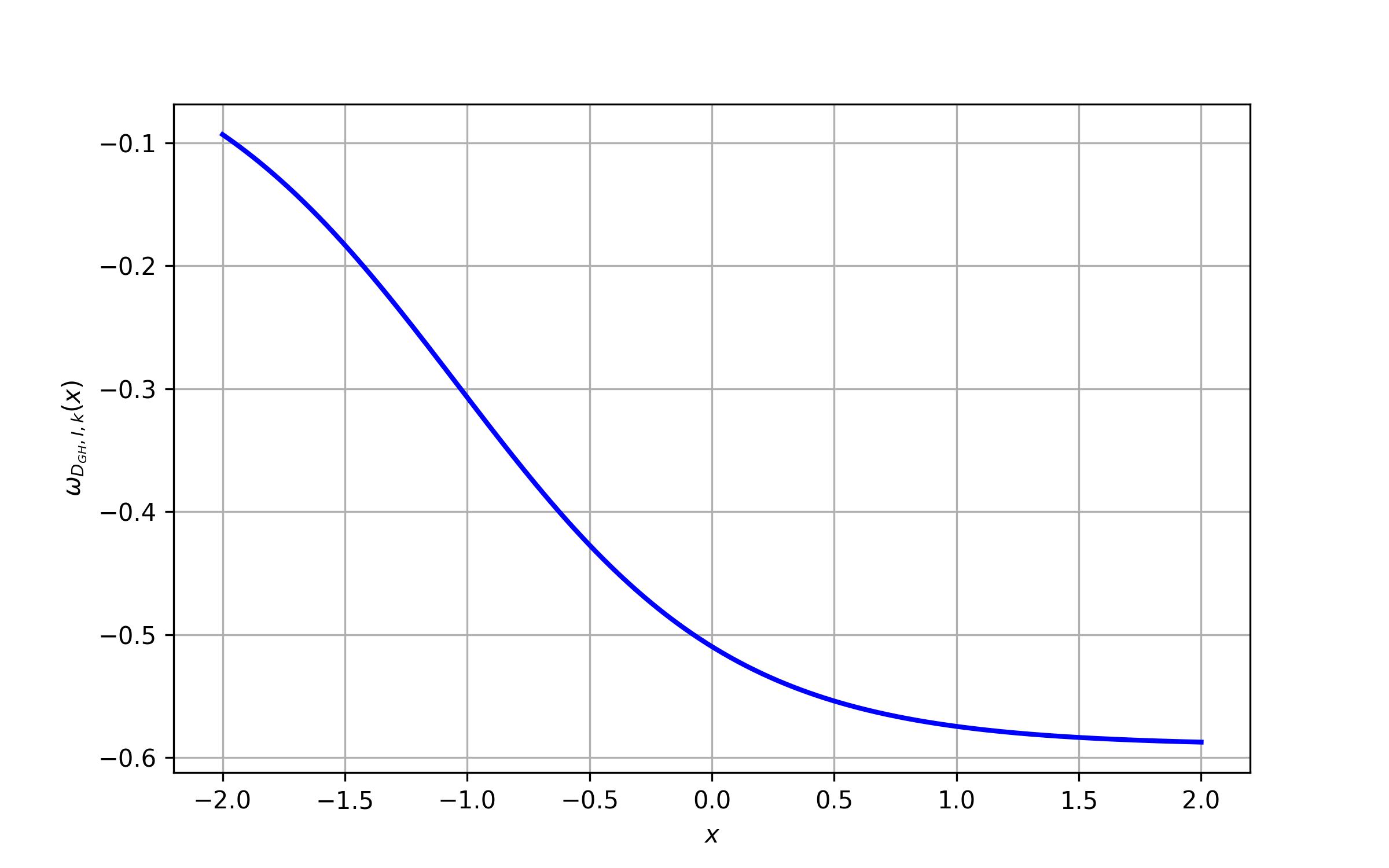}
        \caption{Plot of $\omega_{D_{GH},I,k}(x)$ for $c=0.815$.}
        \label{EoS4-3}
    \end{subfigure}
    \caption{Comparison of the three plots of $\omega_{D_{GH},I,k}(x)$.}
    \label{fig:EoS_all3}
\end{figure}

In Figs. (\ref{EoS4}), (\ref{EoS4-2}) and (\ref{EoS4-3})  we plot the expression of $\omega_{D_{GH},I,k}(x) $ obtained in Eq. (\ref{carolina63}) for $c^2=0.46$, $c=0.579$ and $c=0.815$, respectively.

We obtain the present day values (i.e. the values corresponding to $x=0$) for the EoS for the three different cases: 
\begin{itemize}
    \item $\omega_{D_{GH},I,k,1}(0) \approx -0.849$
    \item $\omega_{D_{GH},I,k,2}(0) \approx -1.261$
    \item $\omega_{D_{GH},I,k,3}(0) \approx -0.504$
\end{itemize}

We now want to calculate the expression of the deceleration parameter $q_{GH,I,k}(x)$.\\
For the model we are considering, we choose the following expression in order to derive $q_{GH,I,k}$: 
\begin{eqnarray}
q_{GH,I,k}(x) &=&  -1 - \frac{1}{2 h_{GH,I,k}^2} \frac{d h_{GH,I,k}^2(x)}{dx}.
\end{eqnarray}
Using the general expression of $h_{GH,I,k}^2$ given in Eq. (\ref{carolina20}) along with the expression of $\frac{d h_{GH,I,k}^2(x)}{dx}$ derived in Eq. (\ref{carolina30}), we can write:
\begin{eqnarray}
q_{GH,I,k} &=& -1+\frac{1}{2}\cdot \left\{ 3(1 - d^2)\Omega_{m0}e^{-3(1 - d^2)x} -2 \Omega_{k0}e^{-2x}\right.\nonumber  \\
&&\left.+3(1 - d^2)\left\{\frac{2 }{  2 \left[1 - c^2 (1 + \epsilon)\right] + 3(1 -  d^2)c^2 \epsilon }-1\right\} \Omega_{m0}e^{-3(1 - d^2)x} \right.\nonumber  \\
&&\left.+2\left[\frac{c^2(\epsilon - 1)}{1-c^2}\right]\Omega_{k0}e^{-2x}  \right. \nonumber \\
&&\left. +2(1-\alpha) \left\{ 1 - \frac{2  \Omega_{m0}}{  2 \left[1 - c^2 (1 + \epsilon)\right] + 3(1 -  d^2)c^2 \epsilon } + \left( \frac{ 1 - \epsilon c^2 }{  1-c^2 }\right)\Omega_{k0} \right\}e^{-2(1-\alpha)x}   \right\}\times \nonumber \\
&&\left\{ \Omega_{m0}e^{-3(1 - d^2)x} - \Omega_{k0}e^{-2x}\right.\nonumber  \\
&&\left.+\left\{\frac{2 }{  2 \left[1 - c^2 (1 + \epsilon)\right] + 3(1 -  d^2)c^2 \epsilon }-1\right\} \Omega_{m0}e^{-3(1 - d^2)x} \right.\nonumber  \\
&&\left.+\left[\frac{c^2(\epsilon - 1)}{1-c^2}\right]\Omega_{k0}e^{-2x} \right. \nonumber \\
&&\left. + \left\{ 1 - \frac{2  \Omega_{m0}}{  2 \left[1 - c^2 (1 + \epsilon)\right] + 3(1 -  d^2)c^2 \epsilon } + \left( \frac{ 1 - \epsilon c^2 }{  1-c^2 }\right)\Omega_{k0} \right\}e^{-2(1-\alpha)x}  \right\}^{-1}. \label{carolina73}
\end{eqnarray}

\begin{figure}[htbp]
    \centering
    \begin{subfigure}{0.8\textwidth}
        \includegraphics[width=0.6\textwidth]{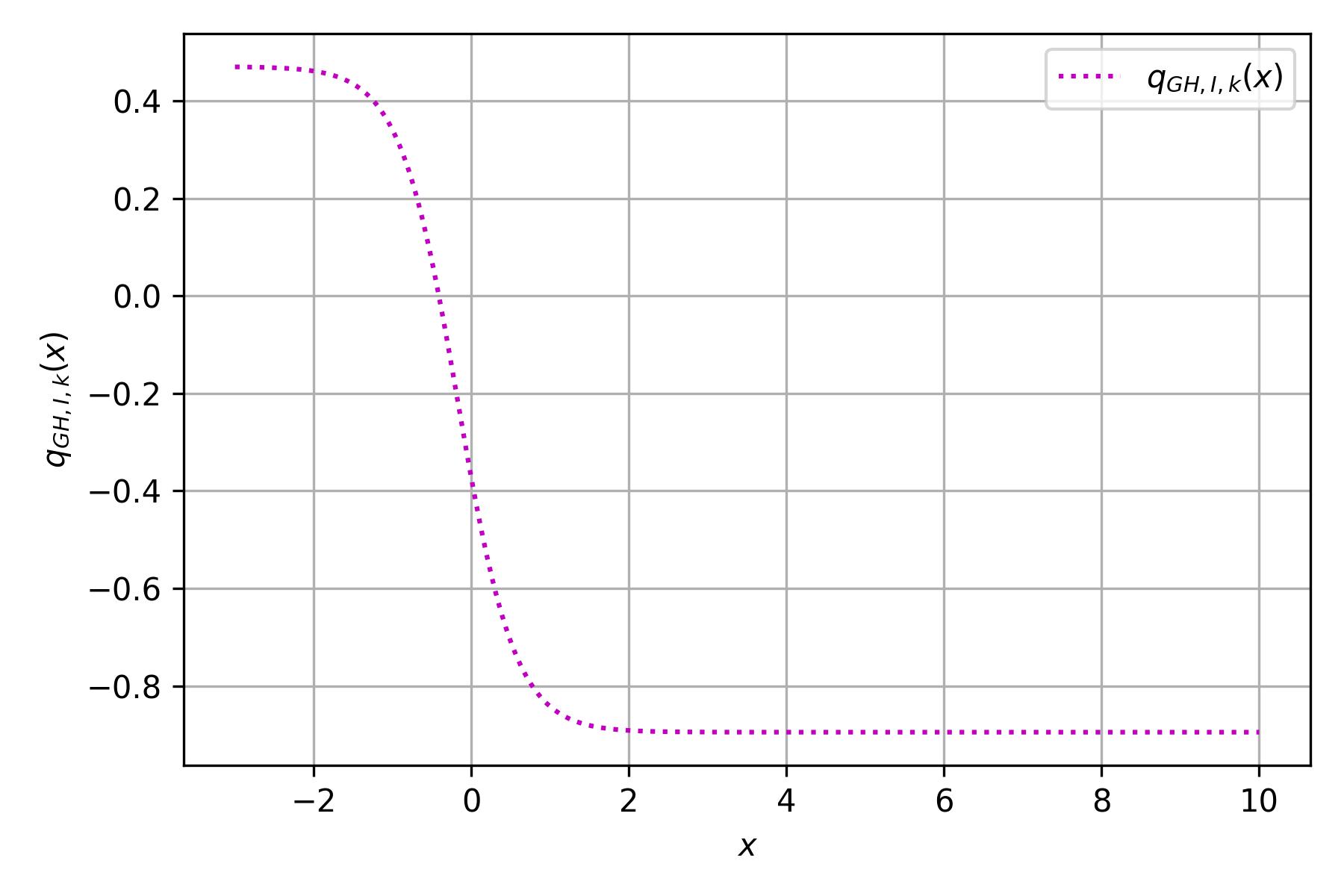}
        \caption{Plot of $q_{GH,I,k}(x)$ for $c^2=0.46$.}
        \label{q4}
    \end{subfigure}\\[0.5cm]
    \begin{subfigure}{0.8\textwidth}
        \includegraphics[width=0.6\textwidth]{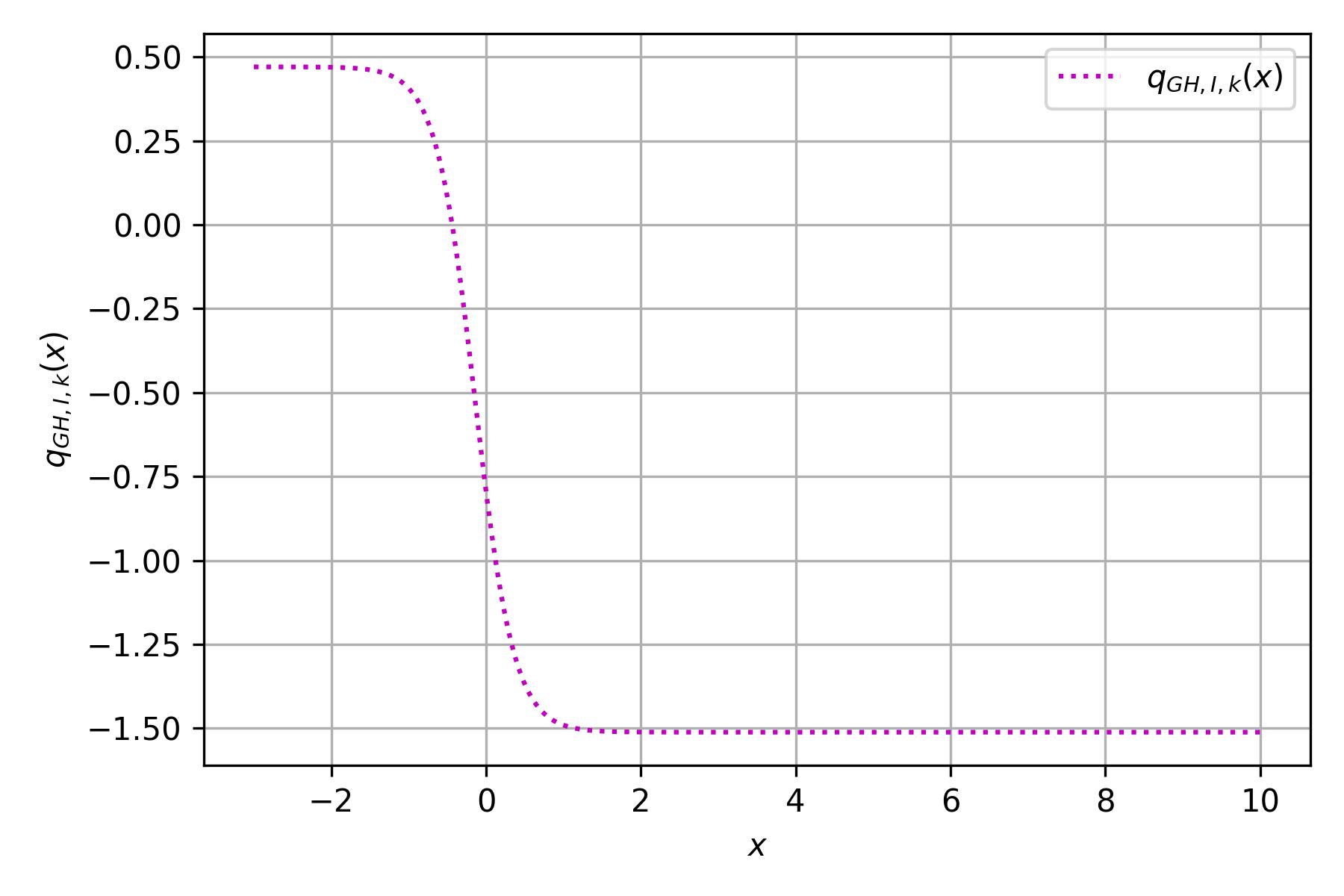}
        \caption{Plot of $q_{GH,I,k}(x)$ for $c=0.579$.}
        \label{q4-2}
    \end{subfigure}\\[0.5cm]
    \begin{subfigure}{0.8\textwidth}
        \includegraphics[width=0.6\textwidth]{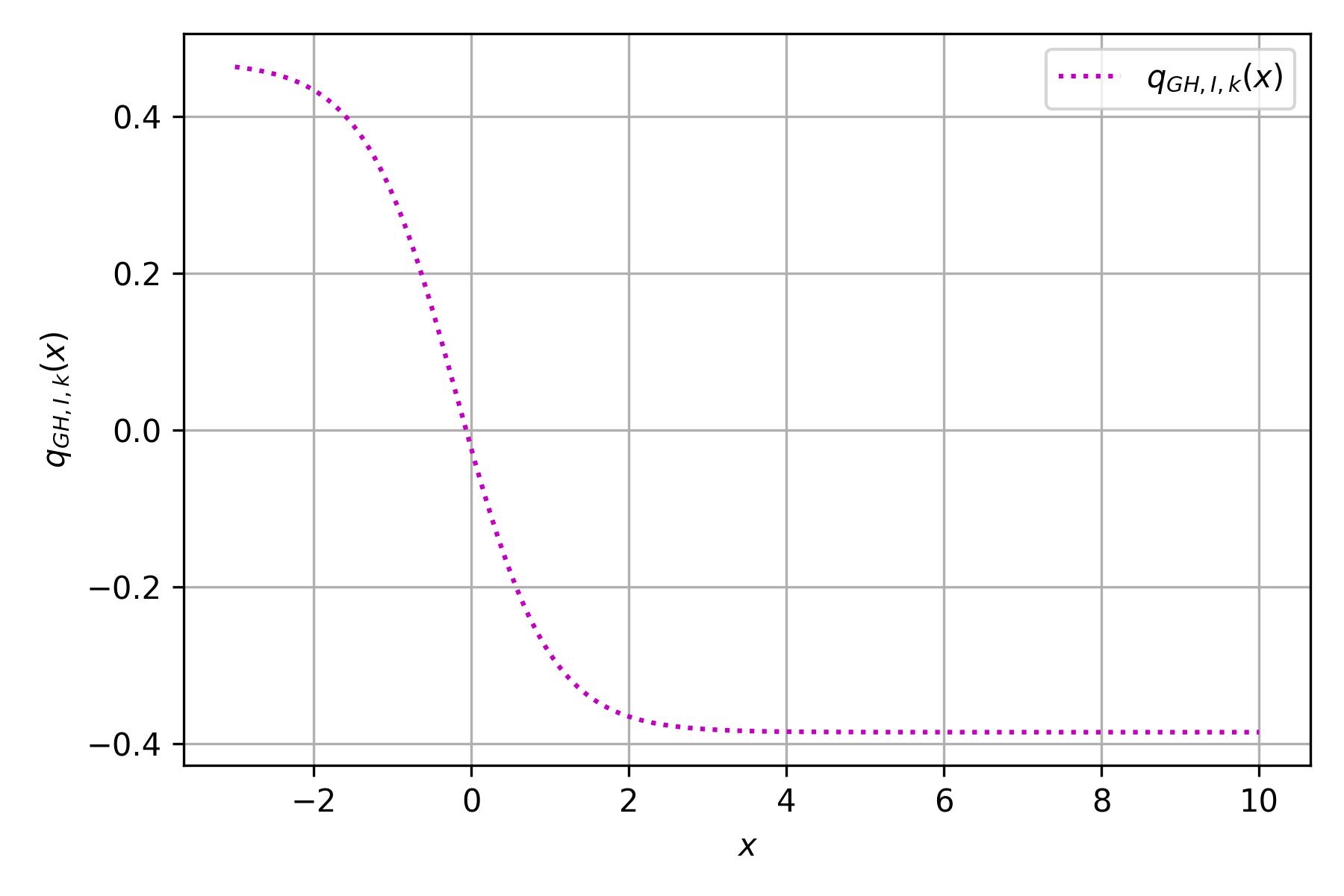}
        \caption{Plot of $q_{GH,I,k}(x)$ for $c=0.815$.}
        \label{q4-3}
    \end{subfigure}
    \caption{Comparison of the three cases for the deceleration parameter $q_{GH,I,k}(x)$.}
    \label{fig:q_all4}
\end{figure}

In Figs. (\ref{q4}), (\ref{q4-2}) and (\ref{q4-3}) we plot the expression of $q_{GH,I,k}(x) $ obtained in Eq. (\ref{carolina73}) for $c^2=0.46$, $c=0.579$ and $c=0.815$, respectively.

We obtain the present day values (i.e. the values corresponding to $x=0$) for $q_{GH,I,k}(x) $ for the three different cases: 
\begin{itemize}
    \item $q_{GH,I,k}(0) \approx -0.373 $
    \item $q_{GH,I,k}(0) \approx -0.796$
    \item $q_{GH,I,k}(0) \approx -0.024$
\end{itemize}
Therefore, all the three cases predict an accelerated expansion of the Universe as predicted by recent observations. 

\subsection{Case 2: $g\left(\frac{H^2}{R}\right)=1-\eta\left(1-\frac{H^2}{R}\right)$}
We now examine the second model in the presence of both spatial curvature and interaction.\\
Also in this case, we obtain that the Friedmann equation given in Eq. (\ref{genfri}) is given by:
\begin{eqnarray}
H^2&=&\frac{1}{3M^{2}_{pl}}\left(\rho_{m}+\rho_{GR}\right)-\frac{k}{a^2}
\end{eqnarray}
The fractional energy density  of the generalized holographic dark energy is given by:
\begin{eqnarray}
\Omega_{D_{GR},I,k}&=&c^{2} \left[ (2-\eta) + H_0^2(1-\eta)\Omega_{k0}(1+z)^2- \frac{(1-\eta)(1+z)}{2H_{{GR},I,k}^{2}} \frac{d H_{{GR},I,k}^{2}}{d x} \right].\label{eq:OmegaGR}
\end{eqnarray}
Therefore, the Friedmann equation can we written as:
\begin{eqnarray}
\frac{H^2_{GR,I,k}(z)}{H_0^2}\left\{\frac{c^2(1-\eta)(1+z)}{2H^2_{GR,I,k}}\frac{d
H^2_{GR,I,k}(z)}{dz}\right. \nonumber \\
\left.+1-c^2(2-\eta)\right\}&=&\Omega_{m0}(1+z)^{3(1-d^2)}\nonumber \\
&&-\left[1- c^2(1-\eta)\right]\Omega_{k0}(1+z)^2,
\end{eqnarray}
which has the following solution:
\begin{eqnarray}
h^2_{GR,I,k}(z) &=&   
\frac{2 \, \Omega_{m0}(1+z)^{3(1-d^2)} }{2 \left[ 1 - c^2 (2-\eta) \right] + 3 c^2 (1-\eta)(1 - d^2)} \nonumber  \\
&&+ \left[ \frac{1 + c^2 (1-\eta)}{1 - c^2} \right]\Omega_{k0}(1+z)^2 \nonumber \\
&& +  \left\{ 1 - \frac{2 \, \Omega_{m0}}{2 \left[ 1 - c^2 (2-\eta) \right] + 3 c^2 (1-\eta)(1 - d^2)}\right.\nonumber \\
&&\left.- \left[ \frac{1 + c^2 (1-\eta)}{1 - c^2}  \right]\Omega_{k0}  \right\} (1+z)^{\lambda}. \label{carolina40}
\end{eqnarray}

We can also write Eq (\ref{carolina40}) as it follows:
\begin{eqnarray}
h^2_{GR,I,k}(z) &=&   \Omega_{m0}(1+z)^{3(1-d^2)} -\Omega_{k0}(1+z)^2\nonumber \\
&&\left\{\frac{2  }{2 \left[ 1 - c^2 (2-\eta) \right] + 3 c^2 (1-\eta)(1 - d^2)}-1\right\}  \Omega_{m0}(1+z)^{3(1-d^2)}\nonumber  \\
&&+ \left[ \frac{1 + c^2 (1-\eta)}{1 - c^2} +1\right]\Omega_{k0}(1+z)^2 \nonumber \\
&& +  \left\{ 1 - \frac{2 \, \Omega_{m0}}{2 \left[ 1 - c^2 (2-\eta) \right] + 3 c^2 (1-\eta)(1 - d^2)}\right. \nonumber \\
&&\left.- \left[ \frac{1 + c^2 (1-\eta)}{1 - c^2}  \right]\Omega_{k0}  \right\} (1+z)^{\lambda}. \label{carolina41}
\end{eqnarray}

The form obtained in Eq.~(\ref{carolina41}) is particularly useful because it allows 
$ h_{GR,I,k}^2(z) $ to be written as the sum of three distinct contributions: 
one from dark matter, one from curvature, and one from dark energy. 
The term $ \Omega_{m0} (1+z)^{3(1-d^2)} $ represents the dark matter component, 
while $ \Omega_{k0} (1+z)^2 $ corresponds to the curvature contribution. 
Consequently, the following expression for $ \rho_{D_{GH},k}(z) $ can be obtained:

\begin{eqnarray}
\rho_{D_{GR},I,k}(z) &=&  \left\{\frac{2  }{2 \left[ 1 - c^2 (2-\eta) \right] + 3 c^2 (1-\eta)(1 - d^2)}-1\right\}  \Omega_{m0}(1+z)^{3(1-d^2)}\nonumber  \\
&&+ \left[ \frac{1 + c^2 (1-\eta)}{1 - c^2} +1\right]\Omega_{k0}(1+z)^2 \nonumber \\
&& +  \left\{ 1 - \frac{2 \, \Omega_{m0}}{2 \left[ 1 - c^2 (2-\eta) \right] + 3 c^2 (1-\eta)(1 - d^2)}\right. \nonumber \\
&&\left.- \left[ \frac{1 + c^2 (1-\eta)}{1 - c^2}  \right]\Omega_{k0}  \right\} (1+z)^{\lambda}.
\end{eqnarray}

We now want to write the expression of $h^2_{GR,I,k}$ and $\rho_{D_{GR},I,k}$ as functions of the parameter $x$ obtaining:
\begin{eqnarray}
h^2_{GR,I,k}(x) &=&   \Omega_{m0}e^{-3(1-d^2)x} -\Omega_{k0}e^{-2x}\nonumber \\
&&+\left\{\frac{2  }{2 \left[ 1 - c^2 (2-\eta) \right] + 3 c^2 (1-\eta)(1 - d^2)}-1\right\}  \Omega_{m0}e^{-3(1-d^2)x}\nonumber  \\
&&+ \left[ \frac{1 + c^2 (1-\eta)}{1 - c^2} +1\right]\Omega_{k0}e^{-2x} \nonumber \\
&& +  \left\{ 1 - \frac{2 \, \Omega_{m0}}{2 \left[ 1 - c^2 (2-\eta) \right] + 3 c^2 (1-\eta)(1 - d^2)}\right. \nonumber \\
&&\left.- \left[ \frac{1 + c^2 (1-\eta)}{1 - c^2}  \right]\Omega_{k0}  \right\} e^{-\lambda x},\label{carolina42} \\
\rho_{D_{GR},I,k}(x) &=&  \left\{\frac{2  }{2 \left[ 1 - c^2 (2-\eta) \right] + 3 c^2 (1-\eta)(1 - d^2)}-1\right\}  \Omega_{m0}e^{-3(1-d^2)x}\nonumber  \\
&&+ \left[ \frac{1 + c^2 (1-\eta)}{1 - c^2} +1\right]\Omega_{k0}e^{-2x} \nonumber \\
&& +  \left\{ 1 - \frac{2 \, \Omega_{m0}}{2 \left[ 1 - c^2 (2-\eta) \right] + 3 c^2 (1-\eta)(1 - d^2)}\right. \nonumber \\
&&\left.- \left[ \frac{1 + c^2 (1-\eta)}{1 - c^2}  \right]\Omega_{k0}  \right\} e^{-\lambda x}.\label{carolina43}
\end{eqnarray}
We now want to calculate the expressions of the fractional energy density of DE $\Omega_{D_{GR},I,k}$, the evolutionary form of the fractional energy density of DE $\Omega'_{D_{GR},I,k}$, the pressure $p_{D_{GR},I,k}$, the EoS parameter $\omega_{D_{GR},I,k}$ and the deceleration parameter $q_{GR,I,k}$ for this case.\\
In order to obtain the final expression of the fractional energy density of DE for this case, we use the relation:
\begin{eqnarray}
\Omega_{D_{GR},I,k}(x)&=&c^{2} \left[ (2-\eta) + H_0^2(1-\eta)\Omega_{k0}e^{-2x}+ \frac{(1-\eta)}{2h_{GR,I,k}^{2}} \frac{d h_{GR,I,k}^{2}}{d x} \right].\label{eq:OmegaGR}
\end{eqnarray}

Using the expression of $h^{2}_{GR,I,k}$ derived in Eq. (\ref{carolina42}), we obtain:
\begin{eqnarray}
    \frac{d h^{2}_{GR,I,k}(x)}{d x} &=& -3(1-d^2)\Omega_{m0}e^{-3(1-d^2)x} +2\Omega_{k0}e^{-2x}\nonumber \\
&&-3(1-d^2)\left\{\frac{2  }{2 \left[ 1 - c^2 (2-\eta) \right] + 3 c^2 (1-\eta)(1 - d^2)}-1\right\}  \Omega_{m0}e^{-3(1-d^2)x}\nonumber  \\
&&-2 \left[ \frac{1 + c^2 (1-\eta)}{1 - c^2} +1\right]\Omega_{k0}e^{-2x} \nonumber \\
&& -\lambda  \left\{ 1 - \frac{2 \, \Omega_{m0}}{2 \left[ 1 - c^2 (2-\eta) \right] + 3 c^2 (1-\eta)(1 - d^2)}\right. \nonumber \\
&&\left.- \left[ \frac{1 + c^2 (1-\eta)}{1 - c^2}  \right]\Omega_{k0}  \right\} e^{-\lambda x}. \label{carolina44}
\end{eqnarray}

Therefore, the final expression of $\Omega_{D_{GR},I,k}(x)$ is given by:
\begin{eqnarray}
\Omega_{D_{GR},I,k}(x)&=&c^{2} \left[ (2-\eta) + H_0^2(1-\eta)\Omega_{k0}e^{-2x} \right]\nonumber \\
&&-\frac{c^2(1-\eta)}{2} \cdot \left\{ 3(1-d^2)\Omega_{m0}e^{-3(1-d^2)x} -2\Omega_{k0}e^{-2x}\right.\nonumber \\
&&\left.+3(1-d^2)\left\{\frac{2  }{2 \left[ 1 - c^2 (2-\eta) \right] + 3 c^2 (1-\eta)(1 - d^2)}-1\right\}  \Omega_{m0}e^{-3(1-d^2)x}\right.\nonumber  \\
&&\left.+2 \left[ \frac{1 + c^2 (1-\eta)}{1 - c^2} +1\right]\Omega_{k0}e^{-2x}\right. \nonumber \\
&& \left.+\lambda  \left\{ 1 - \frac{2 \, \Omega_{m0}}{2 \left[ 1 - c^2 (2-\eta) \right] + 3 c^2 (1-\eta)(1 - d^2)}- \left[ \frac{1 + c^2 (1-\eta)}{1 - c^2}  \right]\Omega_{k0}  \right\} e^{-\lambda x}   \right\} \times \nonumber\\
&&\left\{ \Omega_{m0}e^{-3(1-d^2)x} -\Omega_{k0}e^{-2x}\right.\nonumber \\
&&\left.+\left\{\frac{2  }{2 \left[ 1 - c^2 (2-\eta) \right] + 3 c^2 (1-\eta)(1 - d^2)}-1\right\}  \Omega_{m0}e^{-3(1-d^2)x}\right.\nonumber  \\
&&\left.+ \left[ \frac{1 + c^2 (1-\eta)}{1 - c^2} +1\right]\Omega_{k0}e^{-2x} \right.\nonumber \\
&&\left. +  \left\{ 1 - \frac{2 \, \Omega_{m0}}{2 \left[ 1 - c^2 (2-\eta) \right] + 3 c^2 (1-\eta)(1 - d^2)}\right.\right. \nonumber \\
&&\left.\left.- \left[ \frac{1 + c^2 (1-\eta)}{1 - c^2}  \right]\Omega_{k0}  \right\} e^{-\lambda x}  \right\}^{-1}.
\end{eqnarray}

We now want to calculate the evolutionary form of the fractional energy density of DE.\\
If we put $h^2_{GR,I,k} \equiv A_8$, we can write $\Omega_{D_{GR,I,k}}(x)$ in the following form:
\begin{eqnarray}
\Omega_{D_{GR,I,k}}(x)&=&c^{2} \left[ (2-\eta) + H_0^2(1-\eta)\Omega_{k0}e^{-2x} \right] +  \frac{c^2(1-\eta)}{2} \left( \frac{A_8'}{A_8}  \right) .\label{renato-2}
\end{eqnarray}
Differentiating Eq. (\ref{renato-2}) with respect to $x$, we can write:
\begin{eqnarray}
\Omega_{D_{GR},I,k}'&=& \frac{c^2(1-\eta)}{2}\left[ \frac{A_8''}{A_8}-\left( \frac{A_8'}{A_8}  \right)^2 \right]  -2c^2H_0^2(1-\eta)\Omega_{k0}e^{-2x},\label{renato-2prime}
\end{eqnarray}
where $A_8''= \frac{d^2 h_{GR,I,k}^2(x)}{dx^2}$. \\
Starting from the general expression of $h_{GR,I,k}^2(x)$ provided in Eq.~(\ref{carolina42}), or, equivalently, from the differential form $\frac{d h_{GR,I,k}^2(x)}{dx}$ obtained in Eq.~(\ref{carolina44}), we may express:
\begin{eqnarray}
    \frac{d^2 h^{2}_{GR,I,k}(x)}{d x^2} &=& 9(1-d^2)^2\Omega_{m0}e^{-3(1-d^2)x} -4\Omega_{k0}e^{-2x}\nonumber \\
&&+9(1-d^2)^2\left\{\frac{2  }{2 \left[ 1 - c^2 (2-\eta) \right] + 3 c^2 (1-\eta)(1 - d^2)}-1\right\}  \Omega_{m0}e^{-3(1-d^2)x}\nonumber  \\
&&+4 \left[ \frac{1 + c^2 (1-\eta)}{1 - c^2} +1\right]\Omega_{k0}e^{-2x} \nonumber \\
&& +\lambda^2  \left\{ 1 - \frac{2 \, \Omega_{m0}}{2 \left[ 1 - c^2 (2-\eta) \right] + 3 c^2 (1-\eta)(1 - d^2)}\right.\nonumber \\
&&\left.- \left[ \frac{1 + c^2 (1-\eta)}{1 - c^2}  \right]\Omega_{k0}  \right\} e^{-\lambda x} .
\end{eqnarray}
Therefore, we obtain the following expression for $\Omega_{D_{GR},I,k}(x)'$:
\begin{eqnarray}
\Omega_{D_{GR},I,k}(x)'&=&c^{2} \frac{c^2(1-\eta)}{2} \cdot \left\{ 9(1-d^2)\Omega_{m0}e^{-3(1-d^2)x} -4\Omega_{k0}e^{-2x}\right.\nonumber \\
&&\left.+9(1-d^2)^2\left\{\frac{2  }{2 \left[ 1 - c^2 (2-\eta) \right] + 3 c^2 (1-\eta)(1 - d^2)}-1\right\}  \Omega_{m0}e^{-3(1-d^2)x}\right.\nonumber  \\
&&\left.+4 \left[ \frac{1 + c^2 (1-\eta)}{1 - c^2} +1\right]\Omega_{k0}e^{-2x}\right. \nonumber \\
&& \left.+\lambda ^2 \left\{ 1 - \frac{2 \, \Omega_{m0}}{2 \left[ 1 - c^2 (2-\eta) \right] + 3 c^2 (1-\eta)(1 - d^2)}- \left[ \frac{1 + c^2 (1-\eta)}{1 - c^2}  \right]\Omega_{k0}  \right\} e^{-\lambda x}   \right\} \times \nonumber\\
&&\left\{ \Omega_{m0}e^{-3(1-d^2)x} -\Omega_{k0}e^{-2x}\right.\nonumber \\
&&\left.+\left\{\frac{2  }{2 \left[ 1 - c^2 (2-\eta) \right] + 3 c^2 (1-\eta)(1 - d^2)}-1\right\}  \Omega_{m0}e^{-3(1-d^2)x}\right.\nonumber  \\
&&\left.+ \left[ \frac{1 + c^2 (1-\eta)}{1 - c^2} +1\right]\Omega_{k0}e^{-2x} \right.\nonumber \\
&&\left. +  \left\{ 1 - \frac{2 \, \Omega_{m0}}{2 \left[ 1 - c^2 (2-\eta) \right] + 3 c^2 (1-\eta)(1 - d^2)}- \left[ \frac{1 + c^2 (1-\eta)}{1 - c^2}  \right]\Omega_{k0}  \right\} e^{-\lambda x}  \right\}^{-1} \nonumber \\
&&-\frac{c^2(1-\eta)}{2} \cdot \left\{ 3(1-d^2)\Omega_{m0}e^{-3(1-d^2)x} -2\Omega_{k0}e^{-2x}\right.\nonumber \\
&&\left.+3(1-d^2)\left\{\frac{2  }{2 \left[ 1 - c^2 (2-\eta) \right] + 3 c^2 (1-\eta)(1 - d^2)}-1\right\}  \Omega_{m0}e^{-3(1-d^2)x}\right.\nonumber  \\
&&\left.+2 \left[ \frac{1 + c^2 (1-\eta)}{1 - c^2} +1\right]\Omega_{k0}e^{-2x}\right. \nonumber \\
&& \left.+\lambda  \left\{ 1 - \frac{2 \, \Omega_{m0}}{2 \left[ 1 - c^2 (2-\eta) \right] + 3 c^2 (1-\eta)(1 - d^2)}- \left[ \frac{1 + c^2 (1-\eta)}{1 - c^2}  \right]\Omega_{k0}  \right\} e^{-\lambda x}   \right\}^2 \times \nonumber\\
&&\left\{ \Omega_{m0}e^{-3(1-d^2)x} -\Omega_{k0}e^{-2x}\right.\nonumber \\
&&\left.+\left\{\frac{2  }{2 \left[ 1 - c^2 (2-\eta) \right] + 3 c^2 (1-\eta)(1 - d^2)}-1\right\}  \Omega_{m0}e^{-3(1-d^2)x}\right.\nonumber  \\
&&\left.+ \left[ \frac{1 + c^2 (1-\eta)}{1 - c^2} +1\right]\Omega_{k0}e^{-2x} \right.\nonumber \\
&&\left. +  \left\{ 1 - \frac{2 \, \Omega_{m0}}{2 \left[ 1 - c^2 (2-\eta) \right] + 3 c^2 (1-\eta)(1 - d^2)}- \left[ \frac{1 + c^2 (1-\eta)}{1 - c^2}  \right]\Omega_{k0}  \right\} e^{-\lambda x}  \right\}^{-2}\nonumber \\
&& -2c^2H_0^2(1-\eta)\Omega_{k0}e^{-2x}.
\end{eqnarray}

We now want to obtain the final expression of the pressure $p_{D_{GR},I,k}(x)$. The general expression of $p_{D_{GR},I,k}$ is given by:
\begin{eqnarray}
\label{eq:conserve}
p_{D_{GR}I,k}(x)&=& -\rho_{D_{GR}I,k}(x)-\frac{\rho'_{D_{GR}I,k}(x)}{3}-\frac{Q}{3H}.
\end{eqnarray}
Using the expression of $\rho_{D_{GR}I,k}(x)$ given in Eq. (\ref{carolina43}), we have that $\rho'_{D_{GR},I,k}$ is given by the following relation:
\begin{eqnarray}
\rho'_{D_{GR}I,k}(x)&=&-3(1-d^2) \left\{\frac{2  }{2 \left[ 1 - c^2 (2-\eta) \right] + 3 c^2 (1-\eta)(1 - d^2)}-1\right\}  \Omega_{m0}e^{-3(1-d^2)x}\nonumber  \\
&&-2 \left[ \frac{1 + c^2 (1-\eta)}{1 - c^2} +1\right]\Omega_{k0}e^{-2x} \nonumber \\
&& -\lambda  \left\{ 1 - \frac{2 \, \Omega_{m0}}{2 \left[ 1 - c^2 (2-\eta) \right] + 3 c^2 (1-\eta)(1 - d^2)}\right. \nonumber \\
&&\left.- \left[ \frac{1 + c^2 (1-\eta)}{1 - c^2}  \right]\Omega_{k0}  \right\} e^{-\lambda x}. \label{carolina45}
\end{eqnarray}

Therefore, we obtain the following relation for $p_{D_{GR},I,k} (x)$:
\begin{eqnarray}
p_{D_{GR},I,k} (x)&=& -d^2\left\{\frac{2  }{2 \left[ 1 - c^2 (2-\eta) \right] + 3 c^2 (1-\eta)(1 - d^2)}\right\}  \Omega_{m0}e^{-3(1-d^2)x}\nonumber  \\
&&-\frac{1}{3}\left[ \frac{1 + c^2 (1-\eta)}{1 - c^2} +1\right]\Omega_{k0}e^{-2x} \nonumber \\
&& +\left(\frac{\lambda -3}{3}\right) \left\{ 1 - \frac{2 \, \Omega_{m0}}{2 \left[ 1 - c^2 (2-\eta) \right] + 3 c^2 (1-\eta)(1 - d^2)}\right.\nonumber \\
&&\left.- \left[ \frac{1 + c^2 (1-\eta)}{1 - c^2}  \right]\Omega_{k0}  \right\} e^{-\lambda x}.
\end{eqnarray}

We now want to calculate the EoS parameter for this case.\\
Using the general definition of $\omega_D$, we obtain the following expression for this case:
\begin{eqnarray}
\omega_{D_{GR},I,k}(x) = -1 - \frac{ \rho_{D_{GR},I,k}'(x)  }{3\rho_{D_{GR},I,k}(x) }- \frac{Q}{3H\rho_{D_{GR},I,k}(x)}.
\end{eqnarray}
Using the relations of $\rho_{D_{GR},I,k}(x)$ and $\rho'_{D_{GR},I,k}(x)$ we derived in Eqs. (\ref{carolina43}) and (\ref{carolina45}), we can write:
\begin{eqnarray}
\omega_{D_{GR}I,k} (x)&=& -1 + \frac{1}{3} \cdot \left\{ 3(1-d^2) \left\{\frac{2  }{2 \left[ 1 - c^2 (2-\eta) \right] + 3 c^2 (1-\eta)(1 - d^2)}-1\right\}  \Omega_{m0}e^{-3(1-d^2)x}\right.\nonumber  \\
&&\left.+2 \left[ \frac{1 + c^2 (1-\eta)}{1 - c^2} +1\right]\Omega_{k0}e^{-2x} \right.\nonumber \\
&& \left.+\lambda  \left\{ 1 - \frac{2 \, \Omega_{m0}}{2 \left[ 1 - c^2 (2-\eta) \right] + 3 c^2 (1-\eta)(1 - d^2)}- \left[ \frac{1 + c^2 (1-\eta)}{1 - c^2}  \right]\Omega_{k0}  \right\} e^{-\lambda x} 
\right\}\times \nonumber \\
&&\left\{   \left\{\frac{2  }{2 \left[ 1 - c^2 (2-\eta) \right] + 3 c^2 (1-\eta)(1 - d^2)}-1\right\}  \Omega_{m0}e^{-3(1-d^2)x}\right.\nonumber  \\
&&\left.+ \left[ \frac{1 + c^2 (1-\eta)}{1 - c^2} +1\right]\Omega_{k0}e^{-2x} \right.\nonumber \\
&&\left. +  \left\{ 1 - \frac{2 \, \Omega_{m0}}{2 \left[ 1 - c^2 (2-\eta) \right] + 3 c^2 (1-\eta)(1 - d^2)}- \left[ \frac{1 + c^2 (1-\eta)}{1 - c^2}  \right]\Omega_{k0}  \right\} e^{-\lambda x}              \right\}^{-1}\nonumber \\
&&-d^2  \Omega_{m0}e^{-3(1-d^2)x} \times \nonumber \\
&&\left\{   \left\{\frac{2  }{2 \left[ 1 - c^2 (2-\eta) \right] + 3 c^2 (1-\eta)(1 - d^2)}-1\right\}  \Omega_{m0}e^{-3(1-d^2)x}\right.\nonumber  \\
&&\left.+ \left[ \frac{1 + c^2 (1-\eta)}{1 - c^2} +1\right]\Omega_{k0}e^{-2x} \right.\nonumber \\
&&\left. +  \left\{ 1 - \frac{2 \, \Omega_{m0}}{2 \left[ 1 - c^2 (2-\eta) \right] + 3 c^2 (1-\eta)(1 - d^2)}\right. \right.\nonumber \\
&&\left. \left.- \left[ \frac{1 + c^2 (1-\eta)}{1 - c^2}  \right]\Omega_{k0}  \right\} e^{-\lambda x}              \right\}^{-1}.
\end{eqnarray}

Also in this case, in order to find the final expression of $q_{GR,I,k}$, we use the general expression:
\begin{eqnarray}
q_{GR,I,k}(x) &=&  -1 - \frac{1}{2 h_{GR,I,k}^2(x)} \frac{d h_{GR,I,k}^2(x)}{dx}. \label{deceleration}
\end{eqnarray}
Using the expression of $ \frac{d h_{GR,I,k}^2(x)}{dx}$ we derived in Eq. (\ref{carolina42}) along with the expression of $h_{GR,I,k}^2$ obtained in Eq. (\ref{carolina44}), we can write:
\begin{eqnarray}
q_{GR,I,k}(x) &=&  -1+\frac{1}{2}\cdot \left\{  3(1-d^2)\Omega_{m0}e^{-3(1-d^2)x} +2\Omega_{k0}e^{-2x}\right.\nonumber \\
&&\left.+3(1-d^2)\left\{\frac{2  }{2 \left[ 1 - c^2 (2-\eta) \right] + 3 c^2 (1-\eta)(1 - d^2)}-1\right\}  \Omega_{m0}e^{-3(1-d^2)x}\right.\nonumber  \\
&&\left.+2 \left[ \frac{1 + c^2 (1-\eta)}{1 - c^2} +1\right]\Omega_{k0}e^{-2x} \right.\nonumber \\
&& \left.\lambda  \left\{ 1 - \frac{2 \, \Omega_{m0}}{2 \left[ 1 - c^2 (2-\eta) \right] + 3 c^2 (1-\eta)(1 - d^2)}- \left[ \frac{1 + c^2 (1-\eta)}{1 - c^2}  \right]\Omega_{k0}  \right\} e^{-\lambda x}     \right\}\times \nonumber \\
&& \left\{ \Omega_{m0}e^{-3(1-d^2)x} -\Omega_{k0}e^{-2x}\right.\nonumber \\
&&\left.+\left\{\frac{2  }{2 \left[ 1 - c^2 (2-\eta) \right] + 3 c^2 (1-\eta)(1 - d^2)}-1\right\}  \Omega_{m0}e^{-3(1-d^2)x}\right.\nonumber  \\
&&\left.+ \left[ \frac{1 + c^2 (1-\eta)}{1 - c^2} +1\right]\Omega_{k0}e^{-2x}\right. \nonumber \\
&&\left. +  \left\{ 1 - \frac{2 \, \Omega_{m0}}{2 \left[ 1 - c^2 (2-\eta) \right] + 3 c^2 (1-\eta)(1 - d^2)} \right.\right.\nonumber \\
&&\left. \left.- \left[ \frac{1 + c^2 (1-\eta)}{1 - c^2}  \right]\Omega_{k0}  \right\} e^{-\lambda x}  \right\}^{-1}.
\end{eqnarray}

\section{Statefinder Diagnostic}
The investigation of key cosmological quantities — such as the equation of state (EoS) parameter $ \omega_D $, the Hubble parameter $ H $, and the deceleration parameter $ q $ — has drawn considerable attention in modern cosmology. It is well established that a wide range of dark energy (DE) models typically predict a positive Hubble parameter and a negative deceleration parameter at the present epoch (i.e., $ H > 0 $ and $ q < 0 $ for $ t = t_0 $). As a result, these parameters alone are not sufficient to effectively discriminate among the different DE models under consideration.

To achieve a deeper understanding and more accurate characterization of DE models, it becomes necessary to consider higher-order derivatives of the scale factor $ a(t) $ with respect to cosmic time. In this context, Sahni et al.~\cite{sah} and Alam et al.~\cite{alam} introduced the so-called statefinder diagnostic — a pair of parameters $ \{r, s\} $ involving the third derivative of the scale factor $ a(t) $. The goal of this diagnostic is to go beyond the degeneracy associated with $ H $ and $ q $, providing a more refined tool to distinguish between various DE scenarios, especially at the present stage of cosmic evolution.

The general expressions of $r$ and $s$ can be expressed as functions of different cosmological quantities. \\
We start with the one involving the third derivative of the scale factor $a(t)$.\\
In this case, the expressions of the pair $\left\{ r-s \right\}$  are given, respectively, by:
\begin{eqnarray}
r &=& \frac{\dot{\ddot{a}}} {aH^3},  \label{r1}\\
s &=&   \frac{r -1}{3\left(q-1/2\right)},   \label{s1}
\end{eqnarray}
where $q$ indicates the deceleration parameter.\\
It is evident that the expression for $ s$ depends directly on the expression for $ r $.\\
An useful and alternative method to express the statefinder parameters using the Hubble parameter $H$ and its time derivatives is:
\begin{eqnarray}
r&=& 1 + 3\left(\frac{\dot{H}} {H^2}\right)+ \frac{\ddot{H}} {H^3},  \label{r2}\\
s&=& -\frac{3H\dot{H}+\ddot{H}} {3H\left( 2\dot{H}+3H^2  \right)} \nonumber \\
&=&  -\frac{3\dot{H}+\ddot{H}/H}{3\left( 2\dot{H}+3H^2  \right)}. \label{s2}
\end{eqnarray}

We can also write $r$ and $s$ as functions of the total energy density $\rho=\rho_m + \rho_D+\rho_k$ and of the total pressure $p=p_D$ (and their first derivatives) in the following way:
\begin{eqnarray}
r &=&  1+ \frac{9}{2}\left(\frac{ \rho_{tot} + p_{tot}   }{\rho_{tot} }\right)\left(\frac{\dot{p}_{tot} } {\dot{\rho}_{tot} } \right)\nonumber  \\
&=&1+ \frac{9}{2}\left(\frac{ \rho_{tot} + p_{tot}   }{\rho_{tot} }\right)\left(\frac{p'_{tot} } {\rho'_{tot} } \right), \label{rgen}\\
s &=& \left(\frac{ \rho_{tot}  + p_{tot}   }{p_{tot} }\right)\left(\frac{\dot{p}_{tot} } {\dot{\rho}_{tot} } \right) \nonumber \\
&=& \left(\frac{ \rho_{tot}  + p_{tot}   }{p_{tot} }\right)\left(\frac{p'_{tot} } {\rho'_{tot} } \right), \label{sgen}
\end{eqnarray}
which are equivalent to:
\begin{eqnarray}
r &=& 1+ \frac{9}{2}\left(\frac{ \rho_D + \rho_m +\rho_k+ p_D  }{\rho_D + \rho_m}\right)\left(\frac{\dot{p}_D}{\dot{\rho}_D +\dot{\rho}_m+\dot{\rho}_k}\right)   \nonumber\\
&=&1+ \frac{9}{2}\left(\frac{ \rho_D + \rho_m +\rho_k+ p_D  }{\rho_D + \rho_m}\right)\left(\frac{p'_D}{\rho'_D +\rho'_m+\rho'_k}\right),\label{rgen1-}\\
s &=& \left(\frac{ \rho_D + \rho_m +\rho_k+ p_D  }{p_D}\right)\left(\frac{\dot{p}_D}{\dot{\rho}_D +\dot{\rho}_m++\dot{\rho}_k}\right)    \nonumber \\
&=& \left(\frac{ \rho_D + \rho_m + \rho_k+p_D  }{p_D}\right)\left(\frac{p'_D}{\rho'_D +\rho'_m+\rho'_k}\right). \label{sgen1-}
\end{eqnarray}
In this work, we will use the expressions given in Eqs. (\ref{rgen1-}) and (\ref{sgen1-}) since they involve quantities we already obtained in previous Sections.\\
One of the most important properties of the statefinder parameters $ r $ and $ s $ is that the point with coordinates $ \{r, s\} = \{1, 0\} $ in the $ r\text{–}s $ plane corresponds to the spatially flat $ \Lambda $CDM model \cite{huang}. Therefore, deviations of a given dark energy (DE) model from this fixed point provide a useful measure of the "distance" of that model from the standard $ \Lambda $CDM Cosmology.\\
Moreover, it is important to emphasize that in the $ r\text{–}s $ plane, a positive value of the statefinder parameter $ s $ (i.e., $ s > 0 $) indicates a quintessence-like DE model, while a negative value (i.e., $ s < 0 $) is associated with a phantom-like DE model. Transitions between these regimes—for instance, from phantom to quintessence or vice versa—are characterized by a crossing of the point $ \{r, s\} = \{1, 0\} $ in the $ r\text{–}s $ plane \cite{wu1}.

Several dark energy models — such as braneworld scenarios, the cosmological constant $ \Lambda_{\text{CC}} $, Chaplygin gas, and quintessence — have been thoroughly examined using the statefinder diagnostic in the work by Alam et al.~\cite{alam}, where it was shown that the statefinder pair $ \{r, s\} $ can effectively discriminate among these different cosmologies. In addition, the statefinder diagnostic for the $ f(T) $ modified gravity model was studied by Wu and Yu \cite{wu1}.

We now proceed to investigate the statefinder diagnostic for the models under consideration.

We already derived the expression for $p_D$ and $\rho_D$. For $\rho_m$
we obtain:
\begin{eqnarray}
    \rho_m = \Omega_{m0}e^{-3x},
\end{eqnarray}
which leads to
\begin{eqnarray}
    \rho'_m = -3\Omega_{m0}e^{-3x}. \label{}
\end{eqnarray}

For the interacting case, we have:
\begin{eqnarray}
    \rho_{m,I}= \Omega_{m0}e^{-3(1-d^2)x},
\end{eqnarray}
which leads to
\begin{eqnarray}
    \rho'_{m,I} = -3(1-d^2)\Omega_{m0}e^{-3(1-d^2)x}.
\end{eqnarray}
We start considering the first model of this paper. \\
Considering the expression of $p_{D_{GH}}$ given in Eq. (\ref{murmura}), we obtain:
\begin{eqnarray}
p_{D_{GH}}'(x)&=&-2(1-\alpha)\left[ \frac{2\left(1 - \alpha\right)}{3}-1  \right] \left[1 - \frac{2 \Omega_{m0}}{2 + c^2(\epsilon - 2)} \right] e^{-2\left(1 - \alpha\right)x} .\label{murmuraprima}
\end{eqnarray}
Therefore, we can write:
\begin{eqnarray}
r_{GH} &=&1+ \frac{9}{2}\left(\frac{ \rho_{D_{GH}} + \rho_m + p_{D_{GH}}  }{\rho_{D_{GH}} + \rho_m}\right)\left(\frac{p'_{D_{GH}}}{\rho'_{D_{GH}} +\rho'_m}\right),\label{rgen1}\\
s_{GH} &=& \left(\frac{ \rho_{D_{GH}} + \rho_m + p_{D_{GH}}  }{p_{D_{GH}}}\right)\left(\frac{p'_{D_{GH}}}{\rho'_{D_{GH}} +\rho'_m}\right). \label{sgen1}
\end{eqnarray}

\begin{figure}[htbp]
    \centering
    \begin{subfigure}{0.8\textwidth}
        \includegraphics[width=0.6\textwidth]{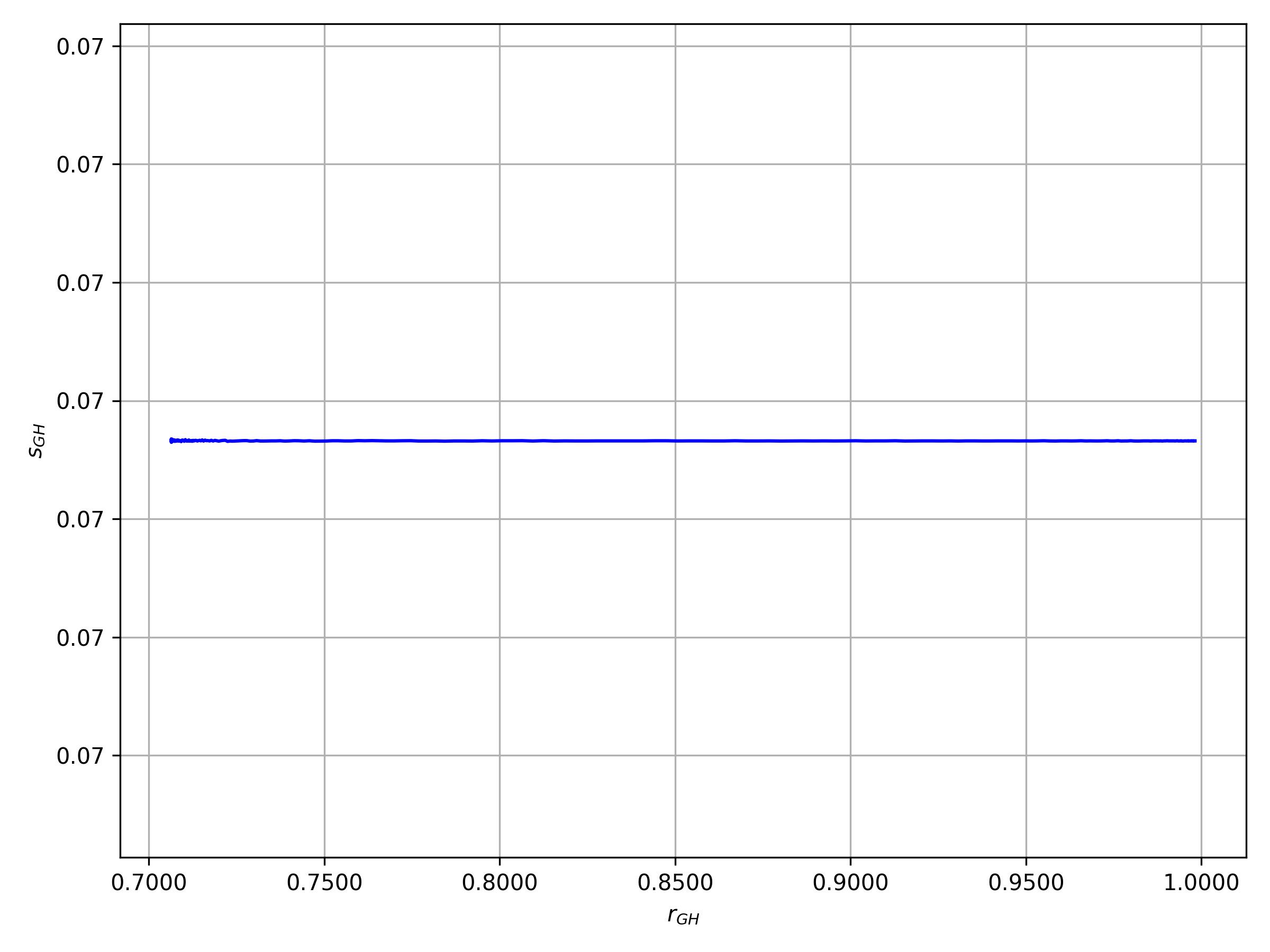}
        \caption{Plot of the $\{ r_{GH}, s_{GH} \}$ pair for $c^2=0.46$.}
        \label{state1}
    \end{subfigure}\\[0.5cm]
    \begin{subfigure}{0.8\textwidth}
        \includegraphics[width=0.6\textwidth]{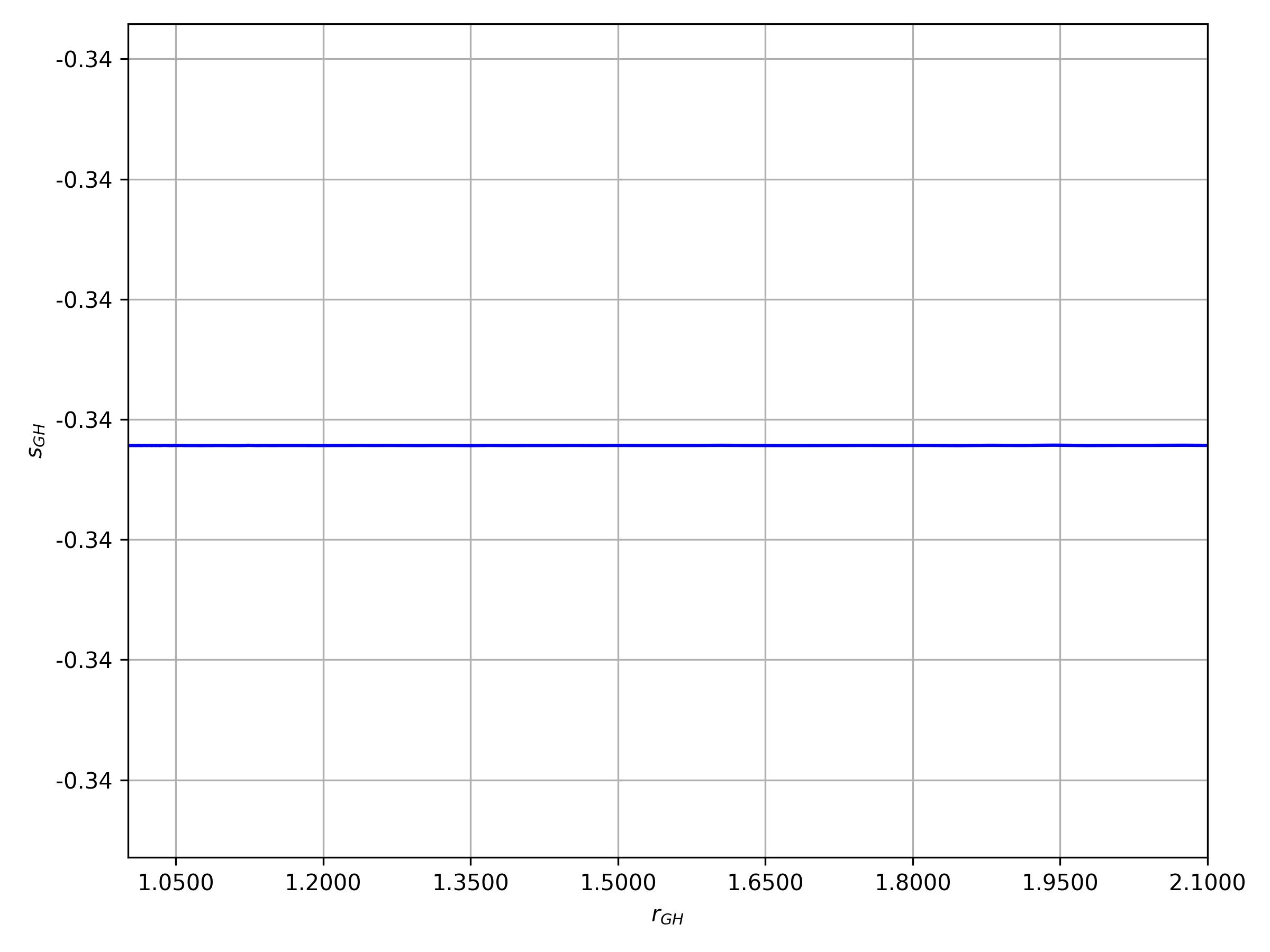}
        \caption{Plot of the $\{ r_{GH}, s_{GH} \}$ pair for $c=0.579$.}
        \label{state1-2}
    \end{subfigure}\\[0.5cm]
    \begin{subfigure}{0.8\textwidth}
        \includegraphics[width=0.6\textwidth]{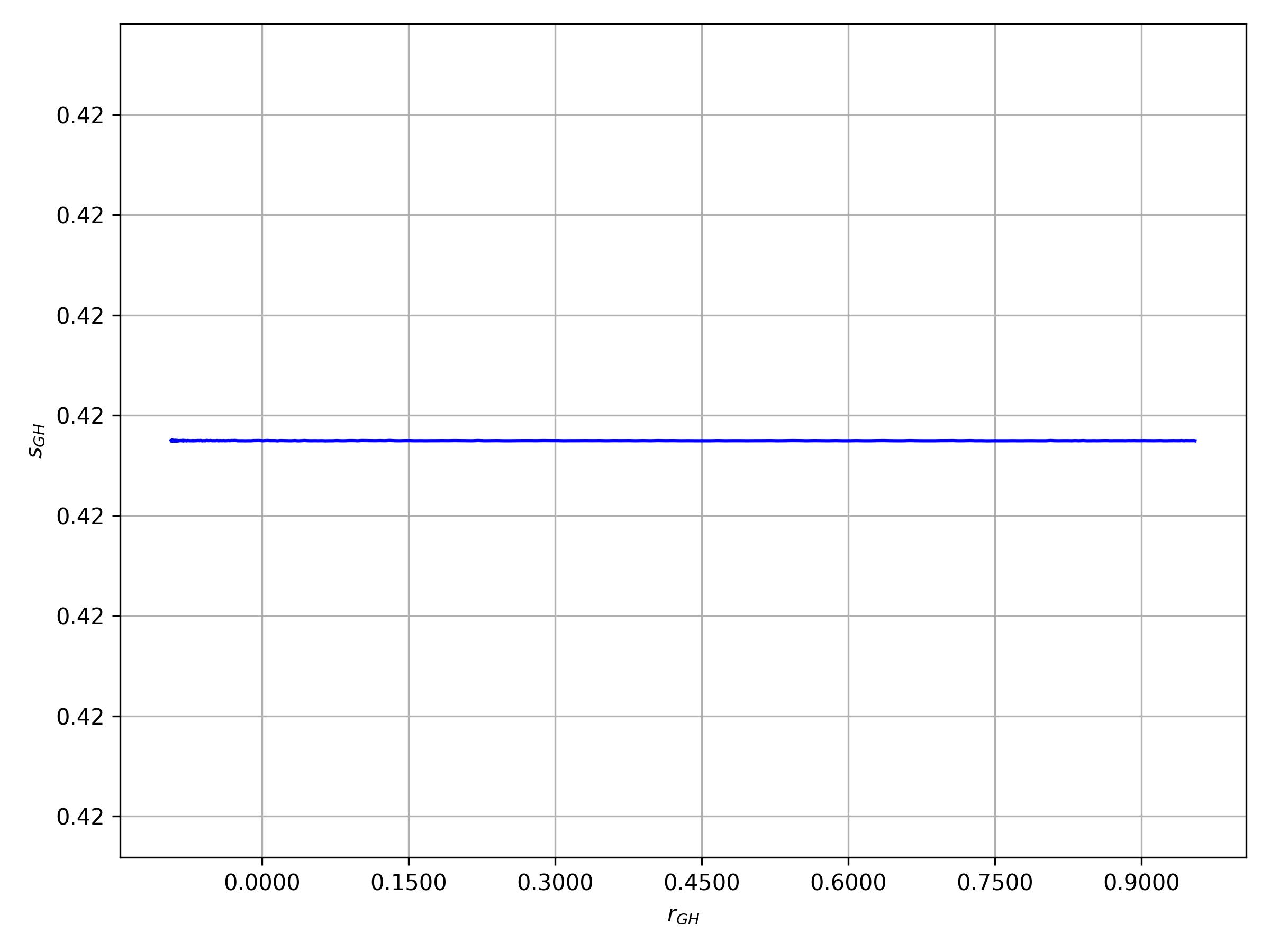}
        \caption{Plot of the $\{ r_{GH}, s_{GH} \}$ pair for $c=0.818$.}
        \label{state1-3}
    \end{subfigure}
    \caption{Comparison of the three cases for the statefinder pair $\{ r_{GH}, s_{GH} \}$.}
    \label{fig:state_all}
\end{figure}

In Figs. (\ref{state1}), (\ref{state1-2}) and (\ref{state1-3}) we plot the expressions of $r_{GH}$ and $s_{GH}$ given in Eqs. (\ref{rgen1}) and (\ref{sgen1}) for $c^2=0.46$, $c=0.579$ and $c=0.818$, respectively.

We obtain that $\left\{ r_{GH,0},s_{GH,0}  \right\} \approx \left\{ 0.8155, 0.0702 \right\}$ for $c^2=0.46$, $\left\{ r_{GH,0},s_{GH,0}  \right\} \approx \left\{ 2.3323,-0.3409 \right\}$ for $c=0.579$ and $\left\{ r_{GH,0},s_{GH,0}  \right\} \approx \left\{0.3525,0.4154  \right\}$ for $c=0.818$.\\
The trajectories shown in (\ref{state1}), (\ref{state1-2}) and (\ref{state1-3}) in the $\left\{r,s\right\}$ plane display a distinctive behavior 
characterized by a constant value of $s \simeq $ while the parameter $r$ 
evolves. This corresponds to a horizontal line in the $\left\{r,s\right\}$ plane, in contrast with the $\Lambda$CDM model, which 
is represented by the fixed point $(r=1, s=0)$. The fact that $s$ remains 
constant throughout cosmic evolution indicates that this model does not 
reproduce $\Lambda$CDM at any epoch, but instead retains a characteristic 
signature in the $(r,s)$ plane. Such a deviation provides a potential 
observational handle to discriminate the model from standard cosmology and 
from alternative dark energy scenarios, since the persistence of $s \neq 0$ 
constitutes a clear departure from the concordance paradigm.

We now study the case where spatial curvature is present.
\\Considering the expression of the pressure $p_{D_{GH},k}(x)$ given in Eq. (\ref{carolina50}), we can write:
\begin{eqnarray}
p_{D_{GH},k}'(x) &=&
 \frac{2}{3}\left[\frac{c^2(\epsilon - 1)}{1-c^2}\right]\Omega_{k0}e^{-2x}\nonumber \\
&&+\frac{2(1-\alpha)(1+2\alpha)}{3}\times\nonumber \\
&&\left[  1 - \frac{2\Omega_{m0}}{2 + c^2(\epsilon - 2)} 
+\left(\frac{1-\epsilon c^2}{1-c^2}\right) \Omega_{k0}(1+z)^2  \right] \,e^{-2\left(1-\alpha \right)} .\label{ciccio1}
\end{eqnarray}
In this case, for $\rho_k$ we can write:
\begin{eqnarray}
    \rho_k= \Omega_{k0}e^{-2x},
\end{eqnarray}
which leads to:
\begin{eqnarray}
    \rho'_k= -2\Omega_{k0}e^{-2x}.\label{omkprime}
\end{eqnarray}
Therefore, we can write:
\begin{eqnarray}
r_{GH,k} &=&1+ \frac{9}{2}\left(\frac{ \rho_{D_{GH,k}} + \rho_m +\rho_k+ p_{D_{GH,k}}  }{\rho_{D_{GH,k}} + \rho_m}\right)\left(\frac{p'_{D_{GH},k}}{\rho'_{D_{GH},k} +\rho'_m+\rho'_k}\right),\label{rgen2}\\
s_{GH,k} &=& \left(\frac{ \rho_{D_{GH},k} + \rho_m +\rho_k+ p_{D_{GH},k}  }{p_{D_{GH},k}}\right)\left(\frac{p'_{D_{GH},k}}{\rho'_{D_{GH},k} +\rho'_m+\rho'_k}\right). \label{sgen2}
\end{eqnarray}

\begin{figure}[htbp]
    \centering
    \begin{subfigure}{0.8\textwidth}
        \includegraphics[width=0.6\textwidth]{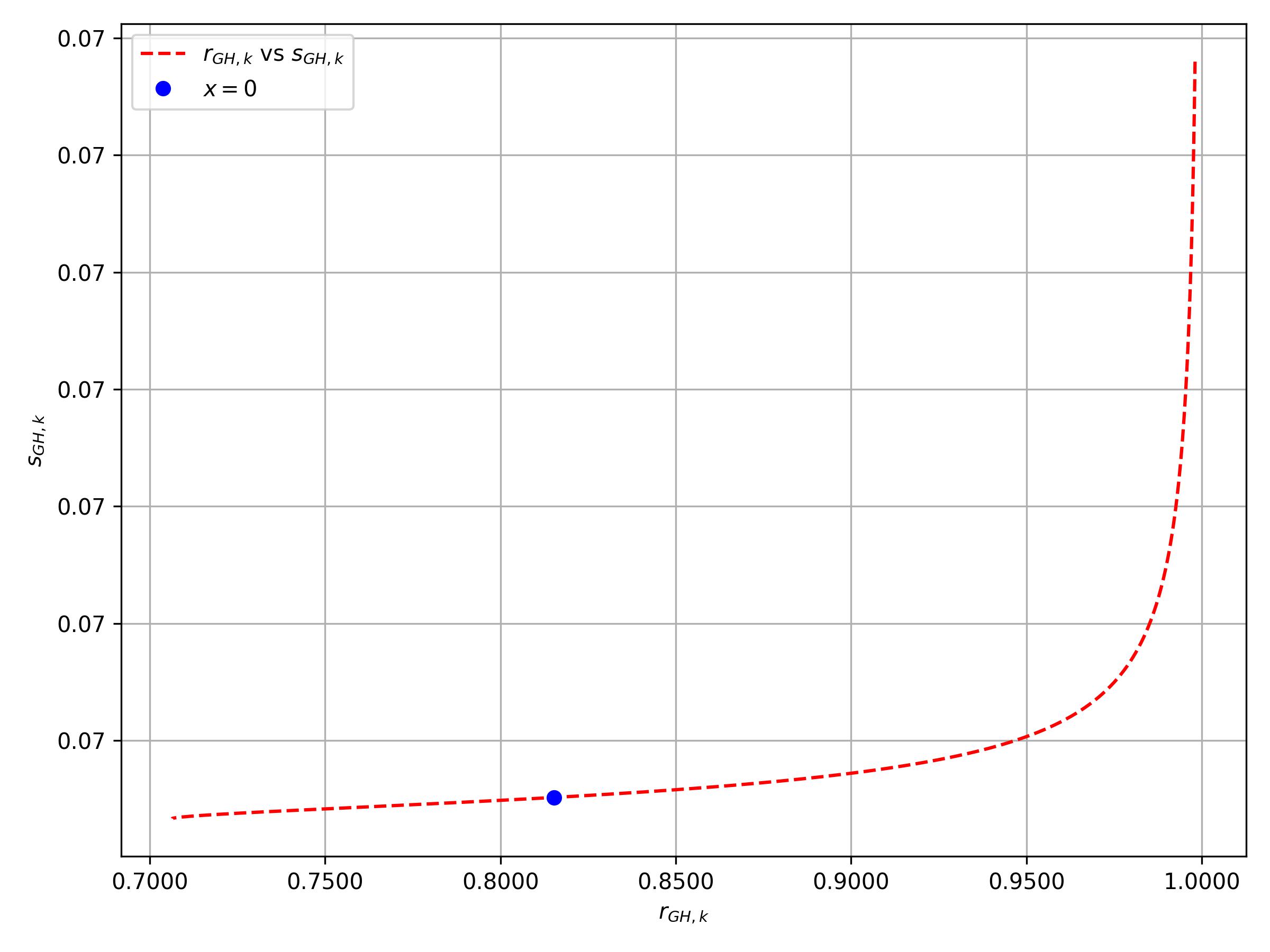}
        \caption{Plot of the $\{ r_{GH,k}, s_{GH,k} \}$ pair for $c^2=0.46$.}
        \label{state2}
    \end{subfigure}\\[0.5cm]
    \begin{subfigure}{0.8\textwidth}
        \includegraphics[width=0.6\textwidth]{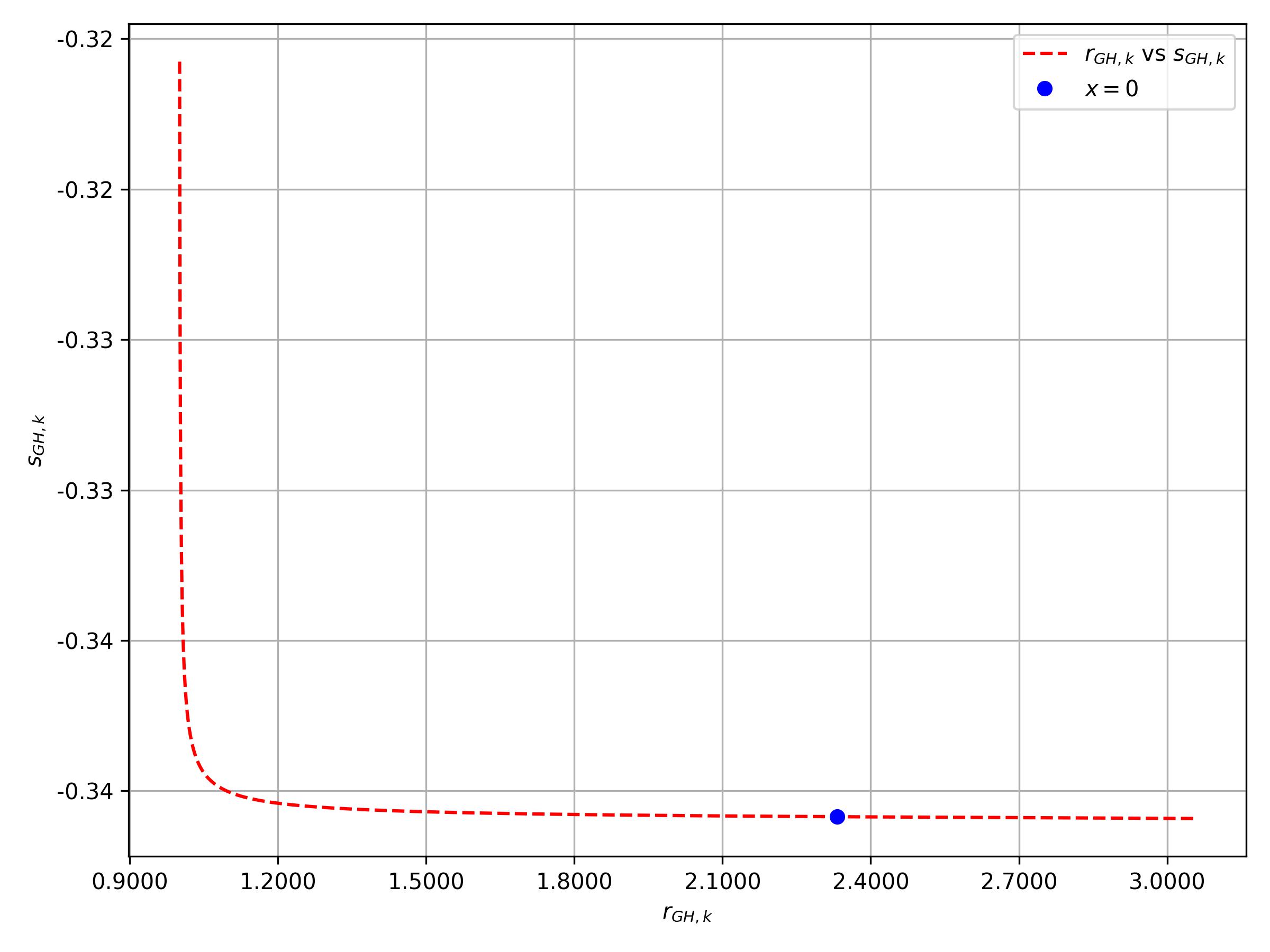}
        \caption{Plot of the $\{ r_{GH,k}, s_{GH,k} \}$ pair for $c=0.579$.}
        \label{state2-2}
    \end{subfigure}\\[0.5cm]
    \begin{subfigure}{0.8\textwidth}
        \includegraphics[width=0.6\textwidth]{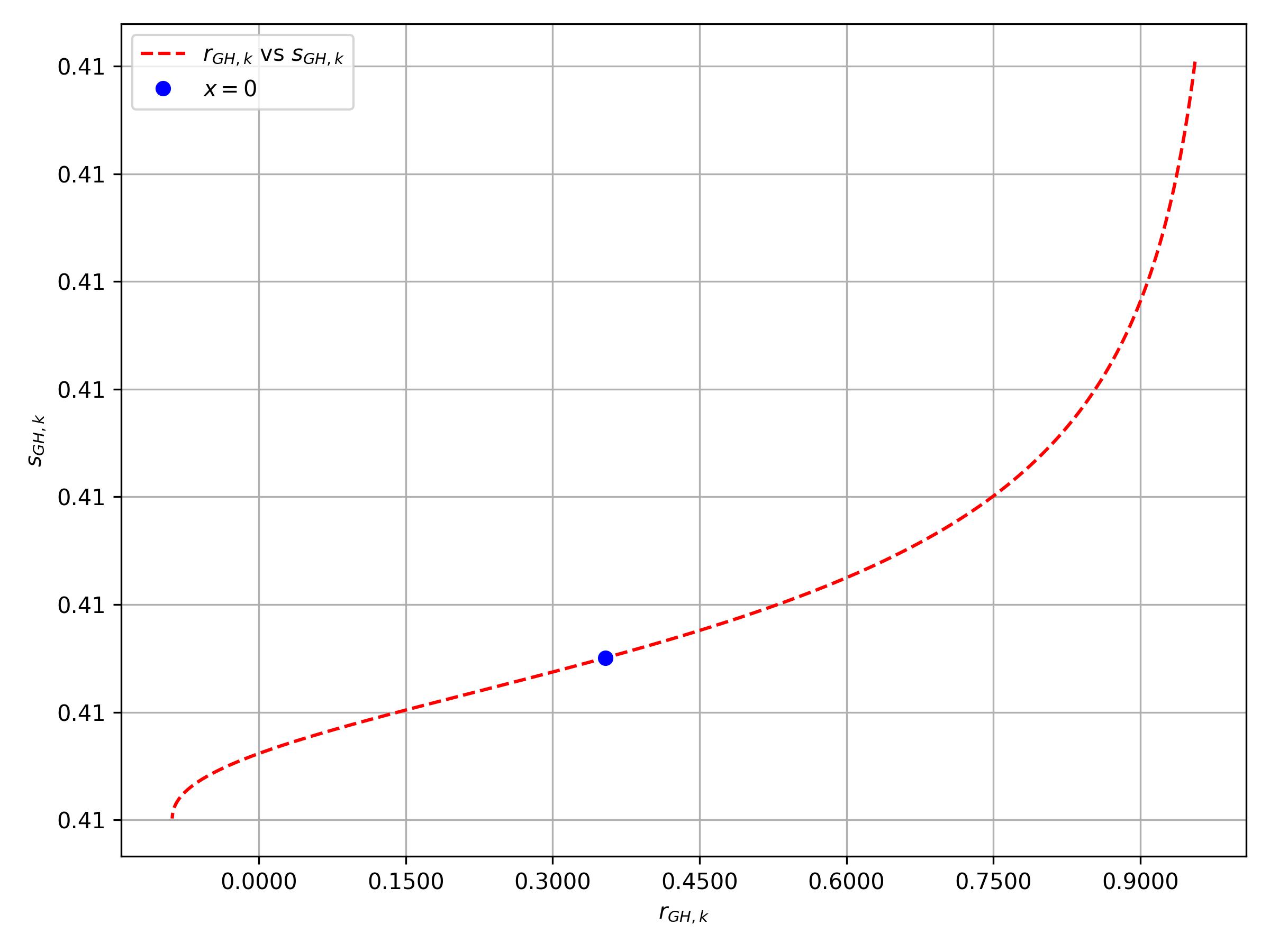}
        \caption{Plot of the $\{ r_{GH,k}, s_{GH,k} \}$ pair for $c=0.815$.}
        \label{state2-3}
    \end{subfigure}
    \caption{Comparison of the three cases for the statefinder pair $\{ r_{GH,k}, s_{GH,k} \}$.}
    \label{fig:state_all2}
\end{figure}

In Figs. (\ref{state2}), (\ref{state2-2}) and (\ref{state2-3}) we plot the expressions of $r_{GH,k}$ and $s_{GH,k}$ given in Eqs. (\ref{rgen2}) and (\ref{sgen2}) for $c^2=0.46$, $c=0.579$ and $c=0.815$, respectively.

We obtain that $\left\{ r_{GH,k,0},s_{GH,k,0}  \right\} \approx \left\{  0.8152,0.0703\right\}  $ for $c^2=0.46$, $\left\{ r_{GH,k,0},s_{GH,k,0}  \right\} \approx \left\{  2.3325,-0.3409\right\} $ for $c=0.579$ and $\left\{ r_{GH,k,0},s_{GH,k,0}  \right\} \approx \left\{0.3537,0.4100 \right\} $ for $c=0.815$.
For $c^2=0.46$, in the $\left\{ r_{GH,k,0},s_{GH,k,0}  \right\}$ plane, our model evolves along an almost horizontal trajectory with  $s \approx 0.07$ nearly constant, while the parameter $r$ increases from $r \simeq 0.7$ towards unity. This indicates that the dynamical evolution drives the model closer to the  $\Lambda$CDM point $(r,s)=(1,0)$, although a small but non-vanishing deviation in the  statefinder parameter $s$ persists. The constancy of $s$ suggests that the underlying relation  between the deceleration and jerk parameters remains fixed throughout the evolution, while the  approach $r \to 1$ reflects the tendency of the model towards a de Sitter-like phase. Overall,  the scenario mimics $\Lambda$CDM at late times but retains a distinct statefinder signature that  could, in principle, be tested with precision observations.

We now pass to the cases where there is interaction between the two Dark Sectors. \\
Considering the expression of the pressure $p_{D_{GH},I}(x)$ given in Eq. (\ref{carolina51}), we can write:
\begin{eqnarray}
p_{D_{GH},I}'(x)&=& 3d^2(1 - d^2)\left\{\frac{2  }{2 \left[1 - c^2 (1+\epsilon)\right] + 3 c^2 \epsilon (1 - d^2)}\right\}\Omega_{m0}e^{-3(1 - d^2)x} \, \nonumber   \\
&&+ \frac{2(1-\alpha)}{3}  \left\{1 - \frac{2 \Omega_{m0}}{2 \left[1 - c^2 (1+\epsilon)\right] + 3 c^2 \epsilon (1 - d^2)} \right\}\, e^{-2(1-\alpha)x} .\label{ciccio2}
\end{eqnarray}
Therefore, we obtain:
\begin{eqnarray}
r_{GH,I} &=&1+ \frac{9}{2}\left(\frac{ \rho_{D_{GH},I} + \rho_{m,I} + p_{D_{GH},I}  }{\rho_{D_{GH},I} + \rho_{m,I}}\right)\left(\frac{p'_{D_{GH},I}}{\rho'_{D_{GH},I} +\rho'_{m,I}}\right),\label{rgen3}\\
s_{GH,I} &=& \left(\frac{ \rho_{D_{GH},I} + \rho_{m,I} + p_{D_{GH},I}  }{p_{D_{GH},I}}\right)\left(\frac{p'_{D_{GH},I}}{\rho'_{D_{GH},I} +\rho'_{m,I}}\right). \label{sgen3}
\end{eqnarray}

\begin{figure}[htbp]
    \centering
    \begin{subfigure}{0.8\textwidth}
        \includegraphics[width=0.6\textwidth]{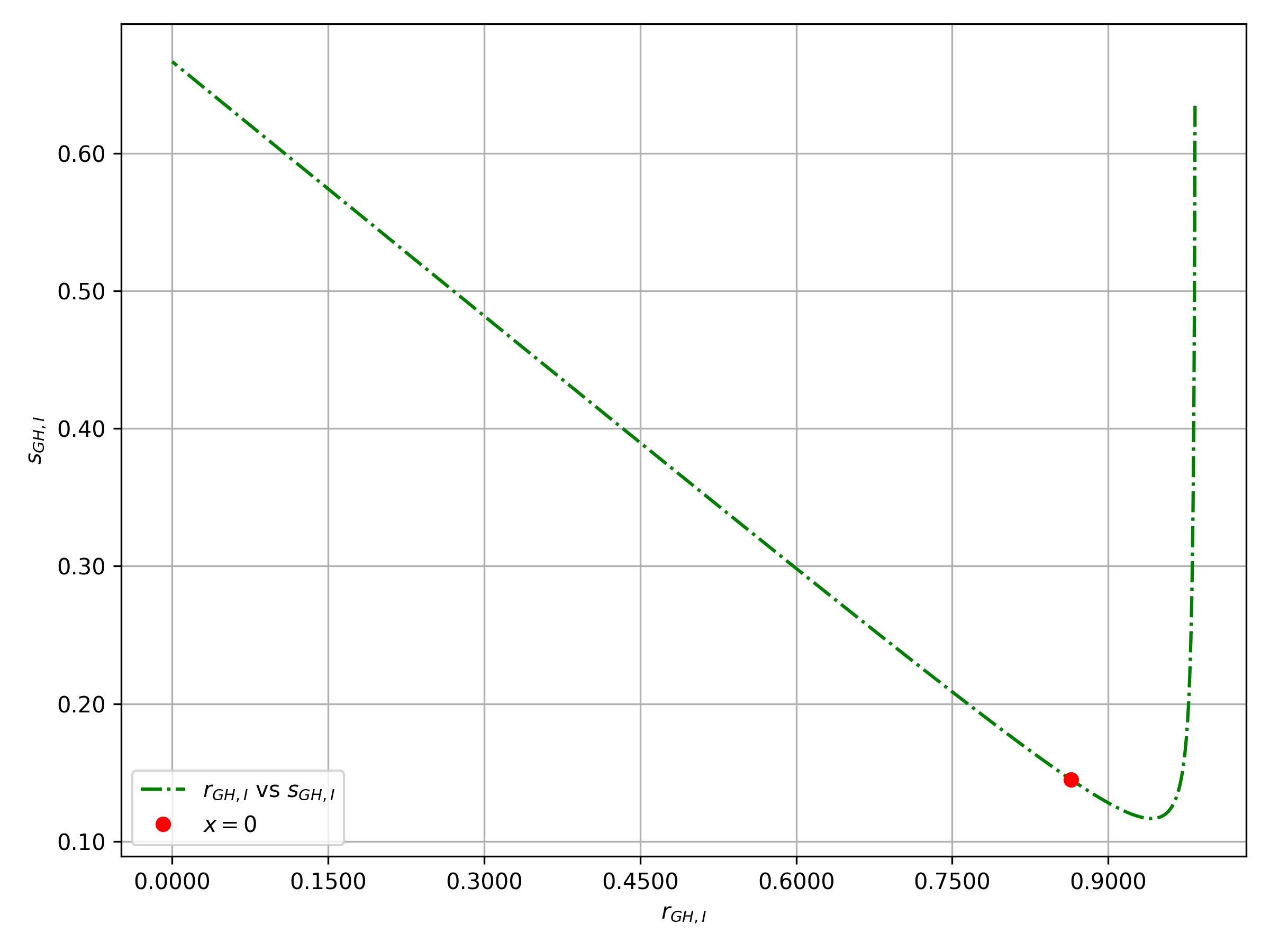}
        \caption{Plot of the $\{ r_{GH,I}, s_{GH,I} \}$ pair for $c^2=0.46$.}
        \label{state3}
    \end{subfigure}\\[0.5cm]
    \begin{subfigure}{0.8\textwidth}
        \includegraphics[width=0.6\textwidth]{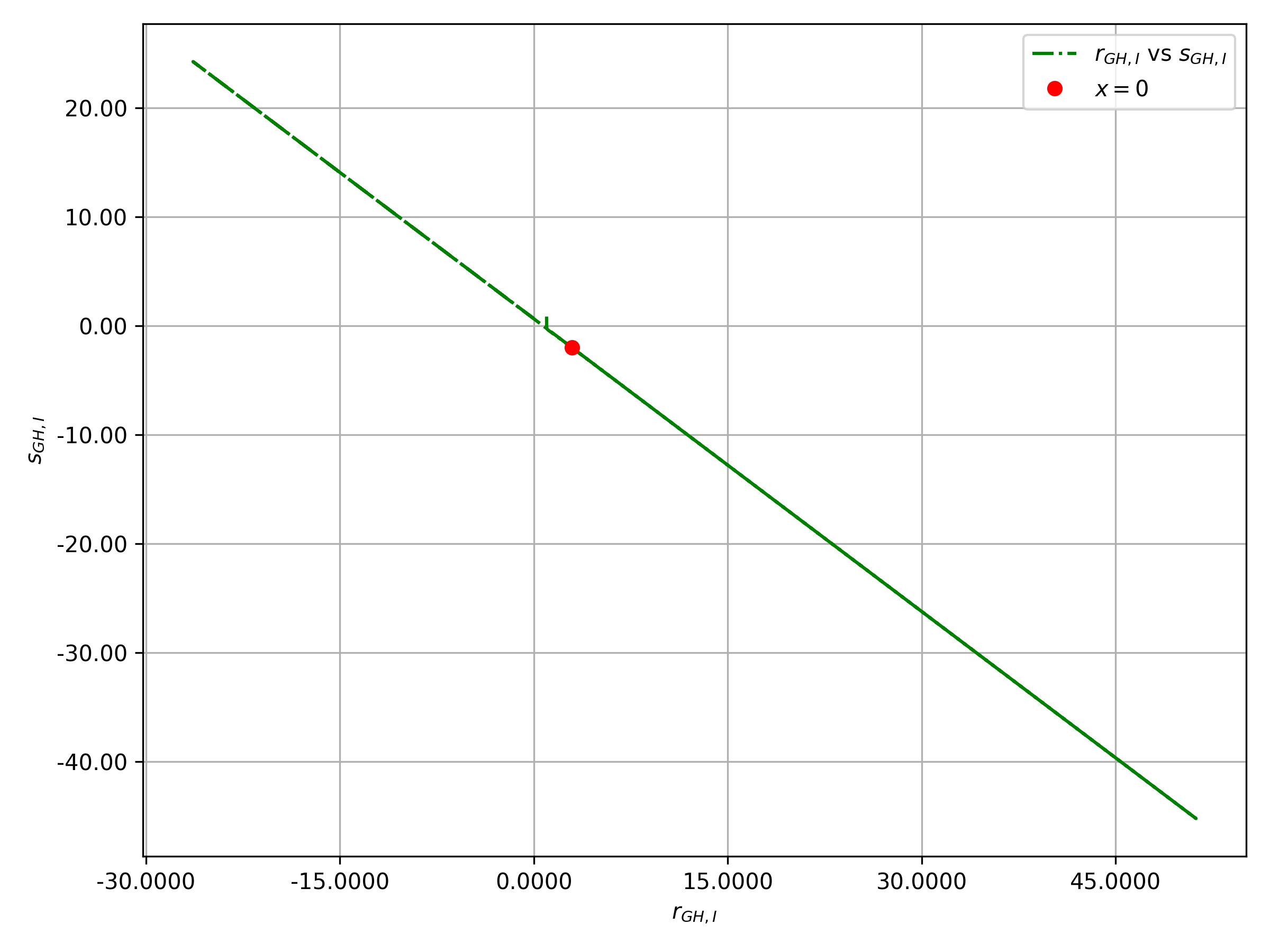}
        \caption{Plot of the $\{ r_{GH,I}, s_{GH,I} \}$ pair for $c=0.579$.}
        \label{state3-2}
    \end{subfigure}\\[0.5cm]
    \begin{subfigure}{0.8\textwidth}
        \includegraphics[width=0.6\textwidth]{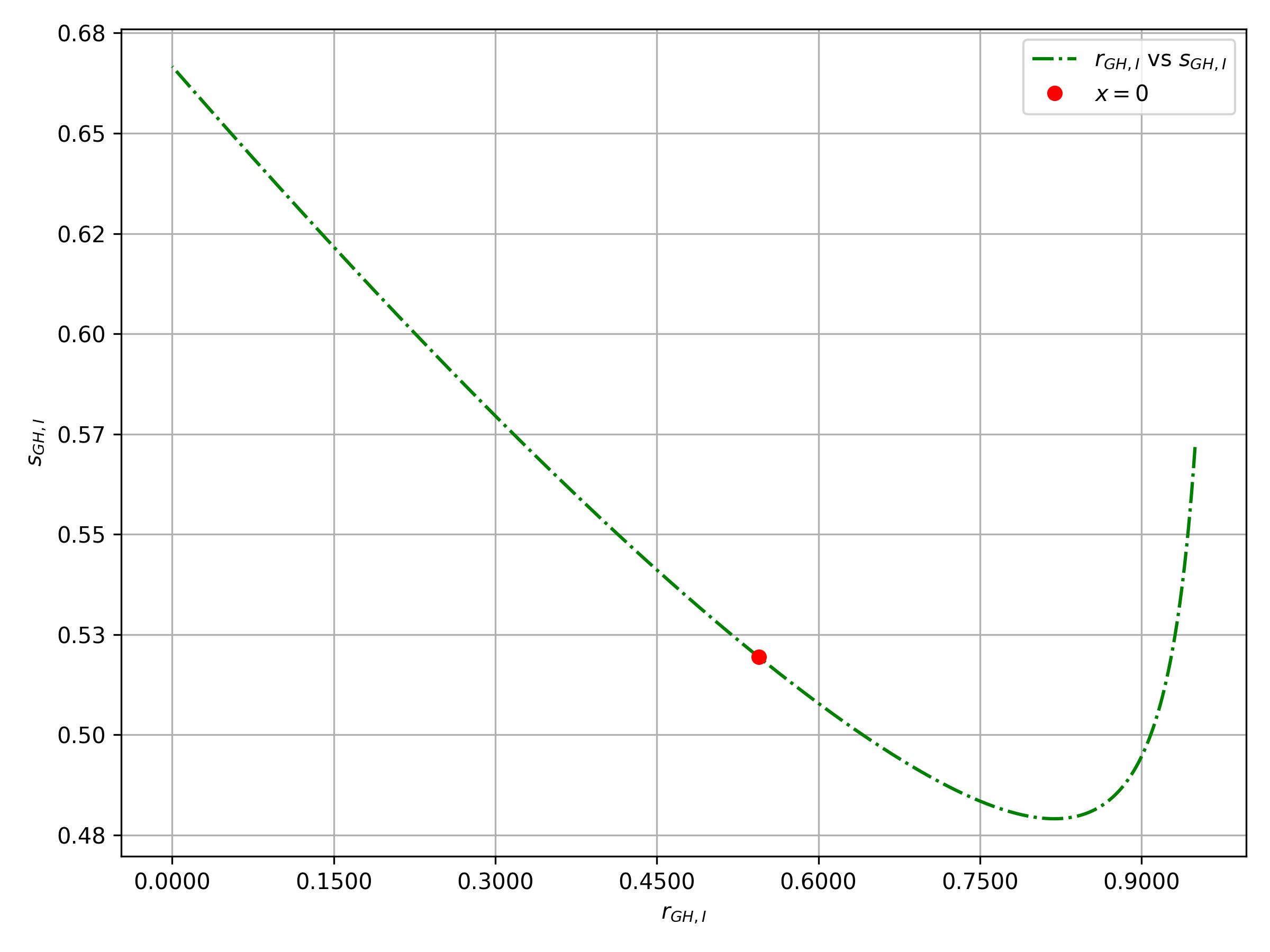}
        \caption{Plot of the $\{ r_{GH,I}, s_{GH,I} \}$ pair for $c=0.818$.}
        \label{state3-3}
    \end{subfigure}
    \caption{Comparison of the three cases for the statefinder pair $\{ r_{GH,I}, s_{GH,I} \}$.}
    \label{fig:state_all3}
\end{figure}

In Figs. (\ref{state3}), (\ref{state3-2}) and (\ref{state3-3}) we plot the expressions of $r_{GH,I}$ and $s_{GH,I}$ given in Eqs. (\ref{rgen3}) and (\ref{sgen3}) for $c^2=0.46$, $c=0.579$ and $c=0.818$, respectively.

We obtain that $\left\{ r_{GH,I,0},s_{GH,I,0}  \right\} \approx \left\{0.885, 0.156 \right\}  $ for $c^2=0.46$, $\left\{ r_{GH,I,0},s_{GH,I,0}  \right\} \approx \left\{ 1.59, -0.7829\right\}$ for $c=0.579$ and $\left\{ r_{GH,I,0},s_{GH,I,0}  \right\} \approx \left\{ 0.650, 0.508 \right\}$ for $c=0.818$.\\
The calculated statefinder parameters $\{r_{GH,I,0}, s_{GH,I,0}\}$ illustrate the strong dependence of the interacting Generalized Holographic (GH) Dark Energy model on the parameter $c$. For $c^2 = 0.46$, we obtain $\{r, s\} \approx \{0.885, 0.156\}$, indicating a small deviation from the standard $\Lambda$CDM scenario $(r,s) = (1,0)$ and suggesting a slightly more dynamic behavior than a pure cosmological constant. For $c = 0.579$, the values $\{r, s\} \approx \{1.59, -0.783\}$ correspond to $r>1$ and $s<0$, a signature of enhanced acceleration typical of quintessence-like or strongly interacting models. Finally, for $c = 0.818$, we find $\{r, s\} \approx \{0.650, 0.508\}$, showing a more pronounced deviation from $\Lambda$CDM, consistent with a softer quintessence-like dynamics. Overall, these results confirm that the GH interacting Dark Energy model is capable of reproducing a wide range of cosmological behaviors, from scenarios closely resembling $\Lambda$CDM to more dynamically evolving dark energy, highlighting the critical role of the parameter $c$ in controlling the model’s evolution.

\begin{figure}[htbp]
    \centering
    \begin{subfigure}{0.8\textwidth}
        \includegraphics[width=0.6\textwidth]{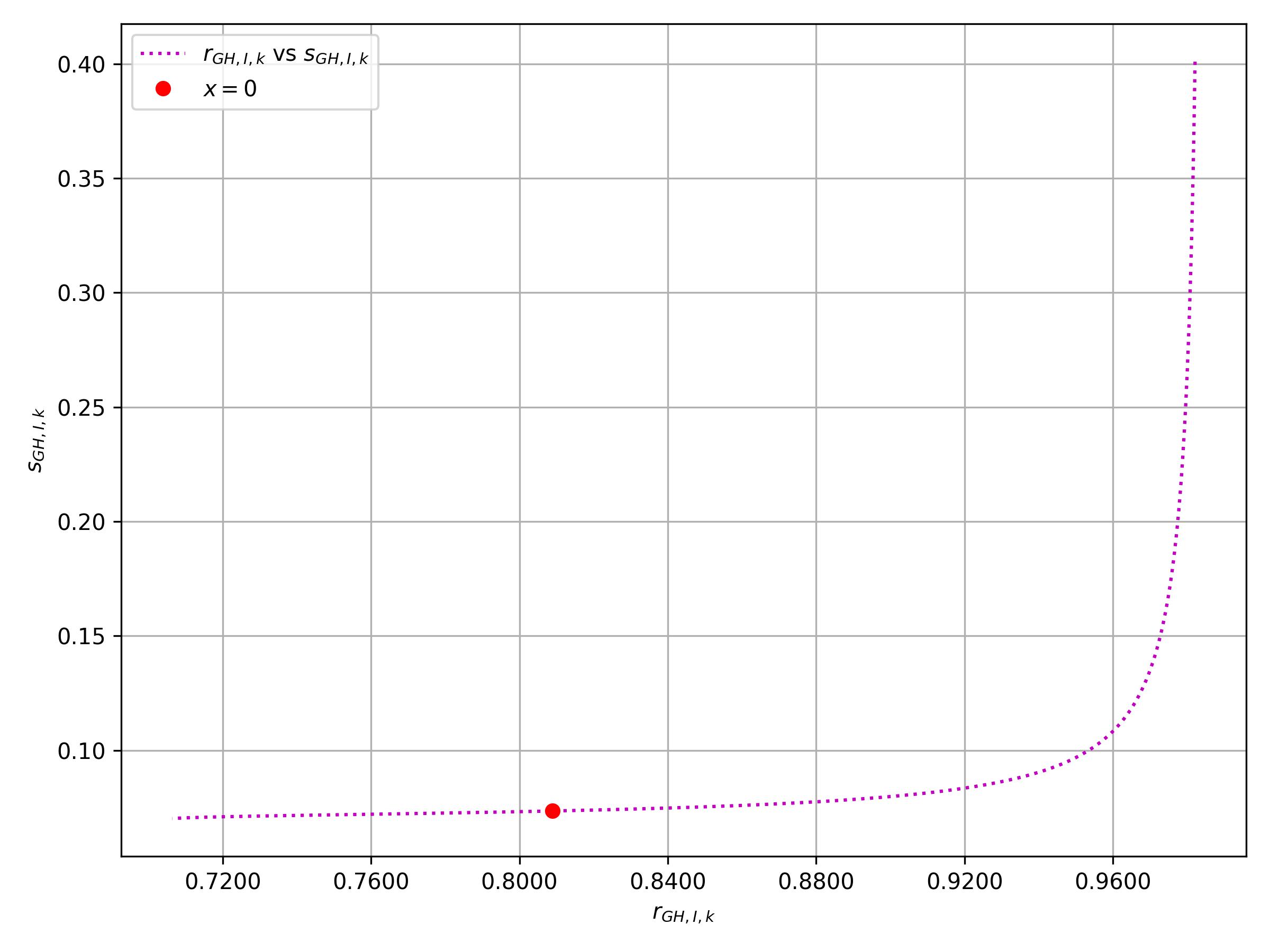}
        \caption{Plot of the $\{ r_{GH,I,k}, s_{GH,I,k}  \}$ pair for $c^2=0.46$.}
        \label{state4}
    \end{subfigure}\\[0.5cm]
    \begin{subfigure}{0.8\textwidth}
        \includegraphics[width=0.6\textwidth]{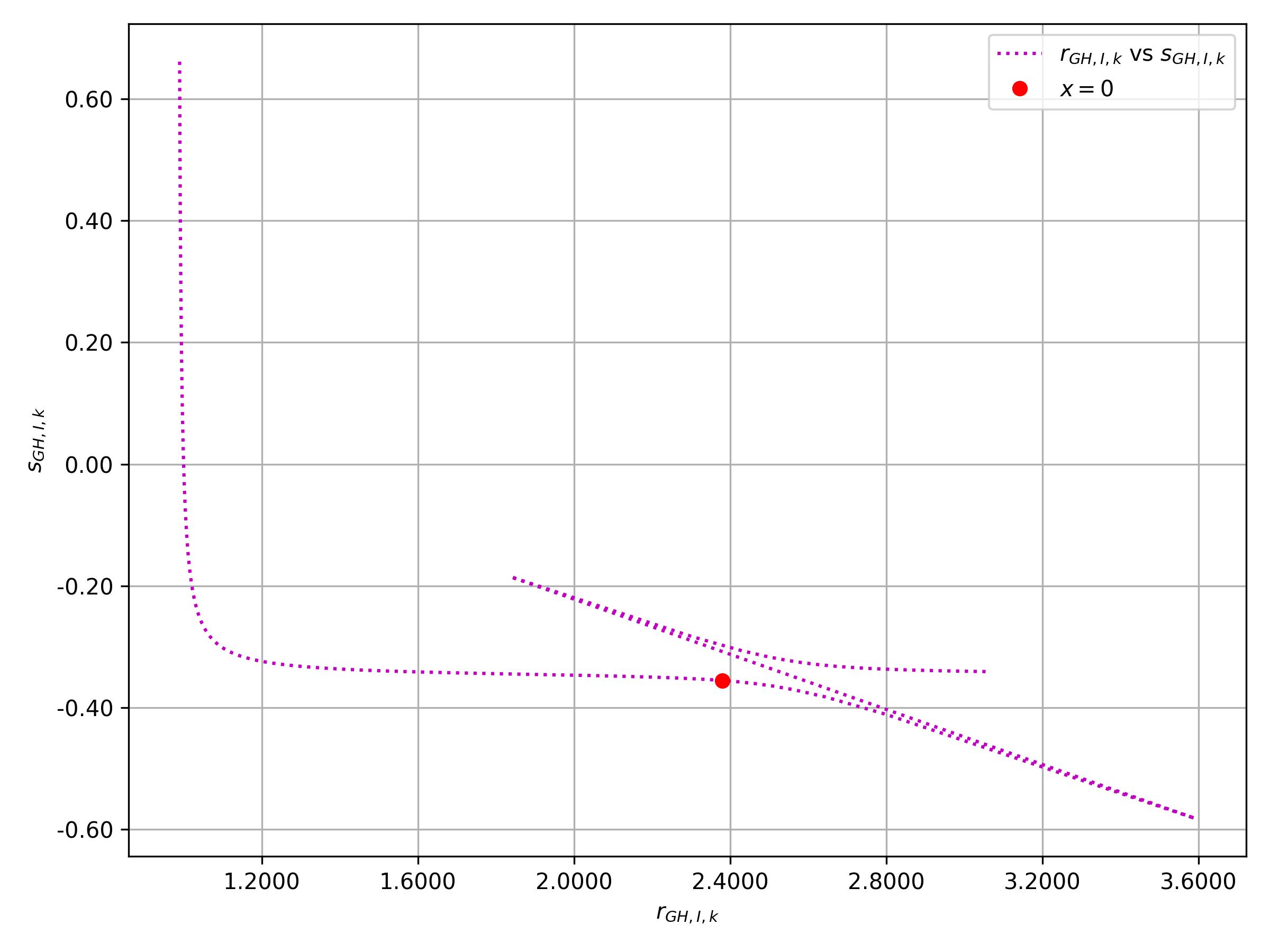}
        \caption{Plot of the $\{ r_{GH,I,k}, s_{GH,I,k}  \}$ pair for $c=0.579$.}
        \label{state4-2}
    \end{subfigure}\\[0.5cm]
    \begin{subfigure}{0.8\textwidth}
        \includegraphics[width=0.6\textwidth]{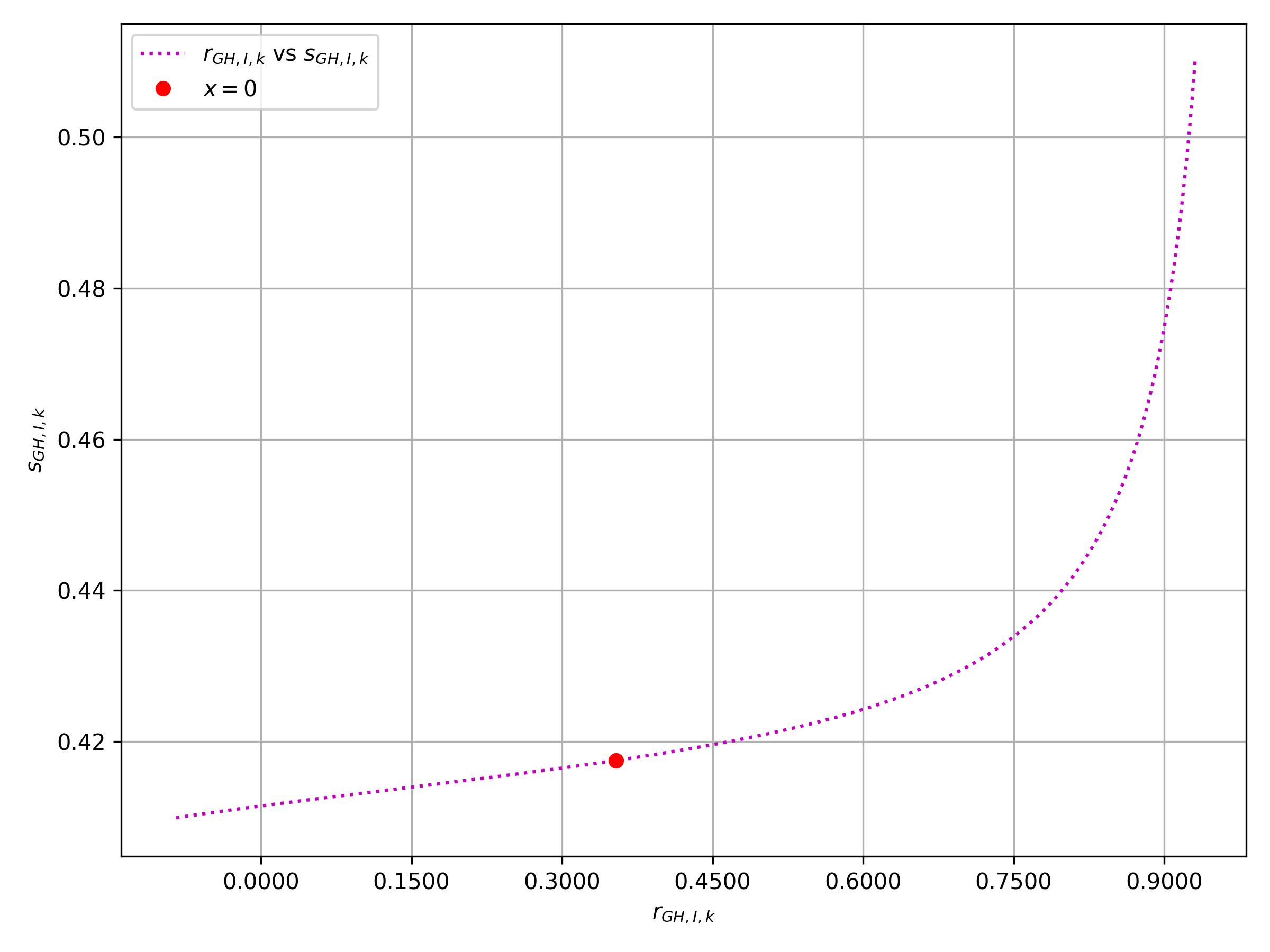}
        \caption{Plot of the $\{ r_{GH,I,k}, s_{GH,I,k}  \}$ pair for $c=0.815$.}
        \label{state4-3}
    \end{subfigure}
    \caption{Comparison of the three cases for the statefinder pair $\{ r_{GH,I,k}, s_{GH,I,k} \}$.}
    \label{fig:state_all4}
\end{figure}

We now consider the case where both spatial curvature and interaction are present.\\
Considering the expression of the pressure $p_{D_{GH},I,k}(x)$ given in Eq. (\ref{carolina52}), we can write:
\begin{eqnarray}
p_{D_{GH},I,k}'(x) &=&3d^2(1 - d^2)\left\{\frac{2 }{  2 \left[1 - c^2 (1 + \epsilon)\right] + 3(1 -  d^2)c^2 \epsilon }\right\} \Omega_{m0}e^{-3(1 - d^2)x} \nonumber  \\
&&+\frac{2}{3}\left[\frac{c^2(\epsilon - 1)}{1-c^2}\right]\Omega_{k0}e^{-2x}  \nonumber \\
&& +\frac{2(1-\alpha)(1+2\alpha)}{3}\times \nonumber \\
&&\left\{ 1 - \frac{2  \Omega_{m0}}{  2 \left[1 - c^2 (1 + \epsilon)\right] + 3(1 -  d^2)c^2 \epsilon } + \left( \frac{ 1 - \epsilon c^2 }{  1-c^2 }\right)\Omega_{k0} \right\}e^{-2(1-\alpha)x}. \label{ciccio3}
\end{eqnarray}
Therefore, we obtain:
\begin{eqnarray}
r_{GH,I,k} &=&1+ \frac{9}{2}\left(\frac{ \rho_{D_{GH},I,k} + \rho_{m,I}+\rho_k + p_{D_{GH},I,k}  }{\rho_{D_{GH},I,k} + \rho_{m,I}+\rho_k}\right)\left(\frac{p'_{D_{GH},I,k}}{\rho'_{D_{GH},I,k} +\rho'_{m,I}+\rho'_k}\right),\label{rgen4}\\
s_{GH,I,k} &=& \left(\frac{ \rho_{D_{GH},I,k} + \rho_{m,I} +\rho_k+ p_{D_{GH},I,k}  }{p_{D_{GH},I,k}}\right)\left(\frac{p'_{D_{GH},I,k}}{\rho'_{D_{GH},I,k} +\rho'_{m,I}+\rho'_k}\right). \label{sgen4}
\end{eqnarray}
In Figs. (\ref{state4}), (\ref{state4-2}) and (\ref{state4-3}) we plot the expressions of $r_{GH,I,k}$ and $s_{GH,I,k}$ given in Eqs. (\ref{rgen4}) and (\ref{sgen4}) for $c^2=0.46$, $c=0.579$ and $c=0.815$, respectively.

We obtain that $\left\{ r_{GH,I,k,0},s_{GH,I,k,0}  \right\} \approx \left\{0.834, 0.08\right\}$ for $c^2=0.46$, $\left\{ r_{GH,I,k,0},s_{GH,I,k,0}  \right\} \approx \left\{1.997, -0.335\right\}$ for $c=0.579$ and $\left\{ r_{GH,I,k,0},s_{GH,I,k,0}  \right\} \approx \left\{0.488, 0.426\right\}$ for $c=0.815$.\\
The statefinder parameters with interaction and including curvature $\{r_{GH,I,k,0}, s_{GH,I,k,0}\}$ show a similar dependence on the parameter $c$ with respect to the case with only interaction. For $c^2 = 0.46$, we find $\{r, s\} \approx \{0.834, 0.08\}$, indicating a mild deviation from the $\Lambda$CDM fixed point $(r,s) = (1,0)$ and a slightly dynamic behavior of the dark energy component. For $c = 0.579$, the values $\{r, s\} \approx \{1.997, -0.335\}$ suggest a stronger acceleration, with $r>1$ and $s<0$, typical of quintessence-like or interacting models. Finally, for $c = 0.815$, we obtain $\{r, s\} \approx \{0.488, 0.426\}$, showing a significant departure from $\Lambda$CDM and indicating a softer, less rapidly evolving dark energy behavior. These results confirm that including curvature modifies the statefinder diagnostics but still preserves the overall trend with respect to $c$, highlighting the role of both the interaction and the curvature in the dynamics of the GH Dark Energy model.

In the limiting case of a flat Dark Dominated Universe, we can write:
\begin{eqnarray}
r_{GH,DD} &=&1+ \frac{9}{2}\left(\frac{ \rho_{D_{GH},DD}  + p_{D_{GH},DD}  }{\rho_{D_{GH},DD} }\right)\left(\frac{p'_{D_{GH},I,k}}{\rho'_{D_{GH},I,k}   }\right),\label{}\\
s_{GH,DD} &=& \left(\frac{ \rho_{D_{GH},DD}  + p_{D_{GH},DD}  }{p_{D_{GH},DD}}\right)\left(\frac{p'_{D_{GH},DD}}{\rho'_{D_{GH},DD} }\right),\label{}
\end{eqnarray}
where:
\begin{eqnarray}
    \rho'_{D_{GH},DD} &=& -2(1 - \alpha)e^{-2(1 - \alpha) x} ,\label{totonno1}\\
    p'_{D_{GH},DD} &=&2\left[\frac{(2\alpha+1)(1 - \alpha)}{3}  \right] e^{-2(1-\alpha)x}.\label{totonno2}
\end{eqnarray}

Therefore, we can write:
\begin{eqnarray}
r_{GH,DD} &=&1- (1-\alpha)(2\alpha+1),\label{}\\
s_{GH,DD} &=& \frac{2(1-\alpha)}{3}. \label{}
\end{eqnarray}
For $c^2=0.46$, we obtain $\left\{  r_{GH,DD},s_{GH,DD}  \right\} \approx  \left\{0.706, 0.070  \right\}$. Instead, for $c=0.579$, we obtain $\left\{  r_{GH,DD},s_{GH,DD}  \right\} \approx \left\{3.057, -0.341  \right\}$. Finally, for $c=0.818$, we obtain $\left\{  r_{GH,DD},s_{GH,DD}  \right\} 
\approx\left\{ -0.093,0.415 \right\}$.\\
Comparing these results with the $\Lambda$CDM fixed point $\left\{ r,s  \right\}= \left\{1,0 \right\} $, it is evident that the parameter $c$ strongly affects the model’s evolution. Cases with $r < 1$ and $s > 0$ (as for $c^2 = 0.46$ and $c = 0.818$) indicate a quintessence-like behavior, moderately deviating from $\Lambda$CDM, whereas the scenario $c = 0.579$ exhibits $r \gg 1$ and $s < 0$, corresponding to a phantom-like behavior. These results demonstrate that varying $c$ allows the model to interpolate between different dark energy dynamics, ranging from quintessence-like to phantom-like evolution.

\section{OM diagnostic}
A useful geometric tool for cosmological model analysis is the \emph{Om} diagnostic. Unlike the statefinder parameters $ r $ and $ s $, which involve higher-order time derivatives of the scale factor $ a(t) $, the \emph{Om} diagnostic relies solely on the first derivative. This is because it is constructed purely from the Hubble parameter $ H(t) $, which itself involves only a first-order time derivative of $ a(t) $. 

As a result, the \emph{Om} diagnostic offers a simpler alternative to the statefinder method for distinguishing cosmological models \cite{Shahalam:2015lra}. It is worth mentioning that this approach has also been successfully employed in the study of Galileon models \cite{Jamil:2013yc,deFromont:2013iwa}.

The Om diagnostic as a function of the redshift $z$ can be performed thanks to the following expression:
\begin{eqnarray}
Om(z)&=&\frac{\left[\frac{H(z)}{H_0}\right]^2-1}{(1+z)^3-1}\nonumber \\
&=& \frac{h^2(z)-1}{(1+z)^3-1}.
\end{eqnarray}
This diagnostic has several attractive features:
\begin{itemize}
    \item \textbf{Model discrimination:} For the standard $\Lambda$CDM cosmology with a cosmological constant, the $ Om(z) $ remains constant with redshift. Deviations from constancy indicate departures from $\Lambda$CDM and can help distinguish between quintessence, phantom dark energy, or modified gravity models.

    \item \textbf{Robustness to observational uncertainties:} Since it relies directly on measurements of $ H(z) $, the Om diagnostic can be estimated using data from cosmic chronometers or baryon acoustic oscillations with fewer model-dependent assumptions.

    \item \textbf{Simplicity and Robustness:} The Om diagnostic depends only on the Hubble parameter $ H(z) $, which involves the first derivative of the scale factor $ a(t) $. This makes it simpler and more robust compared to diagnostics that require higher-order derivatives, such as the statefinder parameters $ r $ and $ s $.
    
    \item \textbf{Model Discrimination:} In the standard $\Lambda$CDM model, $ Om(z) $ remains constant and equals the present matter density parameter $\Omega_{m0}$. Deviations from a constant behavior indicate departures from $\Lambda$CDM, helping to distinguish between various dark energy models including quintessence, phantom energy, or modified gravity.
    
    \item \textbf{Reduced Sensitivity to Measurement Errors:} Since it uses only $ h^2(z) $, the Om diagnostic is less sensitive to the amplification of observational uncertainties compared to methods requiring second or third derivatives.
    
    \item \textbf{Non-parametric Nature:} The diagnostic does not require any explicit assumption about the dark energy equation of state $ \omega_D(z) $ or its parameterization, thus providing a relatively model-independent probe of the expansion history.
    
    \item \textbf{Effectiveness at Low and Moderate Redshifts:} The \emph{Om} diagnostic is especially useful at low to intermediate redshifts where accurate measurements of $ H(z) $ are available, allowing for tight constraints on the nature of dark energy.
    
    \item \textbf{Wide Applicability:} The Om diagnostic has been successfully applied to a range of cosmological models beyond $\Lambda$CDM, including Galileon models, scalar field scenarios, and modified gravity frameworks.
\end{itemize}

For a constant EoS parameter $\omega$,  the expression for $Om(z)$ is given by
\begin{eqnarray}
&&Om(z)= \Omega_{m0} + (1 -\Omega_{m0})\frac{(1+z)^{3(1+\omega)}-1}{(1+z)^3-1}.
\end{eqnarray}
Thus,  we observe that we have different values of  $Om(z) =\Omega_{m0}$  for the  $\Lambda$CDM model,
 quintessence, and phantom cosmological models.\\
 We can easily apply the Om diagnostic to our model since we already calculated the final expressions of $h^2(z)$ for the different cases we studied.

For the GH model, we have:
\begin{eqnarray}
Om(z)_{GH}&=& \frac{h^2_{GH}(z)-1}{(1+z)^3-1}. \label{OM1}
\end{eqnarray}

\begin{figure}[htbp]
    \centering
    \begin{subfigure}{0.8\textwidth}
        \includegraphics[width=0.45\textwidth]{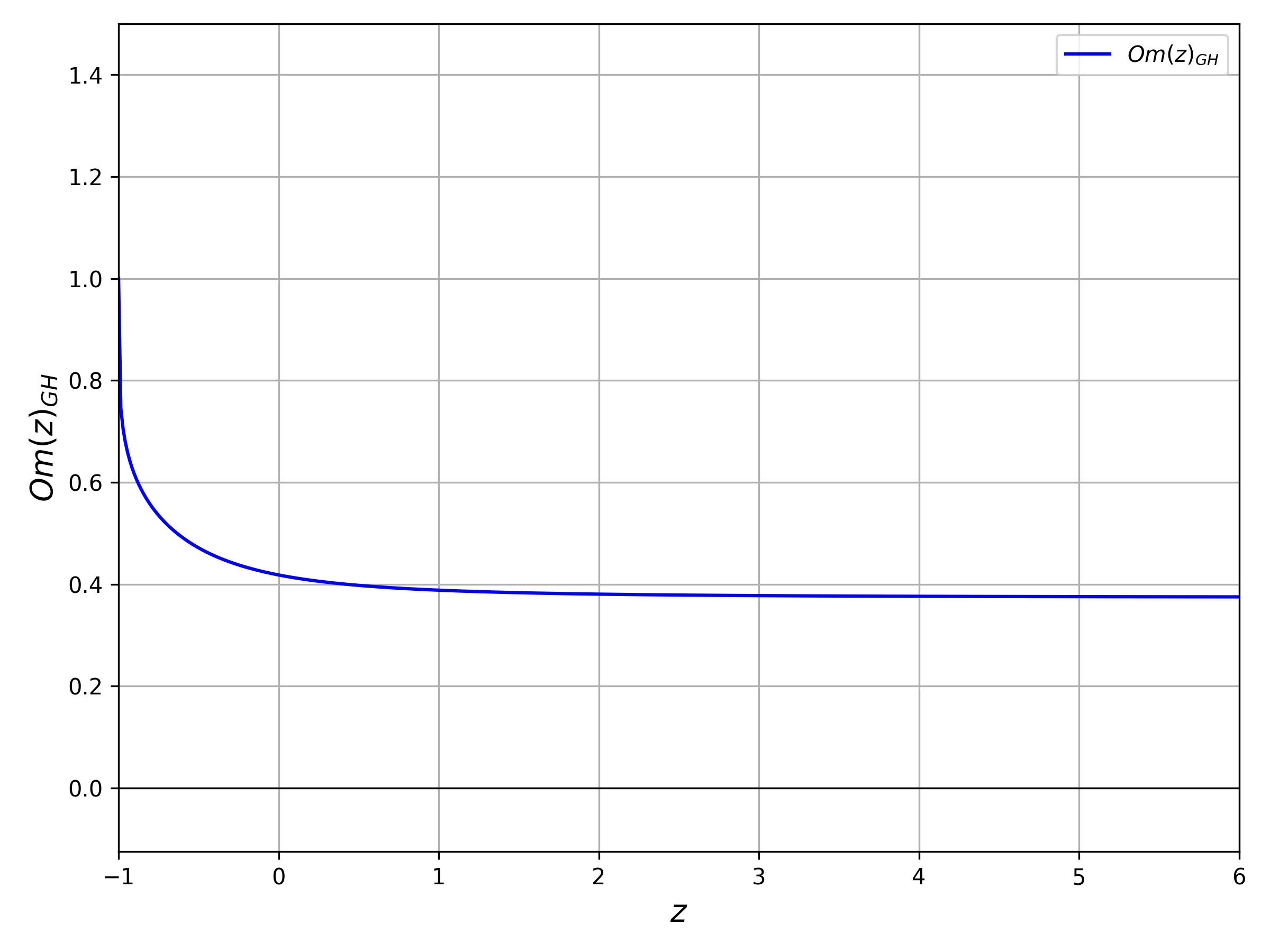}
        \caption{$Om(z)_{GH}$ for  $c^2=0.46$.}
        \label{om1}
    \end{subfigure}\\[0.45cm]
    \begin{subfigure}{0.8\textwidth}
        \includegraphics[width=0.5\textwidth]{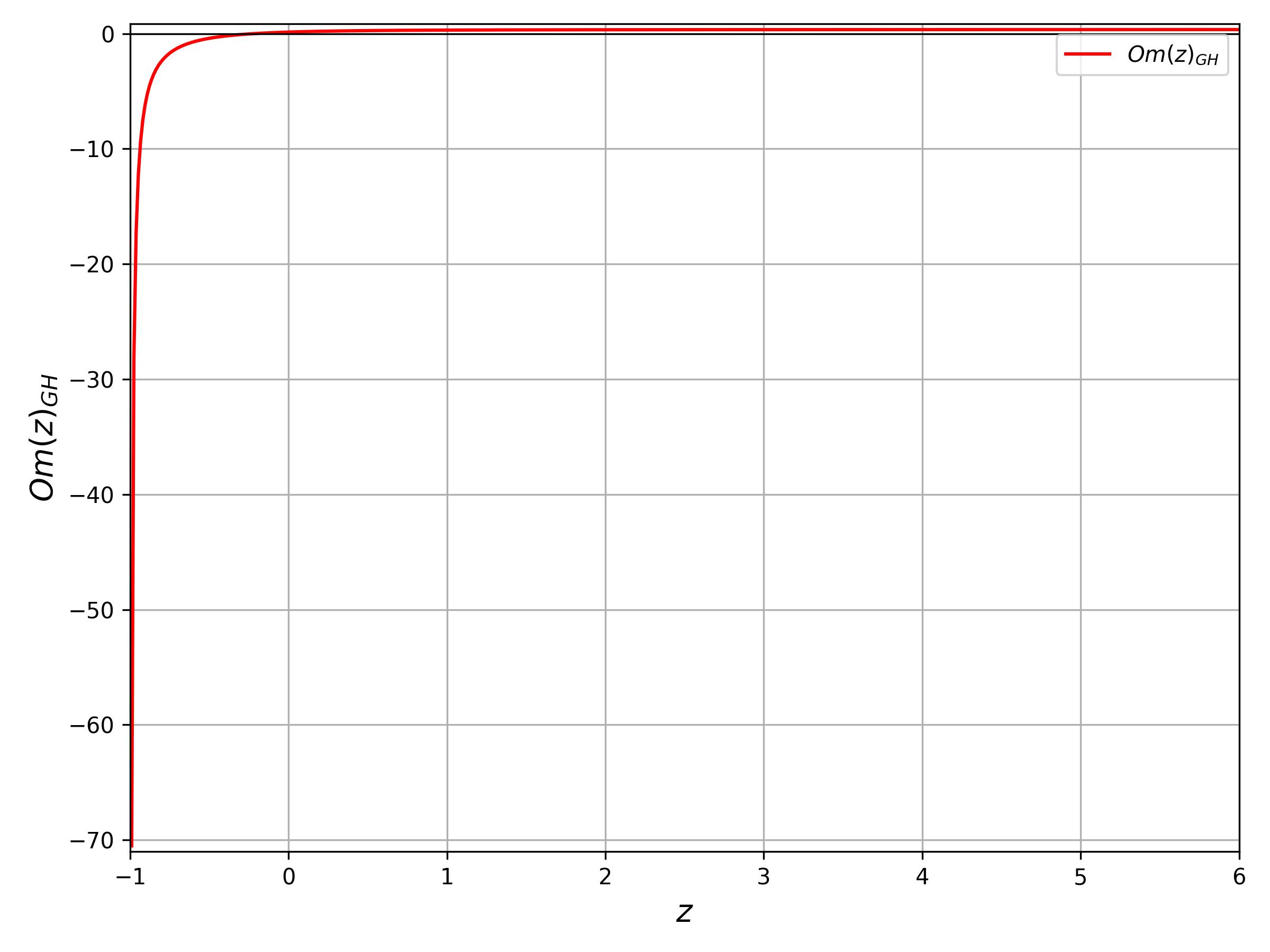}
        \caption{$Om(z)_{GH}$ for  $c=0.579$.}
        \label{om1-2}
    \end{subfigure}\\[0.45cm]
    \begin{subfigure}{0.8\textwidth}
        \includegraphics[width=0.5\textwidth]{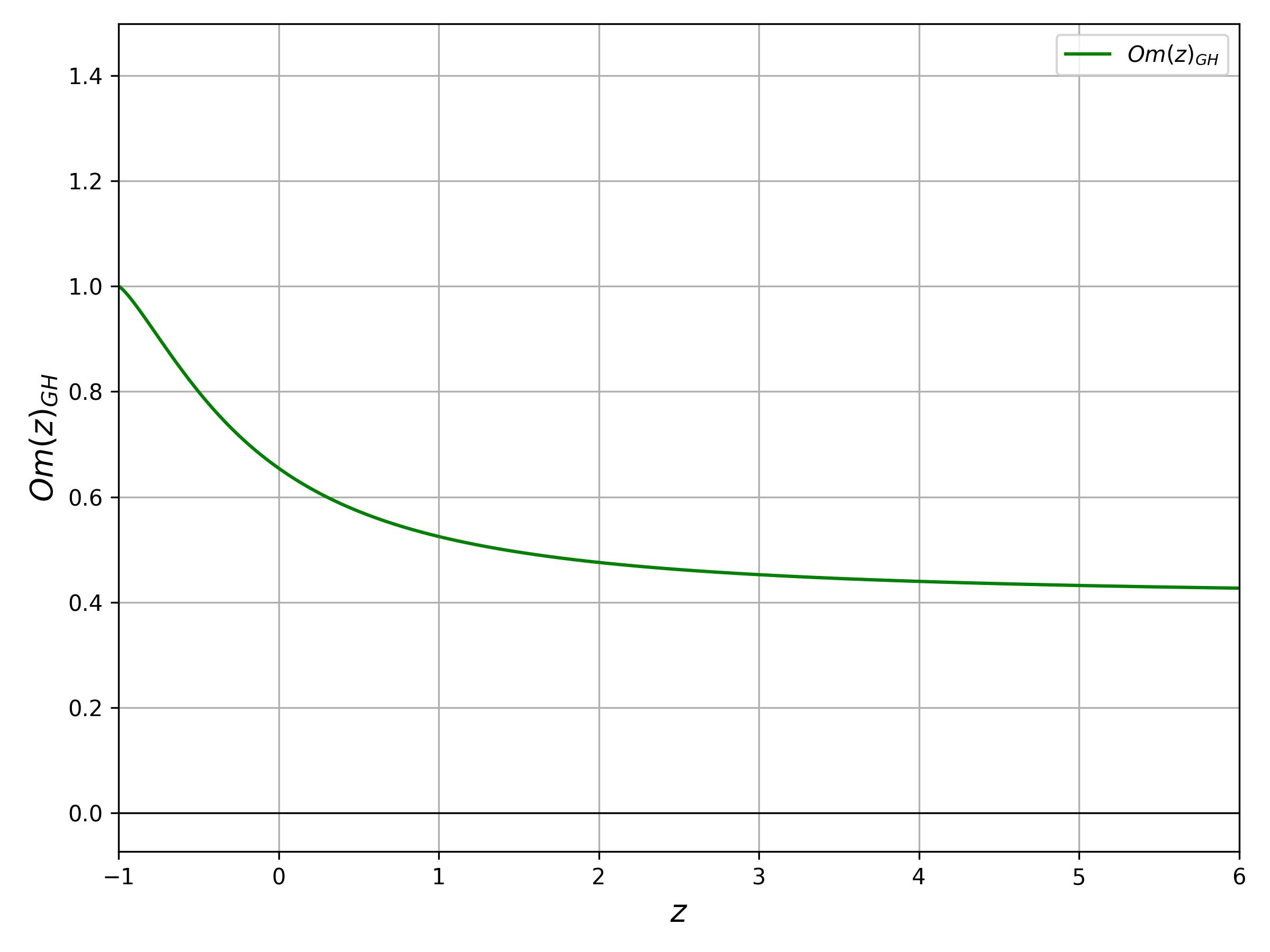}
        \caption{$Om(z)_{GH}$ for  $c=0.818$.}
        \label{om1-3}
    \end{subfigure}
    \caption{Comparison of the three cases for $Om(z)_{GH}$.}
    \label{fig:om_all}
\end{figure}

In Figs. (\ref{om1}), (\ref{om1-2}) and (\ref{om1-3}) we plot the expression of $Om(z)_{GH}$ given in Eq. (\ref{OM1}) for $c^2=0.46$, $c=0.579$ and $c=0.818$, respectively.
For the cases corresponding to $c^2=0.46$ and $c=0.818$, $Om(z)$ is always greater than $\Omega_{m0}$, therefore we have a phantom-like behavior. \\
For $c=0.579$, $Om(z)\equiv \Omega_{m0}$ for $z\approx 1.11$. For $1.11< z<6$, $Om(z)$ has a phantom-like behavior, in the other region a quintessence-like behavior.

If we consider the presence of spatial curvature, we obtain the following expression:
\begin{eqnarray}
Om(z)_{GH,k}&=& \frac{h^2_{GH,k}(z)-1}{(1+z)^3-1}. \label{OM2}
\end{eqnarray}

\begin{figure}[htbp]
    \centering
    \begin{subfigure}{0.8\textwidth}
        \includegraphics[width=0.45\textwidth]{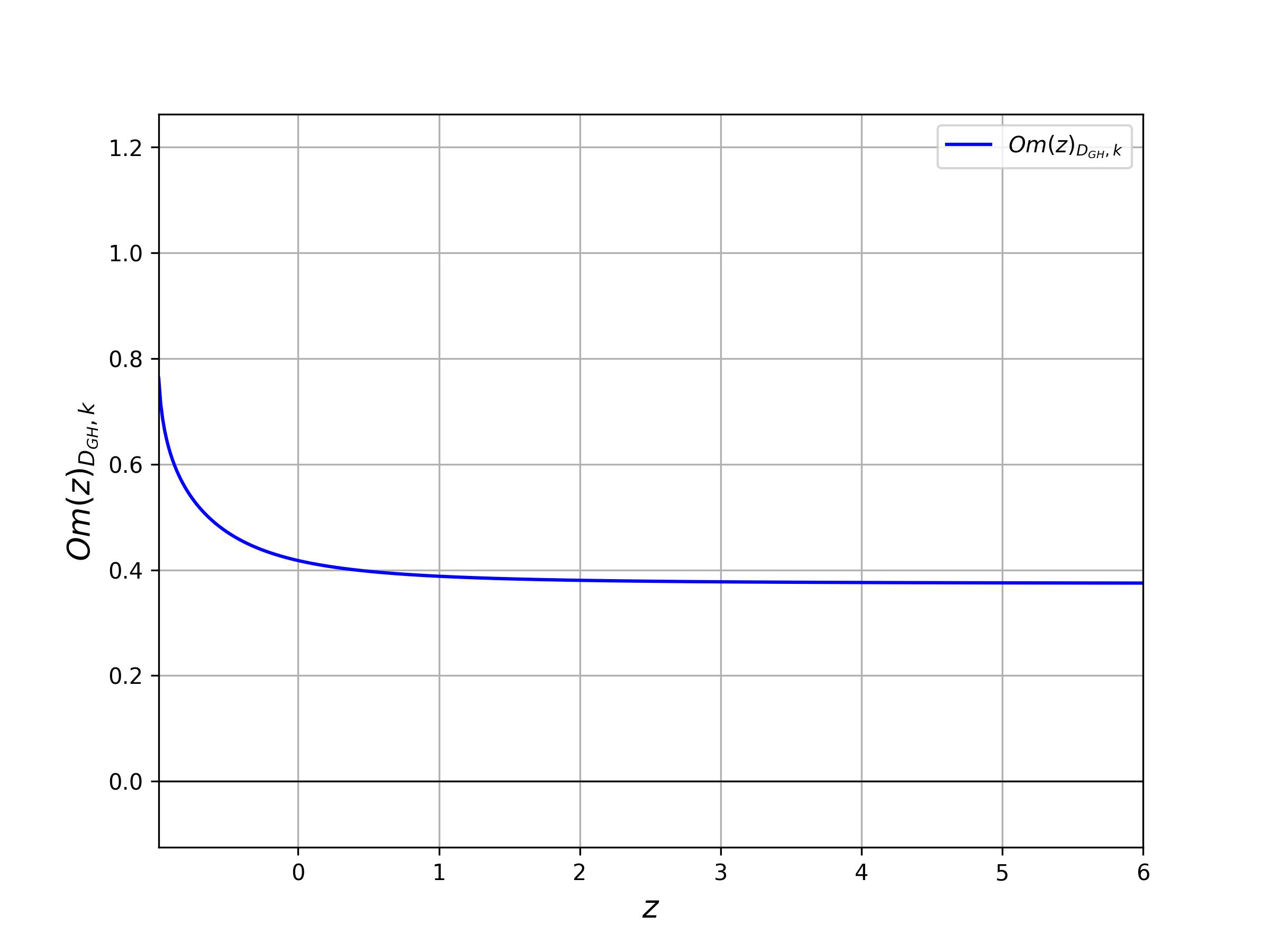}
        \caption{$Om(z)_{GH,k}$ for $c^2=0.46$.}
        \label{om2}
    \end{subfigure}\\[0.45cm]
    \begin{subfigure}{0.8\textwidth}
        \includegraphics[width=0.45\textwidth]{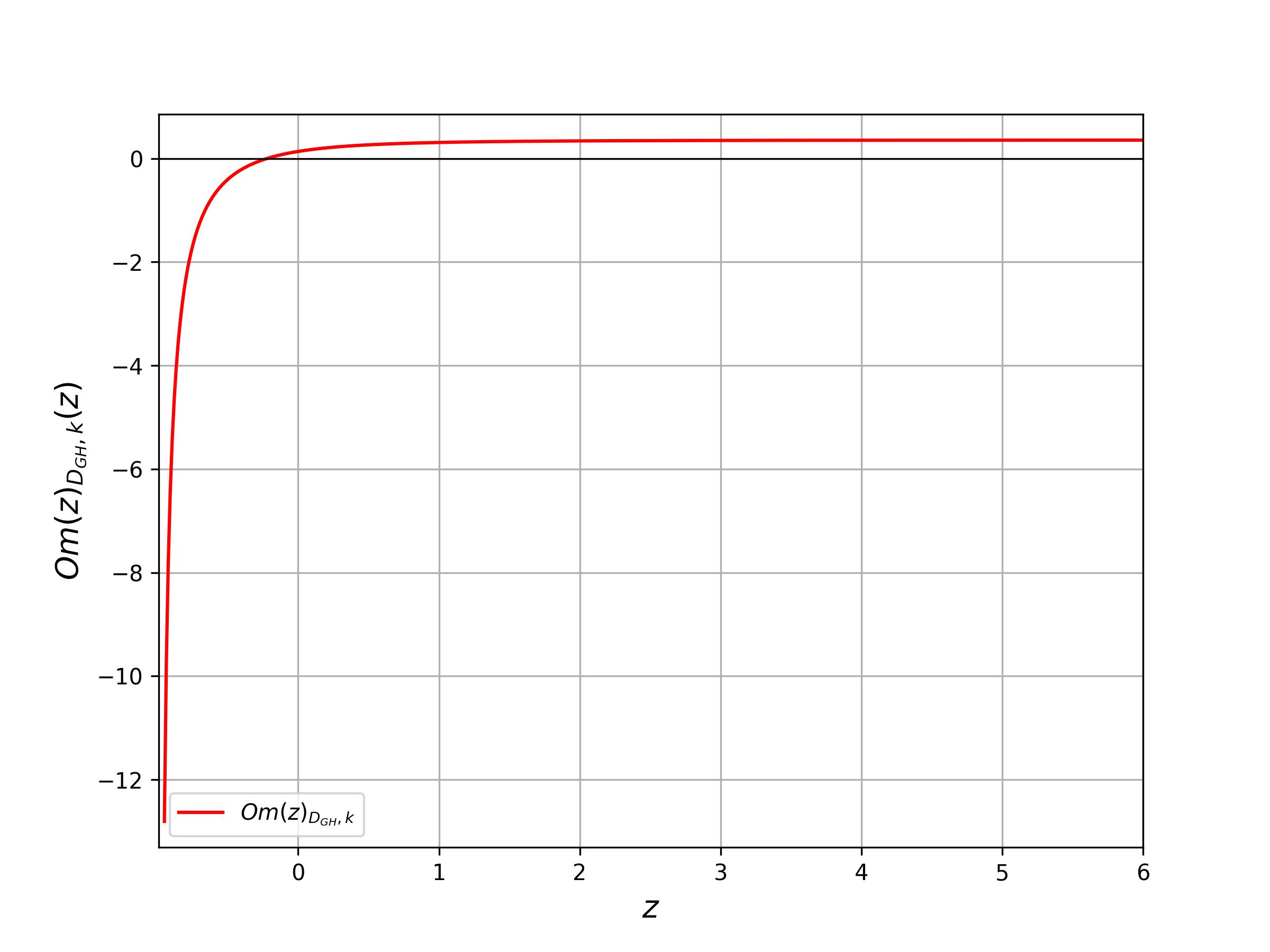}
        \caption{$Om(z)_{GH,k}$ for $c=0.579$.}
        \label{om2-2}
    \end{subfigure}\\[0.45cm]
    \begin{subfigure}{0.8\textwidth}
        \includegraphics[width=0.45\textwidth]{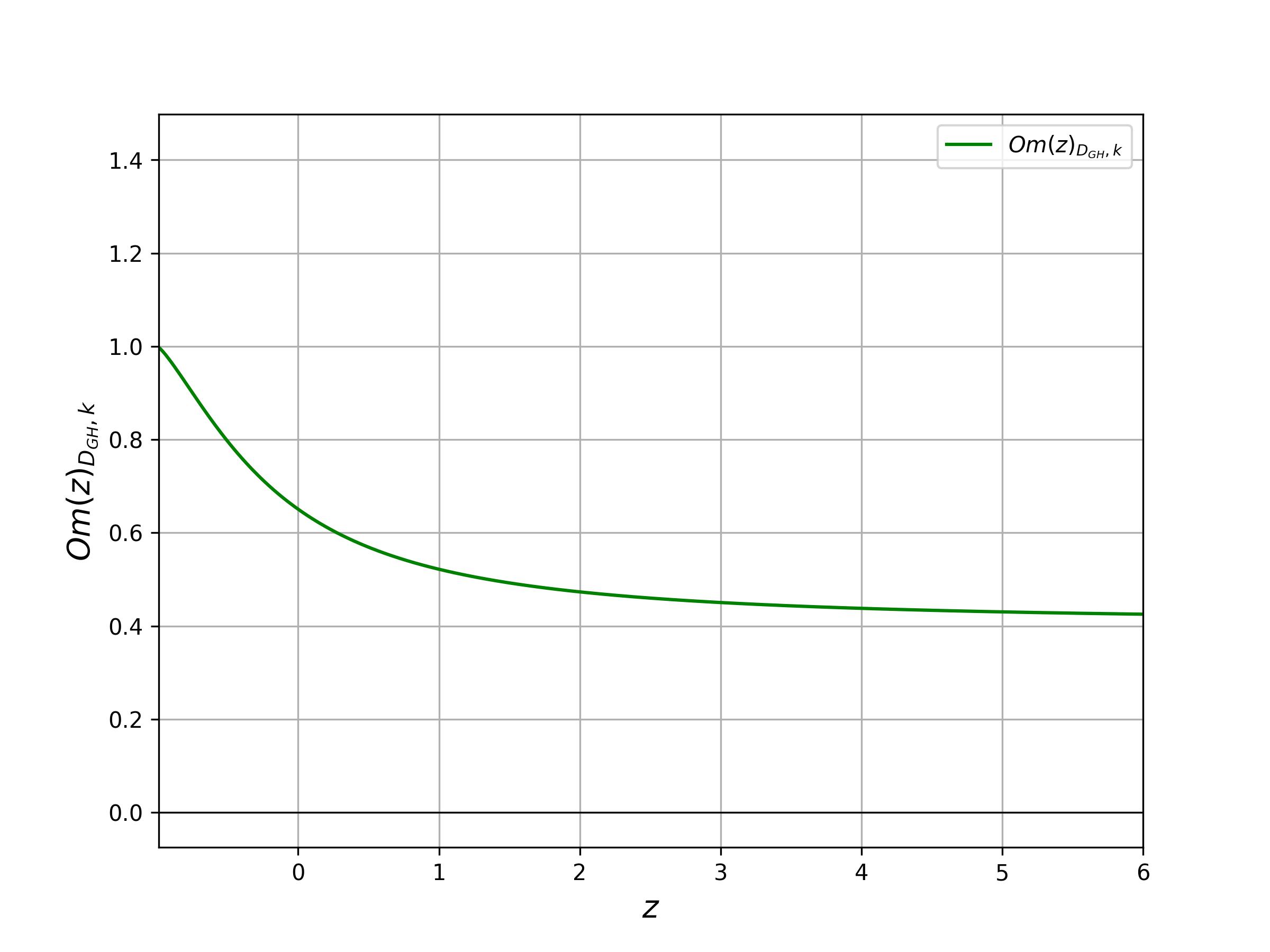}
        \caption{$Om(z)_{GH,k}$ for $c=0.815$.}
        \label{om2-3}
    \end{subfigure}
    \caption{Comparison of the three cases for $Om(z)_{GH,k}$.}
    \label{fig:omk_all2}
\end{figure}

In Figs. (\ref{om2}), (\ref{om2-2}) and (\ref{om2-3}) we plot the expression of $Om(z)_{GH,k}$ given in Eq. (\ref{OM2}) for $c^2=0.46$, $c=0.579$ and $c=0.815$, respectively.
For the cases corresponding to $c^2=0.46$ and $c=0.818$, $Om(z)$ is always greater than $\Omega_{m0}$, therefore we have a phantom-like behavior. \\
For $c=0.579$, $Om(z)\equiv \Omega_{m0}$ for $z\approx 1.12$. For $1.12< z<6$, $Om(z)$ has a phantom-like behavior, in the other region a quintessence-like behavior.

Considering the presence of interaction between the two Dark Sectors, we obtain:
\begin{eqnarray}
Om(z)_{GH,I}&=& \frac{h^2_{GH,I}(z)-1}{(1+z)^3-1}. \label{OM3}
\end{eqnarray}

\begin{figure}[htbp]
    \centering
    \begin{subfigure}{0.8\textwidth}
        \includegraphics[width=0.45\textwidth]{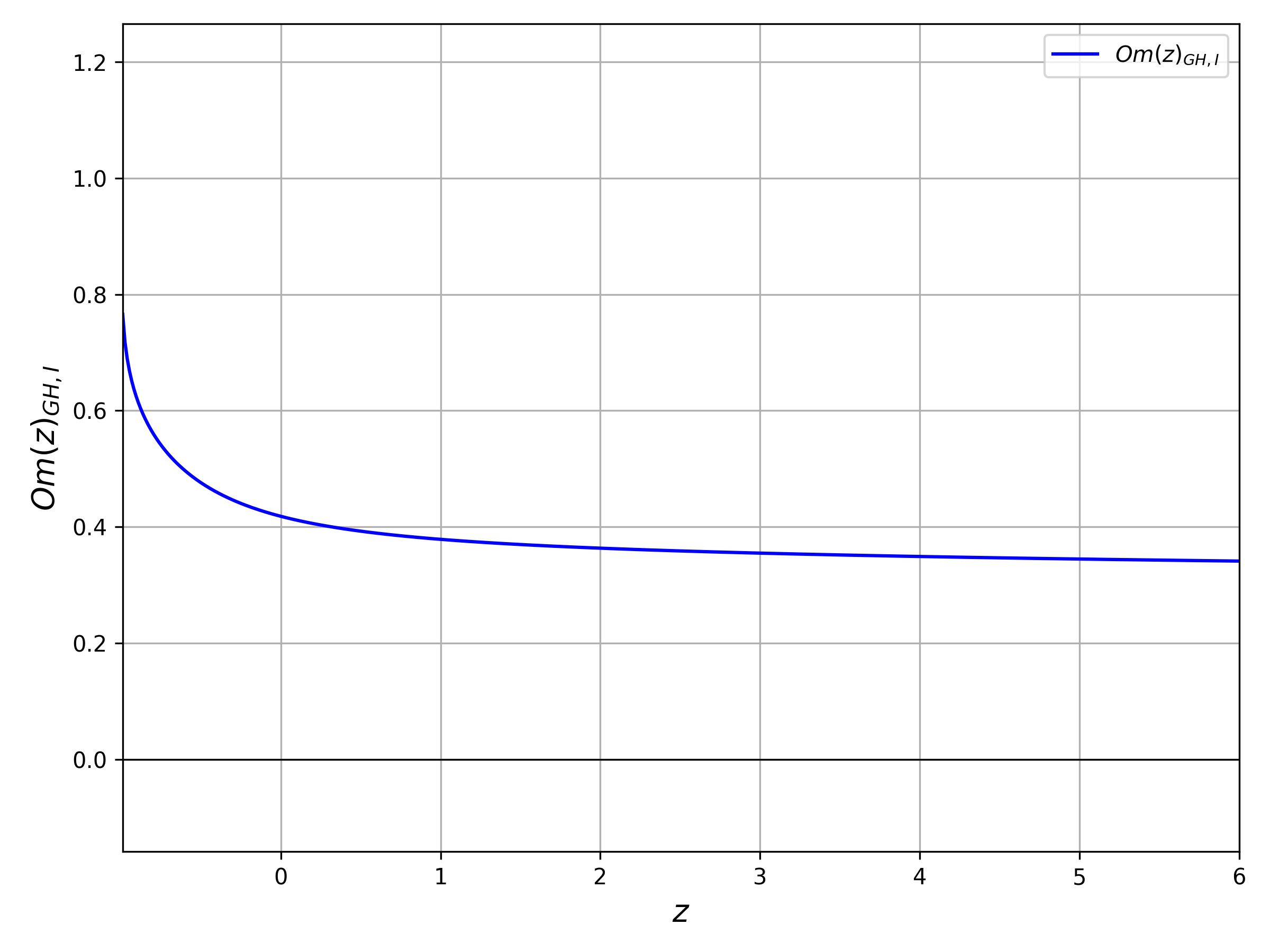}
        \caption{$Om(z)_{GH,I}$ for $c^2=0.46$.}
        \label{om3}
    \end{subfigure}\\[0.45cm]
    \begin{subfigure}{0.8\textwidth}
        \includegraphics[width=0.45\textwidth]{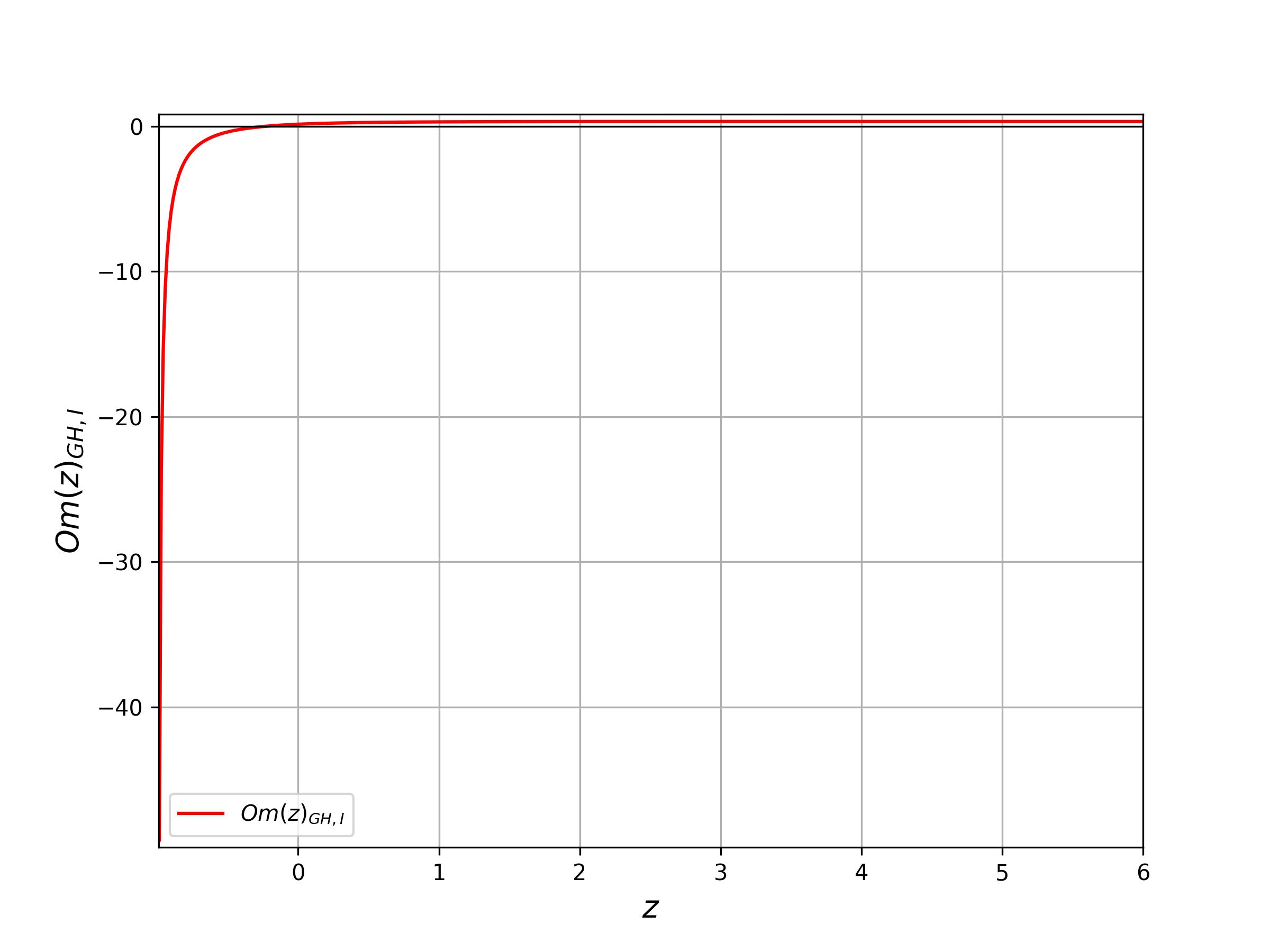}
        \caption{$Om(z)_{GH,I}$ for $c=0.579$.}
        \label{om3-2}
    \end{subfigure}\\[0.45cm]
    \begin{subfigure}{0.8\textwidth}
        \includegraphics[width=0.45\textwidth]{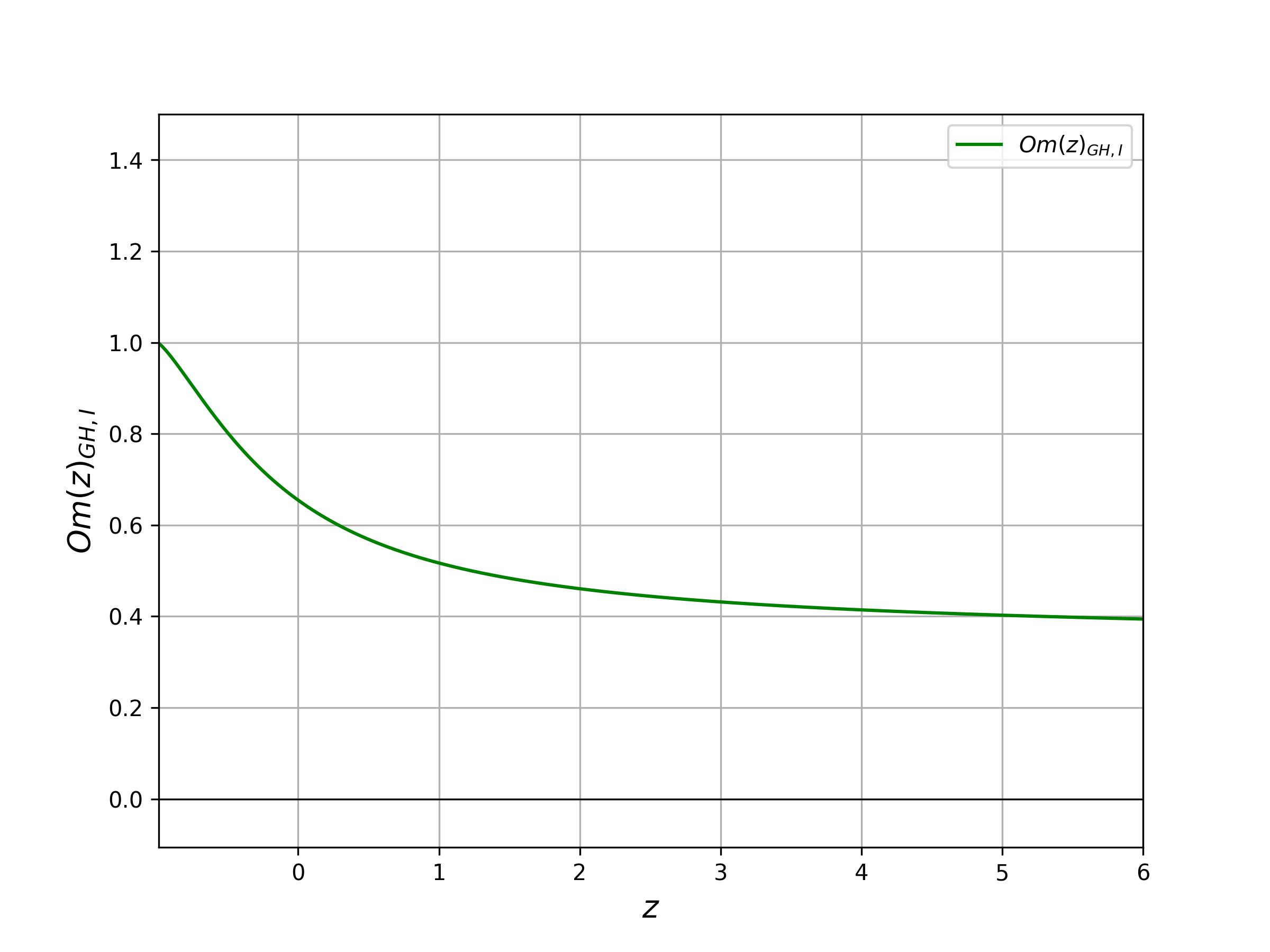}
        \caption{$Om(z)_{GH,I}$ for $c=0.818$.}
        \label{om3-3}
    \end{subfigure}
    \caption{Comparison of the three cases for $Om(z)_{GH,I}$.}
    \label{fig:omhi_all}
\end{figure}

In Figs. (\ref{om3}), (\ref{om3-2}) and (\ref{om3-3}) we plot the expression of $Om(z)_{GH,I}$ given in Eq. (\ref{OM3}) for $c^2=0.46$, $c=0.579$ and $c=0.818$, respectively.
For the cases corresponding to $c^2=0.46$ and $c=0.818$, $Om(z)$ is always greater than $\Omega_{m0}$, therefore we have a phantom-like behavior. \\
For $c=0.579$, $Om(z)\equiv \Omega_{m0}$ for $z\approx 1.53$. For $1.53<z<6$, $Om(z)$ has a phantom-like behavior, in the other region a quintessence-like behavior.

Finally, if both spatial curvature and interaction are present, we obtain the following result:
\begin{eqnarray}
Om(z)_{GH,I,k}&=& \frac{h^2_{GH,I,k}(z)-1}{(1+z)^3-1}.\label{OM4}
\end{eqnarray}

\begin{figure}[htbp]
    \centering
    \begin{subfigure}{0.8\textwidth}
        \includegraphics[width=0.5\textwidth]{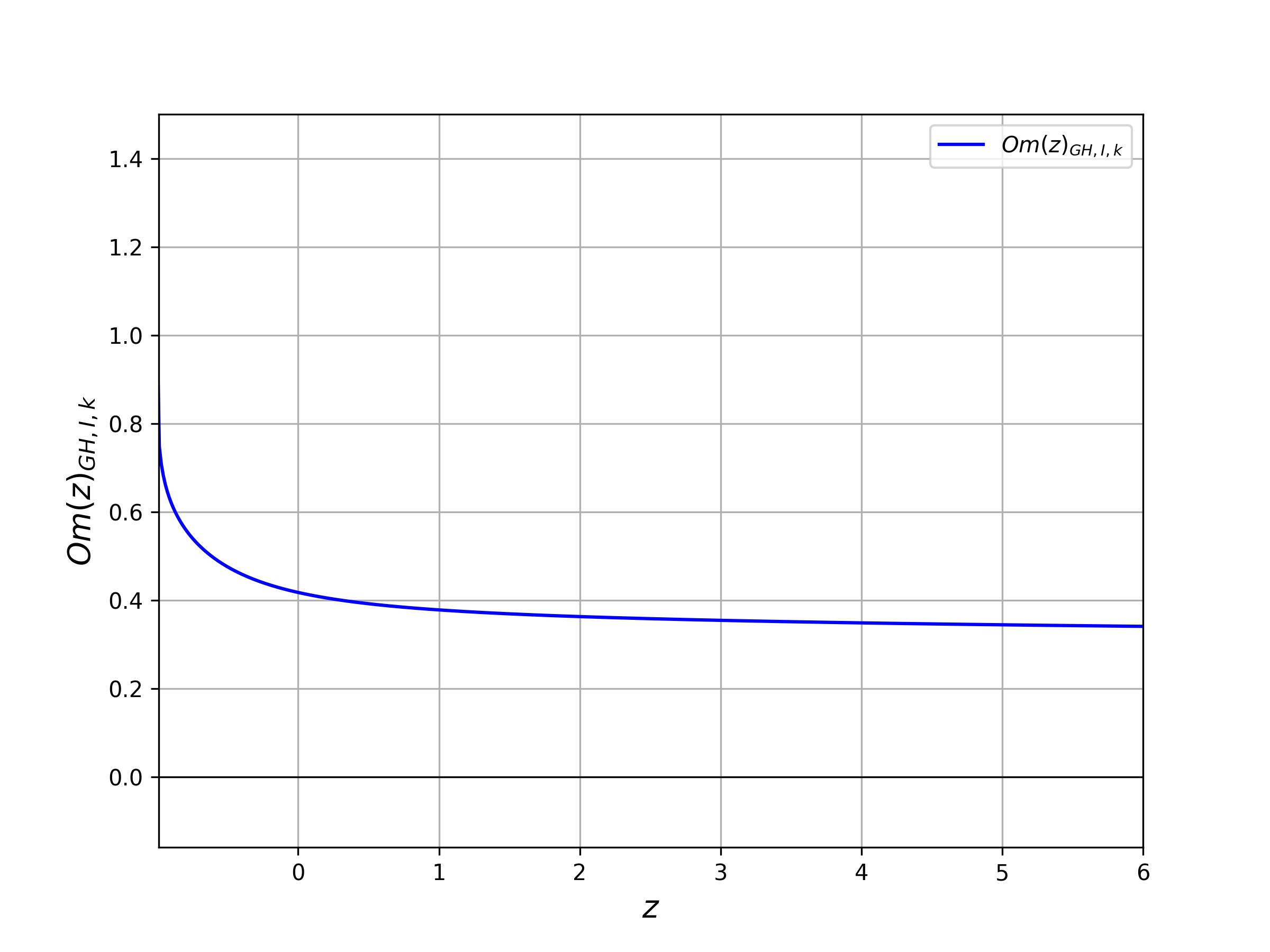}
        \caption{$Om(z)_{GH,I,k}$ for $c^2=0.46$.}
        \label{om4}
    \end{subfigure}\\[0.45cm]
    \begin{subfigure}{0.8\textwidth}
        \includegraphics[width=0.5\textwidth]{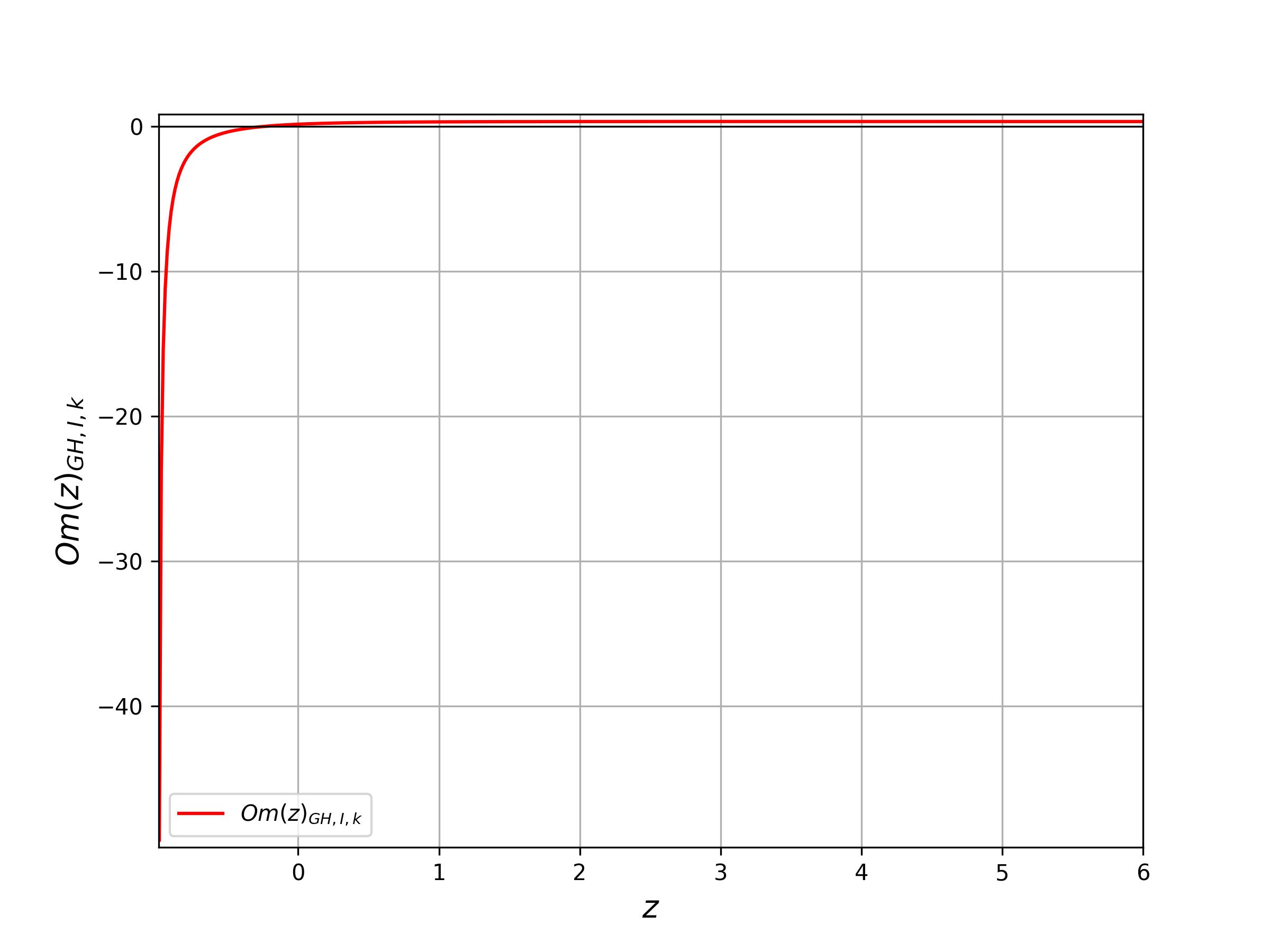}
        \caption{$Om(z)_{GH,I,k}$ for $c=0.579$.}
        \label{om4-2}
    \end{subfigure}\\[0.45cm]
    \begin{subfigure}{0.8\textwidth}
        \includegraphics[width=0.5\textwidth]{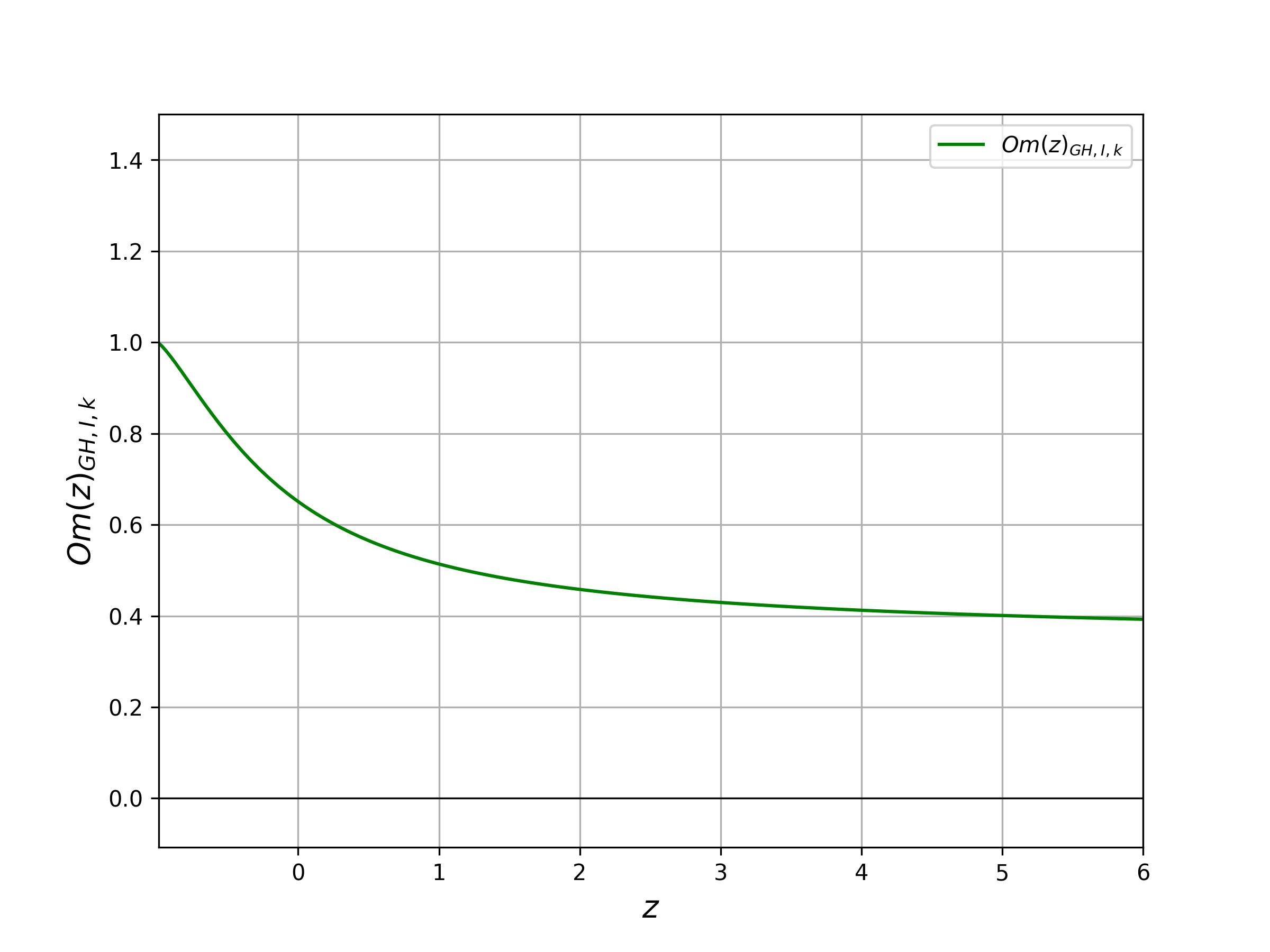}
        \caption{$Om(z)_{GH,I,k}$ for $c=0.815$.}
        \label{om4-3}
    \end{subfigure}
    \caption{Comparison of the three cases for $Om(z)_{GH,I,k}$.}
    \label{fig:omhik_all}
\end{figure}

In Figs. (\ref{om4}), (\ref{om4-2}) and (\ref{om4-3}) we plot the expression of $Om(z)_{GH,I,k}$ given in Eq. (\ref{OM4}) for $c^2=0.46$, $c=0.579$ and $c=0.815$, respectively.
For the cases corresponding to $c^2=0.46$ and $c=0.815$, $Om(z)$ is always greater than $\Omega_{m0}$, therefore we have a phantom-like behavior. \\
For $c=0.579$, $Om(z)\equiv \Omega_{m0}$ for $z\approx 1.56$. For $1.56<z<6$, $Om(z)$ has a phantom-like behavior, in the other region a quintessence-like behavior.

In the limiting case of a Dark Dominated Universe, we have:
\begin{eqnarray}
Om(z)_{GH,DD}&=& \frac{h^2_{GH,DD}(z)-1}{(1+z)^3-1}\nonumber \\
&=&\frac{(1+z)^{2(1 - \alpha)}-1}{(1+z)^3-1}.\label{om5-5}
\end{eqnarray}

\begin{figure}[htbp]
    \centering
    \begin{subfigure}{0.8\textwidth}
        \includegraphics[width=0.45\textwidth]{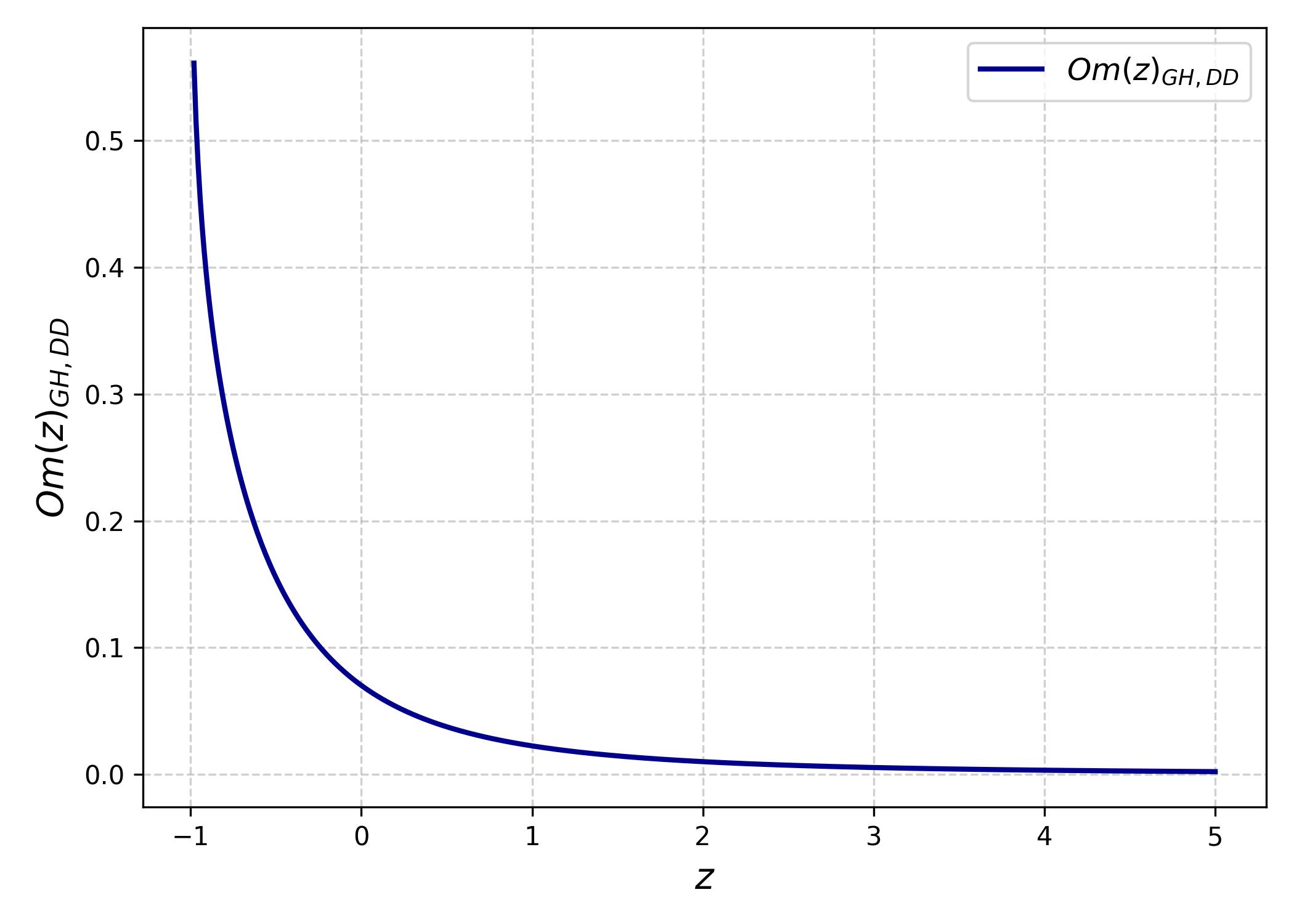}
        \caption{$Om(z)_{GH,DD}$ for $c^2=0.46$.}
        \label{om2DD}
    \end{subfigure}\\[0.45cm]
    \begin{subfigure}{0.8\textwidth}
        \includegraphics[width=0.45\textwidth]{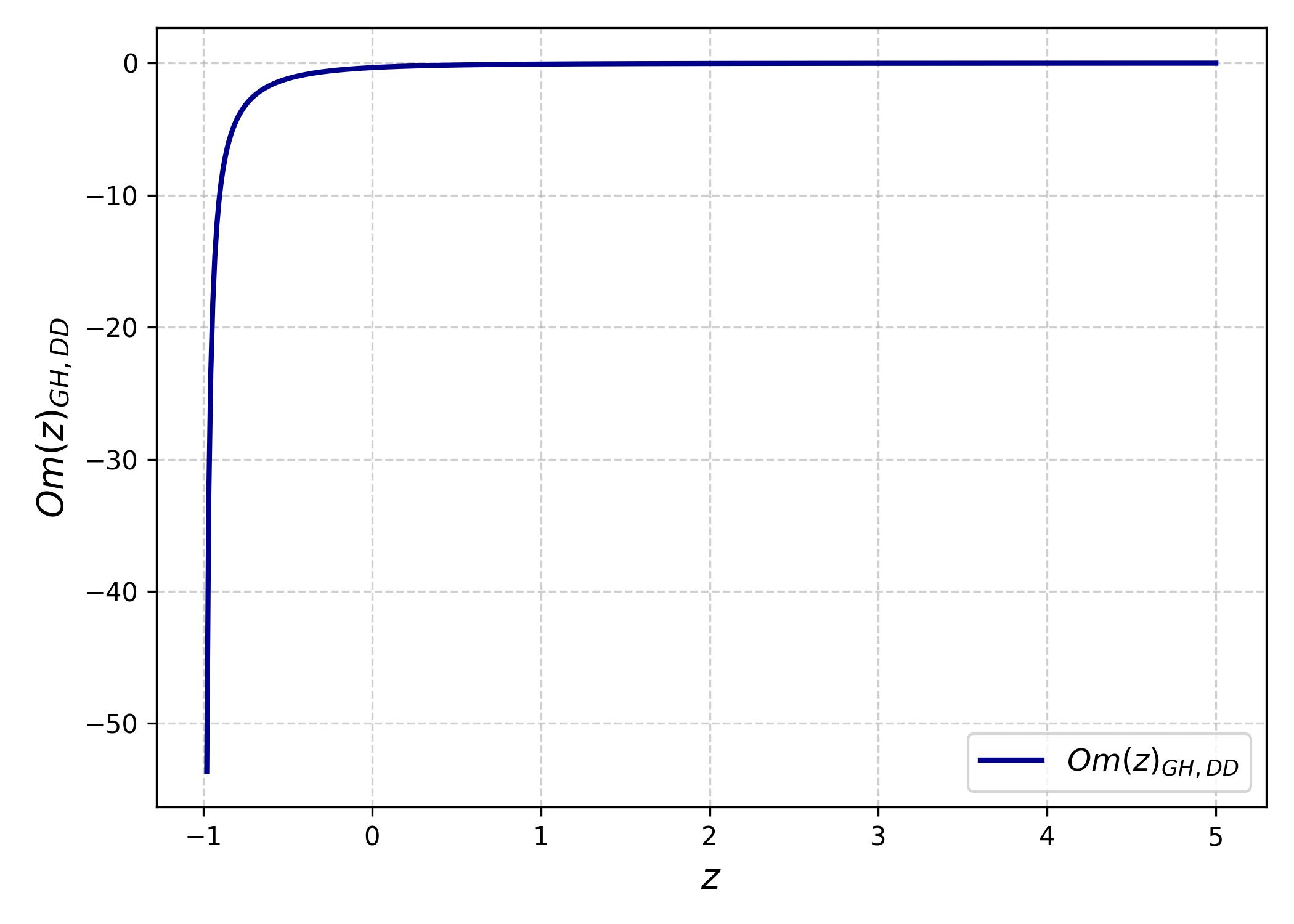}
        \caption{$Om(z)_{GH,DD}$ for $c=0.579$.}
        \label{om2-2DD}
    \end{subfigure}\\[0.45cm]
    \begin{subfigure}{0.8\textwidth}
        \includegraphics[width=0.45\textwidth]{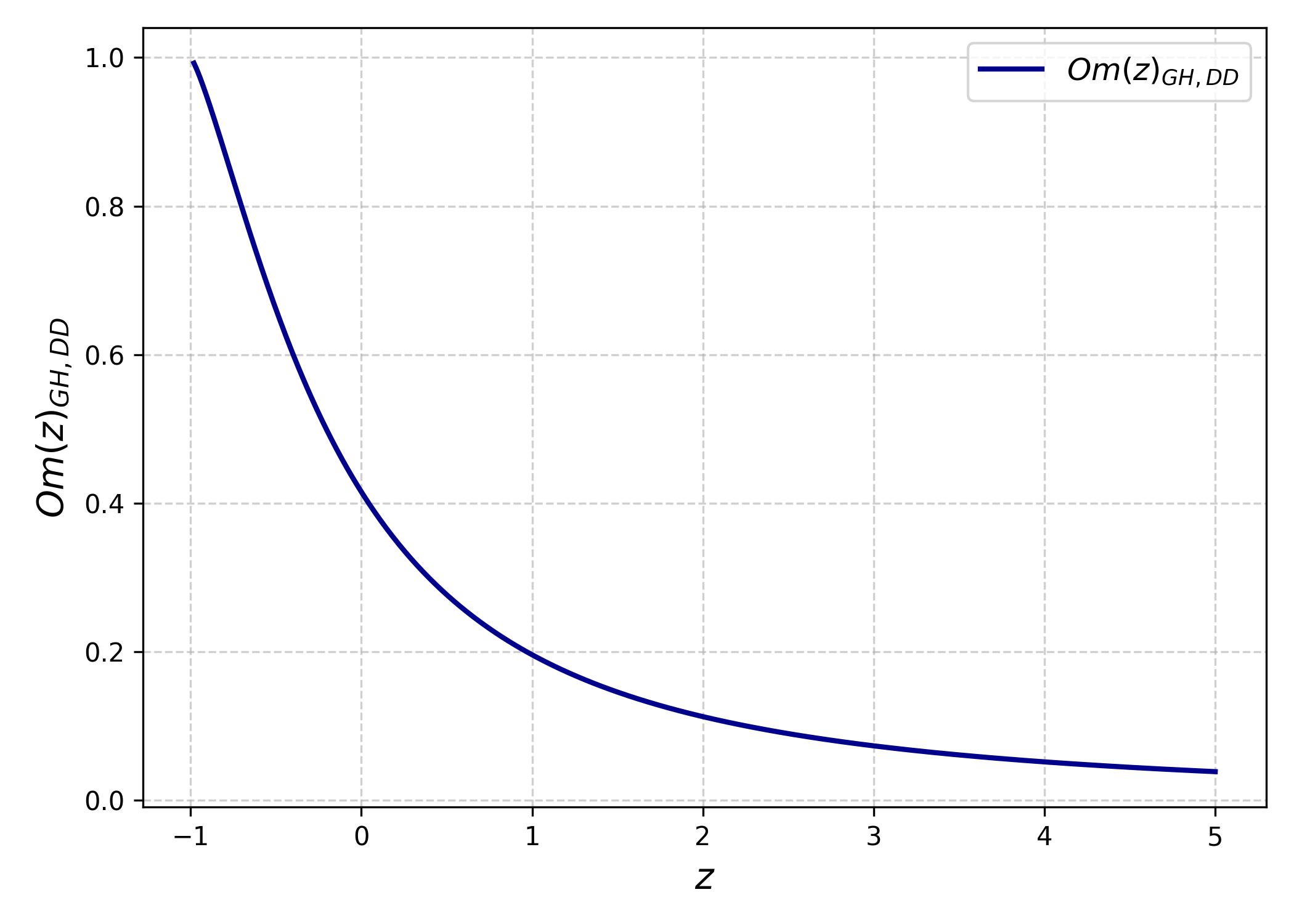}
        \caption{$Om(z)_{GH,DD}$ for $c=0.818$.}
        \label{om2-3DD}
    \end{subfigure}
    \caption{Comparison of the three cases for $Om(z)_{GH,DD}$.}
    \label{fig:omk_all2}
\end{figure}

In Figs. (\ref{om2DD}), (\ref{om2-2DD}) and (\ref{om2-3DD}) we plot the expression of $Om(z)_{GH,DD}$ given in Eq. (\ref{om5-5}) for $c^2=0.46$, $c=0.579$ and $c=0.818$, respectively.\\
For the first and third cases ($c^2 = 0.46$ and $c = 0.818$), the function $Om(z)$ remains positive and monotonically decreasing with redshift, tending to zero at $z \simeq 5$. 
This behavior indicates that the model approaches a cosmological-constant–like regime at high redshift, corresponding to $\omega_D \rightarrow -1$. 
At late times ($z \rightarrow -1$), $Om(z)$ attains higher positive values, showing that the dark-energy contribution becomes dominant as the universe evolves toward its future accelerated phase. 
Conversely, for the intermediate value $c = 0.579$, the $Om(z)$ curve becomes negative for all values of $z$, suggesting an unphysical behavior related to an effective negative matter density parameter.\\ 
Hence, the $Om(z)$ diagnostic effectively distinguishes viable holographic dark-energy models, confirming that only specific values of $c$ yield a consistent cosmic evolution compatible with the observed late-time acceleration.

\section{Squared Speed Of The Sound}
In this section, we aim to gain insights into the behavior of the DE models under consideration in this paper by analyzing the squared speed of sound, indicated by $v_s^2$, a quantity commonly used to assess the stability of various dark energy (DE) models. The squared speed of sound is defined as follows \cite{myung}:
\begin{eqnarray}
v_s^2 = \frac{\dot{p}} {\dot{\rho}}  &=& \frac{\dot{p}_D}{\dot{\rho}_D + \dot{\rho}_m+ \dot{\rho}_k} =\frac{p'_D}{\rho'_D + \rho'_m+\rho_k'}, \label{genvs}
\end{eqnarray}
where $p= p_D$ and $\rho = \rho_D + \rho_m+\rho_k$ indicate, respectively the total pressure and the total energy density of the models.\\
The sign of $v_s^2$ plays a crucial role in analyzing the stability of the background evolution in cosmological models. A negative value of $v_s^2$ indicates the presence of classical instabilities in the evolution of perturbations within the framework of General Relativity \cite{myung,kim}.

Myung \cite{myung} showed that in the Holographic Dark Energy (HDE) model, when the future event horizon is adopted as the infrared (IR) cutoff, the quantity $v_s^2$ remains negative, indicating instability. In contrast, for the Chaplygin gas and tachyon models, $v_s^2$ is found to be non-negative, suggesting classical stability. Kim et al. \cite{kim} further observed that the Agegraphic Dark Energy (ADE) model also leads to a negative $v_s^2$, implying instability in the corresponding perfect fluid description.

More recently, Sharif and Jawad \cite{sharif} demonstrated that the interacting new HDE model is similarly characterized by a negative squared speed of sound. In the context of modified gravity, Jawad et al. \cite{miovs1} found that the $f(G)$ gravity model in an HDE scenario with a power-law scale factor also suffers from classical instability. Likewise, Pasqua et al. \cite{miovs2} showed that a DE model constructed using the Generalized Uncertainty Principle (GUP) and a power-law scale factor $a(t)$ is classically unstable.

\begin{figure}[htbp]
    \centering
    \begin{subfigure}{0.8\textwidth}
        \includegraphics[width=0.45\textwidth]{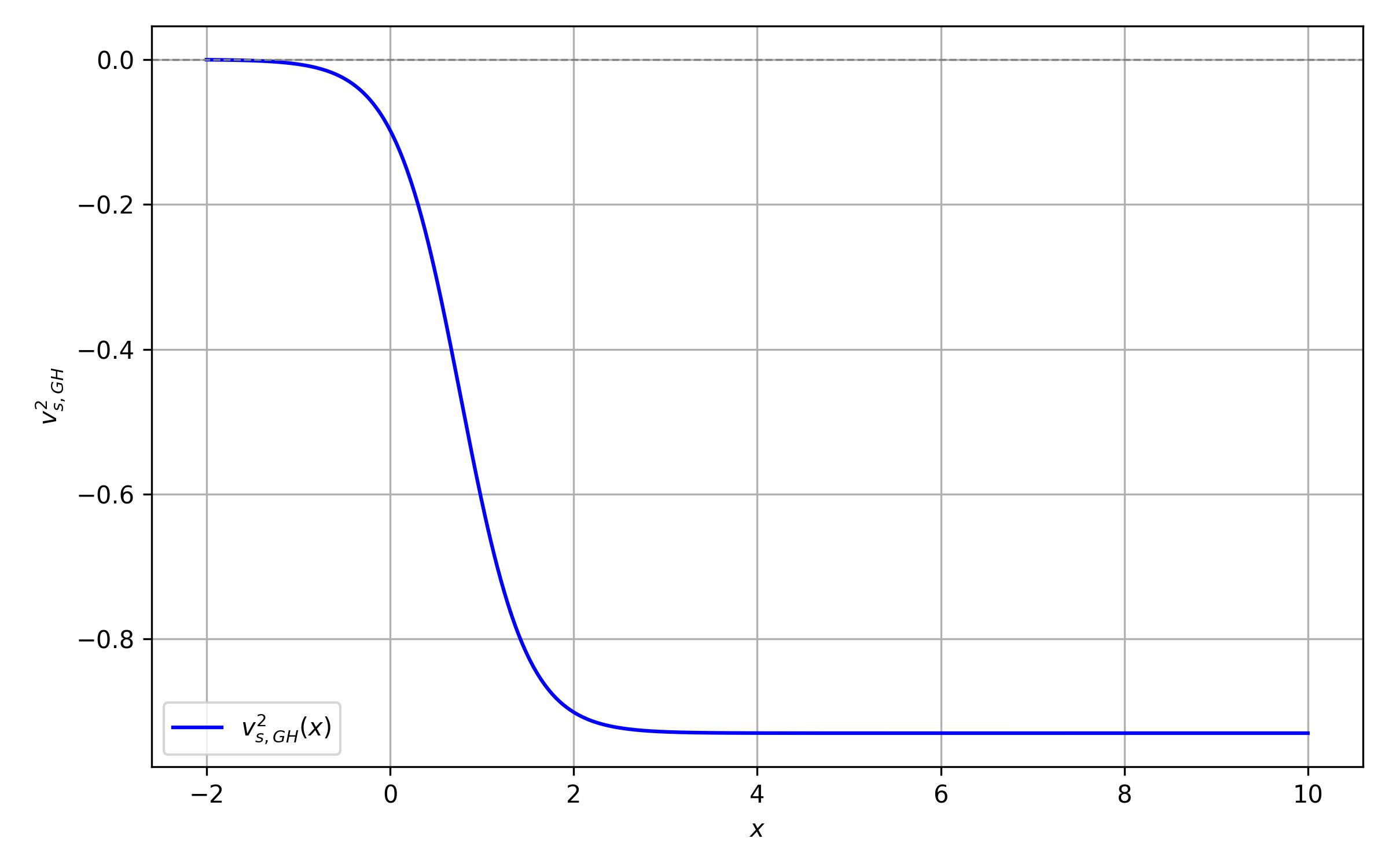}
        \caption{$v_{s,GH}^2$ for $c^2=0.46$.}
        \label{vs1}
    \end{subfigure}\\[0.45cm]
    \begin{subfigure}{0.8\textwidth}
        \includegraphics[width=0.45\textwidth]{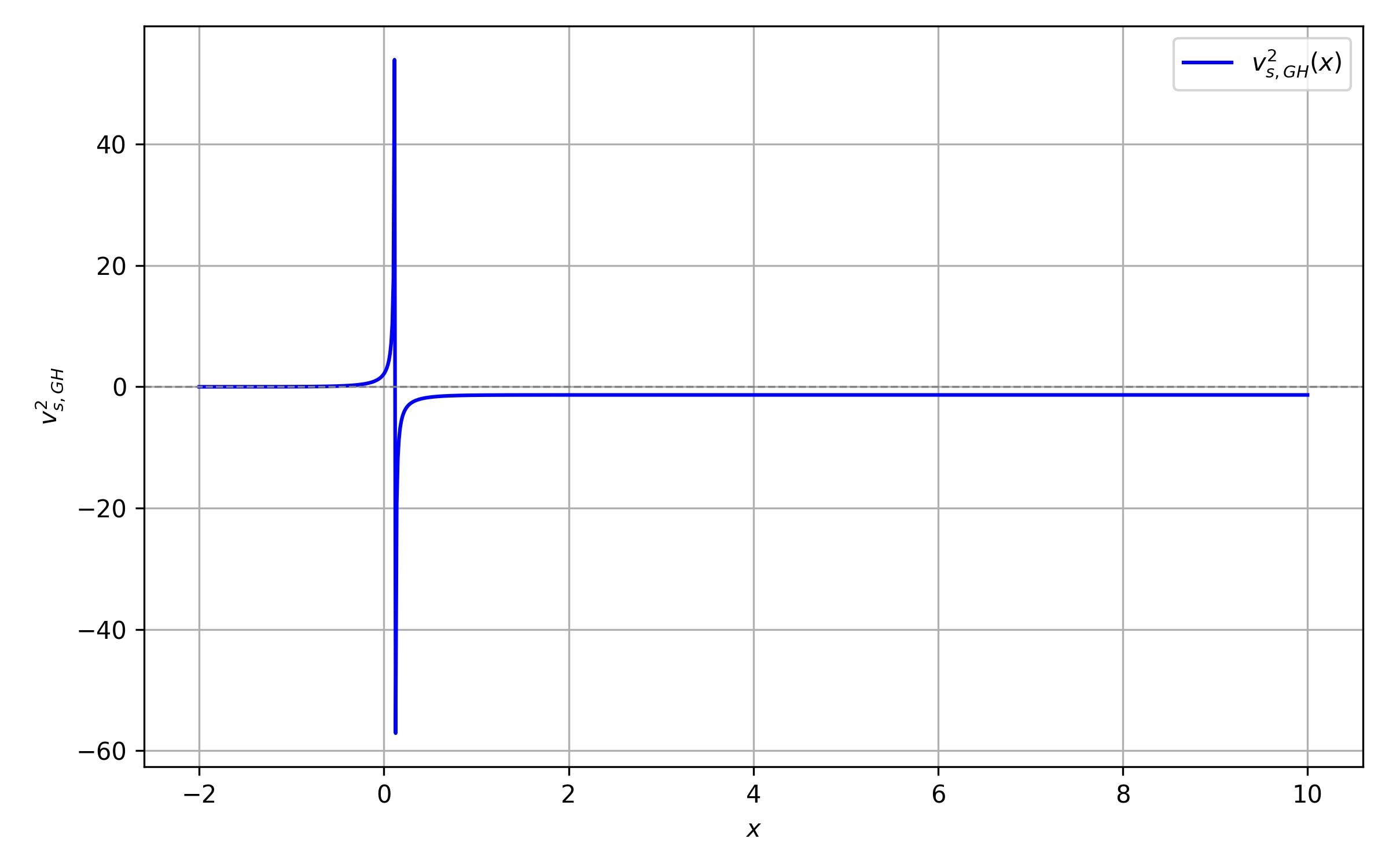}
        \caption{$v_{s,GH}^2$ for $c=0.579$.}
        \label{vs1-2}
    \end{subfigure}\\[0.45cm]
    \begin{subfigure}{0.8\textwidth}
        \includegraphics[width=0.45\textwidth]{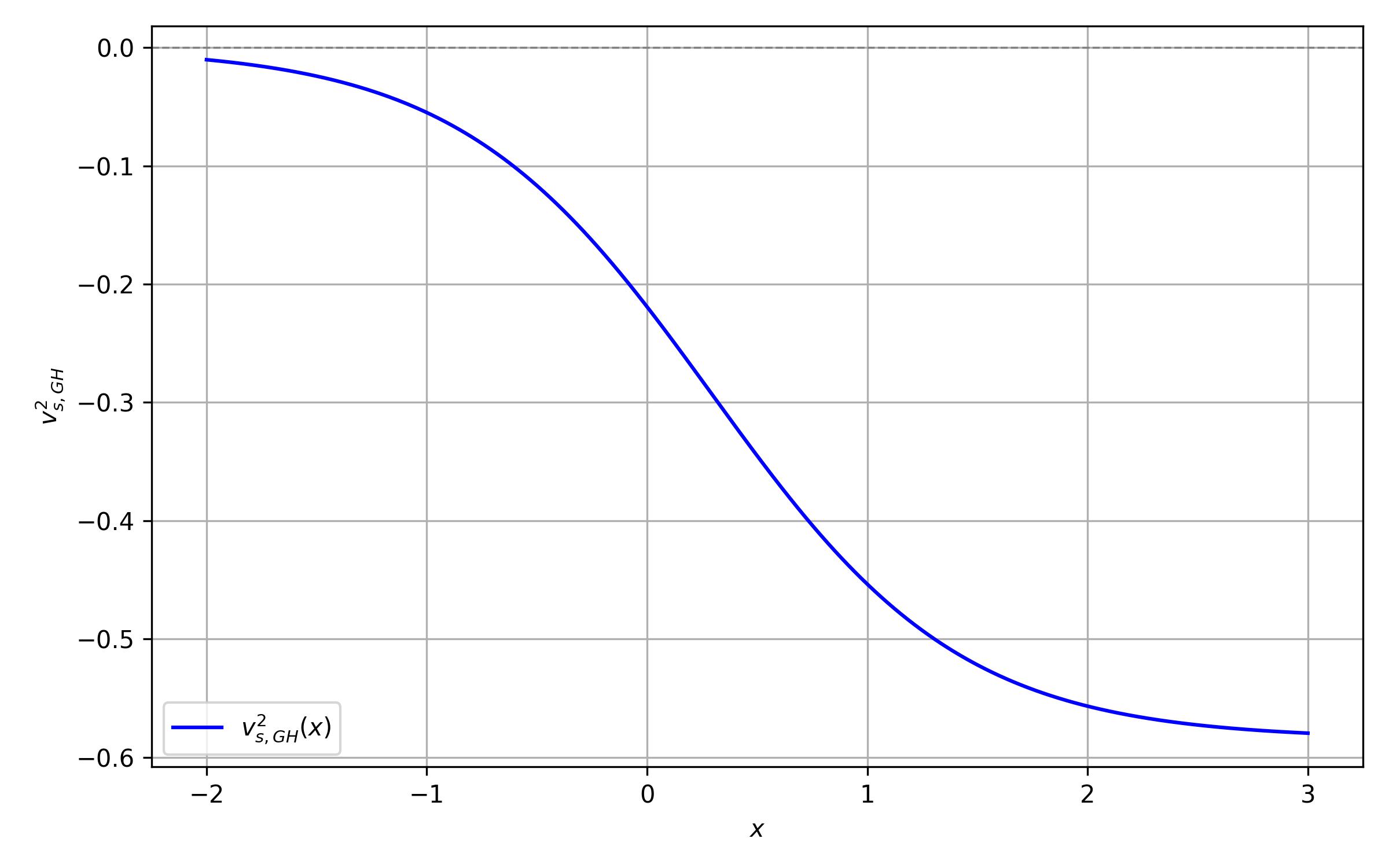}
        \caption{$v_{s,GH}^2$ for $c=0.818$.}
        \label{vs1-3}
    \end{subfigure}
    \caption{Comparison of the three cases for $v_{s,GH}^2$.}
    \label{fig:vsh_all}
\end{figure}

We start considering the first model. \\
For the GH case, we obtain the following expression for $v_{s,GH}^2 (x)$:
\begin{eqnarray}
v_{s,GH}^2 (x)
&=&\frac{(1-\alpha)\left[ \frac{2\left(1 - \alpha\right)}{3}-1  \right] \left[1 - \frac{2 \Omega_{m0}}{2 + c^2(\epsilon - 2)} \right] e^{-2\left(1 - \alpha\right)x}}{\left[1 - \frac{2 \Omega_{m0}}{2 + c^2(\epsilon - 2)} \right] \left(1 - \alpha\right) e^{-2\left(1 - \alpha\right) x} 
+ \left[ \frac{3}{2 + c^2(\epsilon - 2)}\right] \Omega_{m0}e^{-3x}}.  \label{genvs1}
\end{eqnarray}
In Figs. (\ref{vs1}), (\ref{vs1-2}) and (\ref{vs1-3}) we plot the expression of $v_{s,GH}^2 (x)$ given in Eq. (\ref{genvs1}) for $c^2=0.46$, $c=0.579$ and $c=0.818$, respectively.

We find that the model in unstable for $c^2=0.46$ e $c=0.818$ since $v_s^2$ is always negative.\\
In the case with $c=0.579$,  $v_s^2$ exhibits distinctive features that shed light on the stability  properties of the model. At early times (i.e. for $x < 0$),  $v_s^2$ remains positive and close to zero, indicating a classically stable  regime for perturbations. As the Universe evolves toward the present epoch  ($x \approx 0$), the function develops a sharp divergence at $x \approx 0.12$, suggesting the  presence of a transition or singular behavior. The behavior we observe in this case is due to the fact that $\alpha$ is greater than one, which leads to a positive value of the term $-2(1-\alpha)$. In the far future (i.e. $x > 0$), $v_s^2$ settles into negative  values, asymptotically approaching a constant. This signals the onset of  classical instabilities in the perturbative sector, a typical feature in many  dark energy scenarios. Overall, the analysis highlights a transition from  stability in the past to instability in the future, with the present epoch  lying close to a critical point in the sound speed evolution.

\begin{figure}[htbp]
    \centering
    \begin{subfigure}{0.8\textwidth}
        \includegraphics[width=0.45\textwidth]{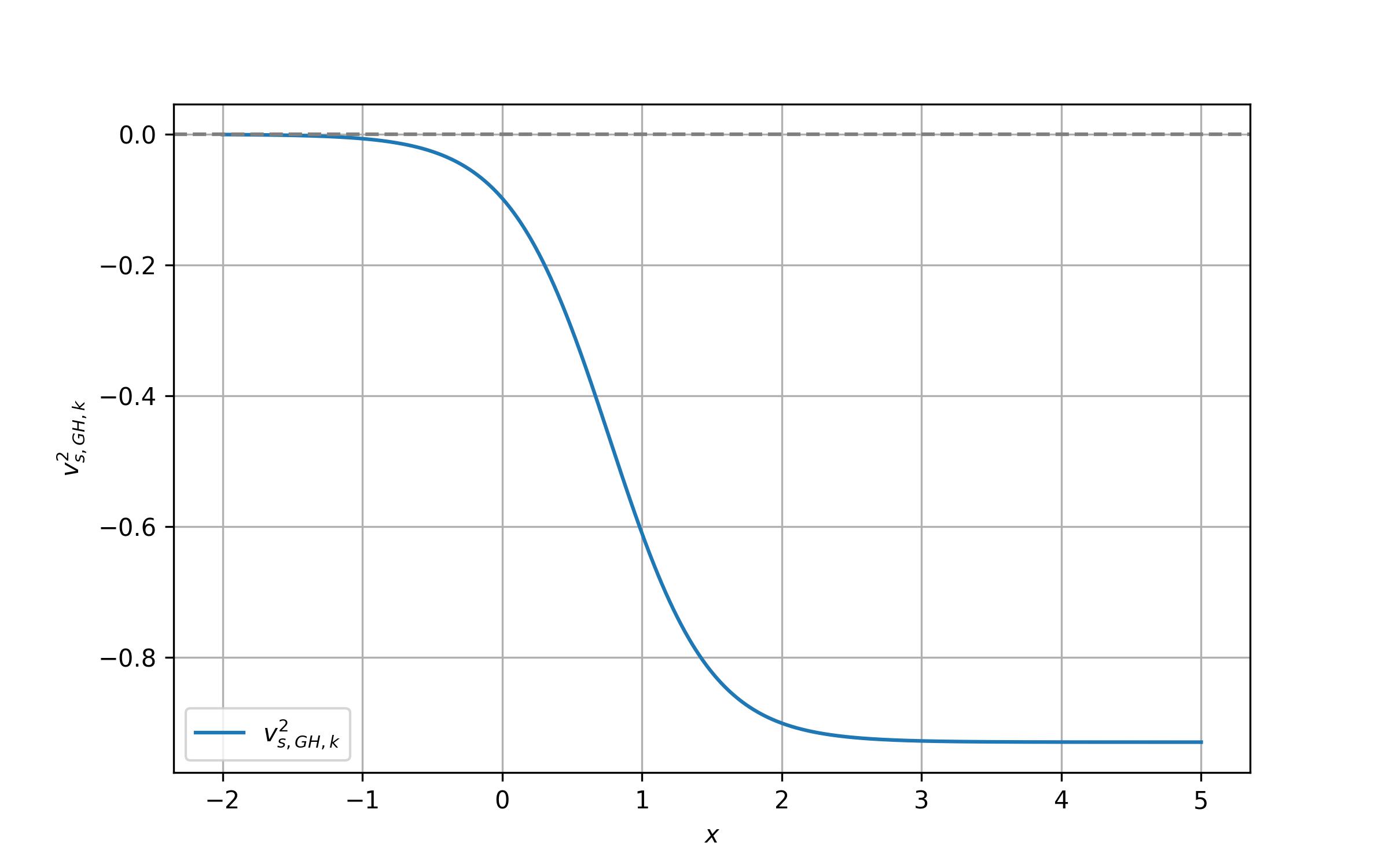}
        \caption{$v_{s,GH,k}^2$ for $c^2=0.46$.}
        \label{vs2}
    \end{subfigure}\\[0.45cm]
    \begin{subfigure}{0.8\textwidth}
        \includegraphics[width=0.45\textwidth]{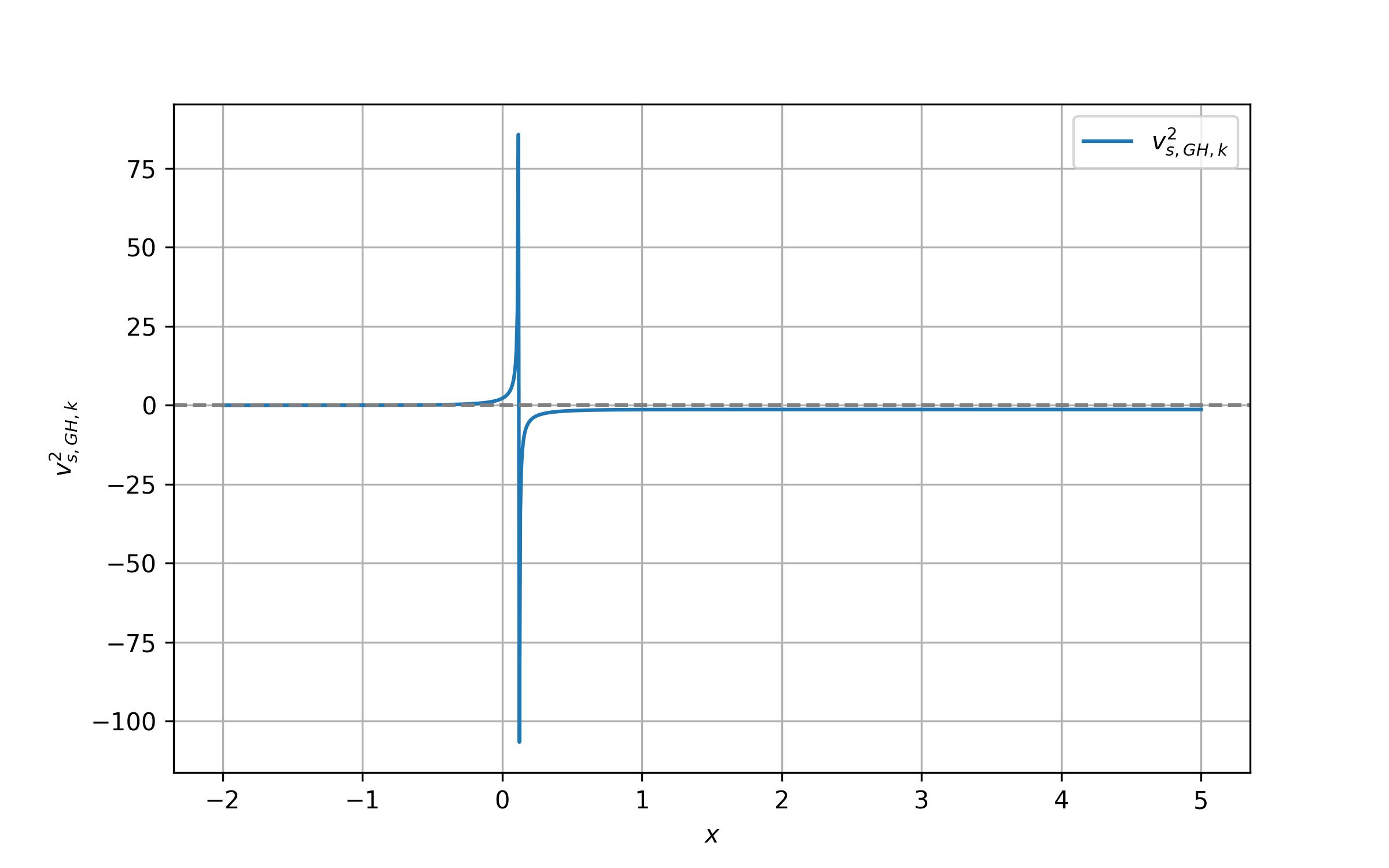}
        \caption{$v_{s,GH,k}^2$ for $c=0.579$.}
        \label{vs2-2}
    \end{subfigure}\\[0.45cm]
    \begin{subfigure}{0.8\textwidth}
        \includegraphics[width=0.45\textwidth]{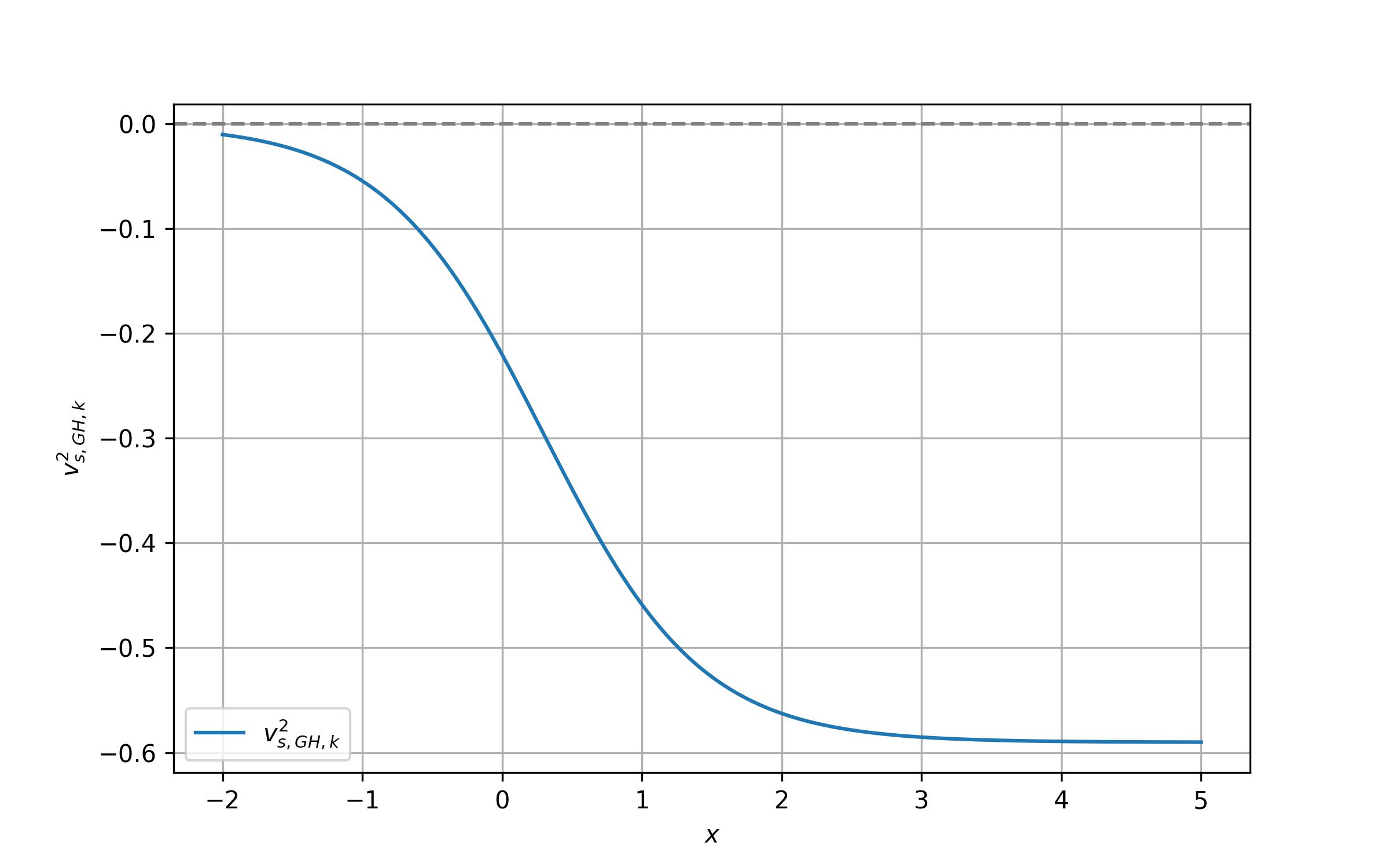}
        \caption{$v_{s,GH,k}^2$ for $c=0.815$.}
        \label{vs2-3}
    \end{subfigure}
    \caption{Comparison of the three cases for $v_{s,GH,k}^2$.}
    \label{fig:vshk_all}
\end{figure}

Considering the presence of spatial curvature, we obtain the following expression for $v_{s,GH,k}^2 (x)$
\begin{eqnarray}
v_{s,GH,k}^2 
&=&\frac{p'_{D_{GH},k}}{\rho'_{D_{GH},k} + \rho'_m+\rho'_k}, \label{genvs2}
\end{eqnarray}
where $\rho'_m$, $\rho'_{D_{GH},k}$, $p'_{D_{GH},k}$  and $\rho'_K$ are defined, respectively, in Eqs. (\ref{boh1}), (\ref{cledi2}), (\ref{ciccio1}) and (\ref{omkprime}).\\
In Figs. (\ref{vs2}), (\ref{vs2-2}) and (\ref{vs2-3}) we plot the expression of $v_{s,GH,k}^2 (x)$ given in Eq. (\ref{genvs2}) for $c^2=0.46$, $c=0.579$ and $c=0.815$, respectively.

We find that the model in unstable for $c^2=0.46$ e $c=0.815$ since $v_s^2$ is always negative.\\
In the case with $c=0.579$,  $v_s^2$ exhibits distinctive features that shed light on the stability  properties of the model. At early times (i.e. for $x < 0$),  $v_s^2$ remains positive and close to zero, indicating a classically stable  regime for perturbations. As the Universe evolves toward the present epoch  ($x \approx 0$), the function develops a sharp divergence at $x \approx 0.119$, suggesting the  presence of a transition or singular behavior. The behavior we observe in this case is due to the fact that $\alpha$ is greater than one, which leads to a positive value of the term $-2(1-\alpha)$. In the far future (i.e. $x > 0$), $v_s^2$ settles into negative  values, asymptotically approaching a constant. This signals the onset of  classical instabilities in the perturbative sector, a typical feature in many  dark energy scenarios. Overall, the analysis highlights a transition from  stability in the past to instability in the future, with the present epoch  lying close to a critical point in the sound speed evolution.

\begin{figure}[htbp]
    \centering
    \begin{subfigure}{0.8\textwidth}
        \includegraphics[width=0.5\textwidth]{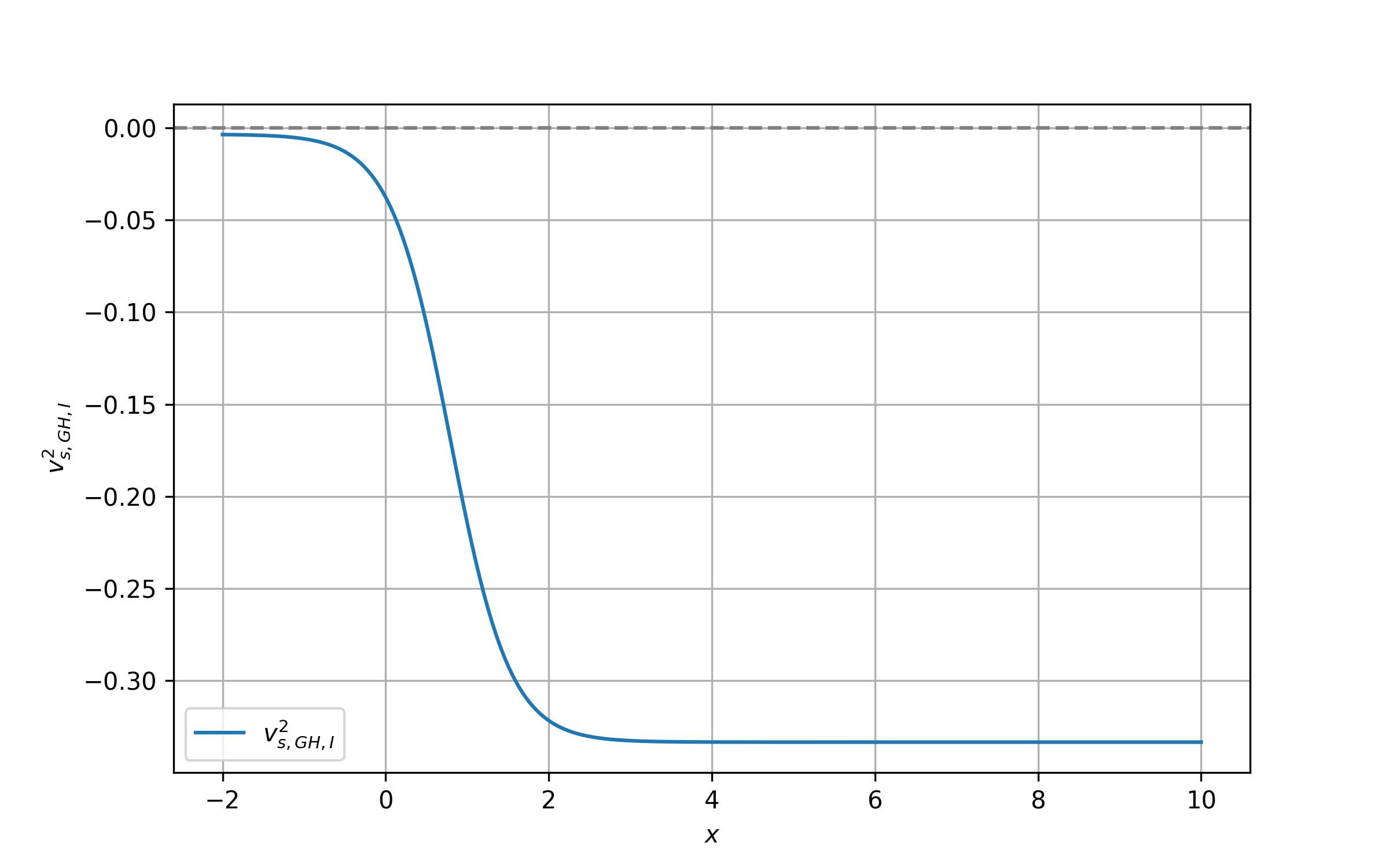}
        \caption{$v_{s,GH,I}^2$ for $c^2=0.46$.}
        \label{vs3}
    \end{subfigure}\\[0.45cm]
    \begin{subfigure}{0.8\textwidth}
        \includegraphics[width=0.5\textwidth]{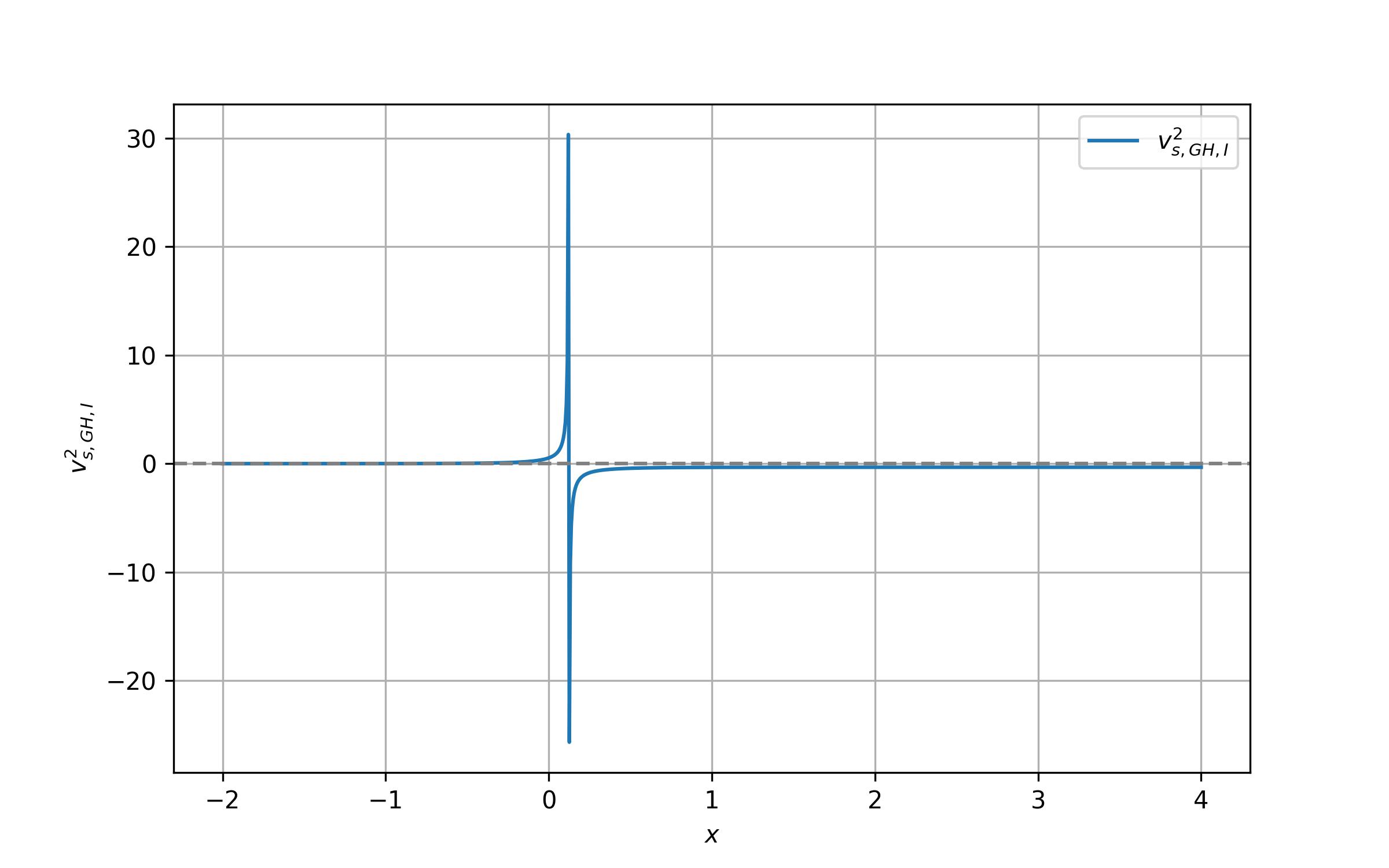}
        \caption{$v_{s,GH,I}^2$ for $c=0.579$.}
        \label{vs3-2}
    \end{subfigure}\\[0.45cm]
    \begin{subfigure}{0.8\textwidth}
        \includegraphics[width=0.5\textwidth]{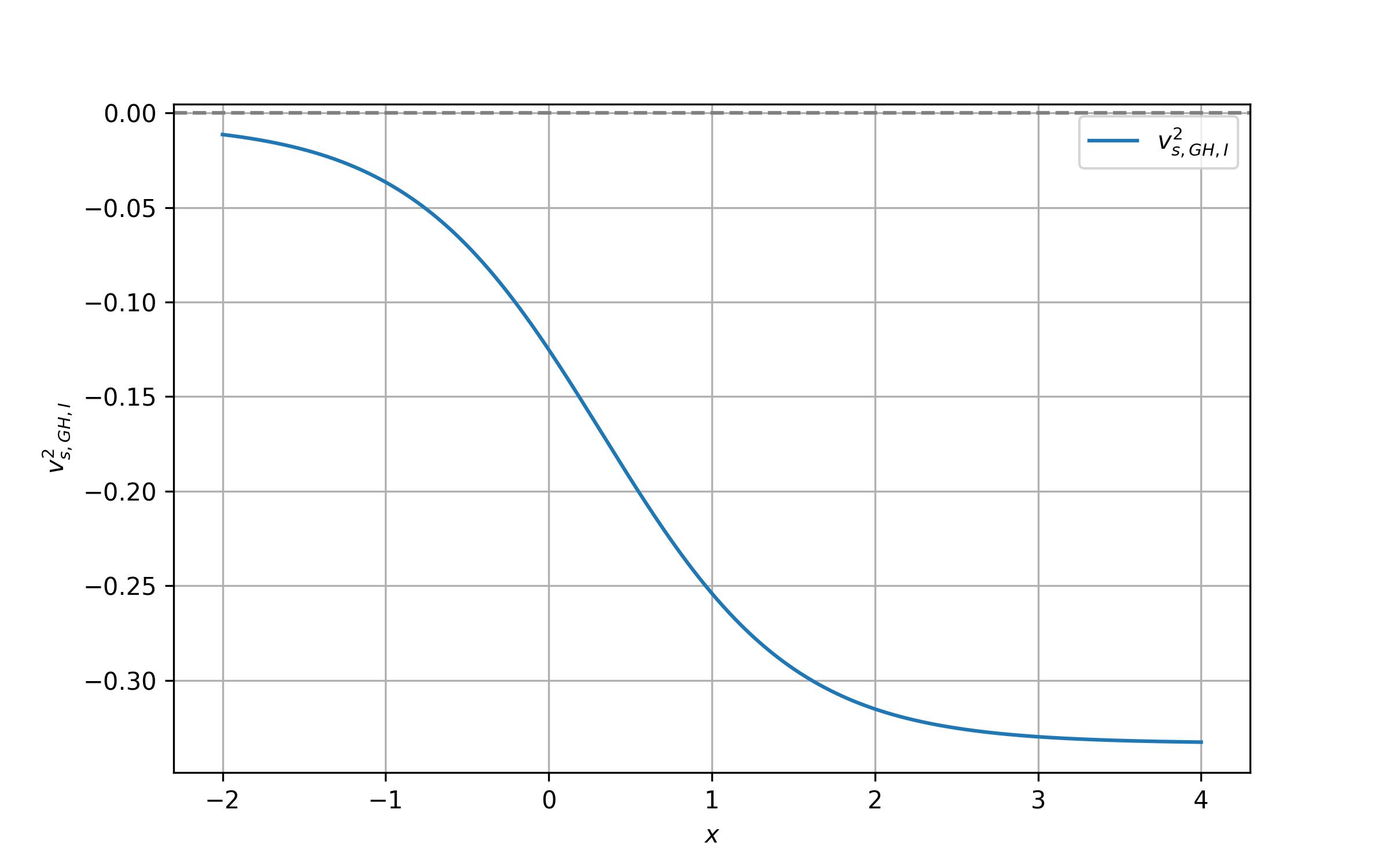}
        \caption{$v_{s,GH,I}^2$ for $c=0.818$.}
        \label{vs3-3}
    \end{subfigure}
    \caption{Comparison of the three cases for $v_{s,GH,I}^2$.}
    \label{fig:vshi_all}
\end{figure}

Taking into account the presence of interacting Dark Sectors, we obtain the following expression for $v_{s,GH,I}^2 (x)$:
\begin{eqnarray}
v_{s,GH,I}^2 
&=&\frac{p'_{D_{GH},I}}{\rho'_{D_{GH},I} + \rho'_{m,I}}, \label{genvs3}
\end{eqnarray}
where $\rho'_{D_{GH},I}$, $p'_{D_{GH},I}$ and $\rho'_{m,I}$ are defined, respectively, in Eqs. (\ref{carolina6}), (\ref{ciccio2}) and (\ref{rhoI}).\\
In Figs. (\ref{vs3}), (\ref{vs3-2}) and (\ref{vs3-3}) we plot the expression of $v_{s,GH,I}^2 (x)$ given in Eq. (\ref{genvs3}) for $c^2=0.46$, $c=0.579$ and $c=0.818$, respectively.

We find that the model in unstable for $c^2=0.46$ e $c=0.818$ since $v_s^2$ is always negative.\\
In the case with $c=0.579$,  $v_s^2$ exhibits distinctive features that shed light on the stability  properties of the model. At early times (i.e. for $x < 0$),  $v_s^2$ remains positive and close to zero, indicating a classically stable  regime for perturbations. As the Universe evolves toward the present epoch  ($x \approx 0$), the function develops a sharp divergence at $x \approx 0.123$, suggesting the  presence of a transition or singular behavior. The behavior we observe in this case is due to the fact that $\alpha$ is greater than one, which leads to a positive value of the term $-2(1-\alpha)$. In the far future (i.e. $x > 0$), $v_s^2$ settles into negative  values, asymptotically approaching a constant. This signals the onset of  classical instabilities in the perturbative sector, a typical feature in many  dark energy scenarios. Overall, the analysis highlights a transition from  stability in the past to instability in the future, with the present epoch  lying close to a critical point in the sound speed evolution.

\begin{figure}[htbp]
    \centering
    \begin{subfigure}{0.8\textwidth}
        \includegraphics[width=0.5\textwidth]{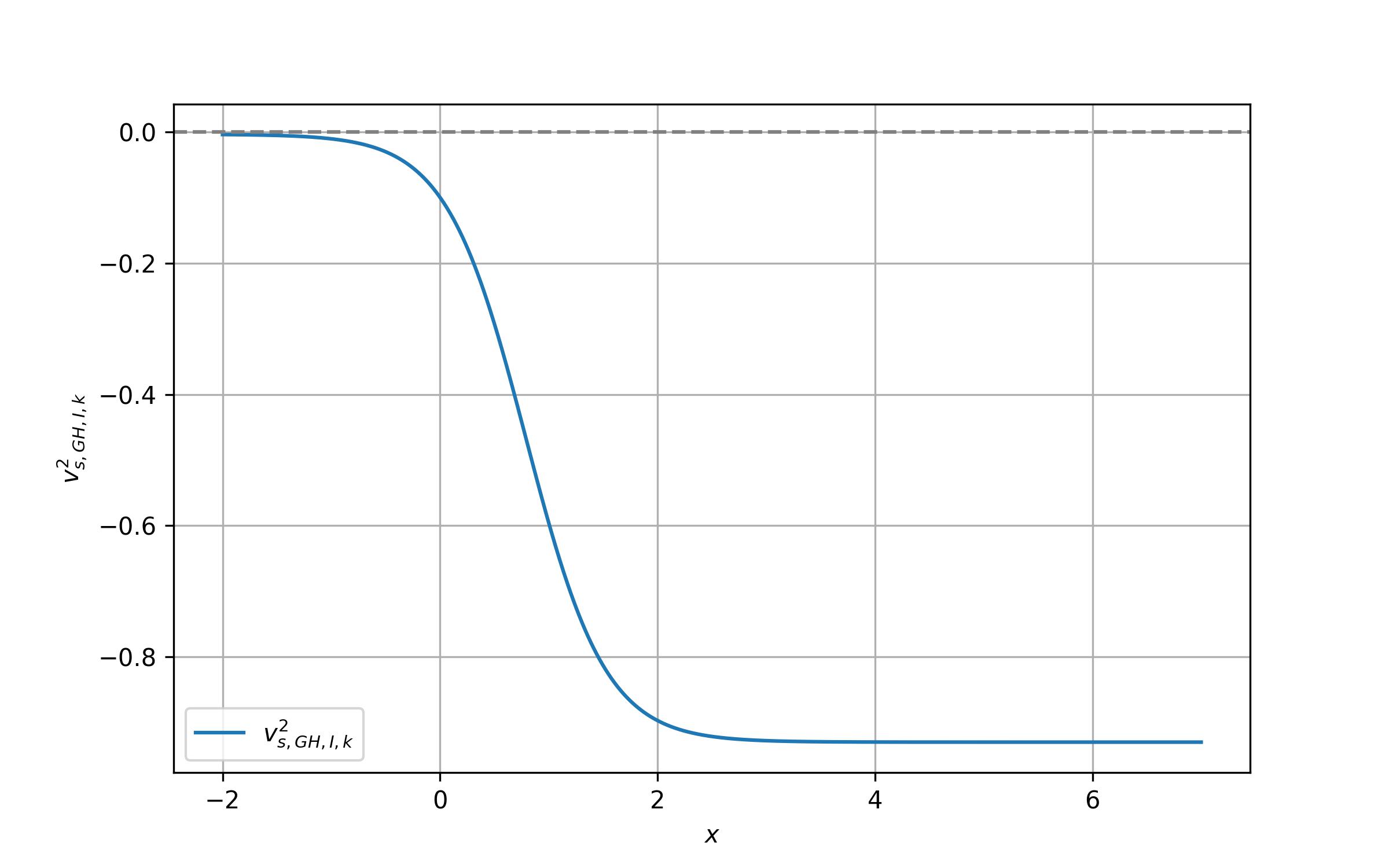}
        \caption{$v_{s,GH,I,k}^2$ for $c^2=0.46$.}
        \label{vs4}
    \end{subfigure}\\[0.45cm]
    \begin{subfigure}{0.8\textwidth}
        \includegraphics[width=0.5\textwidth]{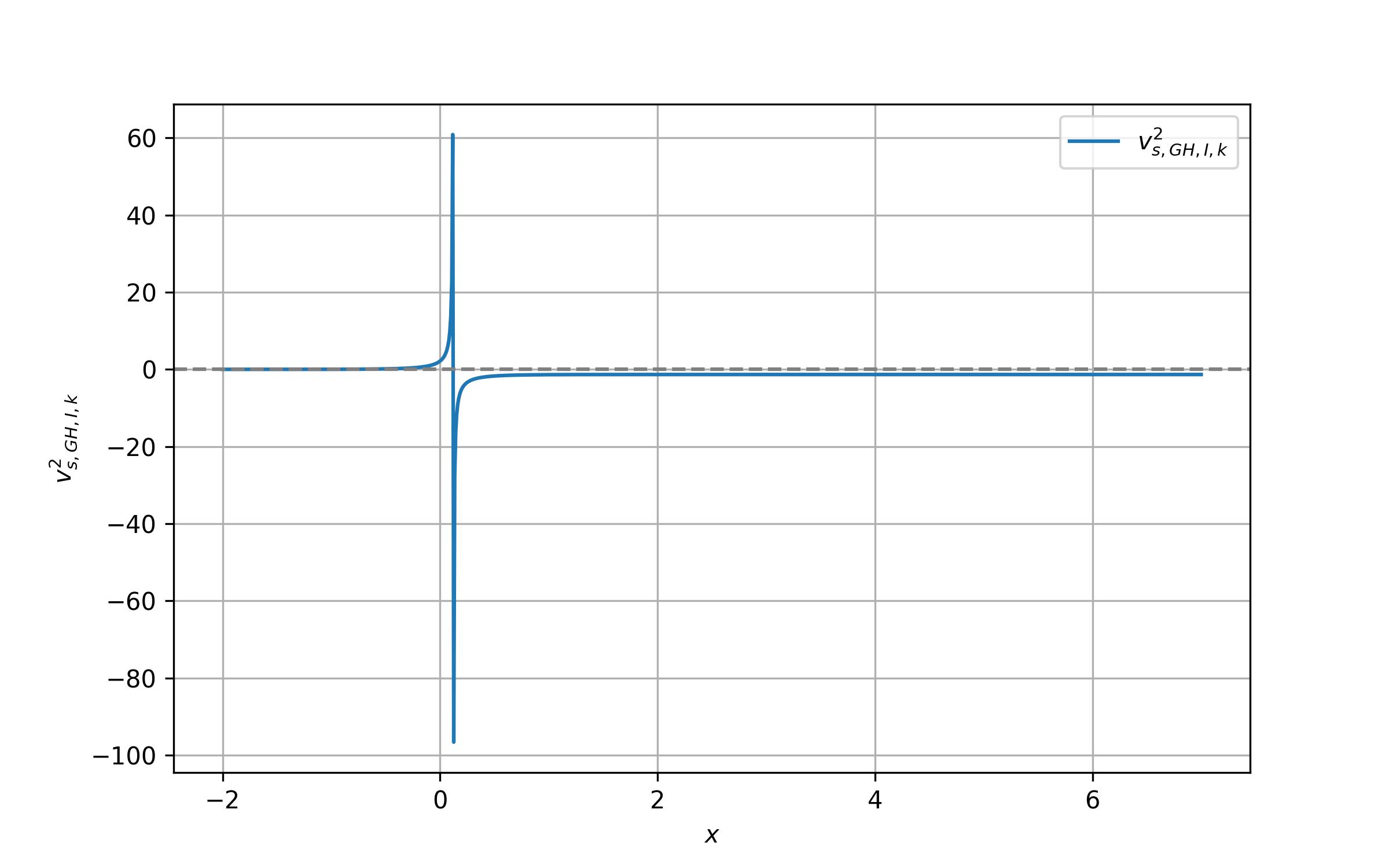}
        \caption{$v_{s,GH,I,k}^2$ for $c=0.579$.}
        \label{vs4-2}
    \end{subfigure}\\[0.45cm]
    \begin{subfigure}{0.8\textwidth}
        \includegraphics[width=0.5\textwidth]{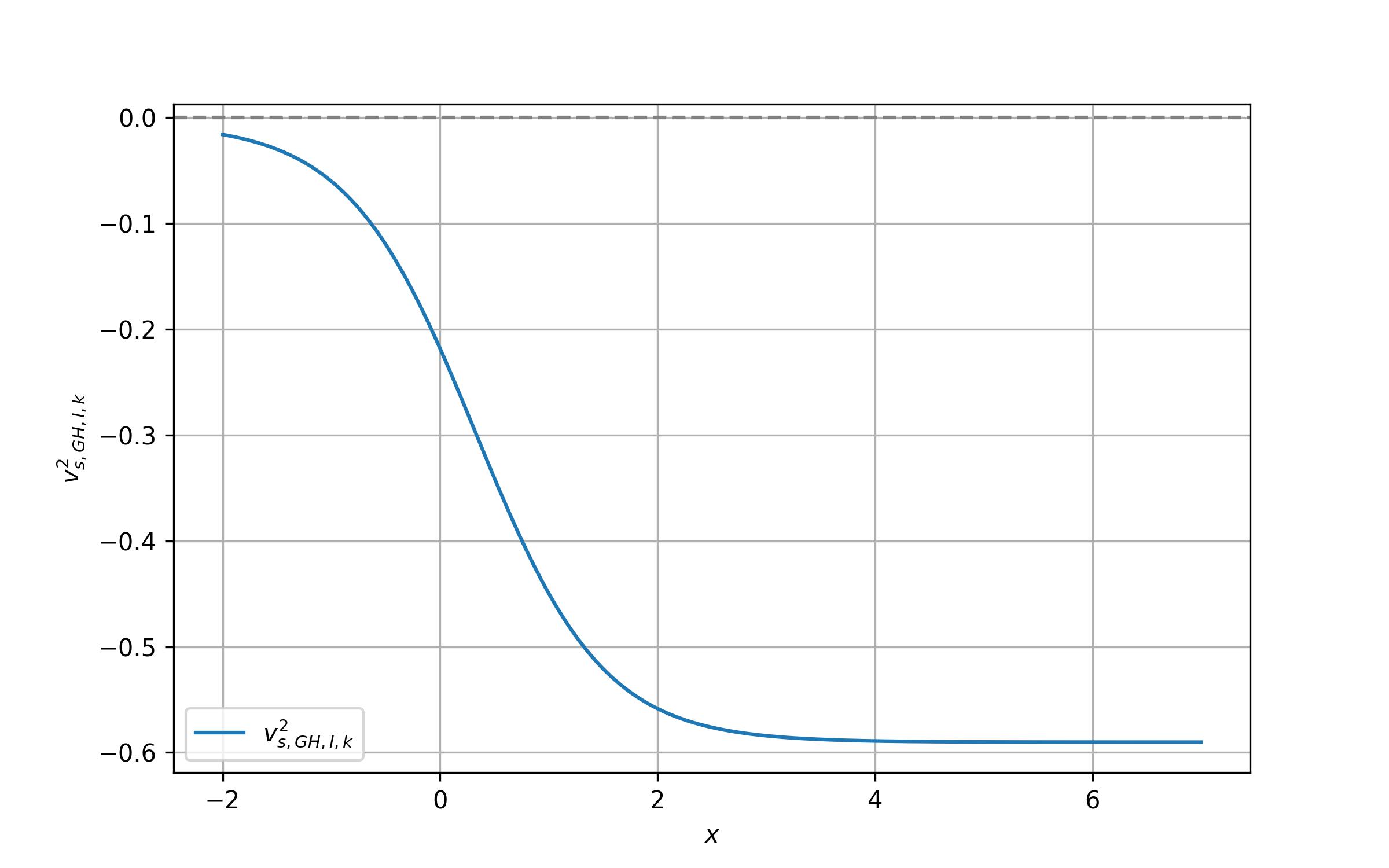}
        \caption{$v_{s,GH,I,k}^2$ for $c=0.815$.}
        \label{vs4-3}
    \end{subfigure}
    \caption{Comparison of the three cases for $v_{s,GH,I,k}^2$.}
    \label{fig:vshik_all}
\end{figure}

Finally, considering the presence of both spatial curvature and interaction, we obtain the following expression for $v_{s,GH,I,k}^2 (x)$:
\begin{eqnarray}
v_{s,GH,I,k}^2 
&=&\frac{p'_{D_{GH},I,k}}{\rho'_{D_{GH},I,k} + \rho'_{m,I}+\rho'_k}, \label{genvs4}
\end{eqnarray}
where $\rho'_{m,I}$, $\rho'_{D_{GH},I,k}$, $p'_{D_{GH},I,k}$  and $\rho'_k$ are defined, respectively, in Eqs.  (\ref{rhoI}), (\ref{carolina31}), (\ref{ciccio3}) and (\ref{omkprime}).\\
In Figs. (\ref{vs4}), (\ref{vs4-2}) and (\ref{vs4-3}) we plot the expression of $v_{s,GH,I,k}^2 (x)$ given in Eq. (\ref{genvs4}) for $c^2=0.46$, $c=0.579$ and $c=0.815$, respectively.

We find that the model in unstable for $c^2=0.46$ e $c=0.815$ since $v_s^2$ is always negative.\\
In the case with $c=0.579$,  $v_s^2$ exhibits distinctive features that shed light on the stability  properties of the model. At early times (i.e. for $x < 0$),  $v_s^2$ remains positive and close to zero, indicating a classically stable  regime for perturbations. As the Universe evolves toward the present epoch  ($x \approx 0$), the function develops a sharp divergence at $x \approx 0.123$, suggesting the  presence of a transition or singular behavior. The behavior we observe in this case is due to the fact that $\alpha$ is greater than one, which leads to a positive value of the term $-2(1-\alpha)$. In the far future (i.e. $x > 0$), $v_s^2$ settles into negative  values, asymptotically approaching a constant. This signals the onset of  classical instabilities in the perturbative sector, a typical feature in many  dark energy scenarios. Overall, the analysis highlights a transition from  stability in the past to instability in the future, with the present epoch  lying close to a critical point in the sound speed evolution.

In the limiting case of a flat Dark Dominated Universe, we obtain:
\begin{eqnarray}
v_{s,GH,DD}^2 
&=&\frac{p'_{D_{GH},DD}}{\rho'_{D_{GH},DD}}. \label{}
\end{eqnarray}
Using the expressions of $\rho'_{D_{GH},DD}$ and $p'_{D_{GH},DD}$ we derived in Eqs. (\ref{totonno1}) and (\ref{totonno2}), we can write:
\begin{eqnarray}
v_{s,GH,DD}^2 =-\frac{2\alpha+1}{3}.
\end{eqnarray}
Therefore, we obtain that $v_{s,GH,DD}^2  \approx -0.930$ for the case with $c^2=0.46$ while $v_{s,GH,DD}^2  \approx -1.341$ for the case with $c=0.579$. Finally, for $c=0.818$, we obtain $v_{s,GH,DD}^2  \approx -0.585$ \\

\section{Cosmographic Parameters}
In this section, we extract important cosmological insights about the model under investigation by analyzing the properties of the cosmographic parameters.

Standard candles, such as Type Ia supernovae (SNe Ia), are powerful tools in modern cosmology, as they can be employed to reconstruct the Hubble diagram—that is, the redshift–distance relation—extending to high redshift values $ z $. A common approach involves constraining a parameterized model using observational data, in order to test its viability and determine its free parameters. However, this method is inherently model-dependent, and ongoing debates persist within the scientific community regarding the robustness and reliability of the constraints it produces on derived cosmological quantities.

To overcome this limitation, one can adopt a model-independent framework known as cosmography, which is based on the Taylor expansion of the scale factor $ a(t) $ with respect to cosmic time $ t $. This expansion yields a distance–redshift relation that does not rely on the specific form of the cosmological model, making cosmography a powerful tool for probing the kinematic behavior of the Universe.

Cosmography can be considered a fundamental benchmark in the study of cosmic dynamics, providing general constraints that any viable cosmological model must be consistent with. In this context, it is useful to introduce the following cosmographic quantities:
\begin{eqnarray}
q &=& -\left(\frac{\ddot{a}} {a}\right)H^{-2}=  -\frac{a^{\left(2\right)}a}{\dot{a}^2}  \label{par1}, \\
j &=& \left(\frac{1}{a}\right)\left(\frac{d^3a}{dt^3}\right)H^{-3} = \frac{a^{\left(3\right)}a^2}{\dot{a}^3}  \label{par2}, \\
s &=&\left( \frac{1}{a}\right)\left(\frac{d^4a}{dt^4}\right)H^{-4}= -\frac{a^{\left(4\right)}a^3}{\dot{a}^4}  \label{par3}, \\
l &=&   \left(\frac{1}{a}\right)\left(\frac{d^5a}{dt^5}\right)H^{-5}= -\frac{a^{\left(5\right)}a^4}{\dot{a}^5}, \label{par4}\\
m &=&   \left(\frac{1}{a}\right)\left(\frac{d^6a}{dt^6}\right)H^{-6}= -\frac{a^{\left(6\right)}a^5}{\dot{a}^6}. \label{par5}
\end{eqnarray}
In general, the $i$-th cosmographic parameter $x^i$ can be obtained thanks to the following general expression:
\begin{eqnarray}
x^{i} &=& \left( -1  \right)^{i+1}\left(\frac{1}{H^{i}}\right) \frac{a^{\left(i\right)}} {a}\nonumber \\
&=& \left( -1  \right)^{i+1} \frac{a^{\left(i\right)}a^{i-1}} {\dot{a}^{i+1}},
\end{eqnarray}
Here, the index $ i $, when enclosed in parentheses, denotes the order of the derivative with respect to cosmic time $ t $, whereas when written without parentheses, it refers to the power-law index of the corresponding quantity.

The quantities defined in Eqs.~(\ref{par1}), (\ref{par2}), (\ref{par3}), and (\ref{par4}) are commonly known as the deceleration, jerk, snap, and lerk parameters, respectively. The parameter $ m $, also known as max-out parameter, has been studied, for example, by Dunsby and Luongo \cite{orlando1} and Aviles et al. \cite{orlando2}. In the previous section, we have already derived the expression of the deceleration parameter $ q $, which can be used as start in order to obtain the expression for the other parameters. \\

The present-day values of these cosmographic parameters, indicated with the subscript 0 (corresponding to $ z=0 $ or equivalently $ t=t_0 $), can be used to characterize the evolutionary state of the Universe. For example, a negative value of $ q_0 $ signals an accelerated expansion, whereas the value of $ j_0 $ helps to discriminate between different accelerating models. Moreover, the jerk parameter $ j_0 $ quantifies the rate of change of the acceleration, with a positive $ j_0 $ implying that the deceleration parameter $ q $ changes sign during the cosmic evolution. \\
Recent studies have also attempted to constrain the snap and lerk parameters. For example, Capozziello and Izzo \cite{capoizzo1} found $ s_0 = 8.32 \pm 12.16 $, while John \cite{capoizzo2, capoizzo3} reported  $ j_0 = 2.7 \pm 6.7 $, $ s_0 = 36.5 \pm 52.9 $ and $ l_0 = 142.7 \pm 320 $. In Luongo and Muccino, it was obtained that $q_0 = -0.55$, $j_0 = 1$ and $s_0 = -0.35$ for the $\Lambda CDM$ model, $q_0 = -0.445$, $j_0 = 0.717$, $s_0 = -0.706$ for the $wCDM$ model and $q_0 = -0.445$, $j_0 = 1.242$, $s_0 = 1.155$ for the $CPL$ parameterization, assuming a value of $\Omega_m=0.3$.
As evident from these results, the uncertainties in the snap and lerk parameters are of order $ 200\% $, indicating that more precise measurements are necessary for meaningful comparisons between observational constraints and theoretical predictions. On the other hand, Aviles et al. \cite{orlando2} provided constraints on the max-out parameter, finding $ m_0 = 71.93^{+382.17}_{-316.76} $. \\

By employing the definitions given in Eqs.~(\ref{par1}), (\ref{par2}), (\ref{par3}), (\ref{par4}), and (\ref{par5}), we can straightforwardly construct the sixth-order Taylor expansion of the scale factor $ a(t) $ as follows:

\begin{eqnarray}
\frac{a\left(t\right)}{a\left(t_0\right)} &=& 1+H_0 \left( t -t_0 \right) - \left(\frac{q_0}{2!}\right)H_0^2 \left( t -t_0 \right)^2 +  \left(\frac{j_0}{3!}\right)H_0^3 \left( t -t_0 \right)^3  \nonumber \\
 &&+\left(\frac{s_0}{4!}\right)H_0^4 \left( t -t_0 \right)^4   +    \left(\frac{l_0}{5!}\right)H_0^5 \left( t -t_0 \right)^5 \nonumber \\
&&+    \left(\frac{m_0}{6!}\right)H_0^6 \left( t -t_0 \right)^6  +O \left[ \left( t-t_0  \right)^7 \right], \label{exp}
\end{eqnarray}
where $t_0 \approx 1/H_0$ is the present day age of the Universe, while $H_0$ indicates the present day value of the Hubble parameter $H$.\\
The deceleration parameter $ q $, previously introduced and analyzed in the other Sections, plays a crucial role in describing the dynamics of cosmic expansion. In particular, we have observed that in order to obtain the crossing divide line crossing it is required to have $ q < 0 $, a condition that aligns with the predictions of contemporary cosmological models.
\\
The jerk parameter $ j $, which is also known as the statefinder parameter $ r $ (which we have already studied in Section...), constitutes a natural extension beyond the Hubble parameter $ H $ and the deceleration parameter $ q $, and provides a more refined diagnostic for distinguishing between different dark energy models \cite{alam,arab}.

The snap parameter $ s $, which involves the fourth time derivative of the scale factor $ a(t) $, is also sometimes referred to as the \emph{kerk} parameter. It has been thoroughly discussed in the works of Dabrowski \cite{dabro}, Dunajski \& Gibbons \cite{duna}, and Arabsalmania \& Sahni \cite{arab}.\\

The lerk parameter, on the other hand, involves the fifth time derivative of the scale factor. Further information and discussion can be found in the paper by Dabrowski \cite{dabro}.\\

Some relations involving the first time derivatives of the cosmographic parameters and the other cosmographic parameters are given by the following expressions:
\begin{eqnarray}
\frac{dq}{dt} &=& -H \left( j -2q^2 -q  \right),  \\
\frac{dj}{dt} &=&H \left[s+j\left( 2+3q  \right)   \right], \\
\frac{ds}{dt} &=&H \left[ l+s \left(3+4q \right)  \right],  \\
\frac{dl}{dt} &=& H \left[m+l\left(4+5q \right)   \right].
\end{eqnarray}
Equivalent expressions involving the cosmographic parameters and the time derivatives of the Hubble parameter $H$ (i.e., the Hubble rate) are the following ones:
\begin{eqnarray}
\frac{d^2H}{dt^2} &=& H^3 \left( j + 3q + 2  \right), \label{}\\
\frac{d^3H}{dt^3} &=& H^4 \left[ s -4j -3q\left( q + 4 \right) -6  \right], \label{}\\
\frac{d^4H}{dt^4} &=& H^5 \left[ l - 5s + 10\left(q+2 \right)j + 30 \left(q+2 \right)q +24  \right], \label{}\\
\frac{d^5H}{dt^5} &=& H^6 \left\{ m - 10j^2 -120j\left(q+1 \right)-3\left[2l + 5\left(24q + 18q^2 + 2q^3-2s-qs +8    \right)   \right]   \right\}.
\end{eqnarray}
From the above Equations, we easily obtain the following expressions for $j$, $s$, $l$ and $m$:
\begin{eqnarray}
j &=& \frac{\ddot{H} }{H^3} - 3q -2,\label{rosa1}\\
s &=& \left(\frac{1}{H^4}\right)\frac{d^3H}{dt^3} +4j +3q\left( q + 4 \right) +6, \label{rosa2}\\
l &=& \left(\frac{1}{H^5}\right)\frac{d^4H}{dt^4} + 5s - 10\left(q+2 \right)j - 30 \left(q+2 \right)q -24, \label{rosa3}\\
m&=& \left(\frac{1}{H^6}\right)\frac{d^5H}{dt^5} + 10j^2 + 120j\left( q+1  \right) + 3 \left[ 2l + 5\left(24q + 18q^2 + 2q^3-2s-qs +8    \right)  \right].\label{rosa4}
\end{eqnarray}
Passing from time derivatives with respect to time to derivatives with respect to $x$, we obtain:
\begin{eqnarray}
\frac{1}{H^3} \frac{d^2 H}{dt^2} 
&=& \left( \frac{1}{h^2} \frac{dh^2}{dx} \right)^2
+ \frac{1}{h^2} \frac{d^2 h^2}{dx^2} \label{rosa5} \\
\frac{1}{H^4} \frac{d^3H}{dt^3}
&=&
\left( \frac{1}{h^2} \frac{dh^2}{dx} \right)^3
+ 3 \left( \frac{1}{h^2} \frac{dh^2}{dx} \right)\left( \frac{1}{h^2} \frac{d^2h^2}{dx^2} \right)
+ \frac{1}{h^2} \frac{d^3h^2}{dx^3} \label{rosa6}\\
\frac{1}{H^5} \frac{d^4 H}{dt^4}
&=&
\left(  \frac{1}{h^2} \frac{dh^2}{dx} \right)^4
+ 6 \left(  \frac{1}{h^2} \frac{dh^2}{dx} \right)^2 \left( \frac{1}{h^2} \frac{d^2h^2}{dx^2}\right)\nonumber \\
&&
+ 4 \left(  \frac{1}{h^2} \frac{dh^2}{dx} \right) \left( \frac{1}{h^2} \frac{d^3h^2}{dx^3}\right)
+ 3 \left( \frac{1}{h^2} \frac{d^2h^2}{dx^2} \right)^2
+ \frac{1}{h^2} \frac{d^4h^2}{dx^4}\label{rosa7}\\
\frac{1}{H^6} \frac{d^5 H}{dt^5}
&=&
\left( \frac{1}{h^2} \frac{dh^2}{dx} \right)^5
+ 10 \left( \frac{1}{h^2} \frac{dh^2}{dx} \right)^3 \left( \frac{1}{h^2} \frac{d^2h^2}{dx^2} \right)
+ 10 \left( \frac{1}{h^2} \frac{dh^2}{dx} \right)^2 \left( \frac{1}{h^2} \frac{d^3h^2}{dx^3}\right)
\nonumber \\
&&+ 15 \left( \frac{1}{h^2} \frac{dh^2}{dx} \right) \left( \frac{1}{h^2} \frac{d^2h^2}{dx^2} \right)^2+ 10 \left( \frac{1}{h^2} \frac{d^2h^2}{dx^2} \right) \left( \frac{1}{h^2} \frac{d^3h^2}{dx^3} \right)\nonumber \\
&&
+ 5 \left( \frac{1}{h^2} \frac{dh^2}{dx} \right) \left( \frac{1}{h^2} \frac{d^4h^2}{dx^4} \right)
+ \frac{1}{h^2} \frac{d^5h^2}{dx^5}. \label{rosa8}
\end{eqnarray}

Defining 
\begin{eqnarray}
W_j&=&\frac{1}{H^3} \frac{d^2 H}{dt^2}, \\
W_s&=&\frac{1}{H^4} \frac{d^3H}{dt^3},\\
W_l&=&\frac{1}{H^5} \frac{d^4 H}{dt^4},\\
W_m&=&\frac{1}{H^6} \frac{d^5 H}{dt^5},
\end{eqnarray}
we can write:
\begin{eqnarray}
j &=& W_j- 3q -2,\label{rosa1}\\
s &=& W_s+4j +3q\left( q + 4 \right) +6, \label{rosa2}\\
l &=& W_l + 5s - 10\left(q+2 \right)j - 30 \left(q+2 \right)q -24, \label{rosa3}\\
m&=& W_m + 10j^2 + 120j\left( q+1  \right) + 3 \left[ 2l + 5\left(24q + 18q^2 + 2q^3-2s-qs +8    \right)  \right],\label{rosa4}
\end{eqnarray}
or equivalently:
\begin{eqnarray}
j(x) &=& \frac{1}{2} \frac{d^2 h^2}{dx^2} + \frac{3}{2} \frac{d h^2}{dx} + h^2,\label{rosa1}\\
s(x) &=& \frac{(h^2)'''}{2 \, h^2} + 2 \, \frac{(h^2)''}{h^2} - \frac{3}{4}\left(\frac{  (h^2)'}{h^2}\right)^2 + 3 \, \frac{(h^2)'}{h^2} + 1 ,\label{rosa2}\\
l(x) &=& \frac{(h^ 2)^{(4)}}{2\,h^2} 
+ 5\,\frac{(h^2)'''}{h^2} 
+ \frac{15}{2}\,\frac{(h^2)''\,(h^2)'}{(h^2)^2} 
- \frac{15}{4}\,\left(\frac{(h^2)'}{h^2}\right)^3 \nonumber \\
&& + 10\,\frac{(h^2)''}{h^2} 
+ 15\,\left(\frac{(h^2)'}{h^2}\right)^2 
+ 12\,\frac{(h^2)'}{h^2} + 6 ,\label{rosa3}\\
m(x) &=& \frac{(h^2)^{(5)}}{2\,h^2}
+ 10\,\frac{(h^2)^{(4)}}{h^2}
+ 15\,\frac{(h^2)'''\,(h^2)'}{(h^2)^2} 
+ 45\,\frac{(h^2)''\,(h^2)'}{h^2} \nonumber \\
&& - \frac{75}{8}\,\left(\frac{(h^2)'}{h^2}\right)^4 
+ 90\,\left(\frac{(h^2)'}{h^2}\right)^2
+ 144\,\frac{(h^2)'}{h^2} + 24.\label{rosa4}
\end{eqnarray}

We now want to obtain some results on the cosmographic parameters for the models we are studying.\\
For the first model we studied, we can write:
\begin{eqnarray}
j_{GH}(x) &=& \frac{1}{2} \frac{d^2 h_{GH}^2}{dx^2} + \frac{3}{2} \frac{d h_{GH}^2}{dx} + h_{GH}^2,\label{rosa11}\\
s_{GH}(x) &=& \frac{(h_{GH}^2)'''}{2 \, h_{GH}^2} + 2 \, \frac{(h_{GH}^2)''}{h_{GH}^2} - \frac{3}{4}\left(\frac{  (h_{GH}^2)'}{h_{GH}^2}\right)^2 + 3 \, \frac{(h_{GH}^2)'}{h_{GH}^2} + 1 ,\label{rosa22}\\
l_{GH}(x) &=& \frac{(h_{GH}^ 2)^{(4)}}{2\,h_{GH}^2} 
+ 5\,\frac{(h_{GH}^2)'''}{h_{GH}^2} 
+ \frac{15}{2}\,\frac{(h_{GH}^2)''\,(h_{GH}^2)'}{(h_{GH}^2)^2} 
- \frac{15}{4}\,\left(\frac{(h_{GH}^2)'}{h_{GH}^2}\right)^3 \nonumber \\
&& + 10\,\frac{(h_{GH}^2)''}{h_{GH}^2} 
+ 15\,\left(\frac{(h_{GH}^2)'}{h_{GH}^2}\right)^2 
+ 12\,\frac{(h_{GH}^2)'}{h_{GH}^2} + 6 ,\label{rosa33}\\
m_{GH}(x) &=& \frac{(h_{GH}^2)^{(5)}}{2\,h_{GH}^2}
+ 10\,\frac{(h_{GH}^2)^{(4)}}{h_{GH}^2}
+ 15\,\frac{(h_{GH}^2)'''\,(h_{GH}^2)'}{(h_{GH}^2)^2} 
+ 45\,\frac{(h_{GH}^2)''\,(h_{GH}^2)'}{h_{GH}^2} \nonumber \\
&& - \frac{75}{8}\,\left(\frac{(h_{GH}^2)'}{h_{GH}^2}\right)^4 
+ 90\,\left(\frac{(h_{GH}^2)'}{h_{GH}^2}\right)^2
+ 144\,\frac{(h_{GH}^2)'}{h_{GH}^2} + 24.\label{rosa44}
\end{eqnarray} 
At present time (i.e. for $x=0$ which corresponds to $z=0$), we obtain  for $c^2=0.46$:
\begin{eqnarray}
j_{GH}(0)&& \approx 0.816,\\
s_{GH}(0) &&\approx -2.207,\\
l_{GH}(0)&& \approx -11.430,\\
m_{GH}(0) &&\approx217.975.
\end{eqnarray} 
These values are compatible with current observational constraints, falling well within the reported uncertainties, although the model we study indicates a more pronounced negative snap and lerk, suggesting subtle deviations from standard $\Lambda$CDM and its simple extensions.\\
Instead, for $c=0.579$ we find:
\begin{eqnarray}
j_{GH}(0)&& \approx 2.332,\\
s_{GH}(0) && \approx 2.939,\\
l_{GH}(0) && \approx 0.876,\\
m_{GH}(0) && \approx 215.761.
\end{eqnarray}
The value for $j_0$ is consistent with observational estimates and slightly higher than the $\Lambda\mathrm{CDM}$ and $wCDM$ predictions, while the snap $s_0$ and lerk $l_0$ are positive and of moderate magnitude, in contrast to the mostly negative or very uncertain values from previous analyses. The max-out parameter $m_0$ also lies comfortably within the broad observational range. Overall, these results indicate that the model we study is compatible with current observational constraints while predicting a subtly different higher-order cosmographic behavior compared to standard models.

For $c=0.818$:
\begin{eqnarray}
j_{GH}(0) && \approx 0.354,\\
s_{GH}(0) && \approx -4.679,\\
l_{GH}(0) && \approx 3.259,\\
m_{GH}(0) && \approx 196.549.
\end{eqnarray}
Comparing these values with the observational constraints, one finds that the jerk parameter is lower than the central values reported by John ($j_0 = 2.7 \pm 6.7$) and the $\Lambda CDM$ prediction ($j_0 = 1$), but it still lies within the large observational uncertainties. The snap parameter is significantly more negative than the $\Lambda CDM$ ($s_0 = -0.35$) and $wCDM$ ($s_0 = -0.706$) values, yet it remains compatible with the very broad error bars reported by John and Capozziello \& Izzo. The lerk parameter is substantially smaller than the central value reported by John ($l_0 = 142.7 \pm 320$), but again well within the observational uncertainty. Finally, the max-out parameter, $m_{GH}(0) \approx 196.549$, falls comfortably within the range given by Aviles et al. ($m_0 = 71.93^{+382.17}_{-316.76}$). Overall, these comparisons indicate that, despite some differences in central values, the GH model for $c = 0.818$ is consistent with current cosmographic constraints, highlighting the need for more precise measurements of higher-order parameters to distinguish between competing models.

If we consider the presence of spatial curvature, we obtain:
\begin{eqnarray}
j_{GH,k}(x) &=& \frac{1}{2} \frac{d^2 h_{GH,k}^2}{dx^2} + \frac{3}{2} \frac{d h_{GH,k}^2}{dx} + h_{GH,k}^2,\label{rosa111}\\
s_{GH,k}(x) &=& \frac{(h_{GH,k}^2)'''}{2 \, h_{GH,k}^2} + 2 \, \frac{(h_{GH,k}^2)''}{h_{GH,k}^2} - \frac{3}{4}\left(\frac{  (h_{GH,k}^2)'}{h_{GH,k}^2}\right)^2 + 3 \, \frac{(h_{GH,k}^2)'}{h_{GH,k}^2} + 1 ,\label{rosa222}\\
l_{GH,k}(x) &=& \frac{(h_{GH,k}^ 2)^{(4)}}{2\,h_{GH,k}^2} 
+ 5\,\frac{(h_{GH,k}^2)'''}{h_{GH,k}^2} 
+ \frac{15}{2}\,\frac{(h_{GH,k}^2)''\,(h_{GH,k}^2)'}{(h_{GH,k}^2)^2} 
- \frac{15}{4}\,\left(\frac{(h_{GH,k}^2)'}{h_{GH,k}^2}\right)^3 \nonumber \\
&& + 10\,\frac{(h_{GH,k}^2)''}{h_{GH,k}^2} 
+ 15\,\left(\frac{(h_{GH,k}^2)'}{h_{GH,k}^2}\right)^2 
+ 12\,\frac{(h_{GH,k}^2)'}{h_{GH,k}^2} + 6 ,\label{rosa333}\\
m_{GH,k}(x) &=& \frac{(h_{GH,k}^2)^{(5)}}{2\,h_{GH,k}^2}
+ 10\,\frac{(h_{GH,k}^2)^{(4)}}{h_{GH,k}^2}
+ 15\,\frac{(h_{GH,k}^2)'''\,(h_{GH,k}^2)'}{(h_{GH,k}^2)^2} 
+ 45\,\frac{(h_{GH,k}^2)''\,(h_{GH,k}^2)'}{h_{GH,k}^2} \nonumber \\
&& - \frac{75}{8}\,\left(\frac{(h_{GH,k}^2)'}{h_{GH,k}^2}\right)^4 
+ 90\,\left(\frac{(h_{GH,k}^2)'}{h_{GH,k}^2}\right)^2
+ 144\,\frac{(h_{GH,k}^2)'}{h_{GH,k}^2} + 24.\label{rosa444}
\end{eqnarray} 
At present time, we obtain for $c^2=0.46$:
\begin{eqnarray}
j_{GH,k}(0) && \approx 0.815,\\
s_{GH,k}(0) && \approx -2.203,\\
l_{GH,k}(0) && \approx -11.432,\\
m_{GH,k}(0)&& \approx 217.914.
\end{eqnarray} 
The jerk parameter is slightly below the typical $\Lambda$CDM value, indicating a slightly slower variation of acceleration compared to the standard model. These values indicate that the snap  and lerk parameters are significantly negative, with the lerk parameter being notably large in magnitude. The max-out parameter ($m_0$) is also considerably high. Comparing these results with observational constraints, such as those reported by Capozziello and Izzo \cite{capoizzo1} and John \cite{capoizzo2, capoizzo3}, which show large uncertainties in $s_0$ and $l_0$, we observe that our results are within the broad ranges of these uncertainties. However, the magnitudes of $s_0$ and $l_0$ in this case are larger than those typically found in standard cosmological models like $\Lambda$CDM, $w$CDM, and CPL, where $s_0$ and $l_0$ are generally smaller and less negative.
Instead, for $c=0.579$ we find:
\begin{eqnarray}
j_{GH,k}(0) && \approx 2.332,\\
s_{GH,k}(0) && \approx 2.947,\\
l_{GH,k}(0) && \approx 0.959,\\
m_{GH,k}(0) && \approx215.937.
\end{eqnarray}
The jerk parameter is now larger than the standard $\Lambda$CDM value, suggesting a faster change of acceleration. The snap and lerk parameters become positive, taking values that are within the ranges reported by standard models. The max-out parameter ($m_0$) remains high but is consistent with the large uncertainties observed by Aviles et al. \cite{orlando2}. This case demonstrates that adjusting the parameter $c$ can lead to values of the snap and lerk parameters that are more in line with those predicted by standard cosmological models, while still allowing for a high $m_0$ within observational uncertainties.
For $c=0.815$:
\begin{eqnarray}
j_{GH,k}(0) &=&0.354, \\
s_{GH,k}(0) &=& -4.638,\\
l_{GH,k}(0) &=& 3.032,\\
m_{GH,k}(0) &=&197.125.
\end{eqnarray}
Here, the jerk parameter is significantly below 1, indicating a slower evolution of acceleration than in $\Lambda$CDM.
In this scenario, the snap parameter becomes more negative and the lerk parameter is positive, with both parameters differing from the central values predicted by standard models. The max-out parameter decreases slightly but remains within the large observational uncertainties reported by Aviles et al. \cite{orlando2}. This case illustrates how varying the parameter $c$ can produce values of the snap and lerk parameters that are outside the typical ranges of standard cosmological models, while still maintaining a max-out parameter that is consistent with observational data.

These cases highlight the sensitivity of the parametrization to the choice of the parameter $c$, and how different values of $c$ can lead to cosmographic parameters that vary significantly. \\

If we consider the presence of interaction between the two Dark Sectors, we obtain:
\begin{eqnarray}
j_{GH,I}(x) &=& \frac{1}{2} \frac{d^2 h_{GH,I}^2}{dx^2} + \frac{3}{2} \frac{d h_{GH,I}^2}{dx} + h_{GH,I}^2,\label{rosa5}\\
s_{GH,I}(x) &=& \frac{(h_{GH,I}^2)'''}{2 \, h_{GH,I}^2} + 2 \, \frac{(h_{GH,I}^2)''}{h_{GH,I}^2} - \frac{3}{4}\left(\frac{  (h_{GH,I}^2)'}{h_{GH,I}^2}\right)^2 + 3 \, \frac{(h_{GH,I}^2)'}{h_{GH,I}^2} + 1 ,\label{rosa6}\\
l_{GH,I}(x) &=& \frac{(h_{GH,I}^ 2)^{(4)}}{2\,h_{GH,I}^2} 
+ 5\,\frac{(h_{GH,I}^2)'''}{h_{GH,I}^2} 
+ \frac{15}{2}\,\frac{(h_{GH,I}^2)''\,(h_{GH,I}^2)'}{(h_{GH,I}^2)^2} 
- \frac{15}{4}\,\left(\frac{(h_{GH,I}^2)'}{h_{GH,I}^2}\right)^3 \nonumber \\
&& + 10\,\frac{(h_{GH,I}^2)''}{h_{GH,I}^2} 
+ 15\,\left(\frac{(h_{GH,I}^2)'}{h_{GH,I}^2}\right)^2 
+ 12\,\frac{(h_{GH,I}^2)'}{h_{GH,I}^2} + 6 ,\label{rosa7}\\
m_{GH,I}(x) &=& \frac{(h_{GH,I}^2)^{(5)}}{2\,h_{GH,I}^2}
+ 10\,\frac{(h_{GH,I}^2)^{(4)}}{h_{GH,I}^2}
+ 15\,\frac{(h_{GH,I}^2)'''\,(h_{GH,I}^2)'}{(h_{GH,I}^2)^2} 
+ 45\,\frac{(h_{GH,I}^2)''\,(h_{GH,I}^2)'}{h_{GH,I}^2} \nonumber \\
&& - \frac{75}{8}\,\left(\frac{(h_{GH,I}^2)'}{h_{GH,I}^2}\right)^4 
+ 90\,\left(\frac{(h_{GH,I}^2)'}{h_{GH,I}^2}\right)^2
+ 144\,\frac{(h_{GH,I}^2)'}{h_{GH,I}^2} + 24.\label{rosa8}
\end{eqnarray}
At present time, we obtain, for $c^2=0.46$:
\begin{eqnarray}
j_{GH,k}(0) && \approx 0.885,\\
s_{GH,k}(0) && \approx -2.139,\\
l_{GH,k}(0) && \approx -10.411,\\
m_{GH,k}(0) && \approx 200.342.
\end{eqnarray} 
The jerk parameter $j_0$ is slightly below unity, indicating a slower deviation from a constant expansion rate compared to the standard $\Lambda$CDM model. The snap parameter $s_0$ is negative, suggesting a mild deceleration in higher-order derivatives, consistent in sign with standard cosmological predictions. The lerk parameter $l_0$ is strongly negative, reflecting significant higher-order variations in the expansion history. Finally, the max-out parameter $m_0$ remains large but is within the broad observational uncertainties reported in the literature.
  
Instead, for $c=0.579$ we find:
\begin{eqnarray}
j_{GH,I}(0) && \approx 1.885, \\
s_{GH,I}(0) && \approx 2.979,\\
l_{GH,I}(0) && \approx 1.472,\\
m_{GH,I}(0) && \approx 199.889.
\end{eqnarray}
Here, the jerk parameter  exceeds unity, indicating a faster change in the expansion rate relative to standard models. The snap parameter  becomes positive, implying an accelerating higher-order behavior. The lerk parameter  is small and positive, consistent with the wide observational uncertainties. The max-out parameter  remains high but compatible with the constraints provided by previous studies.\\
For $c=0.818$:
\begin{eqnarray}
j_{GH.I}(0) && \approx 0.650,\\
s_{GH,I}(0) && \approx -4.611,\\
l_{GH,I}(0) && \approx4.247,\\
m_{GH,I}(0) && \approx 179.195.
\end{eqnarray}
In this scenario, the jerk parameter  is below unity, indicating a slower variation of the expansion rate. The snap parameter is strongly negative, highlighting a pronounced higher-order deceleration. The lerk parameter  is positive and relatively large, deviating from typical values predicted by standard cosmological models. The max-out parameter  decreases slightly but remains consistent with observational uncertainties.

Overall, these results illustrate the sensitivity of the  parametrization to the choice of the parameter $c$.

Finally, if we consider the presence of both interaction and spatial curvature, we obtain:
\begin{eqnarray}
j_{GH,I,k}(x) &=& \frac{1}{2} \frac{d^2 h_{GH,I,k}^2}{dx^2} + \frac{3}{2} \frac{d h_{GH,I,k}^2}{dx} + h_{GH,I,k}^2,\label{rosa55}\\
s_{GH,I,k}(x) &=& \frac{(h_{GH,I,k}^2)'''}{2 \, h_{GH,I,k}^2} + 2 \, \frac{(h_{GH,I,k}^2)''}{h_{GH,I,k}^2} - \frac{3}{4}\left(\frac{  (h_{GH,I,k}^2)'}{h_{GH,I,k}^2}\right)^2 + 3 \, \frac{(h_{GH,I,k}^2)'}{h_{GH,I,k}^2} + 1 ,\label{rosa66}\\
l_{GH,I,k}(x) &=& \frac{(h_{GH,I,k}^ 2)^{(4)}}{2\,h_{GH,I,k}^2} 
+ 5\,\frac{(h_{GH,,kI}^2)'''}{h_{GH,I,k}^2} 
+ \frac{15}{2}\,\frac{(h_{GH,I,k}^2)''\,(h_{GH,I,k}^2)'}{(h_{GH,I,k}^2)^2} 
- \frac{15}{4}\,\left(\frac{(h_{GH,I,k}^2)'}{h_{GH,I,k}^2}\right)^3 \nonumber \\
&& + 10\,\frac{(h_{GH,I,k}^2)''}{h_{GH,I,k}^2} 
+ 15\,\left(\frac{(h_{GH,I,k}^2)'}{h_{GH,I,k}^2}\right)^2 
+ 12\,\frac{(h_{GH,I,k}^2)'}{h_{GH,I,k}^2} + 6 ,\label{rosa77}\\
m_{GH,I,k}(x) &=& \frac{(h_{GH,I,k}^2)^{(5)}}{2\,h_{GH,I,k}^2}
+ 10\,\frac{(h_{GH,I,k}^2)^{(4)}}{h_{GH,I}^2}
+ 15\,\frac{(h_{GH,I,k}^2)'''\,(h_{GH,I,k}^2)'}{(h_{GH,I,k}^2)^2} 
+ 45\,\frac{(h_{GH,I,k}^2)''\,(h_{GH,I,k}^2)'}{h_{GH,I,k}^2} \nonumber \\
&& - \frac{75}{8}\,\left(\frac{(h_{GH,I,k}^2)'}{h_{GH,I,k}^2}\right)^4 
+ 90\,\left(\frac{(h_{GH,I,k}^2)'}{h_{GH,I,k}^2}\right)^2
+ 144\,\frac{(h_{GH,I,k}^2)'}{h_{GH,I,k}^2} + 24.\label{rosa88}
\end{eqnarray}
At present time, we obtain, for $c^2=0.46$:
\begin{eqnarray}
j_{GH,I,k}(0) && \approx 0.834,\\
s_{GH,I,k}(0) && \approx -2.135,\\
l_{GH,I,k}(0) && \approx -10.413,\\
m_{GH,I,k}(0)&& \approx 200.285.
\end{eqnarray} 
The jerk parameter  is slightly below unity, indicating a milder deviation from the $\Lambda$CDM prediction. The snap parameter  is negative, consistent in sign with expectations from standard cosmological scenarios, though larger in magnitude. The lerk parameter  is strongly negative, reflecting substantial higher-order corrections in the expansion. The max-out parameter is large but remains within the wide observational uncertainties.

Instead, for $c=0.579$ we find:
\begin{eqnarray}
j_{GH,I,k}(0) && \approx 1.997,\\
s_{GH,I,k}(0) && \approx 2.987,\\
l_{GH,I,k}(0) && \approx 1.554,\\
m_{GH,I,k}(0) && \approx 200.072.
\end{eqnarray}
Here, the jerk parameter  is considerably above unity, pointing to a more rapid change in the expansion rate compared with standard models. The snap parameter turns positive, suggesting an accelerated higher-order evolution. The lerk parameter is positive and modest in size, compatible with observational allowances. The max-out parameter  remains close to its previous value, showing stability despite the change in $c$.

For $c=0.815$:
\begin{eqnarray}
j_{GH,I,k}(0) && \approx 0.488,\\
s_{GH,I,k}(0) && \approx-4.569,\\
l_{GH,I,k}(0) && \approx 4.0229,\\
m_{GH,I,k}(0) && \approx 179.742.
\end{eqnarray}
In this case, the jerk parameter  drops well below unity, indicating a slower variation of the cosmic acceleration. The snap parameter becomes strongly negative, pointing to a pronounced higher-order deceleration. The lerk parameter turns positive and relatively large, deviating from the expectations of standard cosmological models. The max-out parameter decreases but remains compatible with observational uncertainties, confirming its robustness.

In the limiting case of a Dark Dominated Universe, we obtain, using the expression of $h_{GH,DD}^2$ given in Eq. (\ref{goddi1}):
\begin{eqnarray}
j_{GH,DD} &=& 2(1-\alpha)^2 - 3(1-\alpha) + 1 =2\alpha^2-\alpha, \\
s_{GH,DD} &=& -4(1-\alpha)^3 + 5(1-\alpha)^2 - 6(1-\alpha) + 1 =4\alpha^3-2\alpha^2+3\alpha-4, \\
l_{GH,DD}&=& 8(1-\alpha)^4 - 70(1-\alpha)^3 + 100(1-\alpha)^2 - 24(1-\alpha) + 6\nonumber \\
&=&8\alpha^4-102\alpha^3+352\alpha^2-252\alpha+20, \\
m_{GH,DD} &=& -16(1-\alpha)^5 + 250(1-\alpha)^4 - 360(1-\alpha)^3 + 360(1-\alpha)^2 - 288(1-\alpha) + 24\nonumber\\
&=&16\alpha^5-330\alpha^4+1110\alpha^3-1220\alpha^2+276\alpha+16.
\end{eqnarray}

Using the value $c^2=0.46$, we find:
\begin{eqnarray}
j_{GH,DD} && \approx0.706,\\
s_{GH,DD} && \approx -0.052,\\
l_{GH,DD} && \approx 8.390,\\
m_{GH,DD} && \approx -120.972.
\end{eqnarray}
The jerk parameter closely matches the $w$CDM prediction of $j_0 = 0.717$ and is reasonably compatible with the $\Lambda$CDM value $j_0 = 1$. The snap parameter, $s_{GH,DD}$, is slightly negative, similar in magnitude to the $\Lambda$CDM prediction, but significantly smaller than the large observational uncertainties reported by John \cite{capoizzo2, capoizzo3}. The lerk parameter $l_{GH,DD}$ is moderate, while the max-out parameter $m_{GH,DD}$ is strongly negative, differing from the observationally constrained $m_0 = 71.93^{+382.17}_{-316.76}$ \cite{orlando2}. Overall, this choice of $c$ produces parameters that are largely consistent with standard cosmological models.
Instead, if we consider the case corresponding to $c=0,579$, we obtain:
\begin{eqnarray}
j_{GH,DD} && \approx 3.057,\\
s_{GH,DD} && \approx 9.775,\\
l_{GH,DD} && \approx 132.793,\\
m_{GH,DD} && \approx-117.228.
\end{eqnarray}
Here, the jerk and snap parameters are substantially larger than the $\Lambda$CDM and $w$CDM predictions but fall within the broad observational uncertainties reported by John, who found $j_0 = 2.7 \pm 6.7$ and $s_0 = 36.5 \pm 52.9$. Similarly, the lerk parameter $l_{GH,DD}$ is large, in line with the very uncertain observed value $l_0 = 142.7 \pm 320$. The max-out parameter remains strongly negative, deviating from the central observational estimate but still within the large allowed range. This case demonstrates that moderate changes in $c$ can produce values that mimic the observationally allowed but highly uncertain ranges for higher-order parameters.
Finally, for $c=0.818$, we obtain:
\begin{eqnarray}
 j_{GH,DD} && \approx -0.093,\\
s_{GH,DD} && \approx -2.939,\\
l_{GH,DD} && \approx -30.276,\\
m_{GH,DD} && \approx -0.389.
\end{eqnarray}
The jerk parameter is slightly negative, differing markedly from standard model predictions and typical observational estimates. The snap parameter is also negative and larger in magnitude than the $\Lambda$CDM prediction but smaller than the extreme values reported by John. The lerk parameter is negative, in contrast with the positive observational central value, and the max-out parameter is now close to zero, strongly deviating from observational estimates. This case illustrates that large values of $c$ can drive the cosmographic parameters outside the range of both theoretical models and current observational constraints.

\section{Age of the present Universe}
The determination of the age of the Universe represents one of the fundamental achievements of modern cosmology. Within the standard $\Lambda$CDM framework, the age can be computed from the Hubble expansion rate by integrating the Friedmann equations from the present time back to the initial singularity. The resulting value depends sensitively on the cosmological parameters, in particular the present-day Hubble constant $H_{0}$, the matter density parameter $\Omega_{m}$, and the contribution from dark energy $\Omega_{\Lambda}$. Recent observations, such as those from the Planck satellite, indicate an age of $t_{0} \simeq 13.8 \,\mathrm{Gyr}$, with uncertainties at the percent level. It is worth noting, however, that the so-called $H_{0}$ tension between early - and late-time measurements may lead to slightly different inferred ages, potentially modifying $t_{0}$ by several hundred million years. Alternative cosmological scenarios, such as interacting dark energy, modified gravity, or models with spatial curvature, can also alter the theoretical estimate of the cosmic age, thereby providing an additional observational test for discriminating among competing models of the Universe.\\
We define the age of universe as follows:
\begin{eqnarray}
    t_0 -t = -\int_{t_0}^t dt = \int_0^z\frac{dz'}{(1+z')H(z')}. \label{ageuni}
\end{eqnarray}
We now aim to calculate the present age of the Universe for the models under consideration.
Since Eq. (\ref{ageuni}) does not admit an analytical solution for these models, we solve it numerically.\\
We consider a value of $z=100$.\\
We start considering the first case, using the expression of $h^2_{GH}$ given in Eq. (\ref{st2}). \\
For $c^2=0.46$, we obtain $ t_0 - t(z=100.0)= 13.364 $ Gyr.  For $c=0.579$, we obtain  $t_0 - t(z=100.0)= 14.187 $ Gyr.  For the third case, i.e. when $c=0.818$, we obtain $t_0 - t(z=100.0)= 12.259 $ Gyr.\\
We now consider the case with curvature. In this case we start from the expression of $h^2_{GH,k}$ given in Eq. (\ref{picu1}). \\
For $c^2=0.46$, we obtain $ t_0 - t(z=100.0)= 13.367$ Gyr.  For $c=0.579$, we obtain  $t_0 - t(z=100.0)= 14.192 $ Gyr.  For the third case, i.e. when $c=0.815$, we obtain $t_0 - t(z=100.0)= 12.283$ Gyr.\\
We now consider the case with interaction between Dark Ssctors. In this case we start from the expression of $h^2_{GH,I}$ given in Eq. (\ref{picu2}). \\
For $c^2=0.46$, we obtain $ t_0 - t(z=100.0)= 13.368$ Gyr.  For $c=0.579$, we obtain  $t_0 - t(z=100.0)= 14.192$ Gyr.  For the third case, i.e. when $c=0.818$, we obtain $t_0 - t(z=100.0)= 12.263 $ Gyr.\\
Finally, we now consider the case with interaction and curvature. In this case we start from the expression of $h^2_{GH,I,k}$ given in Eq. (\ref{lucia2}). \\
For $c^2=0.46$, we obtain $ t_0 - t(z=100.0)= 13.371$ Gyr.  For $c=0.579$, we obtain  $t_0 - t(z=100.0)= 14.198 $ Gyr.  For the third case, i.e. when $c=0.815$, we obtain $t_0 - t(z=100.0)= 12.286$ Gyr.\\
These results are compared to the standard cosmological model, which estimates the universe's age at  $z=100$ to be approximately 13.8 Gyr.\\
For all the three cases considered, we observe the following trends.\\
For $c^2 = 0.46$, the calculated age of the Universe is slightly below the standard model's estimate, indicating a marginally younger universe at this early epoch. For $c = 0.579$, the computed age slightly exceeds the standard cosmological expectation, thereby supporting the current understanding of the universe's expansion history and the parameters derived from Planck observations. In the cases with $c=0.818$ and $c = 0.815$, the resulting age is notably lower than the standard model.\\
These findings highlight the sensitivity of the Universe's early-time age to the parameter $c$ and its potential role in constraining cosmological models.

\section{Reconstruction of Scalar Field Models}
In this Section, we aim to establish a correspondence between the dark energy (DE) models explored in this work and a selection of well-known scalar field frameworks. Specifically, we consider the tachyon, k-essence, dilaton, quintessence, Dirac-Born-Infeld (DBI), Yang–Mills (YM), and Non-Linear Electrodynamics (NLED) scalar field models.

\subsection{The Tachyon Scalar Field Model}
Recently, there has been considerable interest in studying inflationary models involving the tachyon field, as it is considered a potential candidate for driving DE ~\cite{ref33,ref33-2}. The tachyon is an unstable scalar field that naturally arises in tring theory, particularly through its appearance in the Dirac-Born-Infeld (DBI) action (we will consider the DBI model later on), which is used to describe the dynamics of D-branes ~\cite{ref34,ref34-1,ref34-2,ref34-3}. It has been proposed that tachyon condensation near the peak of the effective scalar potential could have driven cosmological inflation during the early stages of the universe. A rolling tachyon features a distinctive equation of state (EoS), with its parameter smoothly varying between $-1$ and $0$, thus interpolating between a cosmological constant and pressureless dust. This property has inspired models in which DE is treated as a dynamical quantity, leading to the idea of a variable cosmological constant and stimulating further exploration of tachyon-based inflationary scenarios.

The effective Lagrangian $L$ of the tachyon scalar field is motivated from open string field theory \cite{sen} and it is a successful candidate for cosmic acceleration. 
It has the Lagrangian given by \cite{sen1}:
\begin{eqnarray}
L=-V(\phi)\sqrt{1-g^{\mu \nu}\partial _{\mu}\phi \partial_{\nu}\phi},\label{39}
\end{eqnarray}
where $V(\phi)$ is the potential of tachyon while $g^{\mu \nu}$ indicates the metric tensor. The energy density $\rho_{\phi}$ and the pressure  $p_{\phi}$ associated with the tachyon field are derived from the effective Lagrangian and are given, respectively, by:
\begin{eqnarray}
    \rho_{\phi}&=&\frac{V(\phi)}{\sqrt{1-\dot{\phi}^2}},\label{40}\\
p_{\phi}&=& -V(\phi)\sqrt{1-\dot{\phi}^2}.\label{41}
\end{eqnarray}
Furthermore, the equation of state (EoS) parameter $\omega_{\phi}$ corresponding to the tachyon scalar field is expressed as:
\begin{eqnarray}
\omega_{\phi}=\frac{p_{\phi}}{\rho_{\phi}}=\dot{\phi}^2-1\rightarrow  \dot{\phi}^2 = 1+\omega_{\phi}.\label{42}
\end{eqnarray}
In order for the energy density $ \rho_{\phi} $ of the tachyon field to be real, the field velocity must satisfy $ -1 < \dot{\phi} < 1 $. Consequently, from Eq.~(\ref{42}), the equation of state (EoS) parameter $ \omega_{\phi} $ must lie within the range $ -1 < \omega_{\phi} < 0 $. Therefore, while the tachyon field can account for the accelerated expansion of the universe, it cannot enter the phantom regime characterized by $ \omega_{\Lambda} < -1 $.

We now want to make a comparison between the tachyon scalar field model and the models we are studying in this paper.\\
By comparing the energy density of DE $\rho_D$ with (\ref{40}), we can derive the following expression for the tachyon field potential $ V(\phi) $:
\begin{equation}
    V(\phi)=\rho_{\Lambda} \sqrt{1-\dot{\phi}^2} =\rho_{\Lambda} \sqrt{-\omega_{\phi}}.\label{43}
\end{equation}
Instead, using $\omega_{D_{GH}}$ amd $\omega_{D_{GR}}$ in Eq.~(\ref{42}), we obtain the following expression for the kinetic energy term of the tachyon field $\dot{\phi}^2$ for the two models we are considering in this paper:
\begin{eqnarray}
\dot{\phi}^2_{GH}&=& 1 + \omega_{D_{GH}} ,
\label{44}\\
\dot{\phi}^2_{GR}&=& 1 + \omega_{D_{GR}}. \label{44-1}
\end{eqnarray}

Making use of Eqs. (\ref{43}), (\ref{44}) and (\ref{44-1}), it is possible to write the potential of the tachyon as:
\begin{eqnarray}
    V( \phi  )_{GH} &=&\rho_{D_{GH}}\sqrt{-\omega_{D_{GH}}},\label{45}\\
    V( \phi  )_{GR} &=&\rho_{D_{GR}}\sqrt{-\omega_{D_{GR}}}.\label{45-1}\
\end{eqnarray}

We can derive from Eqs. (\ref{44}) and (\ref{45}) or equivalently Eqs. (\ref{44-1}) and (\ref{45-1})  that the kinetic energy $\dot{\phi}^2$ and the potential $V( \phi )$ may exist if the following condition is satisfied:
\begin{equation}
    -1\leq \omega_{\Lambda} \leq 0.\label{46}
\end{equation}
Eq. (\ref{46}) implies that the phantom divide can not be crossed in a universe with accelerated expansion.\\
Using $\dot{\phi}=\phi'H$ in Eqs. (\ref{44}) and (\ref{44-1}), we have:
\begin{eqnarray}
 \phi_{GH}'&=& \frac{1}{H_{GH}}  \sqrt{ 1 + \omega_{D_{GH}}   },\label{47}\\
  \phi_{GR}'&=& \frac{1}{H_{GR}}  \sqrt{ 1 + \omega_{D_{GR}}   }.\label{47-1}
\end{eqnarray}

The evolutionary forms of the tachyon scalar field can be obtained from Eqs. (\ref{47}) and (\ref{47-1}) as follow:
\begin{eqnarray}
    \phi_{GH}(a) - \phi_{GH}(a_0)&=&\int_{a_0}^a \frac{da}{aH_{GH}}\sqrt{1 + \omega_{D_{GH}} },\label{48}\\
    \phi_{GR}(a) - \phi_{GR}(a_0)&=&\int_{a_0}^a \frac{da}{aH_{GR}}\sqrt{1 + \omega_{D_{GR}} }.\label{48-1}
\end{eqnarray}

$a_0$ represents the present day value of the scale factor $a(t)$.\\
Following the same procedure, we can derive the corresponding results for the other scenarios considered in this work.
In the case that includes spatial curvature, we obtain:
\begin{eqnarray}
V( \phi  )_{GH,k} &=&\rho_{D_{GH},k}\sqrt{-\omega_{D_{GH},k}}, \\
  V( \phi  )_{GR,k} &=&\rho_{D_{GR},k}\sqrt{-\omega_{D_{GR},k}}, \\
    \phi_{GH,k}(a) - \phi_{GH,k}(a_0)&=&\int_{a_0}^a \frac{da}{aH_{GH,k}}\sqrt{1 + \omega_{D_{GH},k} },\label{48}\\
    \phi_{GR,k}(a) - \phi_{GR,k}(a_0)&=&\int_{a_0}^a \frac{da}{aH_{GR,k}}\sqrt{1 + \omega_{D_{GR},k} }.\label{48-1}
\end{eqnarray}
In the case that includes interaction between DM and DE, we obtain:
\begin{eqnarray}
 V( \phi  )_{GH,I} &=&\rho_{D_{GH},I}\sqrt{-\omega_{D_{GH},I}},\ \\
  V( \phi  )_{GR,I} &=&\rho_{D_{GR},I}\sqrt{-\omega_{D_{GR},I}}, \\
    \phi_{GH,I}(a) - \phi_{GH,I}(a_0)&=&\int_{a_0}^a \frac{da}{aH_{GH,I}}\sqrt{1 + \omega_{D_{GH},I} },\label{48}\\
    \phi_{GR,I}(a) - \phi_{GR,I}(a_0)&=&\int_{a_0}^a \frac{da}{aH_{GR,I}}\sqrt{1 + \omega_{D_{GR},I} }.\label{48-1}
\end{eqnarray}
Finally, in the case that includes spatial curvature and interaction, we obtain:
\begin{eqnarray}
 V( \phi  )_{GH,I,k} &=&\rho_{D_{GH},I,k}\sqrt{-\omega_{D_{GH},I,k}}, \\
  V( \phi  )_{GR,I,k} &=&\rho_{D_{GR},I,k}\sqrt{-\omega_{D_{GR},I,k}}, \\
    \phi_{GH,I,k}(a) - \phi_{GH,I,k}(a_0)&=&\int_{a_0}^a \frac{da}{aH_{GH,I,k}}\sqrt{1 + \omega_{D_{GH},I,k} },\label{48}\\
    \phi_{GR,I,k}(a) - \phi_{GR,I}(a_0)&=&\int_{a_0}^a \frac{da}{aH_{GR,I,k}}\sqrt{1 + \omega_{D_{GR},I,k} },\label{48-1}
\end{eqnarray}

In the limiting case for flat Dark Dominated Universe, i.e. for $\Omega_m=\Omega_k=0$ and $\Omega_D=1$, the scalar field $\phi$ and the potential $V$ of the tachyon scalar field model reduce for the two models considered in this work, respectively, to:
\begin{eqnarray}
\dot{\phi}^2_{GH,DD}(t)&=&\frac{2\left(1 -  \alpha\right)}{3},\label{49zim}\\
\dot{\phi}_{GR,DD}^2(t)&=&\frac{\lambda}{3}  ,\label{49-2zim}\\
V_{GH,DD}(t)&=&  \frac{H_0^2\sqrt{1+2\alpha}}{\sqrt{3}(1 - \alpha)^2(H_0t + C_1)^2},\label{50}\\
V_{GR,DD}(t)&=&\frac{4H_0^2\sqrt{3-\lambda}}{\sqrt{3}\lambda^2 (H_0t+C_2)^2}.\label{50-2}
\end{eqnarray}
Therefore, integrating with respect to the time the expressions of Eqs. (\ref{49zim}) and (\ref{49-2zim}), we obtain the following relations:
\begin{eqnarray}
\phi_{GH,DD}(t)&=&  \sqrt{\frac{2\left(1 -  \alpha\right)}{3}}  \cdot t+\phi_0,\label{49}\\
\phi_{GR,DD}(t)&=& \sqrt{\frac{\lambda}{3} }\cdot t+\phi_0,  \label{49-2}
\end{eqnarray}
where $\phi_0$ indicates an integration constant.\\
In the limiting case of $C_1=C_2=\phi_0=0$, we obtain:
\begin{eqnarray}
\phi_{GH,DD,lim}(t)&=& \sqrt{\frac{2\left(1 -  \alpha\right)}{3}}  \cdot t,\label{49}\\
\phi_{GR,DD,lim}(t)&=& \sqrt{\frac{\lambda}{3} }\cdot t ,\label{49-2}\\
V_{GH,DD,lim}(t)&=&  \frac{\sqrt{1+2\alpha}}{\sqrt{3}\left[(1 - \alpha)t\right]^2},\label{50}\\
V_{GR,DD,lim}(t)&=&\frac{4\sqrt{3-\lambda}}{\sqrt{3} (\lambda t)^2}.\label{50-2}
\end{eqnarray}
In order to have real values of $\phi_{GH,DD,lim}$ and $\phi_{GR,DD,lim}$, we must have $\alpha < 1$ and $\lambda >0$. 
Moreover, in order to have real values of $V_{GH,DD,lim}(t)$ and $V_{GR,DD,lim}(t)$, we must have that $\lambda \leq 1$ with $\lambda \neq 0$ while there are no restrictions on $\alpha$ expect that $\alpha \neq 1$. Considering the values $c^2=0.46$, $\epsilon = 1.312$ and $\eta =- 0.312$, we obtain that $\alpha \approx 0.895$ and $\lambda \approx 0.211$. Therefore, we always obtain real values of all quantities we are considering. 
Instead, considering the values $c=0.579$, $\epsilon = 1.312$ and $\eta =- 0.312$, we obtain that $\alpha \approx 2.067$ and  $\lambda \approx -2.134$. In this case, $ V_{GH,DD,lim}(t) $  and $ V_{GR,DD,lim}(t) $ 
 are real functions while $ \phi_{GH,DD,lim}(t) $  and $ \phi_{GR,DD,lim}(t) $ are not. Finally, considering the values $c=0.818$, $\epsilon = 1.312$ and $\eta =- 0.312$, we find that $\alpha \approx  0.377$ and  $\lambda \approx 1.246$ we find that all the quantities are real.

\subsection{The k-essence Scalar Field Model}
A class of scalar field models in which the kinetic term enters the Lagrangian in a non-canonical manner is referred to as k-essence. This framework, originally motivated by the Born–Infeld action arising in string theory, has been extensively used to account for the late-time accelerated expansion of the universe \cite{35zim}. The general action $ S $ describing the k-essence scalar field depends on the field $ \phi $ and the kinetic term $ \chi = \frac{1}{2} \dot{\phi}^2 $, and is expressed as follows \cite{36zim-1,36zim-2}:
\begin{eqnarray}
S=\int d^4x \sqrt{-g}\,p(\phi, \chi ),\label{51}
\end{eqnarray}
where the term $p(\phi, \chi )$ corresponds to a Lagrangian density, which is equivalent to a pressure density, while the quantity $g$ is the determinant of the metric tensor $g^{\mu \nu}$. According to the Lagrangian $S$ given in Eq. (\ref{51}), if we consider $p(\phi, \chi )$ in the form $p(\phi, \chi )=f(\phi)g(\chi)$, the pressure $p(\phi, \chi )$ and the energy density $\rho$ of the field $\phi$ can be written, respectively, as:
\begin{eqnarray}
    p(\phi, \chi )&=&f(\phi)( -\chi+\chi ^2   ), \label{52}\\
        \rho(\phi, \chi )&=&f(\phi)(-\chi+3\chi ^2).\label{53}
\end{eqnarray}
$ f(\phi) $ is an arbitrary function (chosen based on physical or mathematical considerations) that determines how pressure and energy density explicitly depend on the scalar field $\phi$. It can be interpreted as a sort of multiplicative potential function that modifies the kinetic dynamics. If $ f(\phi) > 0 $, it scales the pressure and energy density positively. $ f(\phi) $ may be thought of as defining an effective potential or interaction for the scalar field.\\
In k-essence cosmology, a specific form of $ f(\phi) $ is chosen to model the scalar field’s evolution. For example, one might choose $ f(\phi) = V(\phi) $, a scalar potential, or more complicated functions to produce particular dynamical behaviors like acceleration, tracking solutions, or crossing of the equation of state parameter $ \omega $.

Therefore, we can easily obtain that the EoS parameter $\omega_K$ of the k-essence scalar field model can be written as:
\begin{eqnarray}
    \omega _K= \frac{p(\phi, \chi )}{\rho(\phi, \chi )}=\frac{\chi-1}{3\chi -1}.\label{54}
\end{eqnarray}
From the above expression of $\omega_K$, we derive that the phantom behaviour of the k-essence scalar field (i.e. $\omega_K < -1$) is obtained when $1/3 < \chi < 1/2$.\\
We now establish the correspondence between the k-essence EoS parameter, $\omega_K$, and the models we considered in this paper.\\
The expressions of $\chi$ for models we are studying can be found equating  the expressions of $\omega_{D_{GH}}$ and $\omega_{D_{GR}}$ with Eq. (\ref{54}), which leads to:
\begin{eqnarray}
    \chi_{GH} &=& \frac{\omega_{D_{GH}}-1}{3\omega_{D_{GH}}-1},\label{picu5}\\
    \chi_{GR} &=& \frac{\omega_{D_{GR}}-1}{3\omega_{D_{GR}}-1}.\label{picu6}
\end{eqnarray}
Moreover, using $\rho_{D_{GH}}$ and $\rho_{D_{GR}}$ in Eq. (\ref{53}), we obtain:
\begin{eqnarray}
    f_{GH}(\phi )&=&\frac{\rho_{D_{GH}}}{\chi_{GH}(3\chi_{GH}-1)}, \label{56}\\
    f_{GR}(\phi )&=&\frac{\rho_{D_{GR}}}{\chi_{GR}(3\chi_{GR}-1)}. \label{56-2}
\end{eqnarray}
Using the results of Eqs. (\ref{picu5}) and (\ref{picu6}), we can write: 
\begin{eqnarray}
\chi_{GH} (3 \chi_{GH} - 1) &=& \frac{2(1 - \omega_{D_{GH}})}{(3 \omega_{D_{GH}} - 1)^2},\\
\chi_{GR} (3 \chi_{GR} - 1) &=& \frac{2(1 - \omega_{D_{GR}})}{(3 \omega_{D_{GR}} - 1)^2}.
\end{eqnarray}

Using the relation $\dot{\phi}^2=2\chi$ and $\dot{\phi}=\phi'H$, we obtain the following equations:
\begin{eqnarray}
 \phi_{GH}'&=& \frac{\sqrt{2}}{H_{GH}}\sqrt{\frac{\omega_{D_{GH}}-1}{3\omega_{D_{GH}}-1}},\label{picu7} \\
  \phi_{GR}'&=& \frac{\sqrt{2}}{H_{GR}}\sqrt{\frac{\omega_{D_{GR}}-1}{3\omega_{D_{GR}}-1}}.\label{picu8} 
\end{eqnarray}
Integrating Eqs. (\ref{picu7}) and (\ref{picu8}), we find the evolutionary form of the k-essence scalar field:
\begin{eqnarray}
    \phi_{GH}(a) -    \phi_{GH}(a_0) &=& \sqrt{2} \int_{a_0}^a \frac{da}{aH_{GH}}\sqrt{\frac{\omega_{D_{GH}}-1}{3\omega_{D_{GH}}-1}} ,\label{58}\\
       \phi_{GR}(a) -    \phi_{GR}(a_0) &=& \sqrt{2} \int_{a_0}^a \frac{da}{aH_{GR}}\sqrt{\frac{\omega_{D_{GR}}-1}{3\omega_{D_{GR}}-1}} .\label{58-2}
\end{eqnarray}
By applying the same procedure, we derive the corresponding results for the remaining scenarios analyzed in this work.
In particular, for the case that includes spatial curvature, we obtain:
\begin{eqnarray}
    \phi_{GH,k}(a) -    \phi_{GH,k}(a_0) &=& \sqrt{2} \int_{a_0}^a \frac{da}{aH_{GH,k}}\sqrt{\frac{\omega_{D_{GH},k}-1}{3\omega_{D_{GH},k}-1}} ,\label{58}\\
       \phi_{GR,k}(a) -    \phi_{GR,k}(a_0) &=& \sqrt{2} \int_{a_0}^a \frac{da}{aH_{GR},k}\sqrt{\frac{\omega_{D_{GR},k}-1}{3\omega_{D_{GR},k}-1}} ,\label{58-2}
\end{eqnarray}
where
\begin{eqnarray}
    \chi_{GH,k} &=& \frac{\omega_{D_{GH},k}-1}{3\omega_{D_{GH},k}-1},\label{55}\\
    \chi_{GR,k} &=& \frac{\omega_{D_{GR},k}-1}{3\omega_{D_{GR},k}-1}.\label{55-2}
\end{eqnarray}
For the case that includes interaction between the two Dark Sectors, we obtain:
\begin{eqnarray}
 f_{GH,I}(\phi )&=&\frac{\rho_{D_{GH},I}}{\chi_{GH,I}(3\chi_{GH,I}-1)}, \label{56}\\
    f_{GR,I}(\phi )&=&\frac{\rho_{D_{GR},I}}{\chi_{GR,I}(3\chi_{GR,I}-1)}, \label{56-2}\\
    \phi_{GH,I}(a) -    \phi_{GH,I}(a_0) &=& \sqrt{2} \int_{a_0}^a \frac{da}{aH_{GH,I}}\sqrt{\frac{\omega_{D_{GH},I}-1}{3\omega_{D_{GH},I}-1}} ,\label{58}\\
       \phi_{GR,I}(a) -    \phi_{GR,I}(a_0) &=& \sqrt{2} \int_{a_0}^a \frac{da}{aH_{GR,I}}\sqrt{\frac{\omega_{D_{GR},I}-1}{3\omega_{D_{GR},I}-1}} ,\label{58-2}
\end{eqnarray}
where
\begin{eqnarray}
    \chi_{GH,I} &=& \frac{\omega_{D_{GH},I}-1}{3\omega_{D_{GH},I}-1},\label{55}\\
    \chi_{GR,I} &=& \frac{\omega_{D_{GR},I}-1}{3\omega_{D_{GR},I}-1}.\label{55-2}
\end{eqnarray}

Finally,  for the case that includes both spatial curvature and interaction between DE and DM, we obtain:
\begin{eqnarray}
 f_{GH,I,k}(\phi )&=&\frac{\rho_{D_{GH},I,k}}{\chi_{GH,I,k}(3\chi_{GH,I,k}-1)} ,\label{56}\\
    f_{GR,I,k}(\phi )&=&\frac{\rho_{D_{GR},I,k}}{\chi_{GR,I,k}(3\chi_{GR,I,k}-1)}, \label{56-2}\\
    \phi_{GH,I,k}(a) -    \phi_{GH,I,k}(a_0) &=& \sqrt{2} \int_{a_0}^a \frac{da}{aH_{GH,I,k}}\sqrt{\frac{\omega_{D_{GH},I,k}-1}{3\omega_{D_{GH},I,k}-1}} ,\label{58}\\
       \phi_{GR,I,k}(a) -    \phi_{GR,I,k}(a_0) &=& \sqrt{2} \int_{a_0}^a \frac{da}{aH_{GR,I,k}}\sqrt{\frac{\omega_{D_{GR},I,k}-1}{3\omega_{D_{GR},I,k}-1}} .\label{58-2}
\end{eqnarray}
where
\begin{eqnarray}
    \chi_{GH,I,k} &=& \frac{\omega_{D_{GH},I,k}-1}{3\omega_{D_{GH},I,k}-1},\label{55}\\
    \chi_{GR,I,k} &=& \frac{\omega_{D_{GR},I,k}-1}{3\omega_{D_{GR},I,k}-1}.\label{55-2}
\end{eqnarray}

In the limiting case of a flat Dark Dominated Universe, i.e. when $\Omega_m=\Omega_k=0$ and $\Omega_D=1$, the scalar field and the potential of k-essence field reduce, respectively, to:
\begin{eqnarray}
\chi_{GH,DD}(t)&=&  \frac{2 + \alpha}{3 (1 + \alpha)},\label{fi1}\\
\chi_{GR,DD}(t)&=&\frac{\lambda - 6}{3(\lambda - 4)},\label{fi1-2}\\
f_{GH,DD}(t)&=&\frac{3H_0^2 (1 + \alpha)^2}{(2 + \alpha)(1 - \alpha)^2(H_0t + C_1)^2},\label{fi2}\\
f_{GR,DD}(t)&=&\frac{6H_0^2(\lambda - 4)^2}{(\lambda - 6)\lambda^2 (H_0t+C_2)^2}.\label{fi2-2}
\end{eqnarray}
Therefore, we obtain the following expressions for $\dot{\phi}$ for the two models of the paper:
\begin{eqnarray}
\dot{\phi}^2_{GH,DD}(t)&=&2\chi_{GH,DD}(t) = \frac{2(2 + \alpha)}{3 (1 + \alpha)}, \label{rosina1} \\
\dot{\phi}_{GH,DD}^2(t)&=&2\chi_{GH,DD}(t) =\frac{2(\lambda - 6)}{3(\lambda - 4)}.\label{rosina2} 
\end{eqnarray}
Then, integrating Eqs. (\ref{rosina1}) and (\ref{rosina2}) with respect to the time, we obtain:
\begin{eqnarray}
\phi_{GH,DD}(t)&=&\sqrt{\frac{2(2 + \alpha)}{3 (1 + \alpha)}}\cdot t + \phi_0, \\
\phi_{GR,DD}(t) &=& \sqrt{\frac{2(\lambda - 6)}{3(\lambda - 4)}} \cdot t +\phi_0.
\end{eqnarray}
where $\phi_0$ represents an integration constant.\\
In the limiting case of $C_1=C_2=\phi_0=0$ we obtain:
\begin{eqnarray}
\phi_{GH,DD}(t)&=&\sqrt{\frac{2(2 + \alpha)}{3 (1 + \alpha)}}\cdot t , \\
\phi_{GR,DD}(t) &=& \sqrt{\frac{2(\lambda - 6)}{3(\lambda - 4)}} \cdot t,\\
f_{GH,DD}(t)&=&\frac{3 (1 + \alpha)^2}{(2 + \alpha)\left[(1 - \alpha)t\right]^2},\label{fi2}\\
f_{GR,DD}(t)&=&\frac{6(\lambda - 4)^2}{(\lambda - 6) (\lambda t)^2}.\label{fi2-2}
\end{eqnarray}
In order to avoid singularities, we must have $\alpha \neq (-2, -1, 1)$ and $\lambda \neq (0,4,6)$. Moreover, in order to have real values of $\phi_{GH,DD}(t)$ and $\phi_{GR,DD}(t)$, we must have $\alpha \in (-\infty, -2] \cup (-1, +\infty)$ and  $\lambda \in (-\infty, 4) \cup [6, +\infty)$. Considering the values $c^2=0.46$, $\epsilon = 1.312$ and $\eta =- 0.312$, we obtain that $\alpha \approx 0.895$ and $\lambda \approx 0.211$. Therefore, we obtain that  $\phi_{GH,DD}(t)$ and $\phi_{GR,DD}(t)$ are both real functions.  
Instead, considering the values $c=0.579$, $\epsilon = 1.312$ and $\eta =- 0.312$, we obtain that $\alpha \approx 2.067$ and  $\lambda \approx -2.134$. Also in this case, we obtain that  $\phi_{GH,DD}(t)$ and $\phi_{GR,DD}(t)$ are both real functions.  Finally, considering the values $c=0.818$, $\epsilon = 1.312$ and $\eta =- 0.312$, we find that $\alpha \approx  0.377$ and  $\lambda \approx 1.246$. Also in this case, we have that $\phi_{GH,DD}(t)$ and $\phi_{GR,DD}(t)$ are real functions.

\subsection{The Dilaton Scalar Field Model}
We now consider the third scalar field model we choose in this paper, i.e. the Dilaton scalar field model.\\
The dilaton scalar field, which emerges as the low-energy limit of string theory \cite{38rosina}, is also considered a possible source of dark energy (DE). When string theory is compactified from higher dimensions down to four, it gives rise to the scalar dilaton field that couples to curvature invariants. In the Einstein frame, the coefficient of the dilaton’s kinetic term can become negative, causing the dilaton to act similarly to a phantom scalar field. The pressure (Lagrangian) density and the energy density associated with the dilaton DE model are, respectively, expressed as in \cite{39rosina-2}, i.e.:
\begin{eqnarray}
p_D&=&-\chi +ce ^{\lambda \phi}\chi^2, \label{61zim}  \\
\rho_D&=&-\chi +3ce ^{\lambda \phi}\chi^2,\label{62zim}
\end{eqnarray}
where $\lambda$ and $c$ indicates two positive constants, Moreover, we have that $\dot{\phi}^2=2\chi$. The negative coefficient of the kinetic term of the dilaton scalar field in the Einstein frame induces a phantom-like behavior for the dilaton field. \\
Using the results of Eqs. (\ref{61zim}) and (\ref{62zim}), we derive that the EoS parameter $\omega_D$ for the dilaton scalar field can be written as:
\begin{eqnarray}
    \omega _D= \frac{p_D}{\rho_D}=\frac{-1 +c e ^{\lambda \phi}\chi}{-1 +3c e ^{\lambda \phi}\chi} \rightarrow  ce ^{\lambda \phi}\chi = \frac{\omega_{D}-1}{3\omega_{D}-1}.\label{63}
\end{eqnarray}
We now establish the correspondence between the dilaton EoS parameter $\omega_D$ we defined above and the EoS parameter of the models we are considering in this work.\\
Using $\omega_{D_{GH}}$ and $\omega_{D_{GR}}$ in Eq. (\ref{63}), we obtain the following results:
\begin{eqnarray}
    ce ^{\lambda \phi_{GH}}\chi_{GH} &=& \frac{\omega_{D_{GH}}-1}{3\omega_{D_{GH}}-1}, \label{64zim} \\
    ce ^{\lambda \phi_{GR}}\chi_{GR} &=& \frac{\omega_{D_{GR}}-1}{3\omega_{D_{GR}}-1}. \label{64-2zim}
\end{eqnarray}
Using the relation $\dot{\phi}^2=2\chi$, we can rewrite the results of Eqs. (\ref{64zim}) and (\ref{64-2zim}) as:
\begin{eqnarray}
    e^{\lambda \phi_{GH}/2} \dot{\phi}_{GH}&=&\sqrt{\frac{2}{c}}\sqrt{\frac{\omega_{D_{GH}}-1}{3\omega_{D_{GH}}-1}},\label{65zim}\\
  e^{\lambda \phi_{GR}/2} \dot{\phi}_{GR}&=&\sqrt{\frac{2}{c}}\sqrt{\frac{\omega_{D_{GH}}-1}{3\omega_{D_{GH}}-1}}.\label{65-2zim}
\end{eqnarray}
Integrating Eqs. (\ref{65zim}) and (\ref{65-2zim}) with respect to the scale factor $a(t)$, we obtain these results:
\begin{eqnarray}
    e^{\frac{\lambda \phi_{GH}(a)}{2}} &=& e^{\frac{\lambda \phi_{GH}(a_0)}{2}}+  \frac{\lambda}{\sqrt{2c}}\int_{a_0}^a\frac{da}{aH_{GH}}\sqrt{ {\frac{\omega_{D_{GH}}-1}{3\omega_{D_{GH}}-1}}},\label{66} \\
     e^{\frac{\lambda \phi_{GR}(a)}{2}} &=& e^{\frac{\lambda \phi_{GR}(a_0)}{2}}+  \frac{\lambda}{\sqrt{2c}}\int_{a_0}^a\frac{da}{aH_{GR}}\sqrt{ {\frac{\omega_{D_{GR}}-1}{3\omega_{D_{GR}}-1}}}.\label{66-2} 
\end{eqnarray}

Therefore, we obtain that the evolutionary form of the dilaton scalar field for the two models we consider can be written in the following way:
\begin{eqnarray}
    \phi_{GH}(a)&=&  \frac{2}{\lambda}\ln \left[e^{\frac{\lambda \phi_{GH}(a_0)}{2}} +   \frac{\lambda}{\sqrt{2c}}  \int_{a_0}^a \frac{da}{aH_{GH}}\sqrt{ {\frac{\omega_{D_{GH}}-1}{3\omega_{D_{GH}}-1}} }\right],\label{67}\\
     \phi_{GR}(a)&=&  \frac{2}{\lambda}\ln \left[e^{\frac{\lambda \phi_{GR}(a_0)}{2}} +   \frac{\lambda}{\sqrt{2c}}  \int_{a_0}^a \frac{da}{aH_{GR}}\sqrt{ {\frac{\omega_{D_{GR}}-1}{3\omega_{D_{GR}}-1}} }\right]. \label{67-2}
\end{eqnarray}

Employing the same methodology, we obtain the corresponding results for the other configurations explored in this study.\\
Specifically, for the scenario involving spatial curvature, we find:
\begin{eqnarray}
    \phi_{GH,k}(a)&=&  \frac{2}{\lambda}\ln \left[e^{\frac{\lambda \phi_{GH,k}(a_0)}{2}} +   \frac{\lambda}{\sqrt{2c}}  \int_{a_0}^a \frac{da}{aH_{GH,k}}\sqrt{ {\frac{\omega_{D_{GH},k}-1}{3\omega_{D_{GH},k}-1}} }\right],\label{67}\\
     \phi_{GR,k}(a)&=&  \frac{2}{\lambda}\ln \left[e^{\frac{\lambda \phi_{GR,k}(a_0)}{2}} +   \frac{\lambda}{\sqrt{2c}}  \int_{a_0}^a \frac{da}{aH_{GR,k}}\sqrt{ {\frac{\omega_{D_{GR},k}-1}{3\omega_{D_{GR},k}-1}} }\right]. \label{67-2}
\end{eqnarray}
Moreover, for the scenario involving interaction between DE and DM, we find:
\begin{eqnarray}
    \phi_{GH,I}(a)&=&  \frac{2}{\lambda}\ln \left[e^{\frac{\lambda \phi_{GH,I}(a_0)}{2}} +   \frac{\lambda}{\sqrt{2c}}  \int_{a_0}^a \frac{da}{aH_{GH,I}}\sqrt{ {\frac{\omega_{D_{GH},I}-1}{3\omega_{D_{GH},I}-1}} }\right],\label{67}\\
     \phi_{GR,I}(a)&=&  \frac{2}{\lambda}\ln \left[e^{\frac{\lambda \phi_{GR,I}(a_0)}{2}} +   \frac{\lambda}{\sqrt{2c}}  \int_{a_0}^a \frac{da}{aH_{GR,I}}\sqrt{ {\frac{\omega_{D_{GR},I}-1}{3\omega_{D_{GR},I}-1}} }\right]. \label{67-2}
\end{eqnarray}
Finally, for the scenario involving both interaction between DE and DM and also spatial curvature, we find:
\begin{eqnarray}
    \phi_{GH,I,k}(a)&=&  \frac{2}{\lambda}\ln \left[e^{\frac{\lambda \phi_{GH,I,k}(a_0)}{2}} +   \frac{\lambda}{\sqrt{2c}}  \int_{a_0}^a \frac{da}{aH_{GH,I,k}}\sqrt{ {\frac{\omega_{D_{GH},I,k}-1}{3\omega_{D_{GH},I,k}-1}} }\right],\label{67}\\
     \phi_{GR,I,k}(a)&=&  \frac{2}{\lambda}\ln \left[e^{\frac{\lambda \phi_{GR,I,k}(a_0)}{2}} +   \frac{\lambda}{\sqrt{2c}}  \int_{a_0}^a \frac{da}{aH_{GR,I,k}}\sqrt{ {\frac{\omega_{D_{GR},I,k}-1}{3\omega_{D_{GR},I,k}-1}} }\right]. \label{67-2}
\end{eqnarray}

In the limiting case of a flat Dark Dominated Universe, i.e. for $\Omega_m=\Omega_k=0$ and $\Omega_D=1$, we obtain the following expressions:
\begin{eqnarray}
    ce ^{\lambda \phi_{GH,DD}}\chi_{GH,DD} &=&\frac{2 + \alpha}{3 (1 + \alpha)}, \label{64} \\
    ce ^{\lambda \phi_{GR,DD}}\chi_{GR,DD} &=& \frac{\lambda - 6}{3(\lambda - 4)}. \label{64-2}
\end{eqnarray}

Using the relation $\dot{\phi}^2=2\chi$, Eqs. (\ref{64}) and (\ref{64-2}) lead to the following results:
\begin{eqnarray}
    e^{\lambda \phi_{GH,DD}/2} \dot{\phi}_{GH,DD}=\sqrt{\frac{2}{c}} \sqrt{  \frac{2 + \alpha}{3 (1 + \alpha)}   },\label{65}\\
  e^{\lambda \phi_{GR,DD}/2} \dot{\phi}_{GR,DD}=\sqrt{\frac{2}{c}} \sqrt{ \frac{\lambda - 6}{3(\lambda - 4)}    }.\label{65-2}
\end{eqnarray}

Integrating Eqs. (\ref{65}) and (\ref{65-2}) with respect to the scale factor $a(t)$, we can write:
\begin{eqnarray}
    e^{\frac{\lambda \phi_{GH,DD}(a)}{2}} &=& e^{\frac{\lambda \phi_{GH,DD}(a_0)}{2}}+  \frac{\lambda}{\sqrt{2c}}\int_{a_0}^a\frac{da}{aH_{GH,DD}}\sqrt{ \frac{2 + \alpha}{3 (1 + \alpha)}},\label{66} \\
     e^{\frac{\lambda \phi_{GR,DD}(a)}{2}} &=& e^{\frac{\lambda \phi_{GR,DD}(a_0)}{2}}+  \frac{\lambda}{\sqrt{2c}}\int_{a_0}^a\frac{da}{aH_{GR,DD}}\sqrt{ \frac{\lambda - 6}{3(\lambda - 4)}}.\label{66-2} 
\end{eqnarray}
We must remember that:
\begin{eqnarray}
\int_{a_0}^a\frac{da}{aH} = \int_{t_0}^tdt.
\end{eqnarray}

Therefore, the evolutionary form of the dilaton scalar field for the two models of these papers can be written as:
\begin{eqnarray}
    \phi_{GH,DD}(a)&=&  \frac{2}{\lambda}\ln \left[e^{\frac{\lambda \phi_{GH,DD}(a_0)}{2}} +   \frac{\lambda}{\sqrt{2c}}  \int_{a_0}^a \frac{da}{aH_{GH,DD}}\sqrt{ \frac{2 + \alpha}{3 (1 + \alpha)} }\right]\nonumber\\
    &=&\frac{2}{\lambda}\ln \left[e^{\frac{\lambda \phi_{GH,DD}(a_0)}{2}} +   \frac{\lambda}{\sqrt{2c}}  \sqrt{ \frac{2 + \alpha}{3 (1 + \alpha)} }\cdot t\right],\label{67}\\
     \phi_{GR,DD}(a)&=&  \frac{2}{\lambda}\ln \left[e^{\frac{\lambda \phi_{GR,DD}(a_0)}{2}} +   \frac{\lambda}{\sqrt{2c}}  \int_{a_0}^a \frac{da}{aH_{GR,DD}}\sqrt{ \frac{\lambda - 6}{3(\lambda - 4)} }\right] \nonumber \\
     &=&  \frac{2}{\lambda}\ln \left[e^{\frac{\lambda \phi_{GR,DD}(a_0)}{2}} +   \frac{\lambda}{\sqrt{2c}}  \sqrt{ \frac{\lambda - 6}{3(\lambda - 4)}}\cdot t \right].\label{67-2}
\end{eqnarray}
In order to avoid singularities, from Eqs. (\ref{67}) and (\ref{67-2}), we obtain that $\alpha \in (-\infty, -2] \cup (-1, +\infty)$
 and  $\lambda \in (-\infty, 4) \cup [6, +\infty)$.  Considering the values $c^2=0.46$, $\epsilon = 1.312$ and $\eta = -0.312$, we obtain that $\alpha \approx 0.895$ and $\lambda \approx 0.211$. Therefore, we obtain that both Eqs. (\ref{67}) and (\ref{67-2}) are real functions. Instead, considering the values $c=0.579$, $\epsilon = 1.312$ and $\eta =- 0.312$, we obtain that $\alpha \approx 2.067$ and $\lambda \approx -2.134$. Also in this case, we obtain that both Eqs. (\ref{67}) and (\ref{67-2}) are real functions. Finally, considering the values $c=0.818$, $\epsilon = 1.312$ and $\eta =- 0.312$, we find that $\alpha \approx  0.377$ and  $\lambda \approx 1.246$, we obtain that  both Eqs. (\ref{67}) and (\ref{67-2}) are real functions.

\subsection{The Quintessence Scalar Field Model}

Quintessence is modeled by a homogeneous scalar field $\phi$ that evolves over time and is minimally coupled to gravity. It features a potential $V(\phi)$ responsible for driving the accelerated expansion of the universe. The action $S_Q$ describing the quintessence scalar field model is expressed as \cite{mou1}:
\begin{eqnarray}
    S_Q=\int d^4x \sqrt{-g}\,\Big[-\frac{1}{2}g^{\mu \nu} \partial _{\mu} \phi   \partial _{\nu} \phi - V( \phi )  \Big].\label{69}
\end{eqnarray}
The energy-momentum tensor $T_{\mu \nu}$ of the quintessence scalar  field model can be easily derived by varying the action $S_Q$ given in Eq. (\ref{69}) with respect to the metric tensor $g^{\mu \nu}$:
\begin{eqnarray}
T_{\mu \nu}=\frac{2}{\sqrt{-g}} \frac{\delta S}{\delta g^{\mu \nu}},\label{70}
\end{eqnarray}
which yields to the following result:
\begin{eqnarray}
    T_{\mu \nu}=\partial _{\mu} \phi   \partial _{\nu} \phi - g_{\mu \nu}\Big[\frac{1}{2}g^{\alpha \beta} \partial _{\alpha} \phi   \partial _{\beta} \phi + V( \phi )  \Big].\label{71}
\end{eqnarray}
In a FLRW background, the pressure $p_Q$ and the energy density $\rho_Q$ of the quintessence scalar field model are given, respectively, by:
\begin{eqnarray}
p_Q&=&T_i^i=\frac{1}{2}\dot{\phi}^2-V(\phi),\label{73}\\
    \rho_Q&=&-T_0^0=\frac{1}{2}\dot{\phi}^2+V(\phi).\label{72}
\end{eqnarray}
Therefore, using the results of Eqs. (\ref{73}) and (\ref{72}), we obtain that the Equation of State parameter $\omega_Q$ for the quintessence scalar field model can be written as:
\begin{eqnarray}    \omega_Q=\frac{p_Q}{\rho_Q}=\frac{\dot{\phi}^2-2V(\phi)}{\dot{\phi}^2+2V(\phi)}.\label{74}
\end{eqnarray}
We find from Eq. (\ref{74}) that, when $\omega_Q < -1/3$, the Universe accelerates for $\dot{\phi}^2<V(\phi)$. \\
Taking the variation of the quintessence action $S_Q$, as defined in Eq.~(\ref{69}), with respect to the scalar field $\phi$, leads to the following equation of motion:
\begin{eqnarray}
	\ddot{\phi} + 3H\dot{\phi}+V_{,\phi} = 0,
\end{eqnarray}
We now establish the correspondence between the interacting scenario and the quintessence DE model. Making the correspondences $\omega_Q=\omega_{\Lambda}$ and  $\rho_Q=\rho_{\Lambda}$, we obtain the following relations for $\dot{\phi}^2$ and $ V( \phi )$:
\begin{eqnarray}
    \dot{\phi}^2&=&(1+\omega_{\Lambda})\rho_{\Lambda}, \label{75}\\
    V( \phi ) &=& \frac{1}{2}(1-\omega_{\Lambda})\rho_{\Lambda}.\label{76}
\end{eqnarray}
Substituting $\omega_{D_{GH}}$ and $\omega_{D_{GR}}$ into Eqs. (\ref{75}) and (\ref{76}), the kinetic energy term $\dot{\phi}^2$ and the quintessence potential energy $V( \phi )$ assume the following expressions:
\begin{eqnarray}
    \dot{\phi}_{GH}^2&=&(1+\omega_{D_{GH}})\rho_{D_{GH}},\label{77}\\
    \dot{\phi}_{GR}^2&=&(1+\omega_{D_{GR}})\rho_{D_{GR}},\label{77-2}\\
    V_{GH}( \phi ) &=&\frac{\rho_{D_{GH}}}{2}(1-\omega_{D_{GH}}), \label{78}\\
    V_{GR}( \phi ) &=&\frac{\rho_{D_{GR}}}{2}(1-\omega_{D_{GR}}).\label{78-2}
\end{eqnarray}

Integrating (\ref{77}) and (\ref{77-2}) with respect to the scale factor $a$ and using the relation $\dot{\phi}=\phi' H$, it is possible to obtain the evolutionary form of the quintessence scalar fields as:
\begin{eqnarray}
\phi(a)_{GH} - \phi_{GH}(a_0)&=&
\int_{a_0}^{a}\frac{da}{a}\sqrt{3M_p^2\Omega_{D_{GH}}} \cdot\sqrt{(1+\omega_{D_{GH}})},\label{79}\\
\phi(a)_{GR} - \phi_{GR}(a_0)&=&
\int_{a_0}^{a}\frac{da}{a}\sqrt{3M_p^2\Omega_{D_{GR}}} \cdot\sqrt{(1+\omega_{D_{GR}})}.\label{79-1}
\end{eqnarray}
Using the same approach, we obtain the results for the other cases considered.\\
For the case with spatial curvature, we find:
\begin{eqnarray}
\phi(a)_{GH,k} - \phi_{GH,k}(a_0)&=&
\int_{a_0}^{a}\frac{da}{a}\sqrt{3M_p^2\Omega_{D_{GH},k}} \cdot\sqrt{(1+\omega_{D_{GH},k})},\label{79}\\
\phi(a)_{GR} - \phi_{GR}(a_0)&=&
\int_{a_0}^{a}\frac{da}{a}\sqrt{3M_p^2\Omega_{D_{GR},k}} \cdot\sqrt{(1+\omega_{D_{GR},k})},\label{79-1}\\
V_{GH,k}( \phi ) &=&\frac{\rho_{D_{GH},k}}{2}(1-\omega_{D_{GH},k}), \label{78}\\
    V_{GR,k}( \phi ) &=&\frac{\rho_{D_{GR},k}}{2}(1-\omega_{D_{GR},k}).\label{78-2}
\end{eqnarray}
Instead, for the case with interacting Dark Sectors, we find:
\begin{eqnarray}
\phi(a)_{GH,I} - \phi_{GH,I}(a_0)&=&
\int_{a_0}^{a}\frac{da}{a}\sqrt{3M_p^2\Omega_{D_{GH},I}} \cdot\sqrt{(1+\omega_{D_{GH},I})},\label{79}\\
\phi(a)_{GR,I} - \phi_{GR,I}(a_0)&=&
\int_{a_0}^{a}\frac{da}{a}\sqrt{3M_p^2\Omega_{D_{GH},I}} \cdot\sqrt{(1+\omega_{D_{GR},I})},\label{79-1}\\
V_{GH,I}( \phi ) &=&\frac{\rho_{D_{GH},I}}{2}(1-\omega_{D_{GH},I}), \label{78}\\
    V_{GR,I}( \phi ) &=&\frac{\rho_{D_{GR},I}}{2}(1-\omega_{D_{GR},I}).\label{78-2}
\end{eqnarray}
Finally, for the case with interacting Dark Sectors and spatial curvature, we find:
\begin{eqnarray}
\phi(a)_{GH,I,k} - \phi_{GH,I,k}(a_0)&=&
\int_{a_0}^{a}\frac{da}{a}\sqrt{3M_p^2\Omega_{D}} \cdot\sqrt{(1+\omega_{D_{GH},I,k})},\label{79}\\
\phi(a)_{GR,I,k} - \phi_{GR,I,k}(a_0)&=&
\int_{a_0}^{a}\frac{da}{a}\sqrt{3M_p^2\Omega_{D}} \cdot\sqrt{(1+\omega_{D_{GR},I,k})},\label{79-1}\\
V_{GH,I,k}( \phi ) &=&\frac{\rho_{D_{GH},I,k}}{2}(1-\omega_{D_{GH},I,k}), \label{78}\\
    V_{GR,I,k}( \phi ) &=&\frac{\rho_{D_{GR},I,k}}{2}(1-\omega_{D_{GR},I,k}).\label{78-2}
\end{eqnarray}

In the limiting case of a flat Dark Dominated Universe, i.e. when $\Omega_m=\Omega_k=0$ and $\Omega_D=1$, $\dot{\phi}$ and the potential $V$ of the quintessence scalar field model reduce to:
\begin{eqnarray}
\dot{\phi}_{GH,DD}^2(t) &=&  \frac{2H_0^2}{3(1 - \alpha)(H_0t + C_1)^2},\label{80-11}\\
\dot{\phi}^2_{GR,DD}(t) &=& \frac{4H_0^2}{3\lambda (H_0t+C_2)^2}, \label{80-22}\\
V_{GH,DD}(t) &=&   \frac{(2+\alpha)H_0^2}{3(1 - \alpha)^2(H_0t + C_1)^2},\label{81-1}\\
V_{GR,DD}(t) &=& \frac{2H_0^2(6-\lambda)}{3\lambda^2 (H_0t+C_2)^2}. \label{81-2}
\end{eqnarray}
Therefore, integrating with respect to the time Eqs. (\ref{80-11}) and (\ref{80-22}), we obtain the following results:
\begin{eqnarray}
\phi_{GH,DD}(t) &=&  \sqrt{\frac{2H_0^2}{3(1 - \alpha)}}  \cdot \ln |H_0t + C_1|+\phi_0,\label{80-1}\\
\phi_{GR,DD}(t) &=& \frac{2H_0}{\sqrt{3\lambda} }\cdot \ln |H_0t + C_2| +\phi_0,\label{80-2}
\end{eqnarray}
where $\phi_0$ is an integration constant. \\
In the limiting case of $C_1=C_2=\phi_0=0$, we obtain:
\begin{eqnarray}
\phi_{GH,DD,lim}(t) &=&  \sqrt{\frac{2}{3(1 - \alpha)}}  \cdot \ln |t|,\label{80-1}\\
\phi_{GR,DD,lim}(t) &=& \frac{2}{\sqrt{3\lambda} }\cdot \ln |t |,\label{80-2}\\
V_{GH,DD,lim}(t) &=&   \frac{2+\alpha}{3\left[(1 - \alpha)t \right]^2},\label{81-1}\\
V_{GR,DD,lim}(t) &=& \frac{2(6-\lambda)}{3 (\lambda t)^2}.
\end{eqnarray}
In order to avoid singularities in the results for Dark Dominated Universe, we must have $\alpha <1$ and $\lambda >0$. Considering the values $c^2=0.46$, $\epsilon = 1.312$ and $\eta =- 0.312$, we obtain that  $\alpha \approx 0.895$ and $\lambda \approx 0.211$. Therefore, in this case, we obtain that $\phi_{GH,DD,lim}(t)$ and $\phi_{GR,DD,lim}(t)$ are both real functions. Instead, considering the values $c=0.579$, $\epsilon = 1.312$ and $\eta =- 0.312$, we obtain that $\alpha \approx 2.067$    and $\lambda \approx -2.134$. In this case, we obtain that $\phi_{GH,DD,lim}(t)$ is a real functions while $\phi_{GR,DD,lim}(t)$ is not. Finally, considering the values $c=0.818$, $\epsilon = 1.312$ and $\eta =- 0.312$, we find that $\alpha \approx  0.377$ and  $\lambda \approx 1.246$, we obtain that both $\phi_{GH,DD,lim}(t)$ and $\phi_{GR,DD,lim}(t)$ are real functions.

\subsection{The Dirac-Born-Infeld (DBI) Scalar Field Model}

We now turn our attention to the Dirac–Born–Infeld (DBI) model.\\ Recently, several studies have explored the connection between string theory and inflation. Among these, the concept of branes in string theory has proven especially fruitful. A widely studied scenario involves inflation driven by the open string sector via dynamical Dp-branes, a framework known as Dirac–Born–Infeld (DBI) inflation. This model is part of a specific class of K-inflation theories, characterized by non-standard kinetic terms.

Considering dark energy (DE) to be driven by a DBI scalar field,  the expression of the action $ S_{dbi} $ of the DBI scalar field model is given by the following relation \cite{dbi1,dbi3,dbi5,dbi6,dbi7,dbi8,dbi9,dbi10,dbi11,dbi12}:
\begin{eqnarray}
S_{dbi} = \int d^4x \, a^3(t) \left[ F(\phi) \sqrt{1 - \frac{\dot{\phi}^2}{F(\phi)}} + V(\phi) - F(\phi) \right].\label{murano86}
\end{eqnarray}
The quantity $F(\phi)$ represents the brane tension, while the quantity $V(\phi)$ represents the scalar potential.

From the action $S_{dbi} $ given in Eq. (\ref{murano86}), it is possible to obtain that the pressure $ p_{dbi} $ and the energy density $ \rho_{dbi} $ of the DBI scalar field model are given by the following expressions:
\begin{eqnarray}
p_{dbi} &=& \left( \frac{\gamma - 1}{\gamma} \right) F(\phi) - V(\phi), \label{murano87} \\
\rho_{dbi} &=& (\gamma - 1) F(\phi) + V(\phi). \label{murano88}
\end{eqnarray}
The quantity $\gamma$ represents the analogous of the relativistic Lorentz factor and it can be expressed as:
\begin{eqnarray}
\gamma = \left[ 1 - \frac{\dot{\phi}^2}{F(\phi)} \right]^{-1/2}. \label{murano89}
\end{eqnarray}
Considering the expressions of the pressure $p_{dbi}$ and the energy density $\rho_{dbi}$ we obtained in Eqs. (\ref{murano87}) and (\ref{murano88}), we easily obtain that the Equation of State parameter $\omega_{dbi}$ of the DBI scalar field model can be written in the following way:
\begin{eqnarray}
\omega_{dbi} &=& \frac{\left(\frac{\gamma-1}{\gamma}\right) F\left(\phi\right) -  V\left(\phi\right)}{\left(\gamma-1\right) F\left(\phi\right) +  V\left(\phi\right)} = \frac{\left(\gamma-1\right) F\left(\phi\right) -  \gamma V\left(\phi\right)}{\gamma\left[ \left(\gamma-1\right) F\left(\phi\right) +  V\left(\phi\right)  \right] }.\label{murano90}
\end{eqnarray}

Our goal is now to derive the explicit expressions for the three quantities $F(\phi)$, $\dot{\phi}^2$, and $V(\phi)$.\\
 By adding the results of Eqs.~(\ref{murano87}) and (\ref{murano88}) and employing the general definition of the equation of state parameter $omega_D$, we obtain the following result:
\begin{eqnarray}
F &=& \rho_D\left( \frac{\gamma}{\gamma^2 -1}\right) \left( \omega_D + 1   \right)\label{murano91old}.
\end{eqnarray}
Using the definition of $\gamma$ provided in Eq.~(\ref{murano89}) together with the result from Eq.~(\ref{murano91old}), we arrive at the following expression for $\dot{\phi}$:
\begin{eqnarray}
\dot{\phi} &=& \sqrt{\frac{ \rho_D \left( \omega_D + 1   \right) }{\gamma}}.\label{murano92old}
\end{eqnarray}
Subtracting Eqs.~(\ref{murano87}) and (\ref{murano88}) and applying the general definition of the equation of state parameter $\omega_D$, we obtain the following expression for $V$:

\begin{eqnarray}
V &=& - \left(\frac{\rho_D}{\gamma +1}\right) \left( \gamma \omega_D -1   \right).\label{murano93}
\end{eqnarray}

Therefore, the three terms $F$, $\dot{\phi}$ and $V$ can be expressed in the following way:
\begin{eqnarray}
F &=& \rho_D\left( \frac{\gamma}{\gamma^2 -1}\right) \left(1+ \omega_D \right),\label{murano91}\\
\dot{\phi} &=& \sqrt{\frac{ \rho_D \left( 1+\omega_D    \right) }{\gamma}},\label{murano92}\\
V &=& - \frac{\rho_D}{\gamma +1} \left( \gamma \omega_D  -1   \right) \nonumber \\
&=&  \frac{\rho_D}{\gamma +1} \left(1- \gamma \omega_D     \right).\label{murano93}
\end{eqnarray}

For the DE models we considered in this work, we obtain the following results for  $F$, $\dot{\phi}$ and $V$:
\begin{eqnarray}
F_{GH} &=& \rho_{D_{GH}}\left( \frac{\gamma}{\gamma^2 -1}\right) \left(1+ \omega_{D_{GH}} \right),\label{murano91}\\
F_{GR} &=& \rho_{D_{GR}}\left( \frac{\gamma}{\gamma^2 -1}\right) \left(1+ \omega_{D_{GR}} \right),\label{murano91-}\\
\dot{\phi}_{GH} &=& \sqrt{\frac{ \rho_{D_{GH}} \left( 1+\omega_{D_{GH}}    \right) }{\gamma}},\label{murano92-}\\
\dot{\phi}_{GR} &=& \sqrt{\frac{ \rho_{D_{GR}} \left( 1+\omega_{D_{GR}}    \right) }{\gamma}},\label{murano92}\\
V_{GH} &=&  \frac{\rho_{D_{GH}}}{\gamma +1} \left(1- \gamma \omega_{D_{GH}}     \right),\label{murano93}\\
V_{GR} &=&  \frac{\rho_{D_{GR}}}{\gamma +1} \left(1- \gamma \omega_{D_{GR}}     \right).\label{murano93-}
\end{eqnarray}

Using the same method, we extend our results to the other cases discussed in this paper.\\
For the scenario with spatial curvature, the results are:
\begin{eqnarray}
F_{GH,k} &=& \rho_{D_{GH},k}\left( \frac{\gamma}{\gamma^2 -1}\right) \left(1+ \omega_{D_{GH},k} \right),\label{murano91}\\
F_{GR,k} &=& \rho_{D_{GR},k}\left( \frac{\gamma}{\gamma^2 -1}\right) \left(1+ \omega_{D_{GR},k} \right),\label{murano91-}\\
\dot{\phi}_{GH,k} &=& \sqrt{\frac{ \rho_{D_{GH},k} \left( 1+\omega_{D_{GH},k}    \right) }{\gamma}},\label{murano92-}\\
\dot{\phi}_{GR,k} &=& \sqrt{\frac{ \rho_{D_{GR},k} \left( 1+\omega_{D_{GR},k}    \right) }{\gamma}},\label{murano92}\\
V_{GH,k} &=&  \frac{\rho_{D_{GH},k}}{\gamma +1} \left(1- \gamma \omega_{D_{GH}}     \right),\label{murano93}\\
V_{GR,k} &=&  \frac{\rho_{D_{GR},k}}{\gamma +1} \left(1- \gamma \omega_{D_{GR},k}     \right).\label{murano93-}
\end{eqnarray}
Instead, for the scenario with interaction, the results are:
\begin{eqnarray}
F_{GH,I} &=& \rho_{D_{GH},I}\left( \frac{\gamma}{\gamma^2 -1}\right) \left(1+ \omega_{D_{GH},I} \right),\label{murano91}\\
F_{GR,I} &=& \rho_{D_{GR},I}\left( \frac{\gamma}{\gamma^2 -1}\right) \left(1+ \omega_{D_{GR},I} \right),\label{murano91-}\\
\dot{\phi}_{GH,I} &=& \sqrt{\frac{ \rho_{D_{GH},I} \left( 1+\omega_{D_{GH},I}    \right) }{\gamma}},\label{murano92-}\\
\dot{\phi}_{GR,I} &=& \sqrt{\frac{ \rho_{D_{GR},I} \left( 1+\omega_{D_{GR},I}    \right) }{\gamma}},\label{murano92}\\
V_{GH,I} &=&  \frac{\rho_{D_{GH},I}}{\gamma +1} \left(1- \gamma \omega_{D_{GH},I}     \right),\label{murano93}\\
V_{GR,I} &=&  \frac{\rho_{D_{GR},I}}{\gamma +1} \left(1- \gamma \omega_{D_{GR},I}     \right).\label{murano93-}
\end{eqnarray}
Finally, for the scenario with spatial curvature and interaction, the results are:
\begin{eqnarray}
F_{GH,I,k} &=& \rho_{D_{GH},I,k}\left( \frac{\gamma}{\gamma^2 -1}\right) \left(1+ \omega_{D_{GH},I,k} \right),\label{murano91}\\
F_{GR,I,k} &=& \rho_{D_{GR},I,k}\left( \frac{\gamma}{\gamma^2 -1}\right) \left(1+ \omega_{D_{GR},I,k} \right),\label{murano91-}\\
\dot{\phi}_{GH,I,k} &=& \sqrt{\frac{ \rho_{D_{GH},I,k} \left( 1+\omega_{D_{GH},I,k}    \right) }{\gamma}},\label{murano92-}\\
\dot{\phi}_{GR,I,k} &=& \sqrt{\frac{ \rho_{D_{GR},I,k} \left( 1+\omega_{D_{GR},I,k}    \right) }{\gamma}},\label{murano92}\\
V_{GH,I,k} &=&  \frac{\rho_{D_{GH},I,k}}{\gamma +1} \left(1- \gamma \omega_{D_{GH},I,k}     \right),\label{murano93}\\
V_{GR,I,k} &=&  \frac{\rho_{D_{GR},I,k}}{\gamma +1} \left(1- \gamma \omega_{D_{GR},I,k}     \right).\label{murano93-}
\end{eqnarray}

In the limiting case of a flat Dark Dominated Universe, i.e. for $\Omega_m=\Omega_k=0$ and $\Omega_D=1$, we obtain the following results:
\begin{eqnarray}
F_{GH,DD} &=& \left( \frac{\gamma}{\gamma^2 -1}\right)\frac{2H_0^2}{3(1 - \alpha)(H_0t + C_1)^2} ,\label{murano91}\\
F_{GR,DD} &=& \left( \frac{\gamma}{\gamma^2 -1}\right)\frac{4H_0^2}{3\lambda (H_0t+C_2)^2},\label{murano91-}\\
\dot{\phi}_{GH,DD} &=& \sqrt{\frac{2}{3\gamma(1 - \alpha)}}\cdot \frac{H_0}{(H_0t + C_1)},\label{murano92-}\\
\dot{\phi}_{GR,DD} &=& \sqrt{\frac{ 1 }{3\gamma \lambda}}\cdot\frac{2H_0}{ (H_0t+C_2)},\label{murano92}\\
V_{GH,DD} &=& \frac{H_0^2}{(\gamma+1)(1 - \alpha)^2(H_0t + C_1)^2} \left[1+ \gamma \cdot \left(\frac{2\alpha+1}{3}\right)    \right],\label{murano93}\\
V_{GR,DD} &=& \frac{4H_0^2}{\lambda^2 (\gamma+1)(H_0t+C_2)^2}  \left[1- \gamma \cdot \left(\frac{\lambda -3}{3}\right)    \right].\label{murano93-}
\end{eqnarray}
In the limiting case of $C_1=C_2=0$, we can write:
\begin{eqnarray}
F_{GH,DD,lim} &=& \left( \frac{\gamma}{\gamma^2 -1}\right)\frac{2}{3(1 - \alpha)} \cdot \frac{1}{t^2},\label{murano91}\\
F_{GR,DD,lim} &=& \left( \frac{\gamma}{\gamma^2 -1}\right)\frac{4}{3\lambda }\cdot \frac{1}{t^2},\label{murano91-}\\
\dot{\phi}_{GH,DD,lim} &=& \sqrt{\frac{2}{3\gamma(1 - \alpha)}}\cdot \frac{1}{t },\label{murano92-}\\
\dot{\phi}_{GR,DD,lim} &=& \sqrt{\frac{ 1 }{3\gamma \lambda}}\cdot\frac{2}{t},\label{murano92}\\
V_{GH,DD,lim} &=& \frac{1}{(\gamma+1)(1 - \alpha)^2} \left[1+ \gamma \cdot \left(\frac{2\alpha+1}{3}\right)    \right]\cdot \frac{1}{t^2},\label{murano93}\\
V_{GR,DD,lim} &=& \frac{4}{\lambda^2 (\gamma+1)}  \left[1- \gamma \cdot \left(\frac{\lambda -3}{3}\right)    \right]\cdot \frac{1}{t^2}.\label{murano93-}
\end{eqnarray}
We now consider a particular case of $F$ given by:
\begin{eqnarray}
F\left( \phi  \right) = F_0 \dot{\phi}^2,\label{murano103}
\end{eqnarray}
with $F_0 >0$ being a constant parameter. For this particular case, we can write $\gamma$ in the following way:
\begin{eqnarray}
\gamma = \sqrt{\frac{F_0}{F_0-1}},\label{murano104}
\end{eqnarray}
which  implies that $F_0>1$ in order to have a real value of $\gamma$. Negative values of $F_0$ also lead to positive values of $\gamma$ but they cannot be considered since we previously imposed that $F_0 >0$.

Considering the result obtained in Eq. (\ref{murano104}), we can write Eqs. (\ref{murano91}), (\ref{murano92}) and (\ref{murano93}) in the following forms:
\begin{eqnarray}
F &=& \rho_D\sqrt{F_0\left( F_0-1  \right)} \left( 1+\omega_D   \right),\label{murano105} \\
\dot{\phi} &=&\left(\sqrt{\frac{F_0}{F_0-1}}\right)^{-1/4} \sqrt{ \rho_D \left( 1+\omega_D    \right)},\label{murano106} \\
V &=& - \frac{\rho_D }{\sqrt{\frac{F_0}{F_0-1}} + 1} \left[ \left(\sqrt{\frac{F_0}{F_0-1}}\right) \omega_D -1 \right].\label{murano107}
\end{eqnarray}

For the DE model we are considering in this paper, we can write:
\begin{eqnarray}
F_{GH} &=& \rho_{D_{GH}}\sqrt{F_0\left( F_0-1  \right)} \left( 1+\omega_{D_{GH}}  \right),\label{murano105} \\
F_{GR} &=& \rho_{D_{GR}}\sqrt{F_0\left( F_0-1  \right)} \left( 1+\omega_{D_{GR}}   \right),\label{murano105-} \\
\dot{\phi}_{GH} &=&\left(\sqrt{\frac{F_0}{F_0-1}}\right)^{-1/4} \sqrt{ \rho_{D_{GH}}\ \left( 1+\omega_{D_{GH}}    \right)},\label{murano106} \\
\dot{\phi}_{GR} &=&\left(\sqrt{\frac{F_0}{F_0-1}}\right)^{-1/4} \sqrt{ \rho_{D_{GR}}\ \left( 1+\omega_{D_{GR}}    \right)},\label{murano106-} \\
V_{GH} &=& - \frac{\rho_{D_{GH}}\ }{\sqrt{\frac{F_0}{F_0-1}} + 1} \left[ \left(\sqrt{\frac{F_0}{F_0-1}}\right) \omega_{D_{GH}}\ -1 \right],\label{murano107}\\
V_{GR} &=& - \frac{\rho_{D_{GR}}\ }{\sqrt{\frac{F_0}{F_0-1}} + 1} \left[ \left(\sqrt{\frac{F_0}{F_0-1}}\right) \omega_{D_{GR}}\ -1 \right].\label{murano107-}
\end{eqnarray}

By adopting the same approach, we obtain the relevant results for the remaining configurations examined in this work.\\
Specifically, in the presence of spatial curvature, we obtain:
\begin{eqnarray}
F_{GH,k} &=& \rho_{D_{GH},k}\sqrt{F_0\left( F_0-1  \right)} \left( 1+\omega_{D_{GH},k}  \right),\label{murano105} \\
F_{GR,k} &=& \rho_{D_{GR},k}\sqrt{F_0\left( F_0-1  \right)} \left( 1+\omega_{D_{GR},k}   \right),\label{murano105-} \\
\dot{\phi}_{GH,k} &=&\left(\sqrt{\frac{F_0}{F_0-1}}\right)^{-1/4} \sqrt{ \rho_{D_{GH}}\ \left( 1+\omega_{D_{GH},k}    \right)},\label{murano106} \\
\dot{\phi}_{GR,k} &=&\left(\sqrt{\frac{F_0}{F_0-1}}\right)^{-1/4} \sqrt{ \rho_{D_{GR}}\ \left( 1+\omega_{D_{GR},k}    \right)},\label{murano106-} \\
V_{GH,k} &=& - \frac{\rho_{D_{GH},k}\ }{\sqrt{\frac{F_0}{F_0-1}} + 1} \left[ \left(\sqrt{\frac{F_0}{F_0-1}}\right) \omega_{D_{GH},k}\ -1 \right],\label{murano107}\\
V_{GR,k} &=& - \frac{\rho_{D_{GR},k}\ }{\sqrt{\frac{F_0}{F_0-1}} + 1} \left[ \left(\sqrt{\frac{F_0}{F_0-1}}\right) \omega_{D_{GR},k}\ -1 \right].\label{murano107-}
\end{eqnarray}
In the presence of spatial curvature, we obtain:
\begin{eqnarray}
F_{GH,I} &=& \rho_{D_{GH},I}\sqrt{F_0\left( F_0-1  \right)} \left( 1+\omega_{D_{GH},I}  \right),\label{murano105} \\
F_{GR,I} &=& \rho_{D_{GR},I}\sqrt{F_0\left( F_0-1  \right)} \left( 1+\omega_{D_{GR},I}   \right),\label{murano105-} \\
\dot{\phi}_{GH,I} &=&\left(\sqrt{\frac{F_0}{F_0-1}}\right)^{-1/4} \sqrt{ \rho_{D_{GH},I}\ \left( 1+\omega_{D_{GH}}    \right)},\label{murano106} \\
\dot{\phi}_{GR,I} &=&\left(\sqrt{\frac{F_0}{F_0-1}}\right)^{-1/4} \sqrt{ \rho_{D_{GR}}\ \left( 1+\omega_{D_{GR},I}    \right)},\label{murano106-} \\
V_{GH,I} &=& - \frac{\rho_{D_{GH},I}\ }{\sqrt{\frac{F_0}{F_0-1}} + 1} \left[ \left(\sqrt{\frac{F_0}{F_0-1}}\right) \omega_{D_{GH},I}\ -1 \right],\label{murano107}\\
V_{GR,I} &=& - \frac{\rho_{D_{GR},I}\ }{\sqrt{\frac{F_0}{F_0-1}} + 1} \left[ \left(\sqrt{\frac{F_0}{F_0-1}}\right) \omega_{D_{GR},I}\ -1 \right].\label{murano107-}
\end{eqnarray}
Instead, in the presence of interaction and spatial curvature, we obtain:
\begin{eqnarray}
F_{GH,I,k} &=& \rho_{D_{GH},I,k}\sqrt{F_0\left( F_0-1  \right)} \left( 1+\omega_{D_{GH},I,k}  \right),\label{murano105} \\
F_{GR,I,k} &=& \rho_{D_{GR},I,k}\sqrt{F_0\left( F_0-1  \right)} \left( 1+\omega_{D_{GR},I,k}   \right),\label{murano105-} \\
\dot{\phi}_{GH,I,k} &=&\left(\sqrt{\frac{F_0}{F_0-1}}\right)^{-1/4} \sqrt{ \rho_{D_{GH},I,k}\ \left( 1+\omega_{D_{GH}}    \right)},\label{murano106} \\
\dot{\phi}_{GR,I,k} &=&\left(\sqrt{\frac{F_0}{F_0-1}}\right)^{-1/4} \sqrt{ \rho_{D_{GR},I,k}\ \left( 1+\omega_{D_{GR},I,k}    \right)},\label{murano106-} \\
V_{GH,I,k} &=& - \frac{\rho_{D_{GH},I,k}\ }{\sqrt{\frac{F_0}{F_0-1}} + 1} \left[ \left(\sqrt{\frac{F_0}{F_0-1}}\right) \omega_{D_{GH},I,k}\ -1 \right],\label{murano107}\\
V_{GR,I,k} &=& - \frac{\rho_{D_{GR},I,k}\ }{\sqrt{\frac{F_0}{F_0-1}} + 1} \left[ \left(\sqrt{\frac{F_0}{F_0-1}}\right) \omega_{D_{GR},I,k}\ -1 \right].\label{murano107-}
\end{eqnarray}

In the limiting case of a flat Dark Dominated Universe, i.e. for $\Omega_m=\Omega_k=0$ and $\Omega_D=1$, we obtain the following results:
\begin{eqnarray}
F_{GH,DD} &=& \frac{2H_0^2\sqrt{F_0\left( F_0-1  \right)}}{3(1 - \alpha)(H_0t + C_1)^2} ,\label{murano105} \\
F_{GR,DD} &=& \frac{4H_0^2\sqrt{F_0\left( F_0-1  \right)}}{3\lambda (H_0t+C_2)^2} ,\label{murano105-} \\
\dot{\phi}_{GH,DD} &=&\left(\sqrt{\frac{F_0}{F_0-1}}\right)^{-1/4} \sqrt{  \frac{2}{3(1 - \alpha)} } \frac{H_0}{(H_0t + C_1)} ,\label{picu3} \\
\dot{\phi}_{GR,DD} &=&\left(\sqrt{\frac{F_0}{F_0-1}}\right)^{-1/4} \sqrt{ \frac{1}{3\lambda }}\frac{2H_0}{ (H_0t+C_2)},\label{picu4} \\
V_{GH,DD} &=&  \frac{1 }{\sqrt{\frac{F_0}{F_0-1}} + 1} \frac{H_0^2}{(1 - \alpha)^2(H_0t + C_1)^2} \left[ \left(\sqrt{\frac{F_0}{F_0-1}}\right) \frac{2\alpha+1}{3}  +1 \right],\label{murano107}\\
V_{GR,DD} &=& - \frac{1 }{\sqrt{\frac{F_0}{F_0-1}} + 1} \frac{4H_0^2}{\lambda^2 (H_0t+C_2)^2} \left[ \left(\sqrt{\frac{F_0}{F_0-1}}\right) \frac{\lambda -3}{3} -1 \right].\label{murano107-}
\end{eqnarray}
Integrating Eqs. (\ref{picu3}) and (\ref{picu4})  with respect to the time $t$, we easily derive the following relations:
\begin{eqnarray}
\phi_{GH,DD} &=&\left(\sqrt{\frac{F_0}{F_0-1}}\right)^{-1/4} \sqrt{  \frac{2}{3(1 - \alpha)} }\cdot \ln |H_0t + C_1| + \phi_0 ,\label{murano106} \\
\phi_{GR,DD} &=&2\left(\sqrt{\frac{F_0}{F_0-1}}\right)^{-1/4} \sqrt{ \frac{1}{3\lambda }}\cdot\ln |H_0t+C_2| + \phi_0,\label{murano106-} 
\end{eqnarray}
where $\phi_0$ is an integration constant.\\
In the limiting case of $C_1=C_2 = \phi_0=0$, we obtain:
\begin{eqnarray}
F_{GH,DD,lim} &=& \frac{2\sqrt{F_0\left( F_0-1  \right)}}{3(1 - \alpha)} \cdot \frac{1}{t^2},\label{murano105} \\
F_{GR,DD,lim} &=& \frac{4\sqrt{F_0\left( F_0-1  \right)}}{3\lambda } \cdot \frac{1}{t^2},\label{murano105-} \\
\phi_{GH,DD,lim} &=&\left(\sqrt{\frac{F_0}{F_0-1}}\right)^{-1/4} \sqrt{  \frac{2}{3(1 - \alpha)} } \cdot\ln |t| ,\label{murano106} \\
\phi_{GR,DD,lim} &=&2\left(\sqrt{\frac{F_0}{F_0-1}}\right)^{-1/4} \sqrt{ \frac{1}{3\lambda }}\cdot\ln |t| ,\label{murano106-}  \\
V_{GH,DD\lim} &=&  \frac{1 }{\sqrt{\frac{F_0}{F_0-1}} + 1} \frac{1}{(1 - \alpha)^2} \left[ \left(\sqrt{\frac{F_0}{F_0-1}}\right) \frac{2\alpha+1}{3}  +1 \right]\cdot \frac{1}{t^2},\label{murano107}\\
V_{GR,DD,lim} &=& - \frac{1 }{\sqrt{\frac{F_0}{F_0-1}} + 1} \frac{4}{\lambda^2 } \left[ \left(\sqrt{\frac{F_0}{F_0-1}}\right) \frac{\lambda -3}{3} -1 \right]\cdot \frac{1}{t^2}.\label{murano107-}
\end{eqnarray}
In order to avoid singularities in the above results, we must have $\alpha <1$  and $\lambda >0$. Considering the values $c^2=0.46$, $\epsilon = 1.312$ and $\eta =- 0.312$, we obtain that $\alpha \approx 0.895$ and $\lambda \approx 0.211$. Therefore, both $\phi_{GH,DD,lim}$ and $\phi_{GH,DD,lim}$ are real functions. \\
Instead, considering the values $c=0.579$, $\epsilon = 1.312$ and $\eta =- 0.312$, we obtain that $\alpha \approx 2.067$ and $\lambda \approx -2.134$. In this case, we obtain that $\phi_{GH,DD,lim}$ is a real functiond and $\phi_{GR,DD,lim}$ is not.\\
Finally, considering the values $c=0.818$, $\epsilon = 1.312$ and $\eta =- 0.312$, we find that $\alpha \approx  0.377$ and  $\lambda \approx 1.246$, we find that both $\phi_{GH,DD,lim}$ and $\phi_{GH,DD,lim}$ are real functions.

\subsection{The Yang-Mills (YM) Scalar Field Model}

We now consider the Yang–Mills (YM) scalar field model.\\
This model is inspired by the non-Abelian gauge theory framework and has been explored in the context of dark energy and modified gravity. In what follows, we derive the corresponding expressions for the energy density and pressure associated with the YM field and investigate how it can effectively describe the dynamics of the dark energy component in our cosmological model.\\
Recent published papers suggest that the Yang–Mills (YM) field \cite{ym1,ym2,ym3,ym9-1,ym9-2,ym9-3,ym9-4,ym9-5} can serve as a viable candidate for describing the nature of dark energy (DE). Two main reasons motivate the consideration of the YM field as a source of DE. First, unlike standard scalar field models, the YM field possesses a stronger foundation in fundamental particle physics, being inherently connected to the gauge structure of the Standard Model and its extensions. Second, the YM field naturally allows for scenarios in which the weak energy condition can be violated, thus enabling a rich and varied cosmological evolution that includes accelerated expansion.\\
The Yang-Mills (YM) scalar field model under consideration exhibits several notable features that make it an appealing candidate for modeling dark energy. It constitutes a fundamental component of particle physics, where gauge bosons mediate interactions, and can therefore be naturally embedded within unified theories of fundamental interactions. Furthermore, the equation of state (EoS) parameter associated with the effective Yang-Mills condensate (YMC) deviates significantly from that of standard matter fields and canonical scalar fields. In particular, the YMC can give rise to a wide range of dynamical behaviors, including regimes with $-1 < \omega < 0$, characteristic of quintessence-like dark energy, as well as $\omega < -1$, indicative of phantom-like behavior. This flexibility enables the YM field to accommodate various accelerating scenarios within the cosmic evolution.\\
In the effective Yang-Mills DE model, it is well-known that the effective YM field Lagrangian density $L_{\text{YMC}}$ is given by:
\begin{eqnarray}
L_{YMC} = \frac{bF}{2} \left( \ln \left| \frac{F}{\kappa^2} \right| - 1 \right), \label{murano117}
\end{eqnarray}
where the quantity $\kappa$ is a renormalization scale with dimensions of mass squared, while the term $F$ plays the role of the order parameter of the YM condensate.\\
The quantity $F$ is defined by the gauge-invariant expression:
\begin{eqnarray}
F = -\frac{1}{2} F_{\mu \nu}^\alpha F^{\alpha \mu \nu} = E^2 - B^2, \label{murano118}
\end{eqnarray}
where $F_{\mu \nu}^\alpha$ is the Yang-Mills field strength tensor, and $E$ and $B$ denote the effective electric and magnetic components of the field, respectively.\\
The  pure electric case is recovered in the limiting case of $B = 0$, which leads to the simplification $F = E^2 $.\\
Moreover, the parameter $b$ indicates the Callan-Symanzik coefficient \cite{ym18,ym18-1}, which is, for the gauge group $SU(N)$:
\begin{eqnarray}
b = \frac{11N - 2N_f}{24\pi^2}, \label{murano119}
\end{eqnarray}
where the quantity $N_f$ denotes the number of quark flavors.\\
Considering the gauge group $SU(2)$, the Callan-Symanzik coefficient $b$ takes the value $b = 2 \times \frac{11}{24\pi^2}$ when fermionic contributions are neglected, while the Callan-Symanzik coefficient $b$ assumes the value of $b = 2 \times \frac{5}{24\pi^2}$ when we consider the number of quark flavors $N_f = 6$. In the case of the $SU(3)$ gauge group, the effective Lagrangian presented in Eq.~\eqref{murano117} offers a phenomenological framework to describe the phenomenon of asymptotic freedom exhibited by quarks confined within hadrons~\cite{ym21, ym21-1}.\\
It is crucial to clarify that the $SU(2)$ Yang-Mills scalar field model considered in this work as a possible candidate for dark energy is fundamentally distinct from the QCD gluon fields or the electroweak gauge bosons such as $Z^0$ and $W^{\pm}$. The Yang-Mills Condensate (YMC) relevant for cosmology operates at an energy scale characterized by $\kappa^{1/2} \approx 10^{-3} \, \text{eV}$, which is many orders of magnitude smaller than the energy scales typical of quantum chromodynamics and electroweak processes. This vast difference in scale highlights the phenomenological independence of the YMC dark energy model from standard particle physics gauge fields.\\
The form of the effective YM Lagrangian $L_{YMC}$ described in Eq. \eqref{murano117} can be understood as a 1-loop quantum corrected Lagrangian \cite{ym21,ym21-1}. The classical $SU(N)$ Yang-Mills Lagrangian can be written 
\begin{eqnarray}
L = \frac{1}{2g_0^2} F, \label{murano120}
\end{eqnarray}
where the quantity $g_0$ is the bare coupling constant. Including 1-loop quantum corrections leads to a running coupling constant $g$ that replaces the bare coupling as
\begin{eqnarray}
g_0^2 \rightarrow g^2 = \frac{4 \times 12 \pi^2}{11N \ln\left(\frac{k}{k_0^2}\right)} = \frac{2}{b \ln\left(\frac{k}{k_0^2}\right)}, \label{murano121}
\end{eqnarray}
where the quantity $k$ is the momentum transfer while $k_0$ indicates the corresponding energy scale.\\
To construct an effective theory, the momentum scale $k^2$ is replaced by the field strength $F$ via the substitution
\begin{eqnarray}
\ln\left(\frac{k}{k_0^2}\right) \rightarrow 2 \ln \left| \frac{F}{\kappa^2 e} \right| = 2 \ln \left| \frac{F}{\kappa^2} - 1 \right|. \label{murano122}
\end{eqnarray}
This replacement recovers the effective Lagrangian expression shown in Eq. \eqref{murano117}.\\
Some notable features of the effective Yang-Mills (YM) action include the correct  trace anomaly, asymptotic freedom,  gauge invariance and Lorentz invariance \cite{ym16}.\\
The logarithmic dependence on the field strength in the YMC model’s Lagrangian draws a strong parallel to the Coleman–Weinberg scalar effective potential~\cite{ym19} and similarly echoes features found in the Parker–Raval effective gravity Lagrangian~\cite{ym20}. These analogies highlight the quantum corrections underlying the model and its potential relevance in cosmological contexts.\\
It is important to emphasize that the renormalization scale $\kappa$ is the only adjustable parameter in this effective Yang-Mills model. This stands in stark contrast to scalar field dark energy models, where the potential energy function is often chosen ad hoc to fit observational data. In the Yang-Mills condensate framework, however, the form of the Lagrangian is rigorously fixed by one-loop quantum corrections, which arise naturally from the underlying gauge theory. Consequently, the model possesses a higher degree of theoretical robustness and predictivity, since its dynamics are tightly constrained by fundamental quantum field theory principles rather than phenomenological parameterizations. This feature makes the effective Yang-Mills approach an attractive candidate for a more physically grounded description of dark energy.\\
Starting from the effective Lagrangian presented in  Eq. \eqref{murano117}, we can derive the corresponding expressions for the energy density $\rho_y$ and pressure $p_y$ associated with the Yang-Mills condensate (YMC):
\begin{eqnarray}
\rho_{YMC} &=& \frac{\epsilon E^2}{2} + \frac{b E^2}{2}, \label{murano123}\\
p_{YMC} &=& \frac{\epsilon E^2}{6} - \frac{b E^2}{2}. \label{murano124}
\end{eqnarray}
Here, $\epsilon$ represents the dielectric constant associated with the Yang-Mills field, which plays a crucial role in determining the effective coupling strength within the condensate. It is defined by the relation:
\begin{eqnarray}
\epsilon = 2 \frac{\partial L_{eff}}{\partial F} = b \ln \left| \frac{F}{\kappa^2} \right|. \label{murano125}
\end{eqnarray}
These quantities characterize the macroscopic cosmological behavior of the YMC and are obtained through the standard procedure of varying the Lagrangian with respect to the metric tensor and identifying the energy-momentum tensor components. The resulting formulas are essential for analyzing the dynamics of the dark energy component modeled by the YMC.\\
Equations \eqref{murano123} and \eqref{murano124} can be alternatively rewritten as
\begin{eqnarray}
\rho_y &=& \frac{1}{2} b \kappa^2 (y+1) e^{y}, \label{murano126}\\
p_y &=& \frac{1}{6} b \kappa^2 (y-3) e^{y}, \label{murano127}
\end{eqnarray}
or equivalently in the following form:
\begin{eqnarray}
\rho_y &=& \frac{1}{2} (y+1) b E^2, \label{murano128} \\
p_y &=& \frac{1}{6} (y-3) b E^2. \label{murano129}
\end{eqnarray}
Here, the dimensionless parameter $y$ is introduced and defined as
\begin{eqnarray}
y = \frac{\epsilon}{b} = \ln \left| \frac{F}{\kappa^2} \right| = \ln \left| \frac{E^2}{\kappa^2} \right|. \label{defiy}
\end{eqnarray}
Using the expressions for $\rho_y$ and $p_y$ in Eqs. \eqref{murano126} and \eqref{murano127} (or equivalently in Eqs. \eqref{murano128} and \eqref{murano129}), the equation of state (EoS) parameter $\omega_y$ of the YMC model can be written as
\begin{eqnarray}
\omega_y = \frac{p_y}{\rho_y} = \frac{y - 3}{3(y + 1)}. \label{omegay}
\end{eqnarray}
At the critical point where $\epsilon = 0$ (i.e., $y=0$), we find $\omega_y = -1$, corresponding to a de Sitter expansion of the Universe. In the vicinity of this point, the sign of the dielectric constant $\epsilon$ critically determines the nature of the equation of state (EoS) parameter $\omega_y$. Specifically, when $\epsilon < 0$, the EoS parameter falls below the phantom divide, i.e., $\omega_y < -1$, which corresponds to a phantom-like regime characterized by super-accelerated cosmic expansion. Conversely, if $\epsilon > 0$, the EoS parameter remains above this threshold, $\omega_y > -1$, indicating a quintessence-like behavior where the dark energy density evolves more moderately. This sensitivity highlights the importance of $\epsilon$ in controlling the dynamical properties of the Yang-Mills condensate as a DE candidate.
 Thus, as previously stated, the YMC model naturally realizes the range of EoS values $0 > \omega_y > -1$ as well as $\omega_y < -1$.\\
The expression of $\omega_y$ given in Eq. \eqref{omegay} can be inverted to yield the following relation for the dimensionless parameter $y$:
\begin{eqnarray}
y = - \frac{3(\omega_y + 1)}{3\omega_y - 1}. \label{murano130}
\end{eqnarray}
To guarantee a physically viable model with a positive energy density $\rho_y$, the dimensionless parameter $y$ must satisfy the condition $y > 1 $. This constraint ensures that the Yang-Mills condensate behaves consistently with fundamental energy conditions required in cosmology.\\
 This condition also implies the following condition for $F$:
\begin{eqnarray}
F > \frac{\kappa^2}{e} \approx 0.368\, \kappa^2.
\end{eqnarray}
Before considering a particular cosmological scenario, it is instructive to study $\omega_y$ as a function of the condensate strength $F$. The Yang-Mills condensate (YMC) model exhibits an equation of state (EoS) characteristic of radiation, given by
\begin{eqnarray}
p_y = \frac{1}{2} \rho_y,
\end{eqnarray}
and the corresponding EoS parameter
\begin{eqnarray}
\omega_y = \frac{1}{2}.
\end{eqnarray}
in the limiting case of large values of the dielectric constant $\epsilon$, i.e. for $\epsilon \gg b$ (which is equivalent to $F \gg \kappa^2$).
On the other hand, at the critical point where $\epsilon = 0$ (i.e. $F = \kappa^2$), the YMC behaves like a cosmological constant with
\begin{eqnarray}
\omega_y &=& -1, \nonumber \\
p_y &=& -\rho_y.
\end{eqnarray}
This critical case corresponds to the YMC energy density taking the value
\begin{eqnarray}
\rho_y = \frac{1}{2} b \kappa^2,
\end{eqnarray}
which acts as the critical energy density \cite{ym1,ym3}.\\
An interesting feature of the YMC equation of state is its smooth evolution from a radiation-like behavior, with $\omega_y = \frac{1}{3}$ at high energy densities ($F \gg \kappa^2$), to a cosmological constant–like behavior, with $\omega_y = -1$ in the low-energy limit ($F = \kappa^2$). This gradual transition enables the existence of scaling solutions for the dark energy component within this model~\cite{ym10,ym10-1}. Furthermore, the evolution is smooth and continuous, as the equation of state parameter $\omega_y$ is a well-defined continuous function of the dimensionless variable $y$ over the entire range $( -1, \infty )$.\\
We now examine whether the EoS parameter $\omega_y$ is able to cross the phantom divide $-1$. From Eq. \eqref{omegay}, $\omega_y$ depends solely on the condensate strength $F$. From a theoretical point of view, the condition $\omega_y < -1$ can be obtained when
\begin{eqnarray}
F < \kappa^2,
\end{eqnarray}
and this crossing is also smooth with respect to $F$.\\
However, when the YMC is incorporated into a cosmological model as a dark energy component, alongside matter and radiation, the value of $F$ is no longer arbitrary but evolves dynamically as a function of the cosmic time $t$.\\
Specifically, if the Yang-Mills condensate (YMC) does not decay into radiation or matter, the EoS parameter $\omega_y$ approaches the value of $-1$ asymptotically from above but never crosses this threshold. In this non-interacting scenario, the YMC behaves effectively like a cosmological constant at late times. In contrast, if the YMC decays into matter and/or radiation through an interaction term or coupling, the dynamics of the system change substantially. In such cases, $\omega_y$ can cross below the phantom divide, i.e., $\omega_y < -1$, and evolve into a phantom-like regime. Depending on the strength of the interaction, $\omega_y$ eventually stabilizes at an asymptotic value that can deviate significantly from $-1$. For instance, with moderate coupling strengths, the final value of t.\\
An important advantage of the lower regime $\omega_y < -1$ in the YMC model is that all relevant physical quantities—namely the energy density $\rho_y$, pressure $p_y$, and equation of state parameter $\omega_y$—remain smooth and well-behaved throughout the cosmic evolution. This stands in contrast to many scalar field dark energy models, where entering the phantom regime often leads to instabilities or the emergence of finite-time singularities, such as Big Rip scenarios. In the YMC framework, the phantom-like behavior arises naturally from the gauge field dynamics without introducing pathological features, making the model both theoretically robust and cosmologically viable.\\
By equating the equation of state (EoS) parameter of the Yang-Mills condensate, $\omega_y$, with the EoS parameter of the dark energy model under investigation, one can express the dimensionless variable $y$ in the following form:
\begin{eqnarray}
y = - \frac{3\left(\omega_D+1\right)}{3\omega_D-1}.\label{murano131}
\end{eqnarray}
We obtain from Eq. (\ref{murano131})  $y>1$ if:
\begin{eqnarray}
    -\frac{1}{3}< \omega_{D} <\frac{1}{3}.
\end{eqnarray}
Therefore, inserting the results we derived for the two models considered in this work, we can write:
\begin{eqnarray}
y_{GH} &=& - \frac{3\left(\omega_{D_{GH}}+1\right)}{3\omega_{D_{GH}}-1},\label{murano131!!!}\\
y_{GR} &=& - \frac{3\left(\omega_{D_{GR}}+1\right)}{3\omega_{D_{GR}}-1}.\label{murano131-}
\end{eqnarray}
Using the same method, we extend our results to the other cases discussed in this paper.\\
For the scenario with spatial curvature, the results are:
\begin{eqnarray}
y_{GH,k} &=& - \frac{3\left(\omega_{D_{GH},k}+1\right)}{3\omega_{D_{GH},k}-1},\label{murano131!!!}\\
y_{GR,k} &=& - \frac{3\left(\omega_{D_{GR},k}+1\right)}{3\omega_{D_{GR},k}-1}.\label{murano131-}
\end{eqnarray}
For the scenario with interacting Dark Sectors, the results are:
\begin{eqnarray}
y_{GH,I} &=& - \frac{3\left(\omega_{D_{GH},I}+1\right)}{3\omega_{D_{GH},I}-1},\label{murano131!!!}\\
y_{GR,I} &=& - \frac{3\left(\omega_{D_{GR},I}+1\right)}{3\omega_{D_{GR},I}-1}.\label{murano131-}
\end{eqnarray}
For the scenario with spatial curvature and interacting Dark Sectors, the results are:
\begin{eqnarray}
y_{GH,I,k} &=& - \frac{3\left(\omega_{D_{GH},I,k}+1\right)}{3\omega_{D_{GH},I,k}-1},\label{murano131!!!}\\
y_{GR,I,k} &=& - \frac{3\left(\omega_{D_{GR}}+1\right)}{3\omega_{D_{GR},I,k}-1}.\label{murano131-}
\end{eqnarray}
In the limiting case of a flat Dark Dominated Universe, i.e. for $\Omega_m=\Omega_k=0$ and $\Omega_D=1$, we obtain the following results for $y$ for the models we study:
\begin{eqnarray}
y_{GH,DD}&=& - \frac{\alpha-1}{\alpha+1},\\
y_{GR,DD}&=&  - \frac{\lambda }{\lambda -4}.  
\end{eqnarray}
In order to have $y_{GH,DD}>1$ and $y_{GR,DD}>1$, the conditions $-1 <\alpha <0$ and $2<\lambda<4$ must be satisfied. Taking into account the values $c^2=0.46$, $\epsilon = 1.312$ and $\eta =- 0.312$, we obtain $\alpha \approx 0.895$ and $\lambda \approx 0.211$.  Therefore, we find that condition $y>1$ is not obtained. 
Instead, considering the values $c=0.579$, $\epsilon = 1.312$ and $\eta =- 0.312$, we obtain $\alpha \approx 2.067$ and $\lambda \approx -2.134$.   Therefore, in this case, the condition $y>1$ is not obtained. Finally, if we consider the values $c=0.818$, $\epsilon = 1.312$ and $\eta =- 0.312$, we find that $\alpha \approx  0.377$ and  $\lambda \approx 1.246$, therefore also in this case the condition $y>1$ is not obtained.

\subsection{The Non-Linear Electro-Dynamics (NLED) Scalar Field Model}

We now consider the last model we have chosen to study, i.e., the Non Linear Electro-Dynamics (NLED) scalar field model. \\
Recently, a novel approach has been proposed to address the problem of cosmic singularities by introducing a nonlinear extension of classical Maxwell electrodynamics. In this framework, exact solutions of Einstein's field equations coupled with Nonlinear Electrodynamics (NLED) have demonstrated that nonlinear corrections become significant in the presence of strong gravitational and electromagnetic fields. These corrections can regularize singularities and lead to physically acceptable solutions. Furthermore, the coupling of General Relativity (GR) with NLED has been shown to naturally give rise to an inflationary phase in the early Universe, thereby offering a viable alternative mechanism for primordial inflation within a purely classical field-theoretic setting.\\
In the framework of classical electrodynamics, the dynamics of the free electromagnetic field are governed by the Maxwell Lagrangian density $L_M$, which takes the form \cite{ele1,ele2}:
\begin{equation}
L_M = - \frac{F^{\mu \nu}F_{\mu \nu}}{4\mu}, \label{murano135}
\end{equation}
where $F^{\mu \nu}$ is the electromagnetic field strength tensor, defined as $F^{\mu\nu} = \partial^\mu A^\nu - \partial^\nu A^\mu$ 
and $\mu$ denotes the magnetic permeability of the medium, which enters the field equations when restoring physical units.

We now consider a nonlinear generalization of the Maxwell electromagnetic Lagrangian, expanded to include second-order corrections in the field invariants. The modified Lagrangian density takes the form:
\begin{equation}
L = -\frac{F}{4\mu_0} + \omega F^2 + \eta F^{*2}.\label{murano136}
\end{equation}

Here, $\omega$ and $\eta$ are two free parameters characterizing the strength of the nonlinear corrections, while $F^*$ denotes the electromagnetic pseudo-invariant defined by
\begin{equation}
F^* = F_{\mu\nu} \tilde{F}^{\mu\nu} = \frac{1}{2} \epsilon^{\mu\nu\rho\sigma} F_{\mu\nu} F_{\rho\sigma},
\end{equation}
with $\tilde{F}^{\mu\nu}$ representing the dual of the electromagnetic field strength tensor, and $\epsilon^{\mu\nu\rho\sigma}$ being the totally antisymmetric Levi-Civita tensor.\\
We consider the specific scenario in which the homogeneous electric field $E$ in the plasma rapidly diminishes due to the backreaction of the electric current generated by charged particles. As a consequence, the contribution from the electric field becomes negligible, i.e., $E^2 \approx 0$, and the dynamics are dominated by the magnetic component. Therefore, we can approximate
\begin{eqnarray}
F = 2B^2.
\end{eqnarray}
Thus, $F$ depends only on the value of the magnetic field $B$ squared.\\
The pressure $ p_{\text{NLED}} $ and energy density $ \rho_{\text{NLED}} $ of the Nonlinear Electrodynamics (NLED) field can be derived from the energy-momentum tensor associated with the Lagrangian density $ L_M $. In a spatially homogeneous and isotropic background, these quantities read as
\begin{align}
p_{NLED} &= \frac{B^2}{6\mu}\left(1 - 40 \mu \omega B^2 \right), \label{murano138}\\
\rho_{NLED} &= \frac{B^2}{2\mu}\left(1 - 8 \mu \omega B^2 \right). \label{murano139}
\end{align}
The weak energy condition, $\rho_{NLED} > 0$, holds for values of $B$ satisfying
 \begin{equation}
B < \frac{1}{2\sqrt{2 \mu \omega}}.
  \end{equation}
Conversely, the pressure $p_{NLED}$ becomes negative when
 \begin{equation}
B > \frac{1}{2\sqrt{10 \mu \omega}}.
   \end{equation}
The magnetic field can act as a source of dark energy if the strong energy condition is violated, that is, when
  \begin{equation}
\rho_B + 3 p_B < 0,
   \end{equation}
which occurs for magnetic field strengths satisfying
\begin{equation}
B > \frac{1}{2\sqrt{6 \mu \omega}}.
\end{equation}

The equation of state (EoS) parameter $\omega_{NLED}$ for the Nonlinear Electrodynamics (NLED) field is defined as
\begin{equation}
\omega_{NLED} = \frac{p_{NLED}}{\rho_{NLED}} = \frac{1 - 40 \mu \omega B^2}{3 \left(1 - 8 \mu \omega B^2 \right)}, \label{murano140}
\end{equation}
which can be inverted to express the magnetic field intensity $B^2$ in terms of $\omega_{NLED}$ as
\begin{equation}
B^2 = \frac{1 - 3 \omega_{NLED}}{8 \mu \omega \left(5 - 3 \omega_{NLED}\right)}. \label{murano141}
\end{equation}
By establishing a correspondence between the equation of state parameter of the NLED model and that of the dark energy model under consideration, we obtain
\begin{equation}
B^2 = \frac{1 - 3 \omega_D}{8 \mu \omega \left(5 - 3 \omega_D\right)}. \label{murano142}
\end{equation}

For the model under study, we obtain the following expressions for $B^2$:
\begin{eqnarray}
B_{GH}^2 &=& \frac{1 - 3 \omega_{D_{GH}}}{8 \mu \omega \left(5 - 3 \omega_{D_{GH}}\right)},\\
B_{GR}^2 &=& \frac{1 - 3 \omega_{D_{GR}}}{8 \mu \omega \left(5 - 3 \omega_{D_{GR}}\right)}.
\end{eqnarray}

Applying the same procedure, we obtain the corresponding results for the other scenarios explored in this study.
In particular, for the scenario that includes spatial curvature, we find:
\begin{eqnarray}
B_{GH,k}^2 &=& \frac{1 - 3 \omega_{D_{GH},k}}{8 \mu \omega \left(5 - 3 \omega_{D_{GH},k}\right)},\\
B_{GR,k}^2 &=& \frac{1 - 3 \omega_{D_{GR},k}}{8 \mu \omega \left(5 - 3 \omega_{D_{GR},k}\right)}.
\end{eqnarray}
For the scenario that includes interaction, we find:
\begin{eqnarray}
B_{GH,I}^2 &=& \frac{1 - 3 \omega_{D_{GH},I}}{8 \mu \omega \left(5 - 3 \omega_{D_{GH},I}\right)},\\
B_{GR,I}^2 &=& \frac{1 - 3 \omega_{D_{GR},I}}{8 \mu \omega \left(5 - 3 \omega_{D_{GR},I}\right)}.
\end{eqnarray}
Finally, for the scenario that includes spatial curvature and interaction, we find:
\begin{eqnarray}
B_{GH,I,k}^2 &=& \frac{1 - 3 \omega_{D_{GH},I,k}}{8 \mu \omega \left(5 - 3 \omega_{D_{GH},I,k}\right)},\\
B_{GR,I,k}^2 &=& \frac{1 - 3 \omega_{D_{GR},I,k}}{8 \mu \omega \left(5 - 3 \omega_{D_{GR},I,k}\right)}.
\end{eqnarray}

Moreover, in the limiting case of a flat Dark Dominated Universe, i.e. for $\Omega_m=\Omega_k=0$ and $\Omega_D=0$, we obtain the following results:
\begin{eqnarray}
B^2_{GH,Dark} &=& \frac{1 + \alpha}{8 \mu \omega \left(3 + \alpha\right)},\\
B_{GR,Dark}^2 &=& \frac{4 -\lambda   }{8 \mu \omega \left(8 -\lambda \right)}.
\end{eqnarray}
We must have $\alpha \neq -3$ and $\lambda \neq 8$ in order to avoid singularities.
Moreover, the conditions:
\begin{eqnarray}
B_{GH,Dark} > \frac{1}{2\sqrt{6 \mu \omega}},\\
B_{GR,Dark} > \frac{1}{2\sqrt{6 \mu \omega}},
\end{eqnarray}
are satisfied when $\alpha>0$ and  $\lambda <2$.\\
Considering the values $c^{2} = 0.46$, $\epsilon = 1.312$, and $\eta = -0.312$, we obtain 
$\alpha \approx 0.895$ and $\lambda \approx 0.211$. In this case, the condition 
 $B > \frac{1}{2\sqrt{6 \mu \omega}}$ is fulfilled.  

For $c = 0.579$ (with the same $\epsilon = 1.312$ and $\eta = -0.312$), the corresponding values are  $\alpha \approx 2.067$ and $\lambda \approx -2.134$. The inequality above is satisfied also in this case.  

Finally, for $c = 0.818$, $\epsilon = 1.312$, and $\eta = -0.312$, we find 
$\alpha \approx 0.377$ and $\lambda \approx 1.246$. Once again, the condition 
$
B > \frac{1}{2\sqrt{6 \mu \omega}}$ 
remains satisfied.

\section{Conclusions} 
The present paper reports two generalized versions of
the holographic and Ricci dark energy respectively which in their form can be written as
\begin{eqnarray}
\rho_{GH}&=&3c^2M^{2}_{pl} \left[ 1-\epsilon\left(1-\frac{R}{H^2}\right) \right]H^2,\label{eq:GHD} \nonumber\\
\rho_{GR}&=&3c^2M^{2}_{pl}\left[ 1-\eta\left(1-\frac{H^2}{R}\right) \right]R, \nonumber 
\end{eqnarray}
where $R$ and $H$ denote, respectively, the Ricci scalar and the Hubble parameters, while $\epsilon$ and $\eta$ and two parameters which are related by  $\epsilon = 1-\eta$.\\
For both models, we derived key cosmological quantities, including the Hubble parameter, the dark energy density $\rho_D$, the dark energy pressure $p_D$, as well as explicit expressions for the equation of state (EoS) parameter $\omega_D$ and the deceleration parameter $q$, all expressed as functions of the parameter $x$ and the redshift $z$. We considered four different cases: first, the standard model as presented in \cite{xu2009}; then, we examined the effects of spatial curvature. Subsequently, we extended the analysis done in the two previous cases, including interactions between the dark sectors. We also evaluated these quantities in the limiting case of a Dark Energy Dominated Universe, i.e. for $\Omega_m = \Omega_k= 0$ and $\Omega_D = 1$.

In addition, for the dark energy models under consideration, we carried out a detailed analysis of several diagnostic tools aimed at characterizing their dynamical behavior and distinguishing them from other cosmological scenarios. Specifically, we studied the Statefinder diagnostic, the cosmographic parameters, the $Om(z)$ diagnostic, the squared speed of sound $v_s^2$, which provides insight into the stability of the models under perturbations, and the age of the present Universe.

Furthermore, we developed a correspondence between the considered dark energy models and a variety of scalar field theories, thereby providing a unified framework to describe their dynamics. In particular, we examined the tachyon, k-essence, dilaton, quintessence, DBI, Yang-Mills, and nonlinear electrodynamics (NLED) scalar field models. We also extended this analysis to the limiting case of a Dark Energy Dominated Universe, obtaining when possible, some hints on the parameters involved.

\end{document}